\documentclass[numberedappendix]{emulateapj}
\usepackage{morefloats}	

\newcommand{\fpk}{$F_{\rm pk}$}
\newcommand{\fpkre}{$F_{\rm pk,PRE}$}
\newcommand{\fluen}{$E_b$}	
\newcommand{\fper}{$F_p$}
\newcommand{\fbol}{$F_{\rm bol}$}

\newcommand{\bstart}{$t_0$}
\newcommand{\trise}{$t_{\rm rise}$}
\newcommand{\trec}{$t_{\rm rec}$}	

\newcommand{\eps}{{\rm erg\,s^{-1}}}
\newcommand{\epcs}{{\rm erg\,cm^{-2}\,s^{-1}}}
\newcommand{\epc}{{\rm erg\,cm^{-2}}}
\newcommand{\cts}{{\rm count\,s^{-1}}}
\newcommand{\ledd}{$L_{\rm Edd}$}
\newcommand{\leddh}{$L_{\rm Edd,H}$}
\newcommand{\leddhe}{$L_{\rm Edd,He}$}
\newcommand{\mdotedd}{$\dot{M}_{\rm Edd}$}
\newcommand{\OtoNe}{$^{15}$O$(\alpha,\gamma)~^{19}$Ne}
\newcommand{\malpha}{$\left<\alpha\right>$}

\newcommand{\ans}{{\it ANS}}	
\newcommand{\vfb}{{\it Vela 5-B}}
\newcommand{\xte}{{\it RXTE}}	
\newcommand{\sax}{{\it BeppoSAX}}
\newcommand{\ros}{{\it ROSAT}}

\newcommand{\uhu}{{\it Uhuru}}
\newcommand{\exo}{{\it EXOSAT}}
\newcommand{\ein}{{\it Einstein}}
\newcommand{\asca}{{\it ASCA}}
\newcommand{\osos}{{\it OSO-7}}
\newcommand{\osoe}{{\it OSO-8}}
\newcommand{\sas}{{\it SAS-3}}	
\newcommand{\arv}{{\it Ariel-5}}

\newcommand{\hak}{{\it Hakucho}}
\newcommand{\gin}{{\it Ginga}}
\newcommand{\chandra}{{\it Chandra}}
\newcommand{\igr}{{\it INTEGRAL}}
\newcommand{\hst}{{\it HST}}

\newcommand{\sgamma}{${\mathcal S}\gamma$}
\newcommand{\ssz}{${\mathcal S}S_Z$}
\newcommand{\sdt}{${\mathcal S}\Delta t$}
\newcommand{\sgammah}{${\mathcal S}\gamma_{\rm H}$}
\newcommand{\sgammahe}{${\mathcal S}\gamma_{\rm He}$}
\newcommand{\sszh}{${\mathcal S}S_{Z,{\rm H}}$}
\newcommand{\sszhe}{${\mathcal S}S_{Z,{\rm He}}$}
\newcommand{\sosc}{${\mathcal S}_{\rm osc}$}

\newcommand{\burstnum}{1187}			
\newcommand{\burstpubdate}{2007 June 3}
\newcommand{\newburstsourcedate}{2007 June 3}	
\newcommand{\numbursters}{48}	
\newcommand{\pcarspver}{10.1}	
\newcommand{\lheasoftver}{5.3}	
\newcommand{\lheasoftdate}{2003 November 17}

\newcommand{\numsrc}{35}        
\newcommand{\ngtone}{27}
\newcommand{\numsrcrate}{19}
\newcommand{\numsrburst}{464}	
\newcommand{\numsrcratehe}{6}
\newcommand{\numsrbursthe}{183}

\newcommand{\numszrate}{5}
\newcommand{\numszburst}{259}

\newcommand{\numszratehe}{2}
\newcommand{\numszbursthe}{131}

\newcommand{\burstswithonedecay}{1107}	
\newcommand{\burstswithtwodecays}{811}	
\newcommand{\burstswithonedecayltfifty}{1092}	
\newcommand{\taurelone}{$\tau_1=0.52+0.44\tau$} 
\newcommand{\taurmsone}{2.8}	
\newcommand{\taureltwo}{$\tau_2=4.0+1.21\tau$} 
\newcommand{\taurmstwo}{8.7} 

\newcommand{\preburstswithdecay}{246}	
\newcommand{\pretaugammacorr}{-0.44}	
\newcommand{\pretaugammacorrsig}{$6.9\sigma$}	
\newcommand{\nonpreburstswithdecay}{436}	
\newcommand{\preubgammacorr}{-0.46}	
\newcommand{\preubgammacorrsig}{$7.3\sigma$}	
\newcommand{\preedtgammacorr}{$\rho=-0.31$, $ 4.8\sigma$}	
\newcommand{\preedttwogammacorr}{$\rho=-0.34$, $5.2\sigma$}	
\newcommand{\nonpretaugammacorr}{-0.25}	
\newcommand{\nonpreburstswithsz}{285}	
\newcommand{\preburstswithsz}{163}	
\newcommand{\nonpretauszcorr}{$\rho=-0.77$, $13\sigma$}	

\newcommand{\nvar}{16}	
\newcommand{\meandist}{$(13\pm9)$\%}	
\newcommand{\meanonly}{13\%}
\newcommand{\fstdevmax}{38\%}

\newcommand{\highmean}{$6.4\times10^{-8}\ \epcs$}
\newcommand{\highsig}{7.6\%}
\newcommand{\preburst}{40}

\newcommand{\alphadat}{209}
\newcommand{\alphasrc}{23}

\newcommand{\noscsrc}{16}	
\newcommand{\nsrcsearched}{13}	
\newcommand{\burstsearched}{515}
\newcommand{\burstdet}{194}
\newcommand{\omitted}{12}

\newcommand{\npair}{84}
\newcommand{\npairbad}{10}

\newcommand{\ntriple}{12}

\newcommand{\refbursts}{5}	

\slugcomment{Accepted by ApJS}

\shortauthors{Galloway et al.}
\shorttitle{Thermonuclear bursts observed by RXTE}

\begin{document}

\title{Thermonuclear (type-I) X-ray bursts observed by the 
  \boldmath {\it Rossi X-ray Timing Explorer}}

\author{Duncan K. Galloway\altaffilmark{1,2},
  Michael P. Muno\altaffilmark{3}, 
  Jacob M. Hartman\altaffilmark{4}, 
  Dimitrios Psaltis\altaffilmark{5},
  and Deepto Chakrabarty\altaffilmark{6}. 
}
\affil{
  Kavli Institutute for Astrophysics and Space Research, 
  Massachusetts Institute of Technology,
Cambridge, MA~02139}
\email{Duncan.Galloway@sci.monash.edu.au} 

\altaffiltext{1}{present address:
  School of Physics \& School of Mathematical Sciences, Monash University,
  Victoria 3800, Australia}
\altaffiltext{2}{Monash Fellow}
\altaffiltext{3}{present address: Space Radiation Laboratory,
  California Institute of Technology, Pasadena CA 91125}
\altaffiltext{4}{present address: Space Science Division, Code 7655, Naval
  Research Laboratory, Washington DC 20375}
\altaffiltext{5}{present address: Department of Physics, University of Arizona,
  Tucson AZ 85721}
\altaffiltext{6}{also Department of Physics, Massachusetts Institute of
Technology, Cambridge MA 02139}

\begin{abstract}
We have assembled a sample of \burstnum\ thermonuclear (type-I) X-ray
bursts from observations of \numbursters\ 
accreting neutron stars by the {\it Rossi X-ray Timing Explorer},
spanning more than 
ten years.
The sample contains examples of
two of the three theoretical ignition regimes (confirmed via comparisons
with numerical models) and likely examples of the third. We present a
detailed analysis of the variation of the burst profiles, energetics,
recurrence times, presence of photospheric radius expansion, and presence
of burst oscillations, as a function of accretion rate.

We estimated the distance for \numsrc\ sources exhibiting
radius-expansion bursts, and found that in general the peak flux of
such bursts varies typically by \meanonly,
We classified sources into two main groups based on the
burst properties: both long and short bursts (indicating mixed H/He
accretion), and consistently short bursts (primarily He accretion), and
calculated the mean burst rate as a function of accretion rate for the two
groups. 
The decrease in burst rate observed at 
$>0.06\dot{M}_{\rm Edd}$ ($\ga2\times10^{37}\ \eps$) is associated
with a transition in the persistent spectral state
and (as has been suggested previously) may be related to
the increasing role of steady He-burning.
We found many examples of bursts with recurrence times $<30$~min,
including burst triplets and even quadruplets.

We describe the oscillation amplitudes for 
\nsrcsearched\ of the \noscsrc\
burst oscillation sources,
as well as the
stages and properties of the bursts in which the oscillations are detected.
The burst properties are correlated with the burst oscillation frequency;
sources spinning at $<400$~Hz generally have consistently short
bursts, while the more rapidly-spinning
systems have both long and short bursts. This correlation
suggests
either that shear-mediated mixing dominates the burst properties, or
alternatively
that the nature of the mass donor (and hence the evolutionary history) 
has an influence on the long-term spin evolution.
\end{abstract}


\section{Introduction}
\label{intro}

\begin{figure}
 \epsscale{1.2}
\plotone{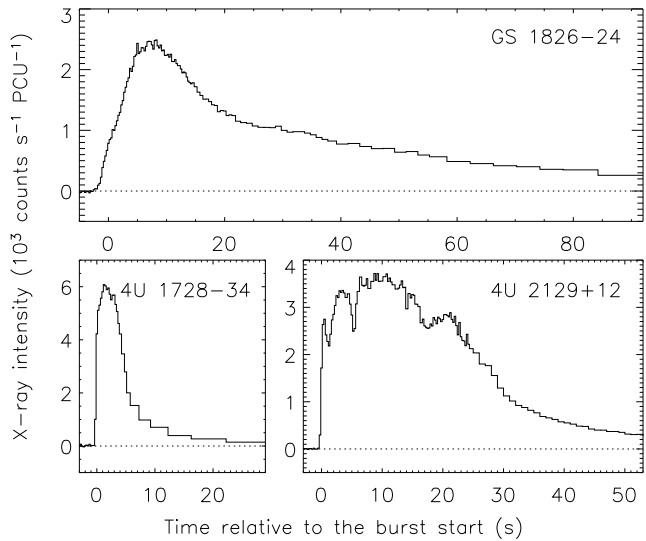}
\caption{Example light curves of bursts observed by \xte. The top panel
shows a long burst from GS~1826$-$24
on 1998 June  8 04:11:45 UT. The lower left panel shows a burst observed
from 4U~1728$-$34
on 1999 June 30 19:50:14 UT, while the burst at lower right was observed
from 4U~2129$+$12 in the globular cluster M15 
on 2000 September 22 13:47:41 UT. The persistent (pre-burst) level has
been subtracted (dotted line). Note the diversity of burst profiles, which
arises in part from variations in the fuel composition; bursts with a slow
rise and decay are characteristic of mixed H/He fuel, while bursts with
much faster rises likely burn primarily He.  Both bursts in the lower
panels exhibited photospheric radius-expansion.
 \label{examples} }
\end{figure}

Thermonuclear (type-I) X-ray bursts manifest as a sudden increase in the X-ray
intensity of accreting neutron stars (NSs),
to many  times brighter than the persistent level.
Typical bursts exhibit rise times of between $\la 1$ and $10$~s,
and last from tens to hundreds of seconds (Fig. \ref{examples}).
These 
events are caused by unstable burning of
accreted H/He on the surface of neutron stars in low-mass X-ray
binary (LMXB) systems \cite[e.g.][]{sb03}, in contrast to type-II bursts,
which are thought to be caused by sudden accretion events
\cite[e.g.][]{lew93}.  
The H/He fuel for type-I bursts is accreted from the binary
companion and accumulates on the surface of the neutron star, forming a
layer several meters thick.
The accreted material is compressed and heated
hydrostatically, 
and if the temperature is sufficiently high any hydrogen present burns
steadily into helium via the ``hot'' ($\beta$-limited)
carbon-nitrogen-oxygen (CNO) process.  After between
$\sim$1 and several tens of 
hours, the temperature and density at the base of the layer become high
enough that the fuel ignites,
burning unstably and spreading rapidly to consume
all the available fuel on the  star in a matter of seconds. 
Such bursts
have been observed to date from more than 70 sources 
\cite[e.g.][]{zand04b}.

The burst X-ray spectrum
is generally consistent with a blackbody of color temperature
$T_{\rm bb}=2$--3~keV.  Time-resolved spectral fits give evidence for an
initial rise in $T_{\rm bb}$ followed by a more gradual decrease following
the burst peak, giving an approximately exponential decay in X-ray
brightness back to the persistent level.  This is naturally interpreted as
heating resulting from the initial fuel ignition, followed by
cooling of the ashes once the available fuel is exhausted.
The primary evidence that the energy source for type-I bursts is
thermonuclear comes from comparisons of the time-integrated persistent and
burst flux.  The ratio $\alpha$ of the integrated persistent flux to the burst
fluence is the standard measure of the relative efficiency of
the two processes.
Early in the study
of type-I bursts it was determined that the energy derived from accretion
was between 40 and a few hundred times greater than the energy liberated
during the bursts.
These values are comparable to those predicted assuming that the burst
energy arises from nuclear burning.

Numerical models of unstable nuclear burning on the surface of a neutron star
reproduce the observed rise times (seconds),
durations (minutes), recurrence times (hours),
and total energies of the bursts \cite[$10^{39}$--$10^{40}$~ergs;
e.g.][]{fhm81, aj82, fl87, fuji87, bil98a,cb00,ramesh03,woos03}.
The frequency, strength, and time  scales of thermonuclear bursts
all depend on the composition of the burning material,
as well as the metallicity 
(here referring to the CNO mass fraction, $Z_{\rm CNO}$)
of the matter accreted onto the neutron
star; the amount of hydrogen burned 
between bursts; and the amount of fuel left-over from the previous burst.
Variations from source to source are also expected because of
differences in the core temperatures of the neutron stars and the  average
accretion rate onto the surface \citep{aj82, fl87,ramesh03}.   

The recently discovered class of extremely long-duration bursts or
``super'' bursts are also thought to arise from thermonuclear processes.
The fuel for these bursts is probably carbon rather than H/He,
giving distinctly different time scales, recurrence times and energetics
\cite[][]{stroh02,cumming04}. 
However, the predicted temperatures in the fuel layer are too low to give
carbon bursts with the observed fluences, suggesting that the cooling in
the crust may be less efficient than previously thought
\cite[]{cumming06}.
Superbursts have been detected from around 10\%
of the Galactic X-ray burst population, with recurrence times estimated at
1.5~yr \cite[]{zand04c}, 
and tend to quench the regular type-I bursts for
weeks--months afterwards.
The connection with intermediate-duration
($\sim30$~min) events observed from 
a few systems is not clear \cite[]{zand04a}.

\subsection{Bursts as a Function of Accretion Rate}

Theoretical ignition models for H- and He-burning thermonuclear bursts
predict how burst properties in an individual system change as the
accretion rate onto the neutron star varies 
\citep[e.g.,][]{fhm81,bil98a,ramesh03}.  
Several regimes of thermonuclear igntion may be identified, depending upon
the local accretion rate ($\dot m$\footnote{Following the usual
convention, we refer to the accretion rate per unit area as $\dot{m}$, and
the total accretion rate integrated over the neutron star as $\dot{M}$.}),
which is usually expressed as a fraction of the local Eddington rate
$\dot{m}_{\rm Edd}$ (
$8.8\times10^4\ {\rm g\,cm^{-2}\,s^{-1}}$, or 
$\equiv1.3\times10^{-8}\ M_\odot\,{\rm yr^{-1}}$ averaged over the
surface of a 10~km NS).
However, the quantitative values of the accretion rates separating these 
regimes of burning are a matter of some debate, as we outline below, 
and summarize in Table~\ref{burstregimes}. 

At the lowest accretion rates \citep[$\lesssim 0.01\dot{m}_{\rm Edd}$,
referred to as case~3 by][]{fhm81},\footnote{See, e..g, \citet{bil98a} for
dependences of these critical accretion rates on the metallicities,
and \citet{ramesh03} for dependences on neutron star compactness.}
the temperature in the burning layer is too low for stable hydrogen
burning; the hydrogen ignites unstably, in turn triggering helium
burning, which produces a type~I X-ray burst in a hydrogen-rich
environment.  At higher accretion rates 
(case~2; $0.01 \lesssim \dot{m}_{\rm Edd}\lesssim 0.1$), 
hydrogen burns stably into helium between bursts, leading to a growing 
pure helium layer at the base of the accreted material. The fuel layer
heats steadily until He ignition occurs and the He burns via the
triple-$\alpha$ process. At these temperatures and
pressures, helium burning is extremely unstable, and a rapid and
intense helium burst follows.  At yet higher accretion rates, 
hydrogen is accreted faster than it can be consumed by steady 
burning (limited by the rate of $\beta$-decays in the CNO
cycle), so that the helium ignites unstably in a H-rich environment
(case 1; $0.1 \lesssim\dot{m}_{\rm Edd} \lesssim 1$). Finally, at the highest 
accretion rates, helium also begins to burn steadily between bursts. 
At just below $\dot{m}_{\rm Edd} (\approx$0.9 in \citealt{hcw07}; see also 
\citealt{ramesh03}), an over-stability may arise that leads to
oscillatory H and He burning, and in turn to intermittent
bursts. 
Once the accretion rate exceeds $\dot{m}_{\rm Edd}$ (``case 0'' in
\citealt{fhm81}), stable helium burning depletes the fuel reserves and
causes bursts to cease altogether.

Although the gross features of these regimes should be robust, several
factors (some poorly-understood) could significantly change the accretion
rates at which the bursting regimes occur. 
First, 
the burst behavior 
may be sensitive to certain individual thermonuclear
reaction rates. 
One reaction that has drawn particular attention is
the ``break-out'' reaction \OtoNe, which 
removes a catalyst from the CNO cycle
\citep[e.g.,][]{fis06}. 
A low rate for this reaction causes more H burning to occur, which
produces a hotter burning layer in which steady helium burning also occurs
\cite[]{cn06}. As a
result, 
unstable burning will cease altogether
at $\dot{m}_{\rm Edd}$$\ga$0.3,
in contrast to predictions using higher
rates 
\cite[]{fhm81,bil98a,hcw07} but in partial agreement with observations.
Recent experimental measurements however favor the original, higher
\OtoNe\ rate \citep{fis07}, and the reduction in the uncertainty means
that this reaction cannot explain the cessation of bursts for most sources
around $\dot{m}_{\rm Edd}$$\ga$0.3.

Second, sedimentation or mixing in the burning layer could change the 
composition at the base of the burning layer. 
At low accretion rates ($\dot{m}_{\rm Edd} \la 0.01$),
CNO elements may settle to the bottom of the burning layer, which may 
prevent unstable H burning from inducing unstable He burning. Therefore,
most case 3 bursts would 
be pure hydrogen. However, a large He layer would also build up that 
would eventually produce a very energetic burst \citep{pbt07}, such as
those seen from ``burst-only'' sources \citep{corn04}.
On the other hand, at high accretion rates ($\dot{m}_{\rm Edd} \ga 0.1$), 
turbulent mixing of accreted fuel into deeper layers could increase the 
amount of steady burning in between bursts \citep{pb07}.

Third, the accreted nuclear fuel may not be
distributed evenly on the neutron star \citep[e.g.,][]{is99,bil00,ps01}. 
If the material is deposited at the equator, a latitudinal gradient 
could develop in the amount of fuel burned steadily between bursts. 
Whether such inhomogeneous distribution of fuel will produce bursts 
that are confined to one hemisphere
\citep{bhatt06a}, or ignite slowly-propagating fires that burn fuel over limited
regions of the neutron star \citep[e.g.,][]{bil95} is uncertain. 
The variation in the effective gravity between the equator and higher
latitudes could also lead to different ignition regimes, depending upon
the spin rate of the neutron star \cite[]{cn07b}.
The spin rate may also affect the spreading via Coriolis forces, which may
give rise to vortices that drift relative to the star as the burning
spreads \cite[]{slu02}.

Understanding these mechanisms is important, because current models 
for X-ray bursts have met with only partial success in explaining
how their rates, energetics, and time scales vary with accretion rates.
The basic predictions of 
Fujimoto et al. (1981; see also Cumming \& Bildsten 2000)
have found validation with the success of models in 
reproducing the energetics of case 3 mixed H/He bursts from
EXO 0748--676 at an accretion rate of $\dot{M} = 0.01\dot{M}_{\rm Edd}$
\citep{boirin07a}, case 2 He bursts bursts from 
SAX J1808.4--3658 at an accretion rate of 
$\dot{M} = 0.06\dot{M}_{\rm Edd}$ \citep{gal06c}, and 
regularly-recurring case 1 mixed H/He bursts from GS 1826--24 at 
$\dot{M} = 0.1\dot{M}_{\rm Edd}$ \citep{gal03d}.

On the other hand, for several sources the burst rate {\it decreases}\/ as
the accretion rate increases.  This decrease typically begins at
$\dot{M}_{\rm Edd} \sim 0.3$, well below the rate at which He-burning is
expected to stabilize.
These sources include 
4U~1636$-$536 \citep{lew87} and
4U~1705$-$44 \citep{langmeier87}, as well as most of the sources in
the sample assembled by \cite{corn03a}.
Furthermore, no correlation was found between persistent flux and burst
recurrence times in Ser~X-1 \citep{szt83} or 4U~1735$-$44
\citep{lew80,vp88}.
These observations may be evidence for ``delayed mixed bursts''
\cite[between cases 1 and 0;][]{ramesh03},
in which helium begins to burn between 
bursts \cite[see also][]{bil95,hcw07}. 
A drop in burst frequency at comparable accretion rates has also been
observed from 3A~1820$-$303 \citep{clark77b}, which, with it's evolved
donor, likely does not accrete any hydrogen.
Alternatively, these observations may indicate that
the accretion rate per unit area (which sets the burst ignition
conditions) is decreasing even though the total accretion rate is
increasing \cite[e.g.][]{bil00}, or perhaps that
the persistent fluxes are not a good measures of the accretion rates in
these sources.

The change in the composition of the fuel layer as 
$\dot{M}$ increases also affects the properties of the bursts. Helium
burning occurs via the triple-$\alpha$
process, which is moderated by the strong nuclear force and proceeds  very
quickly at the temperatures and densities of the burning layer. Hydrogen
burning proceeds more slowly, because it is limited by 
$\beta$-decays moderated by the weak force.
Therefore, faster, more intense bursts characteristic of a helium-rich
burning layer should occur at relatively low accretion rates (case~2),
while hydrogen-rich bursts with slower rise and decay times should occur
at higher rates (case~1).
Surprisingly, most sources behave in the opposite manner.  The decay  time
scales of bursts has been observed to {\it decrease} as the apparent 
$\dot{M}$ increases from 0.01--0.1~\mdotedd for 4U~1608$-$52 
\citep{murakami80b}, 4U~1636$-$536 \citep{lew87}, 4U~1705$-$44 
\citep{langmeier87}, KS 1731$-$260 \citep{muno00}, and in the sample
of \citet[][which includes several of the above LMXBs]{corn03a}.
This discrepancy has been taken as evidence that steady helium burning is
more prolific than expected at $\sim$0.3$\dot{M}_{\rm Edd}$ 
\citep[e.g.][]{ramesh03,cn06}.
Some support for this hypothesis has been found in the appearance
of low-frequency noise and mHz QPOs at these same accretion rates, 
that has been attributed to marginally unstable helium burning 
\citep{bil95,rev01,ramesh03,hcw07}. 

\begin{deluxetable}{llllc}
  \tablewidth{0pt}
  \tablecaption{Bursting Regimes \label{burstregimes} }
  \tablehead{
 & & & \colhead{Burst} & \\
 \colhead{Case} &
 \colhead{$\dot{m}_{\rm Edd}$} &
 \colhead{Steady Burning} &
 \colhead{Composition} &
 \colhead{Ref.} }
 \startdata
3/V & $\la$0.01 & none & mixed H/He & [1,2] \\
2/IV & 0.01 -- 0.1 & stable H & pure He & [3,4] \\
1/III & 0.1 -- 1.0 & stable H & mixed H/He & [4,5,6] \\
-/II & $\sim$1.0 & over-stable H/He & mixed H/He? & [7,8] \\
0/I  & $\ga$1.0 & stable H/He & none & [9] \\
 \enddata
\tablecomments{The names for each of the burst cases are taken from 
\cite[][Arabic numerals]{fhm81} and 
\cite[][Roman numerals]{ramesh03}.
The ranges in accretion rate ($\dot{m}_{\rm Edd}$; 
normalized to the Eddington rate) for each case are taken from \citet{fhm81};
note that \citet{ramesh03} predict lower rates for their cases
II and I ($\sim0.3\dot{m}_{\rm Edd}$). The references represent recent 
examples of calculations and/or comparisons to observations in each regime.}
 \tablerefs{1. \cite{pbt07}; 2. \cite{cn07a}; 3. \cite{gal06c}; 
4. \cite{woos03}; 5. \cite{gal03d}; 6. \cite{fis08};
7. \cite{ramesh03}; 
8. \cite{hcw07}; 
9.  \cite{schatz01}}
 \end{deluxetable}


\begin{figure}
 \epsscale{1.2}
 \plotone{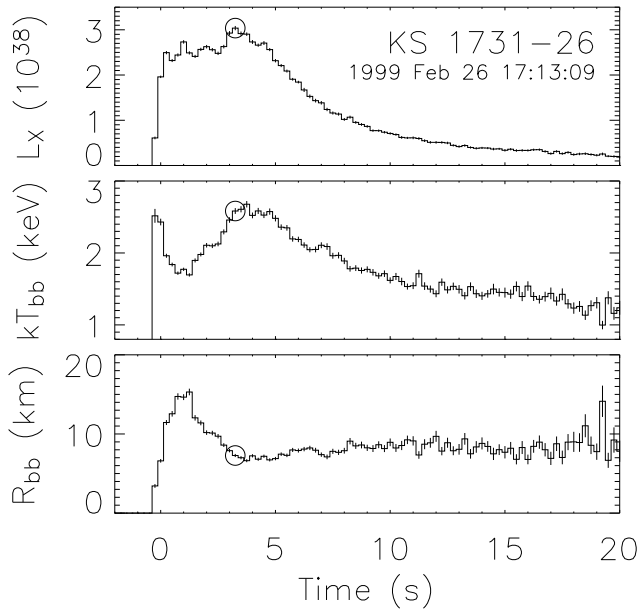}
 \caption{Spectral evolution in a thermonuclear burst exhibiting photospheric
radius-expansion, from KS~1731$-$26. {\it Top panel}\/ 
Burst luminosity $L_X$, in units of $\eps$; 
{\it middle panel}\/ blackbody (color)
temperature $kT_{\rm bb}$; and {\it bottom panel}\/ blackbody radius $R_{\rm
bb}$.
$L_X$ and $R_{\rm Bbb}$ are calculated at an assumed distance of
7.2~kpc (Table \ref{dist}). 	
Note the anticorrelation between $kT_{\rm bb}$ and $R_{\rm bb}$ in the
first few seconds, indicative of the expanding photosphere, and the
approximately constant flux throughout the expansion. The time at which
the flux reaches a maximum is indicated by the open circle; by then the
radius has declined to the asymptotic value in the burst tail, suggesting
that the photosphere has settled (``touched down'') on the NS surface.
 \label{pre_example} }
\end{figure}

\subsection{Bursts as Standard Candles}

The peak flux for very bright bursts can reach 
the Eddington luminosity at the surface of the NS, at which point the
(outward) radiation pressure equals (or exceeds) the gravitational force
binding the outer layers of accreted material to the star.
Such bursts frequently exhibit a characteristic spectral evolution in the
first few seconds, with a local peak in blackbody radius and
at the same time a dip in color temperature, while the flux remains
approximately constant (Fig. \ref{pre_example}). This pattern is thought
to result from expansion of the X-ray emitting photosphere once the burst
flux reaches the Eddington luminosity; the effective temperature must
decrease in order to maintain the luminosity at the Eddington limit, and
excess burst flux is converted into kinetic and gravitational potential
energy in the expanded atmosphere. 

The largest uncertainty in the theoretical Eddington luminosity arises
from possible variations in the photospheric composition. The limiting
flux for a composition with hydrogen present at solar mass fraction will
be a factor of 1.7 below that of a pure helium atmosphere.
Nevertheless, the Eddington luminosities \ledd\ measured for LMXBs with
independently-known distances are generally consistent to within the
uncertainties, at a value estimated as 
$(3.0\pm0.6)\times10^{38}\ {\rm ergs\,s^{-1}}$ by \cite{lew93},
or, more recently,
$(3.79\pm0.15)\times10^{38}\ {\rm ergs\,s^{-1}}$
\cite[cf. with equation \ref{ledd};][]{kuul03a}.  This result is
consistent with the narrow ranges for masses and surface redshifts
expected for the neutron stars in these bursters. 
Consequently, these photospheric radius-expansion (PRE) bursts can be used
as distance indicators \cite[]{bas84}.
Time-resolved spectroscopy of radius-expansion bursts also allow in principle 
measurement of the surface gravitational redshift
\cite[e.g.][]{damen90,smale01},
although this has proved a considerable challenge \cite[see e.g.][for a
more recent study]{ozel06}.

\subsection{A New Diagnostic of Nuclear Burning}

\begin{figure}
 \epsscale{1.2}
\plotone{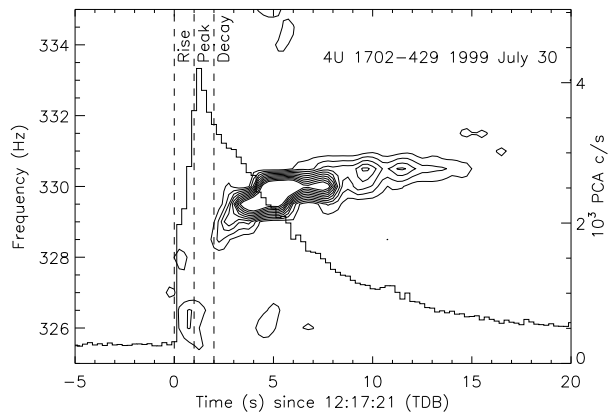}
\caption{A dynamic power spectrum illustrating the typical frequency
evolution of a burst oscillation. Contours of power as a function of
frequency and time were generated from power spectra of 2 s intervals
computed every 0.25 s. A Welch function was used to taper the data to
reduce sidebands in the power spectrum due to its finite length
\citep{nr}. The $n$th contour level has a single-trial probability of
$0.02^n$ of occurring randomly due to noise. The PCA count rate
(histogram) is plotted
referenced to the right axis. The time intervals defined in \S\ref{ibo} as
the rise, peak, and decay are indicated with the vertical dashed lines.
\label{fig:freqev}}
\end{figure}

One of the key capabilities of \xte\/ is for high temporal
resolution X-ray timing studies. Since 1996, this capability has led to
the discovery of several distinct types of kHz variability in LMXBs
\cite[for a recent review, see ][]{gal08b}.
Highly coherent burst oscillations with fractional amplitudes in the
range 5--20\% rms have been detected
in thermonuclear bursts from \noscsrc\ sources to date
(\citealt{stroh96}; see also \citealt{sb03}).
As the burst evolves, these ``nuclear-powered pulsations''
typically increase in frequency by a few Hz, most approaching an
asymptotic value which is stable for a given source from burst to burst
\cite[e.g. Fig. \ref{fig:freqev};][]{muno02b}.  That
the asymptotic frequency traces the NS spin has been confirmed by the
detection of burst oscillations at the spin frequency in the millisecond
pulsars SAX~J1808.4$-$3658 \cite[]{chak03a} and XTE~J1814$-$338
\cite[]{stroh03a}, as well as the prolonged oscillation detected during a
superburst in 4U~1636$-$536 \cite[]{stroh02b}.

Substantial questions remain regarding the mechanism of burst
oscillations,
as well as what conditions determine whether or not the
oscillations will be detectable in a given burst, or a given source.
The oscillations have been suggested to result from initially localized
nuclear burning, which spreads over the surface of the neutron star during
the early stages of the burst \cite[]{stroh96}. However, this explanation
does not account for the oscillations which persist as long as 5--10~s
after the burst rise. 
The frequency drift 
is likely too large to be explained by angular momentum conservation in a
decoupled expanded burning layer \cite[]{cum02},
and may instead result from changes in the  velocity
of a pattern in the surface brightness. 
Slow-moving (in the rotating neutron star frame) anisotropies in the
surface brightness may
originate from hydrodynamic instabilities \citep{slu02} or modes excited
in the neutron star ocean (\citealt{cb00}; see also \citealt{heyl04,pb05}
and references therein).
Recently, the phenomenology has become even more complex with the
observation of intermittent persistent pulsations which appear to be in
some cases related to the occurence of bursts \cite[]{gal07a,altamirano07}
but in other cases not \cite[]{casella07}.

\subsection{The Need for a Global Study}

It is generally unfeasible to 
to study the variation in burst properties over a wide range of accretion
rate using data from a single source.
Typically 
only a narrow range of
accretion rates is observed, and insufficient exposure time is available,
leading to only a small number of
bursts in total. 
A few previous observational studies have focussed on the properties of
bursts from more than one source.
A compilation of  45 measurements in the literature from ten LMXBs
by \cite{vppl88} 
revealed a global decrease in burst duration with increasing
persistent flux, similar to that seen individually for several sources.
They also found that $\alpha$
was correlated with the (normalized)
persistent flux, more strongly than was predicted by numerical models
\cite[e.g.][]{fhm81}. The normalized fluence depended principally on the
burst interval $t_{\rm rec}$, which suggested that continuous stable
burning between bursts is a general phenomenon. 
More recently,
\cite{corn03a} analysed six years of \sax\/ observations of 37 LMXBs,
with a combined sample of 1823 bursts, and identified a transition between
long, H-rich bursts (assumed to result from case~3 ignition) to short,
pure He bursts (case~2), inferring the onset of steady H burning at a
persistent luminosity of
$2\times10^{37}\ \eps$ (equivalent to $0.1\dot{M}_{\rm Edd}$), a factor of
10 higher than predicted by theory. Below this level, bursts
were long, frequent and occurred quasi-periodically, typified by
GS~1826$-$24 and KS~1731$-$26. Above $2\times10^{37}\ \eps$ the burst rate
dropped by a factor of five, and the bursts were short and occurred
irregularly (although short bursts were also observed in the low accretion
rate regime). 
At even higher luminosities, bursts ceased altogether in these sources 
(although are subsequently observed at $\approx\dot{M}_{\rm Edd}$ in two
sources, GX~17+2 and Cyg~X-2).

To date, the {\it Rossi X-ray Timing Explorer}\/ (\xte) has observed
66	
of the known thermonuclear burster sources, and
discovered several new ones.
The \xte\/ data are unparalleled for studies of bursts and bursters,
thanks to the large instrumental effective area and high timing
resolution. 
New data enter the public archive
continually, 
and published analyses rarely take advantage of all the available bursts
in all the public observations, let alone all the bursts from all the
known bursters.
To date, no global
comparisons of theory with these data have been made.
The wealth of observational data 
motivate the present work, which seeks to present a
uniform analysis of all thermonuclear bursts from the bursters
observed by \xte\/ through
\burstpubdate.
By combining the bursts from different sources 
we achieve much larger burst numbers and a larger range of $\dot{M}$ for
global characterization of burst behaviour than is possible for any
individual source.
We also include information on the presence of burst
oscillations, which is only available in the \xte\/ data.
While \xte\/ has also observed several superbursts,
we do not analyse these events in this paper.

We present the contents of the catalog, and the results of our studies, as
follows.
In \S\ref{analysis} we describe the analysis methods
and products, and relate to the physical properties of the bursts.
We summarise the properties of the catalog in \S\ref{results},
and present our detailed analysis in the subsequent sections.
In \S\ref{pflux} we measure the mean peak flux of radius-expansion
bursts, and determine the source distances.
We analyse the properties of the individual bursts in \S\ref{ts}, and
explore the consistency of the burst behaviour of
different sources as a function of accretion rate
in \S\ref{diversity}.
In \S\ref{global} we combine the bursts from various sources in an attempt
to quantify the global burst properties as a function of accretion rate
and compare these properties to predictions from burst theory.
We further discuss the global behaviour of the burst energetics in
\S\ref{secalpha}.
We attempt to place observational limits on the boundaries of the
theoretical ignition regimes in \S\ref{boundary}.
We discuss the properties of the millisecond oscillations in
\S\ref{milosc}.
Finally, we present a number of outstanding theoretical challenges in
\S\ref{challenges},
and summarise our results in \S\ref{summary}.
In appendix \ref{sources}
we present the results for
individual bursters on a case-by-case basis;
we constrain the origin for bursts in crowded fields in appendix
\ref{localize}.

\section{Observations and analysis}
\label{analysis}

Public data from 
\xte\/ observations of thermonuclear burst sources are available through the
High-Energy Astrophysics Science Archive Research Center\footnote{\url
http://heasarc.gsfc.nasa.gov} (HEASARC), dating from
shortly after the launch of the satellite on 1995 December 30. This paper
includes all publicly available data through 
\burstpubdate.
\xte\ carries three instruments sensitive to X-ray photons. The All-Sky
Monitor \cite[ASM;][]{asm96} consists of three scanning shadow cameras
sensitive to photons between 1.5 and 12~keV with a total effective area of
$\approx100\ {\rm cm^2}$, which provide 90~s exposures of most points on
the sky every 96~min.
The High-Energy X-ray Timing Experiment
\cite[HEXTE;][]{hexte96} is comprised of two clusters of NaI/CsI
scintillation detectors sensitive to X-rays between 15 and 250~keV with a
total effective area of $1600\ {\rm cm^2}$.
The Proportional Counter Array \cite[PCA;][]{xte96}  consists of five
identical, co-aligned proportional counter units (PCUs), sensitive to
photons in the energy range 2--60~keV. 
The field-of-view of both the PCA and HEXTE is circular with radius
$\approx1^\circ$.  Photon counts from the PCA are processed
independently by up to 6 Event Analyzers (EAs) in a variety of
configurations.  Two EAs are permanently set to two standard observing
modes, Standard-1 (with 0.125~s time resolution but only one energy
channel) and Standard-2 (16~s binned spectra on 129 energy channels
between 2--60~keV). The remaining EAs may be configured by the observer to
give time resolution down to $1\mu s$ and up to 256 spectral channels.

We extracted 1-s lightcurves covering the full 2--60~keV PCA energy range
from Standard-1 mode data of all public observations covering known burst
sources. The PCA field of view is approximately $1^\circ$ in radius, and
the effective area drops off approximately linearly as a function of
off-axis angle. Thus, we extended our lightcurves to offset pointings of
up to $1\fdg2$, frequently including the end of the satellite's slew to the
source and the beginning of the slew away. We also searched observations
of fields centered less than $1^\circ$ away from known burst
sources. We searched each lightcurve for bursts as follows. For each
observation we calculated the overall mean and standard deviation of the
1-s count rate measurements, and identified burst candidates in bins which
exceeded the mean by more than $4\sigma$. We then visually
inspected the lightcurves to confirm or reject each candidate.
Candidates were rejected if they were attributable to other events which
can produce sharp jumps in the count rate, such as 
detector breakdowns, gamma-ray bursts, or particle events.
For a few weak events, time-resolved spectral analysis (see below) failed
to show significant cooling in the decay; for others, data modes with
sufficient temporal and spectral evolution were not available to undertake
spectral analysis at all. We include these events in the catalog, but they
must be viewed as burst candidates, only.

\subsection{Characterizing the persistent emission}
\label{pers}

In order to coarsely characterise the persistent spectrum,
we computed hard and soft X-ray colors as the ratio of the
background-subtracted detector counts in the (8.6--18.0)/(5.0--8.6) keV
and the (3.6--5.0)/(2.2--3.6) keV energy bands, respectively.  We used
64~s integrations to calculate the colors when the source intensity was
above 100 counts s$^{-1}$, and 256 s integrations otherwise.
We corrected the measured count rates for gain changes over the life of
the mission by normalizing count rates from the Crab Nebula in each
PCU to constant values for each energy band (totaling 2440 counts s$^{-1}$
PCU$^{-1}$ in the 2.2--18.0 keV band) using linear trends.  When this
correction is applied, the hard and soft colors from the Crab Nebula have
values of 0.679 and 1.358, 
with standard deviations of only 0.1\% and 0.5\% respectively.

\begin{figure}
 \epsscale{1.2}
\plotone{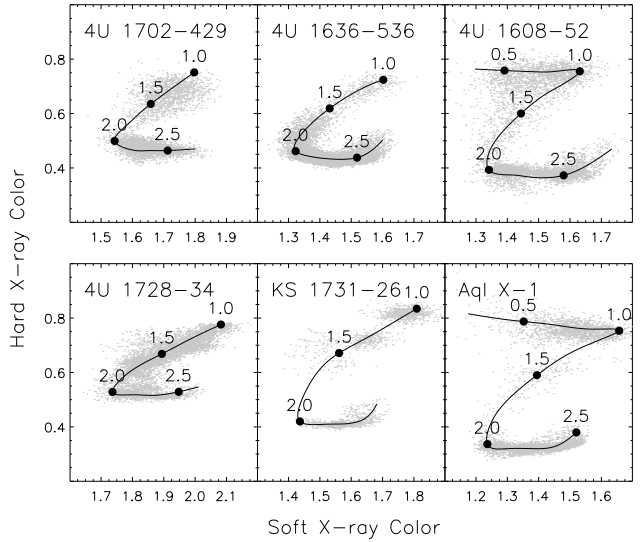}
\caption{Color-color diagrams for six of the nine sources for which it was
possible to define the coordinate $S_Z$ locus.  The soft color is the
ratio of the background-subtracted PCA counts in the energy range
3.6--5.0~keV accumulated in 64 or 256~s, to the counts in the range
2.2--3.6~keV.  The hard color is the ratio of counts in the ranges
8.6--18.0 and 5.0--8.6~keV. Colors are corrected for PCU gain variations.
Filled circles show the points (labeled with their $S_Z$ values) used to
define the overall shape of the curve; solid curves show the spline
interpolation between the points.  The coordinate $S_Z$ is thought to
be proportional to $\dot{M}$.
\label{fig:sz}}
\end{figure}

We show examples of the distribution of source colors 
(color-color diagrams) in Fig. \ref{fig:sz}.
As $\dot{M}$ onto the neutron star increases, a source moves from the
top-left to the bottom-right, roughly tracing a Z-shaped pattern
\citep{hvdk89,muno02,gd02a}. Most bursting LMXBs are classified as ``atoll'' sources, and trace out their full Z-shaped pattern as they vary
in intensity by a factor of $\ga 100$ \citep[see][for a
discussion]{muno02}.  Nine atoll sources were observed on both
the top and bottom portions of their color-color diagrams:
4U~1608$-$52,
4U~1636$-$536,
4U~1702$-$429,
4U~1705$-$44,
4U~1728$-$34,
KS~1731$-$260,
4U~1746$-$37,
XTE~J2123$-$058 and
Aql~X-1.
For those sources, we parameterized the position on the diagram by
defining a curve that followed the middle of the Z-shaped track
\cite[Fig.~\ref{fig:sz};][]{die00}.  We first selected several points 
to define the basic shape of the curve. From these points, we defined
a smooth curve using a spline interpolation. We then assigned $S_Z = 1$ to
the upper-right vertex of the Z-shape and $S_Z = 2$ to the lower-right
vertex, and defined the unit arc-length as the distance along the curve
between these two points. The value of $S_Z$ for any given point on the
color-color diagram by finding the nearest point on the curve, and finding
the value of the arc-length $S_Z$ there. 
We then defined the mean $S_Z$ value from the mean colors for each
observation of the sources listed above. Although $S_Z$ is 
thought to be proportional to $\dot{M}$ \cite[e.g.][]{vrtilek90}, the absolute
calibration is
not well determined.

We also estimated the persistent source flux \fper\/ at the time of the bursts
from spectra extracted from Standard-2 mode data, separately for each PCU
within each observation (excluding a typically $300$~s interval covering each
burst). We fit these spectra over the range 2.5--25~keV with an empirical
model consisting of blackbody and power law components, each attenuated by
neutral absorption with solar abundances.
For many of the spectra,
residuals were present around 6.4~keV which we interpreted as fluorescent Fe
emission, and where these residuals resulted in a reduced-$\chi^2\ga2$ we
added a Gaussian component to improve the fit. For particularly bright
sources such a model did not give a good fit, and for these we used
instead a continuum component describing Comptonisation in a homogeneous
environment \cite[{\tt compTT} in {\sc xspec};][]{tit94}.  The particular
choice of the continuum 
did not significantly affect the measured flux within the energy range
covered by the PCA.
We
then integrated the model over the energy range 2.5--25~keV\footnote{The Crab flux in this
band is $3.3\times10^{-8}\ \epcs$} to estimate the source flux detected by each
PCU.  For each source we calculated the mean PCU-to-PCU offset averaged
over all the public \xte\/ observations, and renormalized the flux
measurements relative to PCU 2.  We adopted the mean and standard
deviation of the rescaled flux measurements as the flux and error for that
observation.

While the majority of the burst flux is emitted in the range 2.5--25~keV,
this is generally  not true for the persistent emission. In order to
estimate the bolometric persistent flux \fbol\ we chose representative
(preferably long) observations for selected sources and undertook combined
fits of each PCU spectra (as described above) along with HEXTE spectra
above 15~keV. We set the upper energy limit for the HEXTE spectra 
individually depending upon the maximum energy to which the source could
be detected (typically 40--80~keV).  Persistent spectra from bursters
frequently exhibit a spectral cutoff between 15 and 50~keV, and so we fit the
broadband spectra with a Comptonisation continuum component attenuated by
neutral absorption, also sometimes with a Gaussian component
representing fluorescent Fe emission around 6.4~keV. We generated an
idealized response covering the energy range 0.1--200~keV with 200
logarithmically spaced energy bins, and integrated the model flux also
over this range. We then calculated a bolometric correction $c_{\rm bol}$
as the ratio between the 0.1--200 and 2.5--25~keV fluxes measured from the
broadband spectral fits.  The error on the bolometric correction was
estimated as the standard deviation of the derived correction over the
available PCUs.
Altogether we estimated bolometric corrections for observations of 
24	
bursting sources, ranging between 1.05 (from a 1997 September observation
of 4U~1728$-$34) and 2.14 (for a 2002 October observation of
SAX~J1808.4$-$3658; Table \ref{cboltable}).  The corrections for the accretion-powered pulsars
tended to be larger than for the non-pulsing burst sources, and we found
the maximum value for the latter sources to be 
1.93.	
In the mean, 
$c_{\rm bol}=1.38$	
for the non-pulsing sources,
and we adopt this value except where we calculated a correction
specifically for that source or observation. The likely error introduced
is thus no more than $\approx40$\%.

%
\begin{deluxetable*}{lllrccccc}
\tablecaption{Bolometric corrections derived from {\it RXTE}\/ observations
  \label{cboltable}
}
\tablewidth{0pt}
\tablehead{
  \colhead{}
 & 
 & 
 & \colhead{Dur.}
 & \colhead{Flux}
 & \colhead{Soft}
 & \colhead{Hard}
 & 
 & 
\\
  \colhead{Source}
 & \colhead{Obs ID}
 & \colhead{Start time (UT)}
 & \colhead{(ks)}
 & \colhead{($10^{-9}\ \epcs$)\tablenotemark{a}}
 & \colhead{color}
 & \colhead{color}
 & \colhead{$S_Z$}
 & \colhead{$c_{\rm bol}$}
}
\startdata
EXO~0748$-$676 & 10108-01-07-01  & 1996 Aug 15 21:24:00 & 1.86 & $0.277\pm0.005$ & 2.30 & 0.98 & \nodata & $1.93\pm0.02$ \\
1M~0836$-$425 & 70031-03-01-00  & 2003 Jan 24 06:10:03 & 6.00 & $1.85\pm0.06$ & 1.99 & 0.892 & \nodata & $1.82\pm0.02$ \\
4U~1254$-$69 & 60044-01-02-00  & 2001 Dec  7 00:02:04 & 44.8 & $0.830\pm0.011$ & 1.55 & 0.440 & \nodata & $1.13\pm0.03$ \\
4U~1323$-$62 & 20066-02-01-00  & 1997 Apr 25 22:00:02 & 21.1 & $0.2477\pm0.0014$ & 2.14 & 0.894 & \nodata & $1.67\pm0.05$ \\
4U~1608$-$52 & 60052-03-01-01  & 2001 Nov 20 23:48:03 & 3.78 & $2.175\pm0.013$ & 1.56 & 0.694 & 1.19 & $1.77\pm0.04$ \\
4U~1636$-$536 & 40028-01-13-00  & 2000 Jan 22 01:29:03 & 16.1 & $8.08\pm0.04$ & 1.58 & 0.449 & 2.63 & $1.118\pm0.017$ \\
 & 60032-05-15-00  & 2002 Feb 28 13:43:04 & 25.9 & $5.1\pm0.2$ & 1.53 & 0.464 & 2.57 & $1.20\pm0.06$ \\
4U~1702$-$429 & 20084-02-01-00  & 1997 Jul 19 08:50:03 & 20.5 & $1.339\pm0.012$ & 1.63 & 0.460 & 2.28 & $1.099\pm0.013$ \\
 & 40025-04-01-01  & 1999 Feb 22 02:44:03 & 4.44 & $1.820\pm0.014$ & 1.55 & 0.520 & 1.93 & $1.117\pm0.010$ \\
4U~1705$-$44 & 20073-04-01-00  & 1997 Apr  1 13:25:02 & 13.8 & $2.02\pm0.15$ & 1.86 & 0.712 & 1.15 & $1.413\pm0.018$ \\
XTE~J1710$-$281 & 60049-01-01-00  & 2001 Aug 12 07:05:03 & 24.8 & $0.1095\pm0.0017$ & 1.40 & 0.488 & \nodata & $1.42\pm0.13$ \\
XTE~J1723$-$376 & 40705-01-03-00  & 1999 Feb  3 21:53:03 & 10.4 & $1.398\pm0.017$ & 1.95 & 0.380 & \nodata & $1.05\pm0.02$ \\
4U~1728$-$34 & 20083-01-02-01  & 1997 Sep 21 15:42:02 & 13.9 & $4.4\pm0.4$ & 1.87 & 0.513 & 2.33 & $1.050\pm0.006$ \\
 & 50030-03-06-02  & 2001 Jul 22 12:09:03 & 4.92 & $1.59\pm0.02$ & 1.83 & 0.634 & 1.67 & $1.373\pm0.011$ \\
KS~1731$-$260 & 30061-01-02-01  & 1998 Oct  2 13:04:02 & 3.60 & $1.35\pm0.12$ & 1.67 & 0.731 & 1.30 & $1.58\pm0.04$ \\
4U~1735$-$44 & 20084-01-02-03  & 1997 Sep  1 08:52:03 & 8.76 & $4.88\pm0.04$ & 1.66 & 0.561 & \nodata & $1.099\pm0.012$ \\
XTE~J1739$-$285 & 91015-03-04-04  & 2005 Nov  7 03:26:04 & 10.1 & $0.942\pm0.006$ & 1.54 & 0.473 & \nodata & $1.30\pm0.06$ \\
SAX~J1748.9$-$2021 & 60035-02-02-03  & 2001 Oct  8 07:44:04 & 10.6 & $3.19\pm0.04$ & 1.49 & 0.403 & \nodata & $1.18\pm0.05$ \\
EXO~1745$-$248 & 50054-06-03-00  & 2000 Aug  6 12:56:04 & 4.62 & $3.93\pm0.05$ & 2.05 & 0.840 & \nodata & $1.53\pm0.02$ \\
4U~1746$-$37 & 10112-01-01-00  & 1996 Oct 25 00:13:00 & 22.0 & $0.165\pm0.003$ & 1.65 & 0.767 & 0.751 & $1.45\pm0.05$ \\
SAX~J1808.4$-$3658 & 70080-01-02-00  & 2002 Oct 18 02:09:03 & 23.2 & $2.056\pm0.006$ & 1.42 & 0.688 & \nodata & $2.14\pm0.03$ \\
XTE~J1814$-$338 & 80418-01-01-09  & 2003 Jun 12 13:19:04 & 9.12 & $0.49\pm0.05$ & 1.56 & 0.883 & \nodata & $1.86\pm0.03$ \\
GX~17+2 & 30040-03-02-00  & 1998 Nov 18 06:41:02 & 20.8 & $14.8\pm0.3$ & 1.84 & 0.377 & \nodata & $1.083\pm0.017$ \\
GS~1826$-$24 & 20089-01-01-02  & 1997 Nov  5 21:09:02 & 15.6 & $1.085\pm0.009$ & 1.66 & 0.857 & \nodata & $1.70\pm0.03$ \\
Ser~X-1 & 70027-04-01-01  & 2002 May 27 15:05:04 & 5.76 & $4.8\pm0.4$ & 1.48 & 0.383 & \nodata & $1.24\pm0.08$ \\
4U~1916$-$053 & 30066-01-02-08  & 1998 Jul 23 11:27:02 & 13.2 & $0.382\pm0.004$ & 1.57 & 0.667 & \nodata & $1.37\pm0.09$ \\
XTE~J2123$-$058 & 30511-01-05-00  & 1998 Jul 22 05:11:02 & 26.5 & $1.017\pm0.008$ & 1.49 & 0.471 & 2.24 & $1.19\pm0.06$ \\
\enddata
\tablenotetext{a}{Flux in the 2.5--25~keV band, averaged over the entire
observation (excluding any bursts present).}
\end{deluxetable*}

From the persistent
flux \fper\ and the distance $d$ (derived from the peak flux of
radius-expansion bursts), we may also estimate the accretion rate per unit area at the
neutron star surface,
$\dot{m}$. We assume that the X-ray luminosity is
\begin{equation}
  L_{X,\infty} = \frac{4\pi R_{\rm NS}^2\dot{m} Q_{\rm grav}}{(1+z)}
            = 4\pi d^2 F_{\rm p} c_{\rm bol}
\end{equation}
where 
$R_{\rm NS}$ is the NS radius, and
$Q_{\rm grav}$ is the energy released per nucleon during accretion
($=c^2z/(1+z)\approx GM_{\rm NS}/R_{\rm NS}$).
Here we assume implicitly that the accreted fuel covers the neutron star
surface evenly, and that the persistent emission is isotropic.
Because the neutron star has such a strong gravitational field,
the luminosity measured by an distant observer is significantly
lower than at the NS surface due to gravitational redshift. Thus, we
correct the 
quantities at the NS surface by a factor $(1+z)$, where
$z$ is the surface redshift; $1+z=(1-2GM/R_{\rm NS}c^2)^{-1/2}=1.31$ for a
NS with mass $M_{\rm NS}=1.4M_\odot$ and radius $R_{\rm NS}=10$~km. Both
mass measurements \cite[]{tc99} and predictions from a range of plausible
equations of state \cite[e.g.][]{lp01} suggest that the masses and radii
of neutron stars (and hence the compactness $M_{\rm NS}/R_{\rm NS}$) span
relatively narrow ranges.  A surface redshift has only been
tentatively measured (via redshifted absorption lines) in one burster,
EXO~0748$-$676, at $z=0.35$ \cite[]{cott02}; subsequent analyses have
failed to confirm this result \cite[]{cott07}. Thus, our assumption of a
constant redshift of $z=0.31$ for all the bursters in our sample
unavoidably introduces a small systematic error when combining burst
measurements from different sources (see \S\ref{energetics}).
Then
\begin{eqnarray}
  \dot{m} & = & 6.7\times10^3\ \left(\frac{F_p c_{\rm bol}}{10^{-9}\ \epcs}\right)
            \left(\frac{d}{10\ {\rm kpc}}\right)^2
            \left(\frac{M_{\rm NS}}{1.4\ M_\odot}\right)^{-1} 
\nonumber \\ & & \times\  
                      \left(\frac{1+z}{1.31}\right)
            \left(\frac{R_{\rm NS}}{10\ {\rm km}}\right)^{-1}\
         {\rm g\,cm^{-2}\,s^{-1}}
\label{mdot}
\end{eqnarray}
It is generally thought that $L_{X,\infty}$ is proportional to
$\dot{m}$ within intervals of $~$days, but that the absolute calibration
can shift substantially on longer timescales \cite[e.g.][]{mend01}.

\subsection{Temporal and spectral analysis of individual bursts}
\label{temporal}

Once each burst was located, high time- and spectral resolution data
(where available) from the PCA covering the
burst (100--200~s) were downloaded and processed to provide a range of
analysis products.  For most bursts, multiple spectral channels were
available with time resolution of 0.25~s or better.  We extracted
2--60~keV spectra within intervals of 0.25--2~s covering the entire burst.
We set the initial integration time for the spectra at 0.25, 0.5, 1 or 2~s
depending upon the peak count rate of the burst ($>6000$, 3000--6000,
1500--3000 or $<1500\ \cts$ respectively, neglecting the pre-burst persistent
emission).  Each time the count rate following the peak
decreased by a factor of $\sqrt2$ we doubled the spectral time bin size.
Since the evolution of the burst flux is slower in the tail, this increase
in time bin size does not adversely affect the data quality.

We fitted each burst spectrum with a blackbody model multiplied by a
low-energy cutoff, representing interstellar absorption 
using the cross-sections of \cite{wabsxc} and solar
abundances from \cite{wabsabund}. 
A spectrum extracted from a (typically) 16~s interval prior to the burst
was subtracted as the background; this approach is well-established as a
standard procedure in X-ray burst analysis \cite[e.g.][]{vpl86,kuul02a}.
The observations span multiple PCA gain epochs, which
are defined by instances where the gain was manually re-set by the
instrument team (on 1996 March 21, 1996 April 15, 1999 March 22 and 2000
May 13).  In addition to these abrupt changes more gradual variation in
the instrumental response is known to occur, due to a number of factors.
To take into account these gain variations we generated a separate
response matrix for each burst using {\sc pcarsp} version
\pcarspver\footnote{We note that the geometric area of the PCUs was
changed for this release for improved consistency between PCUs and with
(e.g.) canonical models of calibration sources, particularly the Crab
pulsar.  These changes have the effect of reducing the measured flux
compared to analyses using previous versions of the response generating
tools, by 12--14\%.  See
\cite{xtecal06} for more
details.}, which is 
included as part of {\sc lheasoft} version
\lheasoftver\ (\lheasoftdate).
The initial fitting was performed with the absorption column
density $N_{\rm H}$ free to vary; subsequently it was fixed at the mean
value over the entire burst to estimate the bolometric flux.
The bolometric flux at
each timestep $i$ was calculated according to
\begin{eqnarray}
  F_i & = & \sigma T_{{\rm bb},i}^4 \left( \frac{R}{d} \right)_i^2 \nonumber \\
      & = & 1.076\times10^{-11}\ 
          \left(\frac{kT_{{\rm bb},i}}{1\ {\rm keV}}\right)^4 K_{{\rm bb},i}\ 
\epcs
\label{flux}
\end{eqnarray}
where $T_{\rm bb}$ is the blackbody temperature,
$R$ is the effective radius of the emitter, $d$ is the distance to the
source, and $K_{\rm bb}$ is the normalisation of the blackbody component
(we assume isotropic emission for the burst flux throughout, unless stated
otherwise).
For bursts observed in slews or offset pointings, we rescaled the measured
peak flux and fluence by $1/\Delta\theta$, where $\theta$ is the offset
between the pointing angle and the source position (see appendix
\ref{localize}).
It is important to note that the apparent blackbody temperature
for a distant observer $T_{\rm bb}\equiv T_{\rm bb,\infty}$ and the
apparent temperature measured at the surface differ by a factor of
$(1+z)$.
Furthermore, spectral hardening arising from radiation transfer effects
in the atmosphere increase the apparent surface
temperature compared to the effective temperature
\cite[e.g.][]{lth86,tit94b,madej04}.
Unless otherwise stated, we make no correction for the effects of redshift or
spectral hardening, and quote the observed parameters for distant
observers only.

\begin{figure}
 \epsscale{1.2}
\plotone{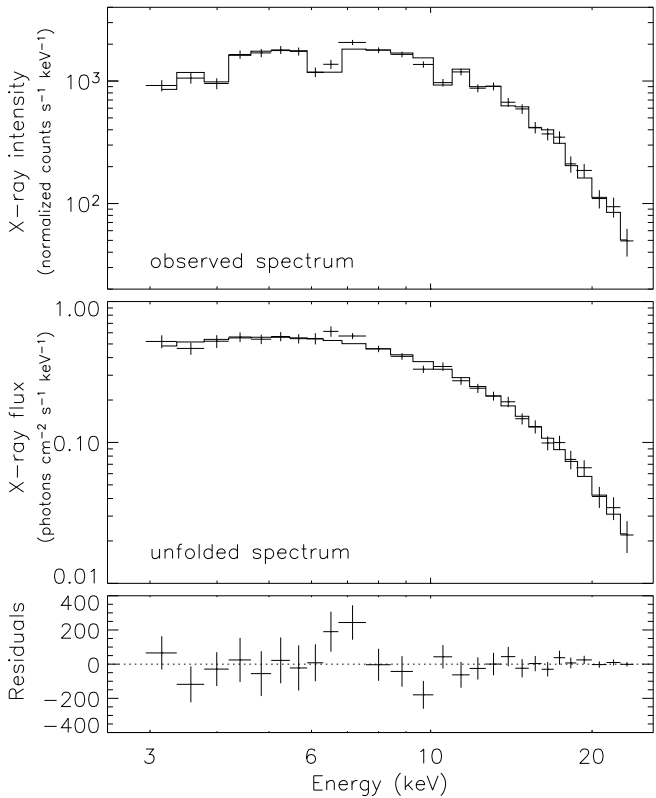}
\caption{Example 0.25-s spectrum from the peak of a radius-expansion
burst observed from
4U~1728$-$34 on 1999 June 30 19:50:14 UT by \xte.  The top panel shows the
observed spectrum (after subtracting the pre-burst persistent
emission), while the middle panel shows the inferred burst
spectrum after correcting (``unfolding'') for the instrumental response.
The histogram in both panels shows the best model fit, in this case a
blackbody with color temperature $kT=2.99\pm0.04$~keV and radius
$5.09_{-0.12}^{+0.13}$~km (assuming $d=5.2$~kpc; Table \ref{dist})
absorbed by neutral material with column density of $6.36\times10^{21}\
{\rm cm^{-2}}$ (the mean value derived from spectral fits over the entire
burst). The corresponding unabsorbed bolometric flux is
$(8.2\pm0.2)\times10^{-8}\ \epcs$. Although the measured  radius is
smaller than expected for a typical neutron-star equation of state, it is
important to note that this is an apparent radius which is reduced by
biases in the color temperature measurement.  The bottom panel shows the
residual counts for the fit, with $\chi^2=20.5$ for 25 degrees of freedom
indicating a statistically good fit.
The most noticeable deviations from zero are between 6 and 7~keV, and
may originate from fluorescent Fe K$\alpha$ emission from material
surrounding the neutron star. 
 \label{exspec} }
\end{figure}

Implicit in equation~(\ref{flux}) is the bolometric correction to the
burst flux measured in the PCA bandpass;
this correction adds $\simeq 7$\%
to the peak 2.5--25~keV PCA flux of radius expansion bursts.  Should the
emitted spectrum deviate significantly from a blackbody outside the PCA
passband,
equation~(\ref{flux}) will not give the correct bolometric flux.
Reassuringly, the blackbody model gave a good fit to the vast majority of
the burst spectra (e.g. Fig. \ref{exspec}),
although we consider that systematic errors of order as large as the
bolometric correction may yet be present in the flux estimates presented
here.
The model fits tended to result in poor $\chi^2_\nu$ values preferentially
at low fluxes, in the burst tail; 13\% of the burst spectra with fluxes
$<0.25$ of the peak in that burst had $\chi^2_\nu>2$, while only 4.6\% of
the spectra with flux $>0.75$ of the peak had $\chi^2_\nu>2$. For a
$\chi^2$ distribution with 24 degrees of freedom (which is the typical
number for the fits) we expect only 2.5\% of reduced-$\chi^2$ values in
excess of 2, indicating that the $\chi^2$ distribution even for the 
bright spectra was skewed to higher values. The largest values of
$\chi^2_\nu$ were obtained in bright bursts with extreme radius expansion,
like those from 4U~1724$-$307 (see \S\ref{s1724}) and 4U~2129+12
(\S\ref{s2129}; see also \S\ref{pflux}).

Fixing the $N_{\rm H}$ at the mean $\left<N_{{\rm H},i}\right>$ derived
over each burst may introduce additional errors into the burst flux and
fluence, if
this value is substantially different from the true
column towards the source at the time of the burst.  Due to its modest
low-energy response, the PCA can generally accurately determine the
$N_{\rm H}$ only when it is $\ga10^{22}\ {\rm cm^{-2}}$.
Alternative approaches, such as fixing the absorption at the measured
Galactic line-of-sight column density, gave poorer fits overall and
hence less reliable fit parameters. 
Furthermore, the local contribution to the line-of-sight $N_{\rm H}$ can
vary with time in LMXBs due to changes in the local distribution of
matter, and in the absence of contemporaneous measurements by instruments
with better low-energy response we must rely on the values measured by
the PCA.
Errors in the fluence from incorrect $N_{\rm H}$ values for individual
time-resolved spectra are likely to average out in the sum, so that the
remaining parameter most likely to be affected by this source of error is
the peak flux \fpk.  The magnitude of the introduced error is
$\approx10^{-9}\ \epcs$ per $10^{22}\ {\rm cm^{-2}}$; that is, for every
$10^{22}\ {\rm cm^{-2}}$ we overestimate the column for the spectral fits,
we calculate an unabsorbed flux $\approx10^{-9}\ \epcs$ larger.
We can estimate the magnitude of the error by comparing the peak flux
determined from the spectral fits with $N_{\rm H}$ fixed, to those where
it is free to vary. We find that these values are consistent for 90\% of
the bursts, and for the remainder the peak flux with $N_{\rm H}$ fixed is
consistently less than the peak flux with $N_{\rm H}$ free to vary. This
is because the fitted $N_{\rm H}$ value for a few low signal-to-noise
spectra in some (typically very long) bursts are much higher than the
mean, resulting in
an erroneously large peak (unabsorbed) flux.
Thus, through our approach we 
avoid these erroneously high peak fluxes by
re-fitting with the $N_{\rm H}$ frozen at the mean, and the additional
error introduced to the \fpk\ are likely comparable to our estimated
uncertainty on those values.

We defined the burst start time \bstart\ to be the time when the burst flux
first exceeded 25\% of the peak flux \fpk\ (see e.g. Fig.
\ref{fig:freqev}).
The rise time \trise\ is the interval from \bstart\ to when the burst flux
exceeds 90\% of \fpk.
These definitions were chosen for ease of implementation and insensitivity
to Poisson or systematic variations in the burst rises during, for
example, strong radius expansion bursts. 
In order to describe the entire lightcurve quantitatively, we also fitted
an exponential curve with decay constant $\tau_1$ to the bolometric flux
from where the flux first dropped below 90\% of \fpk\ through the decay.
For many bursts, the evolution was not consistent with a single
exponential decay, and so we fitted a subsequent exponential curve, with
an independent decay constant $\tau_2$ until the end of the 
interval over
which we extracted burst spectra (128~s by default\footnote{We extracted
data over longer windows for sources with typically long bursts, e.g.
GS~1826$-$24 and GX~17+2.}).
Note that the decay curves for most bursts were not statistically
consistent with this ``broken'' exponential model, mainly due to
variations on smaller time-scales (for this reason we do not
quote uncertainties for the decay constants $\tau_1$ and $\tau_2$).
However, we chose the time ranges to fit the exponential segments so as to
qualitatively described the burst decay with as few parameters as
possible, even if the actual fit was poor.

We measured the fluence \fluen\ by summing the measured fluxes over the
burst, and integrating the final exponential curve to account for any
additional flux beyond the data window. We found significant emission
compared to the pre-burst level at the end of the window particularly for
long bursts.
We extrapolated the second (or first, in cases where only one
exponential was used to fit the lightcurve) decay curve beyond the end of the
data window, and integrated to estimate the
burst flux missed by truncating the high-resolution data.  Where this
extrapolated contribution to the fluence was greater than the propagated
error, we adopted it as the uncertainty instead of the propagated error.
We also calculated for each
burst a simpler, less model-dependent time scale $\tau=E_b/F_{\rm pk}$
traditionally used to characterise burst evolution \cite[e.g.][]{vppl88}
for which we estimated the uncertainties by propagating the errors on $E_b$
and \fpk.

From the observed (bolometric) integrated burst flux $E_b$, we estimate
the column depth $y$ at which the burst is ignited as
\begin{eqnarray}
y & = & \frac{L_b d^2 (1+z)}{R_{\rm NS}^2 Q_{\rm nuc}}\ 
                                                            \nonumber \\
  & = & 3.0\times10^8 \left(\frac{E_{\rm b}}{10^{-6}\ \epc}\right)
                      \left(\frac{d}{10\ {\rm kpc}}\right)^2\
                      \left(\frac{R_{\rm NS}}{10\ {\rm km}}\right)^{-2}
\nonumber \\ & & \times\  
             \left(\frac{Q_{\rm nuc}}{4.4\ {\rm MeV/nucleon}}\right)^{-1}
                      \left(\frac{1+z}{1.31}\right)
                      {\rm g\,cm^{-2}}
\label{column}
\end{eqnarray}
where $L_b=4\pi d^2E_b$ is the total burst luminosity,
and $Q_{\rm nuc}$ the energy
released, $4.4$~MeV/nucleon for material with solar
abundances.
For sub-solar hydrogen fraction $X$, 
$Q_{\rm nuc}=1.6+4\left<X\right>$ where (strictly speaking) $X$ is
averaged over the burning layer; 
this expression assumes $\approx 35$\% energy loss due to neutrinos during
the rp process (e.g. \citealt{fuji87}; see also
\citealt{schatz99,schatz01}).
As with the persistent emission, we assume that the burst emission is
isotropic.

For bursts where the recurrence time could be measured
unambiguously, we calculated the ratio of the integrated persistent flux
to the burst fluence:
\begin{equation}
\alpha = \frac{F_p c_{\rm bol} \Delta t}{E_b}
 \label{alpha}
\end{equation}
We propagated the errors on each of the observeable parameters (excluding
$\Delta t$, for which the fractional measurement errors were negligible)
to calculate the error on $\alpha$.
By substituting expressions \ref{column} and \ref{mdot} into
the simple equality $y=\dot{m}\Delta t$ (assuming implicitly that all the
accreted fuel is burnt during the burst) we
obtain the expected value of $\alpha$, which depends upon the
compactness of the star and the burst fuel composition:
\begin{eqnarray}
\alpha & = & \frac{Q_{\rm grav}}{Q_{\rm nuc}}\,(1+z) \nonumber \\
       & = & 44\ \left(\frac{M}{1.4M_\odot}\right)
                 \left(\frac{R}{10\ {\rm km}}\right)^{-1}
             \left(\frac{Q_{\rm nuc}}{4.4\ {\rm MeV/nucleon}}\right)^{-1}
 \label{alphatheory}
\end{eqnarray}
We note that if the burst fuel is not completely consumed,
the observed fluence $E_b$ will be lower than expected (given
the available fuel), and the measured $\alpha$ (equation \ref{alpha}) will
thus be in excess of the expected value.

\subsection{Photospheric radius-expansion}
\label{pre}

The time-resolved spectral analyses result, for each burst with high
temporal and spectral data coverage, in time-series of blackbody
temperature $kT_{\rm bb}$ (units of keV) and normalization $K_{\rm bb}$
(units of $({\rm km}/10\,{\rm kpc})^2$) throughout the burst.
We examined the spectral variation throughout each of the bursts in the
catalog and classified them according to the following criteria.
We considered that radius expansion occurred
when 1) the blackbody normalization $K_{\rm bb}$ reached a (local) maximum
close to the time of peak flux; 2) lower values of $K_{\rm bb}$ were
measured following the maximum, with the decrease significant to $4\sigma$
or more; and 3) there was evidence of a (local) minimum in the fitted
temperature $T_{\rm bb}$ at the same time as the maximum in $K_{\rm bb}$.
Bursts where just one or two of these criteria were satisfied we refer to
as ``marginal'' cases, in which the presence of PRE could not be
conclusively established\footnote{We note, however, that in the
case of 4U~1728$-$34, the marginal cases have an identical flux
distribution as the confirmed radius expansion bursts \cite[]{gal03b}; so 
excluding these bursts from the sample of PRE bursts may be overly
conservative.}.

For spherically symmetric emission, the Eddington luminosity measured
by an observer at infinity is given by \cite[]{lew93}
\begin{eqnarray}
  L_{\rm Edd,\infty} & = & \frac{8\pi G m_p M_{\rm NS} c
  [1+(\alpha_{\rm T}T_{\rm e})^{0.86}]} {\sigma_T(1+X)[1+z(R)]} 
       \nonumber \\
  & = & 2.7\times10^{38} \left(\frac{M_{\rm NS}}{1.4M_\odot}\right)
 \frac{1+(\alpha_{\rm T}T_{\rm e})^{0.86}}{(1+X)}
\nonumber \\ & & \times\  
    \left[\frac{1+z(R)}{1.31}\right]^{-1}\ 
              \eps
  \label{ledd}
\end{eqnarray}
where
$T_{\rm e}$ is the
effective temperature of the atmosphere, $\alpha_{\rm T}$ is a coefficient
parametrizing the temperature dependence of the electron scattering
opacity \cite[$\simeq 2.2\times10^{-9}$~K$^{-1}$;][]{lew93},
$m_p$ is the mass of the proton, $\sigma_T$ the Thompson scattering
cross-section, and $X$ is the
mass fraction of hydrogen in the atmosphere ($\approx0.7$ for cosmic
abundances).
The final factor in
square brackets represents the gravitational redshift 
at the photosphere $1+z(R)=(1-2GM_{\rm NS}/R
c^2)^{-1/2}$, which may be elevated significantly above the NS
surface (i.e. $R\ge R_{\rm NS}$). 
\cite{kuul03a} analysed all bursts detected from 
the 12 bursters in globular clusters, for which independent
distance estimates are available, in order to rigorously test whether the
radius-expansion bursts reached a ``standard candle'' luminosity. They
found that for about two-thirds of the sources the radius-expansion bursts
reached $3.79\pm0.15\times10^{38}\ \eps$. This value is consistent with
equation \ref{disteq} only for H-poor material and where the radius expansion
drives the photosphere to very large radii; we note however that the spectral
evidence generally does not support the latter condition
\cite[e.g.][]{seh84}. 

Given the observed peak flux of a
PRE burst \fpkre, we estimated the distance as
\begin{eqnarray}
 d & = & \left(\frac{L_{\rm Edd,\infty}}{4\pi F_{\rm pk, RE}}\right)^{1/2} \nonumber \\
   & = & 8.6
	\left( \frac{F_{\rm pk, RE}}{3\times10^{-8}\ \epcs} \right)^{-1/2}
       	\left(\frac{M_{\rm NS}}{1.4M_\odot}\right)^{1/2}
\nonumber \\ & & \times\  
	\left[\frac{1+z(R)}{1.31}\right]^{-1/2}
	(1+X)^{-1/2}\ {\rm kpc}
 \label{disteq}
\end{eqnarray}
We discuss the properties of the radius-expansion bursts in \S\ref{pflux}.

\subsection{Burst oscillations}
\label{ibo}
\label{bosc}

We searched for burst oscillations in data recorded with $2^{-13}$~s
(122~$\mu$s) time
resolution.  We computed fast Fourier transforms of each 1~s interval of
data for the first 16~s of the burst, and searched for signals in bursts
from sources with previously-detected burst oscillations within 5~Hz
of the known oscillation frequencies.
We considered a signal to be a detection if it had less than a 1\% chance
of occurring due to noise given the 160 trial frequencies searched for
each burst. A signal was considered significant if it passed any of three
tests: (1) having a probability of $< 6\times10^{-5}$ that it
was produced by noise in a single trial, (2) persisting for two adjacent
(independent)
time and frequency bins with a chance probability of $<
(6\times10^{-5})^{1/2}/6 = 1.3\times10^{-3}$, or (3) occurring in the
first second of a burst with a chance probability of $<10^{-3}$.

If oscillations were detected during a burst, we then determined whether
they were observed during the rise, peak or decay of the burst.  We
defined these characteristic times during the burst using data with 0.25~s
time resolution. The rise of the burst was defined to start 0.25~s before
the first time bin in which the count rate rose above 25~\% of the peak
count rate, and ended in the last time bin for which the count rate was
less than 90\% of the peak count rate. The peak of the burst was defined
to last from the end of the rise until the count rate dropped back below
90\% of the peak count rate. The decay of the burst commenced at the end
of the peak, and lasted 
through to the end of the high-time resolution data (typically $\la
200$~s duration).

Figure~\ref{fig:freqev}\ illustrates how these times were defined for a
burst from 4U~1702--429. Oscillations were detected during the rise and
decay of this burst, but not during the peak.
Note that
the over-sampled dynamic power spectrum displayed in this figure was not
used to search for oscillations; only non-overlapping power spectra were
used for the oscillation search.
Where oscillations were not detected in any of the 1-s intervals, we also 
computed FFTs of 4-s intervals covering the burst.
We used a probability threshold of $10^{-3}$ to determine when the
oscillations occurred, corresponding to a 1\% chance of observing a
spurious signal during each 1~s interval (10 trial frequencies).

We computed the amplitudes of the oscillations according to
\begin{equation}
A =
\left({{P}\over{I_\gamma}} 
\right)^{1/2}{{I_\gamma}\over{I_\gamma-B_\gamma}},
\label{eq:acont}
\end{equation}
where $P$ is the power from the Fourier spectrum, $I_{\gamma}$ is the
total number of counts in the profile, and $B_\gamma$ is the estimated
number of background counts. We estimated the background using the mean
count rate 16 seconds prior to the start of the burst.  Since the
detection threshold is a fixed power, the minimum detectable amplitude
depends on the number of counts produced by a burst.
We discuss the global properties of the bursts from burst oscillation
sources in \S\ref{milosc}.

\subsection{Combined burst samples}
\label{energetics}

We combined samples of bursts from different sources 
in order to analyse larger samples than would be possible for a single
source. Throughout this paper we refer to these combined samples as
``${\mathcal S}${\it label}\/'' where {\it label}\/ corresponds to the
selection criteria. The details of each subsample discussed in the text
are listed in Table \ref{samples}.

We assembled two principal samples
using the two 
measures of the accretion rate $\dot{M}$ (see \S\ref{pers}). 
First, we calculated the mean peak flux $\left<F_{\rm pk,PRE}\right>$ of
the radius-expansion bursts from each source that exhibited at least one.
We identify this value as the flux corresponding to the Eddington
luminosity for that source, $F_{\rm Edd}$. For each observation we then
calculated the dimensionless persistent flux $\gamma=F_p/F_{\rm
Edd}$,
rescaling the measured persistent flux (see \S\ref{pers}) by
the $F_{\rm Edd}$ for that source \cite[this approach
follows][]{vppl88}\footnote{For reference, $\gamma=0.1$ corresponds to
between 1.6--$3.5\times10^{37}\ \eps$ (depending upon the maximum radius
$R$ and atmospheric composition $X$ in Equation \ref{ledd}).}.
We excluded sources for which the PCA field of view contains other
active sources, since we cannot reliably estimate the persistent flux in
those cases (GRS~1741.9$-$2853, 2E~1742.9$-$2929 and SAX~J1747.0$-$2853).
It is possible that the Eddington luminosity 
varies from source to source, e.g. due to variations in the composition of
the photosphere, and this may introduce a bias when combining data
based on $\gamma$ of up to a factor of 1.7.
We also neglect the 
precise bolometric correction $c_{\rm bol}$ for each observation, which
introduces an error up to a factor of two 
(see \S\ref{pers}).
Ideally, $\gamma$ is 
approximately equal to the accretion rate as a fraction of the
Eddington rate, i.e.  $\gamma\approx\dot{M}/\dot{M}_{\rm Edd}$. 
This approach has the advantage of being independent of assumptions about
exactly what is the value of the Eddington limit reached by the bursts.
The principal drawback 
is that it is only possible for those sources with at least one detected
PRE burst, although these sources represent the majority of the bursts
contributing to our sample. We refer to this sample as \sgamma\ (see Table
\ref{samples} for a summary).
We also calculated the normalized peak burst flux $U_p=F_{\rm pk}/F_{\rm Edd}$
and fluence $U_b=E_b/F_{\rm Edd}$ in order to measure the combined
distribution of those parameters in \S\ref{pflux}.

\begin{figure}
 \epsscale{1.2}
 \plotone{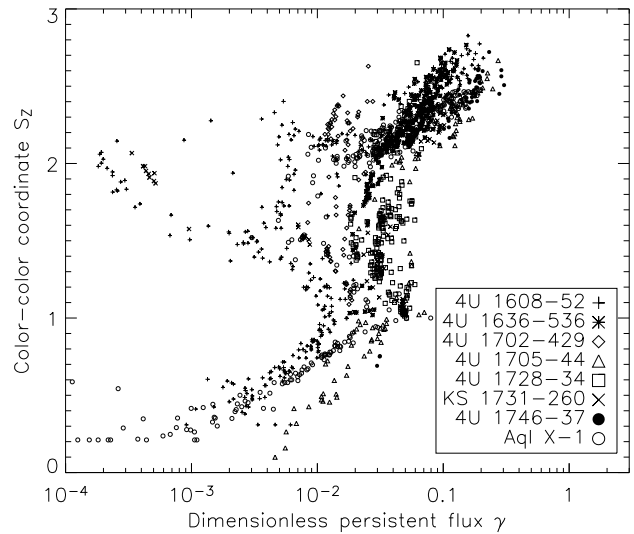}
 \figcaption[]{Comparison of the two parameters used in \S\ref{global} as
proxies for $\dot{M}$ to combine bursts from different sources.  The
position along the color-color track, $S_Z$ (averaged over each 
observation) is plotted against the normalised persistent flux
$\gamma\equiv F_{\rm p}/F_{\rm Edd}$ for eight of the nine sources with
well-defined color-color diagrams.  
XTE~J2123$-$058 is excluded since no
PRE bursts were detected, and thus $\gamma$ cannot be determined.
The two parameters are roughly proportional at high and low $S_Z$ for all
sources (in the upper and lower branches of the color-color diagrams in
Fig.  \ref{fig:sz}), but in the range $S_Z\simeq1$--2 $\gamma$ is
approximately constant or varies over a wide range, depending upon the
source.
  \label{comparison} }
\end{figure}

There is substantial evidence that \fper\ may not strictly track
$\dot{M}$ \cite[e.g.][]{hvdk89}, so that $\gamma$ may not be the best
available measure of $\dot{M}$.  Thus, we 
assembled a second sample of bursts
from only those sources with a well-defined color-color diagram, in this
case
adopting the position along the color-color diagram $S_Z$ as a proxy for
$\dot{M}$ (see \S\ref{pers}; we refer to this sample as \ssz, Table
\ref{samples}).
We show in Fig. \ref{comparison} the comparison between the
observation-averaged $S_Z$ and $\gamma$ for eight of the nine sources with
both well-defined color-color diagrams.
For most of the sources, $S_Z$ was proportional to $\gamma$ when $S_Z<1$
(upper or ``island'' horizontal branch in Fig. \ref{fig:sz}) and $S_Z>2$
(lower or ``banana'' branch). However, between these two branches 
$\gamma$ is essentially constant at between 0.006 and 0.06, indicating
that the transition between the two branches takes place at a roughly
constant flux. For two sources, 4U~1608$-$52 and KS~1731$-$26, the
relationship is more complex, and some observations with very hard spectra
($S_Z\approx2$) actually have very low $\gamma$. 

\begin{figure*}[t]
 \plotone{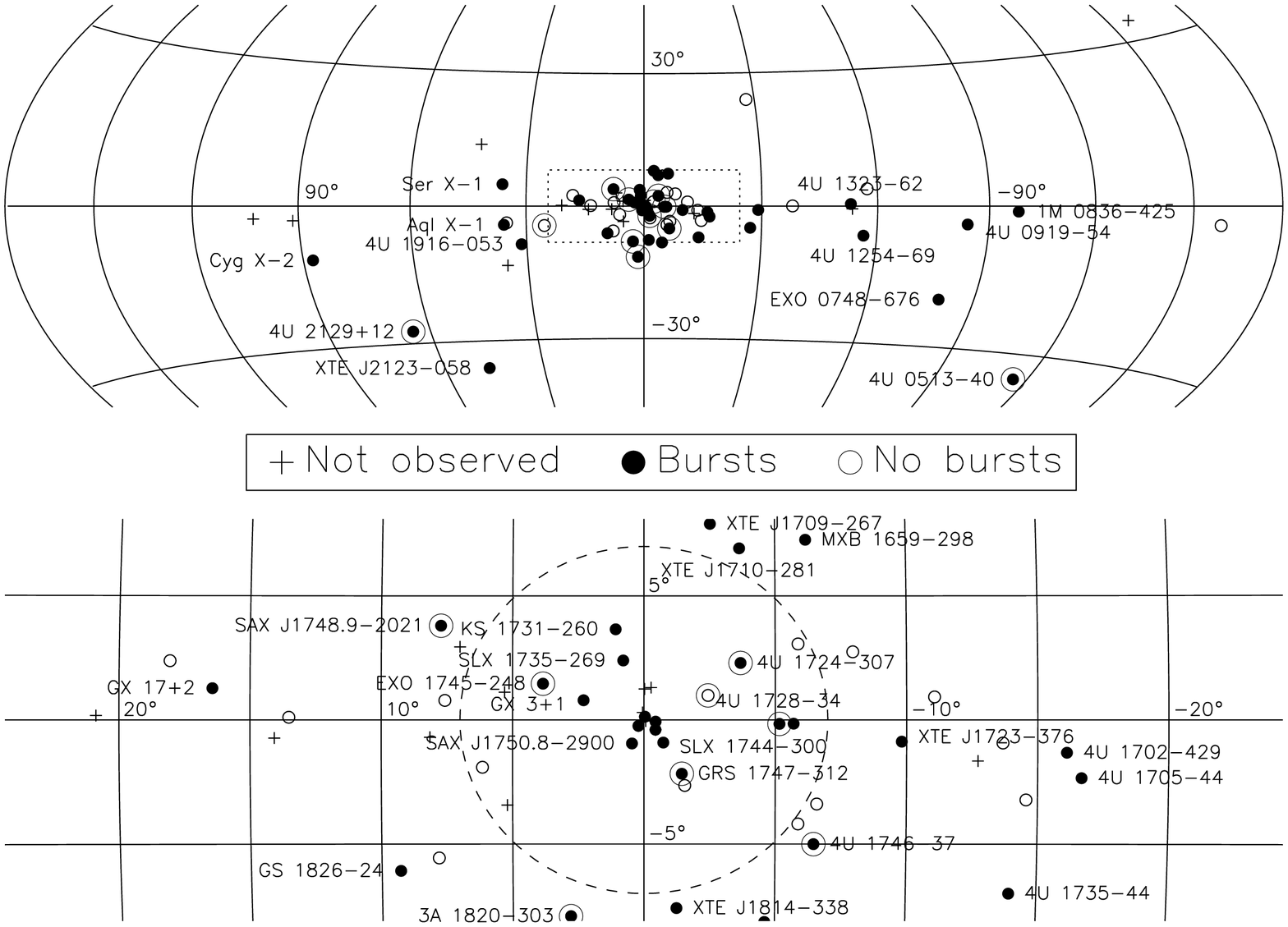}
 \figcaption[f1.eps]{Sky distribution of bursters showing those
observed by \xte, as well as those from which bursts were detected.
Sources within globular clusters are additionally indicated by a larger
concentric circle.  The lower panel shows the region around the Galactic
center. The four (unlabeled) sources closest to the Galactic origin are
(clockwise from lower left) are
SAX~J1747.0$-$2853,
GRS~1741.9$-$2853,
KS~1741$-$293
and
2E~1742.9$-$2929.
The dashed line shows the approximate projected radius of the
Galactic bulge.
  \label{skydist} }
\end{figure*}

We measured mean burst rates for the combined burst samples \sgamma\ and
\ssz\ as a function of accretion rate (by proxy).
In order to guarantee reasonable statistics, we
defined bins
such that some minimum number of bursts fell into each bin. We then
calculated the mean burst rate within each bin as the number of bursts
divided by the total duration of observations which also fell into that
bin. 
We also measured recurrence times \trec\ from a third sample \sdt\
comprising successive pairs of bursts with recurrence times $\la 10$~hr.
In general 
the measured separation of a pair of bursts is an upper limit
on the recurrence time, due to the possibility of missing intervening
events in data gaps arising from Earth occultations. Additional data gaps arise
from satellite passages through the South Atlantic Anomaly, during
which the PCA is turned off to protect the electronics from damage.  
We can estimate the recurrence time with more confidence when the burst
is observed within a high duty-cycle observing interval,
or where the source is bursting regularly\footnote{We note that this
sample is not complete, but is limited to sources of particular interest;
see Table \ref{samples}.}.  For each pair of bursts
matching this criteria we calculated the instantaneous burst rate as
$1/t_{\rm rec}$.
We note that \sdt\ also includes bursts with very short separations
$<30$~min, for which the occurrence is episodic and thus not
indicative 
of a steady recurrence time (see \S\ref{dblbursts}).

We also calculated mean $\alpha$-values (which we designate \malpha) for
the combined samples \sgamma\ and \ssz.
%
For the bursts arising from observations
which fall in a bin $j$ (defined as a range in $\gamma$ or $S_Z$, 
for samples \sgamma\ and \ssz, respectively) we calculated
\begin{equation}
\left<\alpha\right>_j=\frac{\sum \gamma_i t_i}{\sum U_b}
\end{equation} 
where $t_i$ is the duration of each observation.
We note that \malpha\ calculated in this manner is an underestimate of
the true value, since $\gamma$ is calculated from the 2.5--25~keV flux,
which broad-band spectral fits indicate contributes typically 60--90\% of the
bolometric persistent flux (depending upon the source and the spectral
state).
We measured the $\alpha$
values directly for pairs of bursts from sample \sdt.
Here we determined the bolometric correction appropriate for the
observations(s) in which the bursts occurred (see
\S\ref{pers}) and calculated $\alpha$
according to equation \ref{alpha}.
We discuss the variations in burst properties, including \malpha, as a
function of $\dot{m}$ in \S\ref{global}.

\begin{deluxetable*}{lccccccccc}
\tabletypesize{\scriptsize}
\tablecaption{LMXBs with type-I X-ray bursts detected by \xte\/ up to \burstpubdate
  \label{bursters}
}
\tablewidth{0pt}
\tablehead{
  \colhead{}
 & 
 & 
 & 
 & \colhead{$P_{\rm orb}$}
 & 
 & \colhead{Time}
 & 
 & \colhead{Mean burst}
 & 
\\
  \colhead{Source}
 & \colhead{Type\tablenotemark{a}}
 & \colhead{RA}
 & \colhead{Dec}
 & \colhead{(hr)}
 & \colhead{$n_{\rm obs}$}
 & \colhead{(ks)}
 & \colhead{$n_{\rm burst}$}
 & \colhead{rate (hr$^{-1}$)}
 & \colhead{$n_{\rm RE}$}
}
\startdata
4U~0513$-$40 & G & 05 14 06.6 & -40 02 37 & \nodata & 555 & 1200 & 7 & 0.021 & 1 \\
EXO~0748$-$676 & TD & 07 48 33.8 & -67 45 08 & 3.82 & 489 & 1390 & 94 & 0.24 & 3 \\
1M~0836$-$425 & T & 08 37 22.7 & -42 53 08 & \nodata & 29 & 143 & 17 & 0.43 & \nodata \\
4U~0919$-$54 & \nodata & 09 20 26.8 & -55 12 24 & \nodata & 79 & 387 & 4 & 0.037 & 1 \\
4U~1254$-$69 & DS & 12 57 37.2 & -69 17 20 & 3.93 & 26 & 279 & 5 & 0.064 & \nodata \\
4U~1323$-$62 & D & 13 26 36.1 & -62 08 10 & 2.93 & 15 & 264 & 30 & 0.41 & \nodata \\
4U~1608$-$52 & TAOS & 16 12 42.9 & -52 25 22 & \nodata & 546 & 1650 & 31 & 0.068 & 12 \\
4U~1636$-$536 & AOS & 16 40 55.5 & -53 45 05 & 3.8 & 454 & 2790 & 172 & 0.22 & 46 \\
MXB~1659$-$298 & TDO & 17 02 06.3 & -29 56 45 & 7.11 & 80 & 356 & 26 & 0.26 & 12 \\
4U~1702$-$429 & AO & 17 06 15.2 & -43 02 09 & \nodata & 196 & 1300 & 47 & 0.13 & 5 \\
4U~1705$-$44 & A & 17 08 54.6 & -44 06 02 & \nodata & 140 & 605 & 47 & 0.28 & 3 \\
XTE~J1709$-$267 & T & 17 09 30.2 & -26 39 27 & \nodata & 39 & 145 & 3 & 0.075 & \nodata \\
XTE~J1710$-$281 & DT & 17 10 12.4 & -28 07 54 & \nodata & 63 & 231 & 18 & 0.28 & 1 \\
XTE~J1723$-$376 & T & 17 23 38.0 & -37 39 42 & \nodata & 4 & 34.1 & 3 & 0.32 & \nodata \\
4U~1724$-$307 & GA & 17 27 33.2 & -30 48 07 & \nodata & 93 & 512 & 3 & 0.021 & 2 \\
4U~1728$-$34 & AO & 17 31 57.3 & -33 50 04 & \nodata & 346\tablenotemark{b} & 1940\tablenotemark{b} & 106 & 0.20 & 69 \\
Rapid~Burster & GT & 17 33 24.0 & -33 23 16 & \nodata & 344\tablenotemark{b} & 1900\tablenotemark{b} & 66 & 0.13 & \nodata \\
KS~1731$-$260 & TOS & 17 34 13.0 & -26 05 09 & \nodata & 78 & 483 & 27 & 0.20 & 3 \\
SLX~1735$-$269 & T & 17 38 16.0 & -27 00 18 & \nodata & 53\tablenotemark{b} & 287\tablenotemark{b} & 1 & 0.013 & \nodata \\
4U~1735$-$44 & AS & 17 38 58.2 & -44 26 59 & 4.65 & 82 & 454 & 11 & 0.087 & 6 \\
XTE~J1739$-$285 & T & 17 39 54.0 & -28 29 00 & \nodata & 27\tablenotemark{c} & 135\tablenotemark{c} & 6 & 0.16 & \nodata \\
KS~1741$-$293 & T & 17 44 49.2 & -29 21 06 & \nodata & 452\tablenotemark{c} & 2090\tablenotemark{c} & 1 & 0.0017 & \nodata \\
GRS~1741.9$-$2853 & O & 17 45 00.6 & -28 54 06 & \nodata & 440\tablenotemark{c} & 2070\tablenotemark{c} & 8 & 0.014 & 6 \\
2E~1742.9$-$2929 & T? & 17 46 06.2 & -29 31 05 & \nodata & 440\tablenotemark{c} & 2070\tablenotemark{c} & 84 & 0.15 & 2 \\
SAX~J1747.0$-$2853 & T & 17 47 02.6 & -28 52 59 & \nodata & 433\tablenotemark{c} & 2090\tablenotemark{c} & 23 & 0.040 & 10 \\
IGR~17473$-$2721 & \nodata & 17 47 18.1 & -27 20 39 & \nodata & 114\tablenotemark{b} & 560\tablenotemark{b} & 2 & 0.013 & \nodata \\
SLX~1744$-$300 & \nodata & 17 47 25.9 & -30 02 30 & \nodata & 39\tablenotemark{c} & 290\tablenotemark{c} & 3 & 0.037 & \nodata \\
GX~3+1 & A & 17 47 56.0 & -26 33 48 & \nodata & 106\tablenotemark{b} & 539\tablenotemark{b} & 2 & 0.013 & 1 \\
1A~1744$-$361 & TD? & 17 48 14.0 & -36 07 30 & \nodata & 15 & 40.6 & 1 & 0.089 & \nodata \\
SAX~J1748.9$-$2021 & TG & 17 48 53.4 & -20 21 43 & \nodata & 27 & 149 & 16 & 0.39 & 6 \\
EXO~1745$-$248 & TG & 17 48 55.7 & -24 53 40 & \nodata & 51 & 148 & 22 & 0.54 & 2 \\
4U~1746$-$37 & GA & 17 50 12.6 & -37 03 08 & 5.7 & 51 & 445 & 30 & 0.24 & 3 \\
SAX~J1750.8$-$2900 & TO & 17 50 24.0 & -29 02 18 & \nodata & 42\tablenotemark{c} & 161\tablenotemark{c} & 4 & 0.090 & 2 \\
GRS~1747$-$312 & GDT & 17 50 45.5 & -31 17 32 & 12.36 & 114\tablenotemark{b} & 670\tablenotemark{b} & 7 & 0.038 & 3 \\
XTE~J1759$-$220 & D? & 17 59 42.0 & -22 01 00 & \nodata & 17 & 107 & 1 & 0.034 & \nodata \\
SAX~J1808.4$-$3658 & PT & 18 08 27.5 & -36 58 44 & 2.01 & 316 & 1390 & 6 & 0.016 & 5 \\
XTE~J1814$-$338 & TP & 18 13 40.0 & -33 46 00 & 4.27 & 90 & 447 & 28 & 0.23 & \nodata \\
GX~17+2 & ZS & 18 16 01.3 & -14 02 11 & \nodata & 147 & 917 & 12 & 0.047 & 2 \\
3A~1820$-$303 & GAS & 18 23 40.5 & -30 21 40 & 0.19 & 186 & 1230 & 5 & 0.015 & 5 \\
GS~1826$-$24 & T & 18 29 27.0 & -23 47 29 & \nodata & 127 & 929 & 65 & 0.25 & \nodata \\
XB~1832$-$330 & GT & 18 35 44.1 & -32 59 29 & \nodata & 13 & 102 & 1 & 0.035 & 1 \\
Ser~X-1 & S & 18 39 57.5 & +05 02 08 & \nodata & 41 & 251 & 7 & 0.10 & 2 \\
HETE~J1900.1$-$2455 & TP & 19 00 08.6 & -24 55 14 & 1.39 & 194 & 664 & 2 & 0.011 & 3 \\
Aql~X-1 & TAO & 19 11 15.9 & +00 35 06 & 19.0 & 411\tablenotemark{b} & 1650\tablenotemark{b} & 57 & 0.12 & 9 \\
4U~1916$-$053 & D & 19 18 47.9 & -05 14 08 & 0.83 & 51 & 412 & 14 & 0.12 & 12 \\
XTE~J2123$-$058 & TA & 21 23 16.1 & -05 47 30 & 5.96 & 5 & 67.2 & 6 & 0.32 & \nodata \\
4U~2129+12 & G & 21 29 58.3 & +12 10 02 & 17.1 & 32 & 343 & 1 & 0.010 & 1 \\
Cyg~X-2 & Z & 21 44 41.2 & +38 19 18 & 236.2 & 379 & 1910 & 55 & 0.10 & 8 \\
\tableline
 Total (48 sources) & & & & & & 1187 & &  247 \\
 \enddata
%
\tablenotetext{a}{Source type, adapted from \cite{lmxb01}; A = 
atoll source, D = ``dipper'', G = globular cluster
association, O = burst oscillation, P = pulsar, 
S = superburst, T = transient, Z = Z-source. We omit the ``B''
designation indicating a burst source.}
\tablenotetext{b}{For sources with a neighbor within $1\arcdeg$, 
we combine all observations which include this source within the
field of view (possibly including observations of the neighbor).}
\tablenotetext{c}{Similarly, for sources towards the Galactic 
center, we combine all observations which include the source in
the field of view to calculate the mean burst rate.}
\end{deluxetable*}


\section{Thermonuclear X-ray bursts observed by RXTE}
\label{results}

We detected a total of \burstnum\ thermonuclear bursts from \numbursters\
sources in public \xte\/ observations up to \burstpubdate.
We summarise the numbers of bursts from individual sources in Table
\ref{bursters}.  The sky distribution of burst sources is shown in Fig.
\ref{skydist}.
Our burst search included all known thermonuclear burst sources as at
\newburstsourcedate. Recently discovered examples include the faint {\it
INTEGRAL}\/ source IGR~J17364$-$2711 \cite[]{chelov06}; the 
transients IGR~J17464$-$2811/XMMU~J174716.1$-$281048 \cite[]{delsanto07} and
IGR~J17473$-$2721/XTE~J1747$-$274 \cite[see \S\ref{xmmu1747};][]{greb05b}
and the 294~Hz burst oscillation source
IGR~J17191$-$2821 \cite[]{kw07,mark07a}.
Eighteen	
sources previously found to exhibit thermonuclear bursts were
observed by \xte\/ but with no detected bursts (Table \ref{nobursts}).

For each source with bursts observed by \xte\/ we list the relevant
analysis results in Table \refbursts.
The superscript to the burst number indicates a range of potential
analysis issues, as follows:
\begin{enumerate}
 \renewcommand{\labelenumi}{\alph{enumi}}
\item The burst was observed during a slew, and thus offset from the
source position. 
\item The observation was offset from the source position.
In cases (a) and (b) we scaled the flux and fluence by
the mean collimator response 
appropriate for the position of the source in the field of view, as
described in appendix \ref{localize}. 
\item The origin of the burst is uncertain; the burst may have been from
another source in the field-of-view (we rescaled the flux and
fluence, if necessary, based upon the assumed origin); 
\item Buffer overruns (or some other instrumental effect) caused gaps in
the high time-resolution data; 
\item The burst was so faint that only the peak flux could be measured,
and not the fluence or other parameters; 
\item An extremely faint burst or possibly problems with the background
subtraction, resulting in no fit results; 
\item the full burst profile was not observed, so that the event can be
considered an unconfirmed burst candidate.
Typically in these cases the initial burst rise is missed, so that the
measured peak flux and fluence are lower limits only; 
\item High-time resolution datamodes did not cover the burst. 
\end{enumerate}
Column (2) lists the ID for the observation during which the burst was
observed; (3) the burst start time in UT and MJD (we neglect corrections
to give the time at the solar-system barycenter);
(4) the peak flux \fpk, in
units of $10^{-9}\ \epcs$; (5) the burst fluence \fluen, in units of $10^{-6}\ \epc$;
(6) the presence of radius expansion; (7) rise time (s); (8) peak count
rate, per PCU;
(9,10) the exponential decay constants $\tau_1$, $\tau_2$ describing the
decay of the burst, where available;
(11) burst time scale ($E_{\rm b}/F_{\rm pk}$); and
(12) burst fluence normalized by the mean peak
flux of the PRE bursts 
$F_{\rm Edd}$, where available
\cite[$U_b$ in][]{vppl88}.
From analysis of the persistent spectrum, excluding the bursts (see
\S\ref{pers}) we list in column 
(13) the persistent flux level \fper\ prior to the burst (2.5--25~keV,
units of $10^{-9}\ \epcs$);
(14) the persistent flux normalised by 
$F_{\rm Edd}$ \cite[$\gamma$ in][]{vppl88};
(15,16) soft and hard color prior to the burst; and
(17) position on the color-color diagram $S_Z$, where available. 
From estimates of the recurrence times of regular and/or approximately
contemporaneous bursts from a subset of \alphasrc\ sources, we compiled a set of
\alphadat\ \trec\ values (this we refer to as sample \sdt; see Table
\ref{samples}) from which we derived
(18) the inferred burst recurrence time $\Delta t$; using this value and
(19) the correction $c_{\rm bol}$ (see also Table \ref{cboltable}) used to estimate the bolometric flux from
the measured 2.5--25~keV persistent flux, we derived 
(20) the corresponding $\alpha$-value, calculated according to equation
\ref{alpha}.
Finally, in column (21) we list references to previously published
analyses of the burst.
Note that we do not list values for columns 12--17 for those observations
where we expect that more than one source is active in the field (see
appendix \ref{localize}).
We assembled various subsets of the detected bursts, as described in
\S\ref{energetics}, for the subsequent
analyses; these samples are summarized in Table \ref{samples}.

\begin{figure*}[!h]
 \epsscale{1.15}
 \plotone{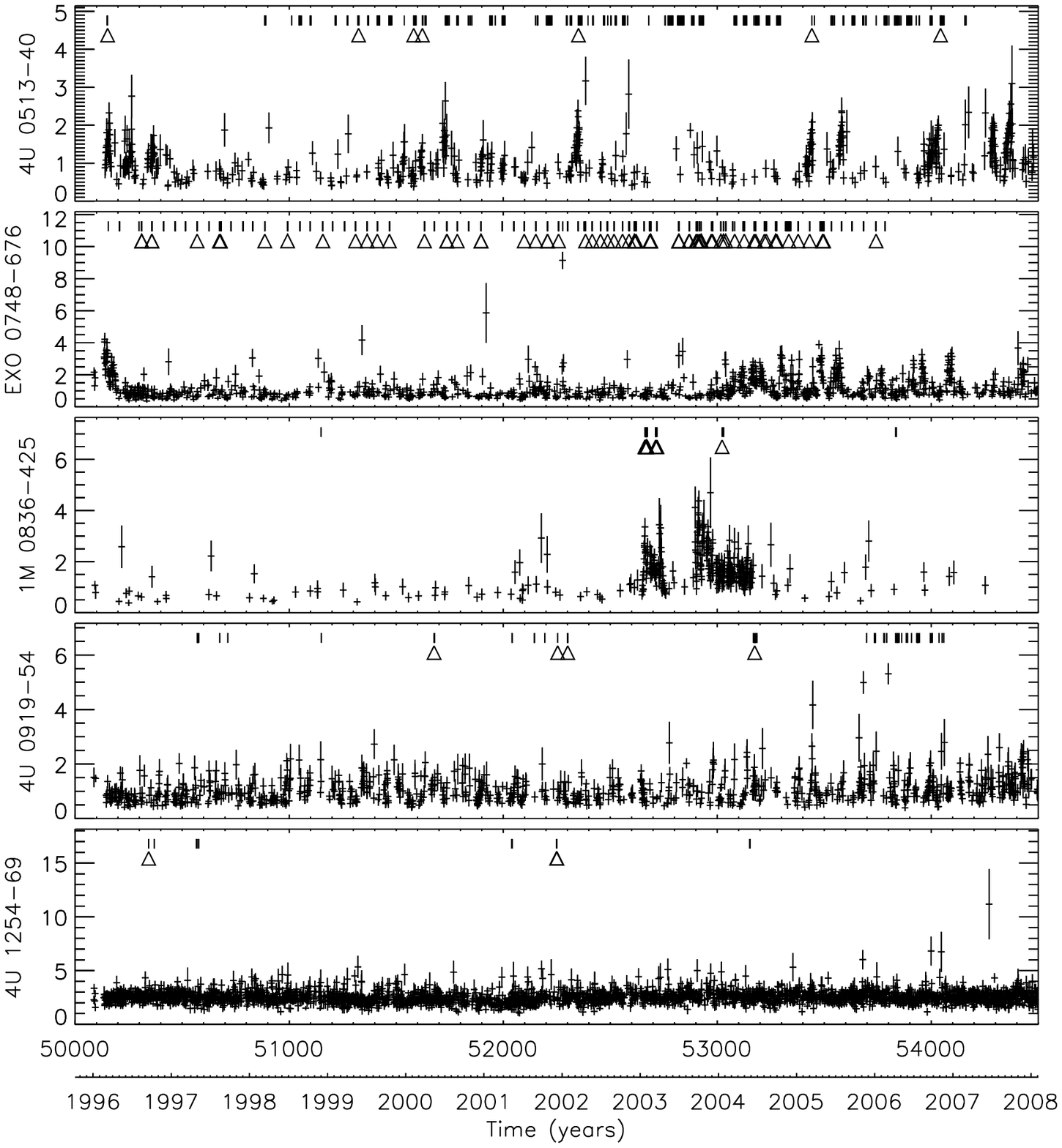}
 \figcaption[asmlc1.eps]{1-d averaged ASM light curves of selected sources
for which X-ray bursts have been observed by \xte\/ (see Table
\ref{bursters}). The times of the PCA observations are shown by the
vertical lines at the top of each plot, while the burst times are shown by
the open triangles. The error bars show the $1\sigma$ uncertainties on
each measurement; we exclude measurements where 
the significance of the detection is $<3\sigma$.
{\it All pages of this
figure can be found in the on-line version of the paper; this version
contains only the first.}
 \label{asmlc} }
\end{figure*}

\begin{figure*}[!hp]
  \epsscale{1.1}
  \plotone{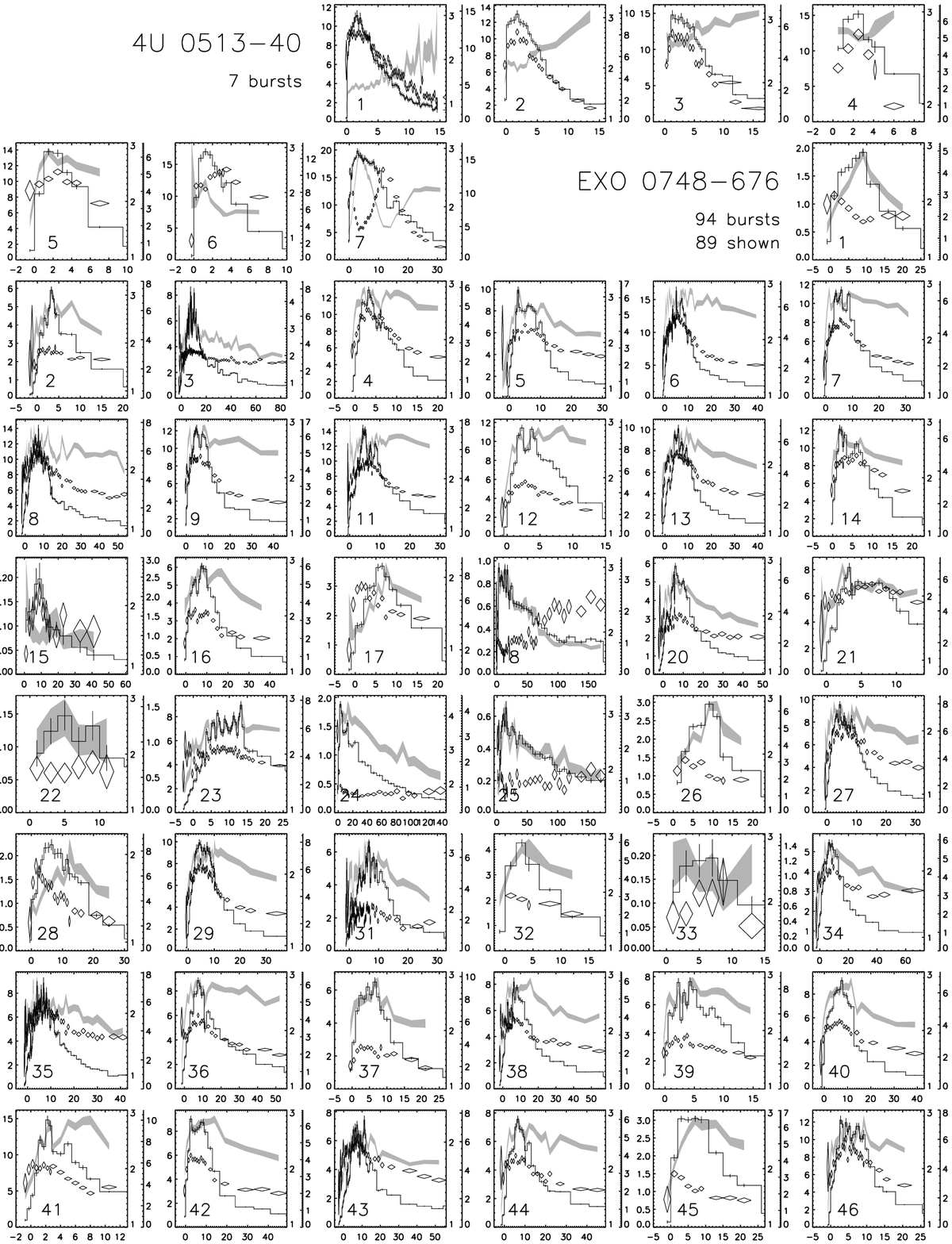}	
 \figcaption{Time-resolved spectral parameters for bursts listed in Table
\refbursts.
The histogram in each panel shows the (bolometric) burst flux (left-hand
$y$-axis) in units of $10^{-9}\ {\rm erg\,cm^{-2}\,s^{-1}}$, with error
bars indicating the $1\sigma$ uncertainties. The grey
ribbon shows the $1\sigma$ limits of the blackbody radius (outer
right-hand $y$-axis) in ${\rm km}/d_{\rm 10kpc}$. The diamonds show
the $1\sigma$ error region for the blackbody temperature (inner right-hand
$y$-axis) in keV. 
%
{\it All pages of this
figure can be found in the on-line version of the paper; this version
contains only the first.}
  \label{profiles} }
\end{figure*}

We plot the ASM intensity and burst activity of selected sources in Fig.
\ref{asmlc}. We also plot lightcurves and spectral evolution for
individual bursts for all sources
in Fig.  \ref{profiles}. Note that only the bursts with time-resolved
spectral information are plotted; this excludes bursts flagged with
superscripts $e$, $f$, $g$, $h$, or $i$
in Table \refbursts.

We searched \burstsearched\ bursts from sources with previously detected
burst oscillations, and detected
oscillations from \burstdet\ in this manner\footnote{
Our analyses led to the discovery of 267~Hz burst oscillations in
4U~1916$-$053 \cite[]{1916burst}, as well as 620~Hz oscillations in 
4U~1608$-$52 (Hartman et al. 2008, in preparation).}; fewer than 10 should be
spurious detections of noise signals according to our selection criteria.
We omitted \omitted\ bursts from our analysis,
principally because they lacked data with sufficiently high time resolution
to search for oscillations.
We also omitted bursts from EXO~0748$-$676 from our analysis, since the
oscillations in that source have only been detected in summed FFTs from
many bursts \cite[]{villarreal04}.
The bursts from the vicinity of the Galactic center exhibiting 589~Hz
oscillations were originally thought to originate from MXB~1743$-$29,
although we attribute them instead to GRS~1741.9$-$2853, a
newly-discovered
source at the time (\S\ref{gcbo}; \S\ref{s17419}).

\begin{deluxetable*}{lcccccccl}
\tabletypesize{\scriptsize}
\tablecaption{Type-I X-ray burst sources observed by \xte\/ with no detected bursts
  \label{nobursts}
}
\tablewidth{0pt}
\tablehead{
  \colhead{}
 & 
 & 
 & 
 & \colhead{$P_{\rm orb}$}
 & 
 & \colhead{Time}
 & \colhead{$\left<F_p\right>$}
 & 
\\
  \colhead{Source}
 & \colhead{Type\tablenotemark{a}}
 & \colhead{RA}
 & \colhead{Dec}
 & \colhead{(hr)}
 & \colhead{$n_{\rm obs}$\tablenotemark{b}}
 & \colhead{(ks)}
 & \colhead{($10^{-9}\,\epcs$)}
 & \colhead{Ref.}
}
\startdata
4U~0614+09 & AS & 06 17 07.3 & +09 08 13 & \nodata & 339(300) & 2161.3 & 0.60--2.7 & [1,2] \\
4U~1246$-$58 & \nodata & 12 49 36.0 & -59 07 18 & \nodata & 15(13) & 21.2 & 0.16--0.52 & [3] \\
Cen~X-4 & T & 14 58 22.0 & -31 40 07 & 15.1 & 4 & 10.4 & $0.0091\pm0.0014$ & [4,5] \\
Cir~X-1 & TAD & 15 20 40.8 & -57 10 00 & 398.4 & 599(467) & 2667.3 & 0.050--65 & [6] \\
1E~1603.6+2600 & D? & 16 05 45.8 & +25 51 45 & 1.85 & 12(6) & 30.1 & $0.0132\pm0.0016$ & [7] \\
4U~1705$-$32 & \nodata & 17 08 54.4 & -32 18 57 & \nodata & 1 & 3.6 & 0.11 & [8] \\
4U~1708$-$40 & \nodata & 17 12 23.0 & -40 50 36 & \nodata & 14 & 61.1 & $0.57\pm0.13$ & [9] \\
SAX~J1712.6$-$3739 & T & 17 12 34.0 & -37 38 36 & \nodata & 2 & 3.8 & $0.278\pm0.008$ & [10] \\
2S~1711$-$339 & T & 17 14 17.0 & -34 03 36 & \nodata & 10(9) & 42.1 & 0.019--0.84 & [11] \\
3A~1715$-$321 & \nodata & 17 18 47.3 & -32 10 40 & \nodata & 13(5) & 100.2 & 0.020--6.7 & [12] \\
GPS~1733$-$304 & G & 17 35 47.6 & -30 28 55 & \nodata & 7 & 56.6 & $0.079\pm0.017$ & [13] \\
IGR~J17364$-$2711~alt.~pos. & T? & 17 38 05.0 & -37 49 05 & \nodata & 1 & 3.8 & 0.086 & [14] \\
SAX~J1752.3$-$3138\tablenotemark{c} & \nodata & 17 52 24.0 & -31 37 42 & \nodata & 2 & 2.8 & \nodata & [15] \\
SAX~J1806.8$-$2435 & TA & 18 06 51.0 & -24 35 06 & \nodata & 19(15) & 62.9 & 0.043--14 & [16] \\
GX~13+1 & A & 18 14 31.0 & -17 09 25 & 577.6 & 65(37) & 560.8 & $8.6\pm1.1$ & [17] \\
4U~1812$-$12 & A & 18 15 12.0 & -12 04 59 & \nodata & 27 & 187.6 & $0.64\pm0.14$ & [18] \\
AX~J1824.5$-$2451 & \nodata & 18 24 30.0 & -24 51 00 & \nodata & 19(10) & 177.2 & $0.013\pm0.002$ & [19] \\
4U~1850$-$08 & GA & 18 53 04.8 & -08 42 19 & 0.343 & 11 & 65.1 & 0.22--1.4 & [20,21] \\
1A~1905+00\tablenotemark{d} & \nodata & 19 08 27.0 & +00 10 07 & \nodata & 13 & 71.9 & \nodata & [22] \\
\tableline
Total (19 sources) & & & & & 1173 &   6290. \\
 \enddata
\tablerefs{
 1. \cite{swank78};
 2. \cite{brandt92};
 3. \cite{piro97};
 4. \cite{bce72};
 5. \cite{mats80};
 6. \cite{tfs86a};
 7. \cite{hakala05};
 8. \cite{zand04b};
 9. \cite{mig03};
10. \cite{cocchi01d};
11. \cite{corn02b};
12. \cite{mak81b};
13. \cite{mak81};
14. \cite{chelov06};
15. \cite{cocchi01b};
16. \cite{muller98b};
17. \cite{matsuba95};
18. \cite{murakami83};
19. \cite{gk97};
20. \cite{swank76};
21. \cite{hoff80};
22. \cite{lhd76c}. }
\tablenotetext{a}{Source type, as for Table \ref{bursters}}
\tablenotetext{b}{Number in parentheses is the number of observations
analysed to determine the flux range in column 8.}
\tablenotetext{c}{Flux measurements are unreliable due to contamination
from the nearby ($0\fdg49$ away) transient GRS~1747$-$312}
\tablenotetext{d}{Flux measurements are unreliable due to contamination
from the nearby ($0\fdg82$ away) transient Aql~X-1}
\tablecomments{This table does not include observations towards the
Galactic center which cover multiple sources}.
\end{deluxetable*}

\begin{table*}
 \epsscale{1.3}
  \plotone{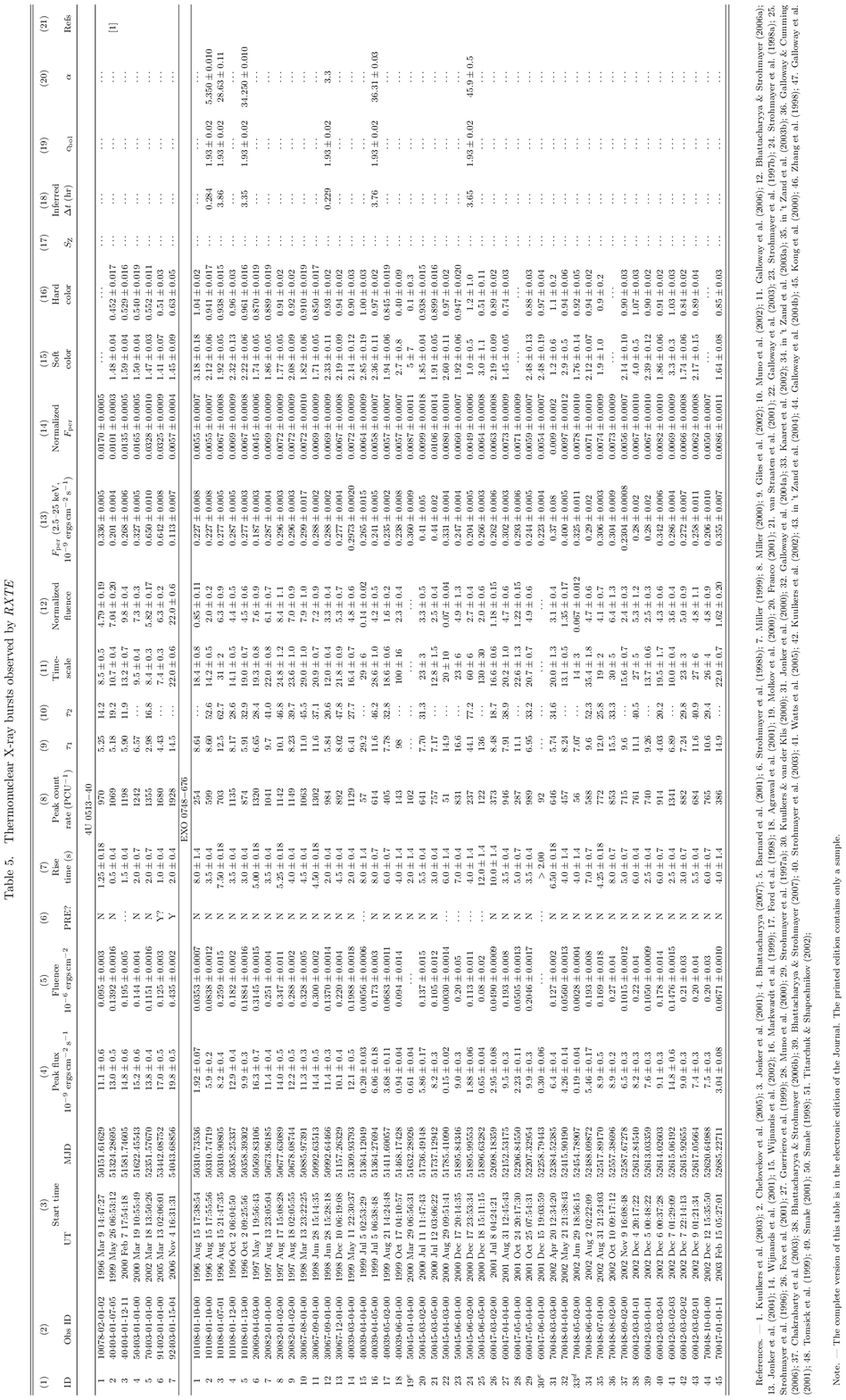}
\end{table*}
\setcounter{table}{5}

We searched for, but did not detect, oscillations at 410~Hz in
SAX~J1748.9$-$2021 \cite[as claimed by][]{kaaret03}; this source has
subsequently exhibited intermittent persistent pulsations at 442~Hz
\cite[]{gavriil07,altamirano07}, strongly suggesting the earlier detection was
spurious.
We also made a targeted search around 1122~Hz for the bursts from
XTE~J1739-285, without success \cite[cf. with][]{kaaret07a}. 
This may be attributable to a different choice of time windows for the
4-s FFTs; in our search the windows do not overlap (to ensure they are
independent), while 
the earlier search adopted overlapping windows beginning
each 0.125~s.

We summarise the millisecond oscillation search in Table \ref{tab:osc},
and list the results for individual bursts in Table \ref{osctbl}. Where
sufficient signal was available to detect oscillations independently in
the rise, peak and decay of the burst, we indicate the detection in the
``Location'' column (R, P and D, respectively). We also list the maximum
Leahy power and the maximum fractional rms for the oscillation.

\subsection{Photospheric radius expansion and source distances}
\label{pflux}

\begin{figure}
 \epsscale{1.2}
 \plotone{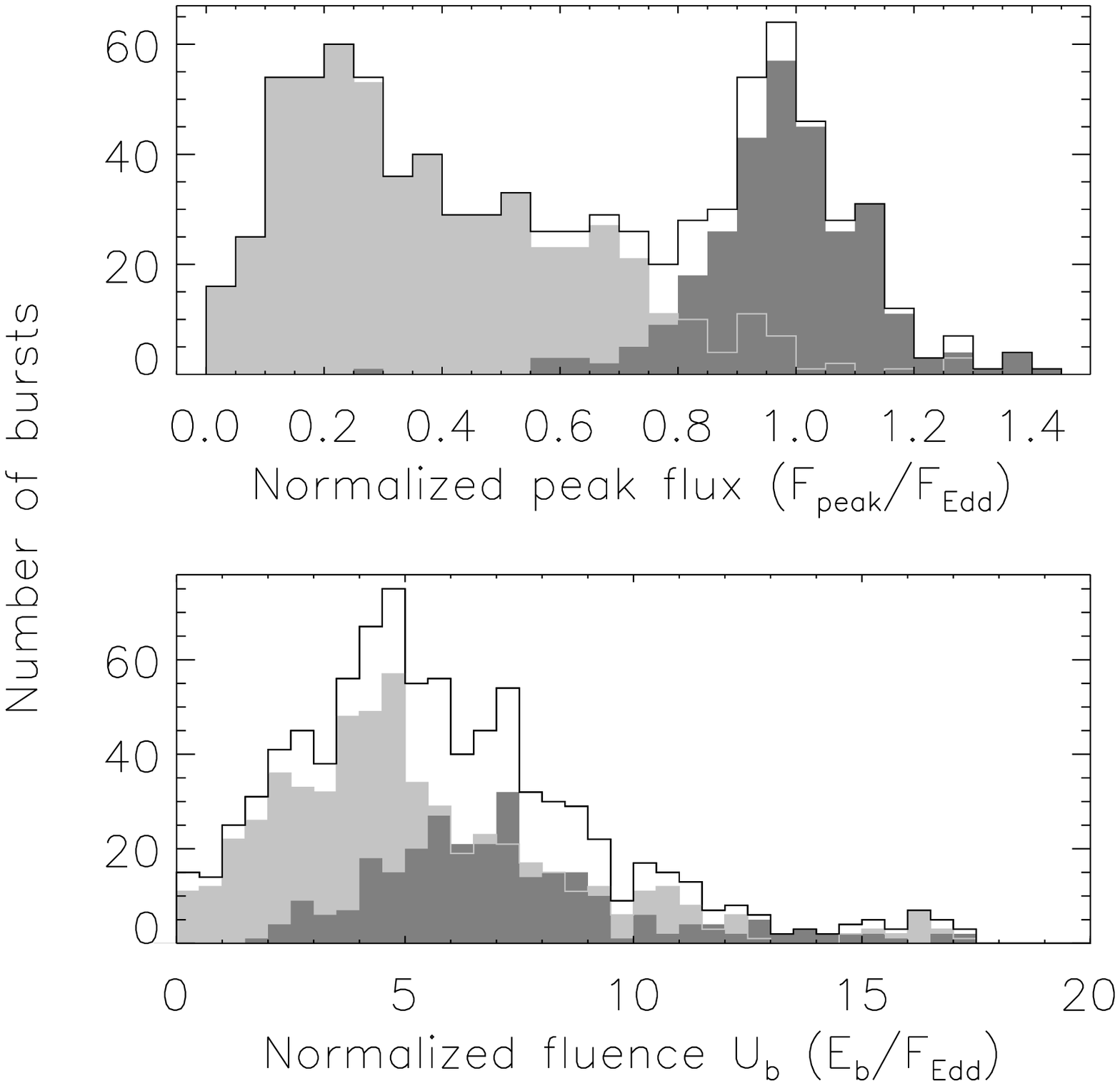}
 \figcaption[f8.eps]{{\it Top panel} Distribution of (normalized) peak burst
flux $F_{\rm pk}/F_{\rm Edd}$ for radius-expansion ({\it dark gray}) and
non-radius expansion ({\it
light gray}) bursts.
The distribution of peak fluxes of the radius-expansion bursts is broad,
with standard deviation 0.14. The radius-expansion burst with the
lowest peak flux $\simeq0.3F_{\rm Edd}$ is from 4U~1636$-$536 (see
also \S\ref{s1636}).
The black histogram shows the combined distribution. 
{\it Bottom panel} Distribution of normalized fluence $U_b=E_b/F_{\rm
Edd}$ for both types of bursts. There is significant overlap between the
two distributions, suggesting that the amount of accreted fuel is
relatively unimportant in determining whether bright bursts exhibit radius
expansion or not. 
Not shown are 
18 extremely energetic bursts with $U_b>20$~s, all exhibiting
radius-expansion, from
4U~0513$-$40,	
4U~1608$-$52, 4U~1636$-$536, 4U~1724$-$307, GRS~1741.9$-$2853 (2),
GRS~1747$-$312, GX~17+2 (8), XB~1832$-$330,
HETE~J1900.1$-$2455 and 4U~2129+12. 
 \label{npflux} }
\end{figure}

Photospheric radius-expansion (PRE) bursts are a key phenomenon which
allow measurement of the distance to bursting sources, as well as (in
principle) determination of neutron star masses and radii
\cite[]{damen90,ozel06}. For these bursts, the large collecting area of
\xte\/ permitted time-resolved spectral analysis, allowing us to
identify radius-expansion episodes from the overwhelming majority of
bursts in which they occurred.
We were able to satisfy our criteria for radius-expansion (see
\S\ref{pre}) for bursts peaking at fluxes as low as $5\times10^{-9}\
\epcs$; the brightest PRE bursts in our sample reached more than
$1.7\times10^{-7}\ \epcs$.
We thus identified photospheric radius-expansion bursts from \numsrc\ of the
\numbursters\ sources with bursts detected by \xte\/
(Table \ref{dist}).

We adopt the mean peak flux of
radius-expansion bursts as the Eddington luminosity for that source,
$F_{\rm Edd}=\left<F_{\rm pk,PRE}\right>\equiv L_{\rm Edd}/4\pi d^2$.
Comparisons of bursts from different sources based on this value may be
biased, for two reasons.
First, 
PRE bursts from individual sources 
may not all reach consistent peak fluxes; and
second, the Eddington luminosities \ledd\/ may vary from source to
source.
We observed highly significant ($>5\sigma$) variations in the peak PRE burst
fluxes for
\nvar\ of the \ngtone\ sources with more than one PRE burst.
For individual bursts, the estimated (statistical) uncertainty on the
measured peak flux was typically $\sim2$\%,
while the fractional variation of peak fluxes from all PRE bursts from a
given source was typically 5--10\%.
The mean fractional standard deviation over all \ngtone\ sources was
\meandist, with a maximum of \fstdevmax\ for GRS~1747$-$312.

The peak fluxes of the combined sample of radius-expansion bursts,
normalized by the $F_{\rm Edd}$ value for each source, 
were in the range 0.26--1.4 (Fig. \ref{npflux}, upper panel). We note that
the PRE burst sample includes those with marginal evidence for
radius-expansion (see \S\ref{pre}). The burst with the smallest
normalized peak flux in
this class was from 4U~1636$-$536 (0.26), although \cite[as discussed
by][]{gal06a} the spectral variation in this burst may have instead arisen
from the same mechanism that gave rise to the double-peaked bursts (see
\S\ref{s1636}).
We also plot the distribution of the normalized fluence, $U_b=E_b/F_{\rm
Edd}$, in Fig.  \ref{npflux}, lower panel. There was much greater overlap
in $U_b$
between the distributions for the PRE and non-PRE bursts, than for the
normalized peak flux. 
The 
18	
bursts with the highest normalized fluence ($U_b>20$, not shown) all
exhibited PRE.  These include extreme radius expansion bursts from
4U~1724$-$307
(\S\ref{s1724}) and 4U~2129+12 (\S\ref{s2129}), amongst others, as well as
the very long bursts from GX~17+2 (\S\ref{sgx17p2}).

Several factors apparently contribute to the scatter in peak PRE burst
fluxes for individual sources.
Faint, symmetric bursts observed from 4U~1746$-$37 and GRS~1747$-$312
appeared to exhibit PRE but
reached significantly sub-Eddington fluxes.
The three PRE bursts from 4U~1746$-$37 reached peak fluxes of
(4.5--$6.3)\times10^{-9}\ \epcs$. At the estimated distance to the host
cluster NGC~6441 \cite[$d=11$~kpc;][]{kuul03a} this flux indicates a peak
isotropic luminosity of (7--$9)\times10^{37}\ \eps$,  well
below the expected luminosity even for the (lower) Eddington
limit for H-rich material, $\approx1.6\times10^{38}\ \eps$ (equation
\ref{ledd} with $X=0.7$).
Similar results were found for this source by
\cite{szt87}.
One of the three PRE bursts observed from GRS~1747$-$312 was similarly
underluminous,
reaching a peak flux of $1.0\times10^{-9}\ \epcs$ which \cite[for the
estimated distance to Terzan~6 of 9.5~kpc;][]{kuul03a} corresponds to an
isotropic luminosity of
$1.1\times10^{38}\ \eps$.
Two other PRE bursts were detected from GRS~1747$-$312, 
both reaching much higher peak
fluxes of 1.7 and
$2.2\times10^{-9}\ \epcs$.
The corresponding range of PRE burst peak fluxes for this source was the
highest of all the sources in the \xte\/ sample, with the brightest PRE
burst 2.2 times brighter than the faintest.
The spectral evolution of these faint PRE bursts was distinctly
different from the PRE bursts from other sources
\cite[see also][]{gal04a,gal08a}, and resembled that of the short, anomalous
bursts from Cyg~X-2 (see \S\ref{scygx2}).
These underluminous PRE bursts 
define a distinct class of bursts which appear to
exhibit radius expansion but which reach peak luminosities significantly
below the Eddington limit, 
in some cases resulting in a much greater variation
in their peak fluxes (for an individual source) than for 
typical PRE bursts. 

\begin{figure}
 \epsscale{1.2}
 \plotone{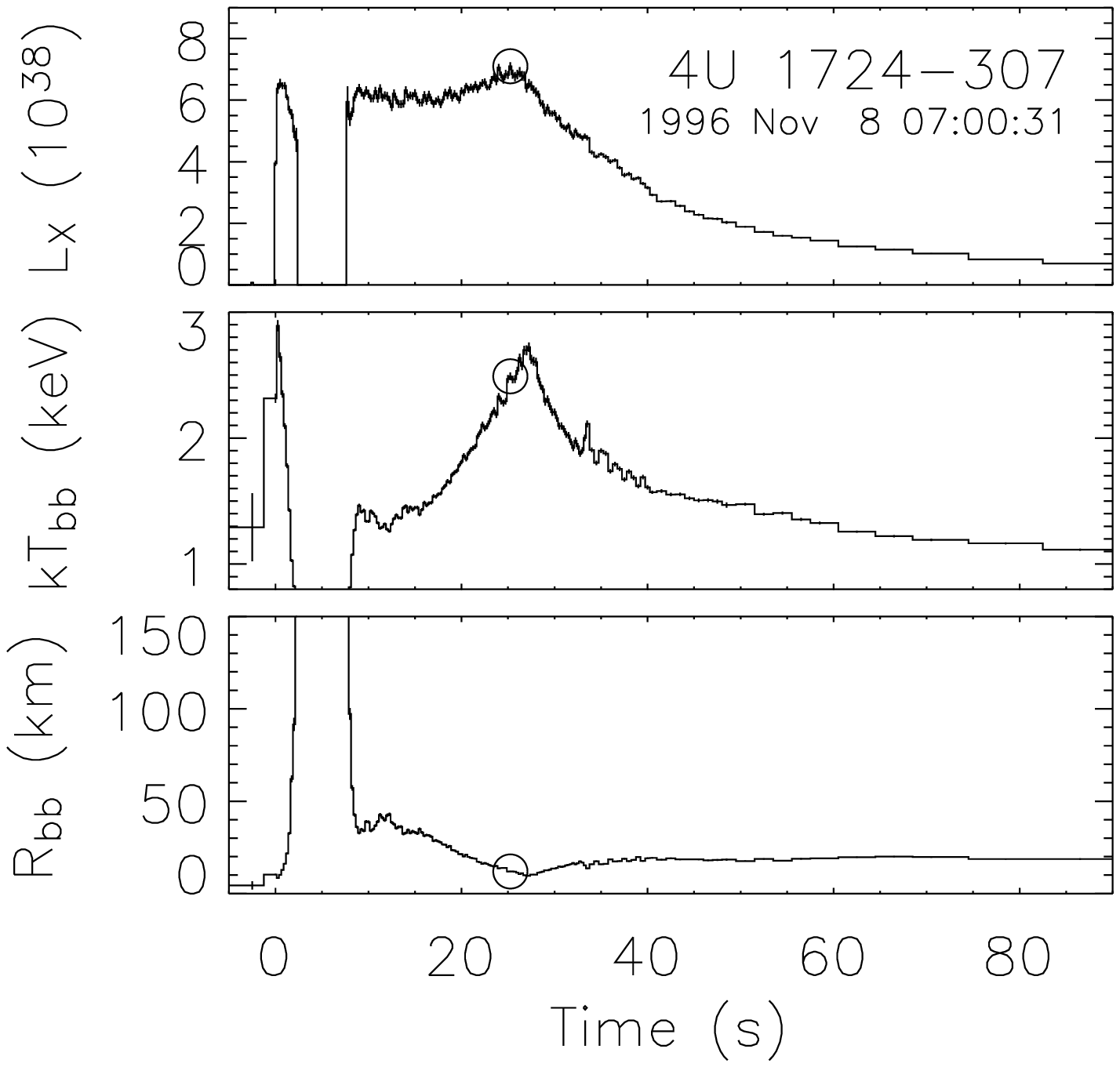}
 \caption{An example of an extremely strong photospheric radius-expansion
burst observed from 4U~1724$-$307 in the globular cluster Terzan 2 by
\xte.
{\it Top panel}\/ 
Burst luminosity (in units of $10^{38}\ \eps$; {\it middle panel}\/
blackbody (color) temperature $kT_{\rm bb}$; and {\it bottom panel}\/
blackbody radius $R_{\rm bb}$. $L_X$ and $R_{\rm bb}$ are calculated
assuming a distance to the host globular cluster Terzan~2 of 9.5~kpc 
\cite[]{kuul03a}.
The time at which the flux reaches its maximum value is indicated by the
open circle.
Note the gap in the first 10~s of this burst, preceded by an abrupt
increase in the apparent blackbody radius to very large values. This gap
was caused not by an
interruption in the data but because the radius-expansion was sufficiently
extreme to drive the peak of the spectrum below the PCA's energy range. In
such cases we expect the luminosity is maintained at approximately the
Eddington limit, although it is no longer observable by \xte.
\label{b1724} }
\end{figure}

Conversely,
the intrinsically most energetic events, exemplified by bursts from
4U~2129+12 and 4U~1724$-$307, tended to reach luminosities significantly in excess of
the Eddington luminosity \ledd. These ``giant'' bursts
exhibited unusually large fluences (and hence long
timescales), and expansion to large radii (e.g. Fig.  \ref{b1724}). At the
estimated distances to the host globular clusters of these sources, the
implied isotropic luminosities reached by these bursts were a factor
of 2--3 larger than the expected Eddington luminosity, even for $X=0$.
As with the other PRE bursts, we applied the gravitational redshift
correction in equation \ref{ledd} with $R=R_{\rm NS}$. If the photospheric
radius is larger at the flux maximum, the intrinsic luminosity will be
overestimated.
However, examination of the spectral evolution of
these two bursts indicates that the blackbody radius at flux maximum is
similar to the asymptotic value in the tail of the burst. Thus, the
redshift correction alone cannot explain the unusually large peak fluxes.
Apparently super-Eddington PRE bursts were also observed from
EXO~1745$-$248 (see \S\ref{s1745}). One explanation for these
discrepancies is that the globular cluster distances are systematically in
error, and the sources are actually significantly closer \cite[note
the smaller distance estimate for the host cluster
of EXO~1745$-$248 made by][]{ortolani07}.
However, extremely energetic bursts observed from sources
outside globular clusters reached peak fluxes significantly larger than
less energetic PRE bursts from the same source.
For example,
the brightest burst from GRS~1741.9$-$2853, on 1996 July, reached a peak
flux 25\% higher than the next brightest PRE burst. The 1996 July burst 
had $U_b=65$, compared to the next highest
value of 23. Similarly, the first burst observed by \xte\/ from the
millisecond accretion-powered pulsar HETE~J1900.1$-$2455 had a peak flux
20\% greater than the second, again with a much higher $U_b=55$ compared
to 15.

While these two factors played a significant role in the overall
variation of PRE burst peak fluxes, smaller variations were observed from
other sources without notably under- or over-luminous PRE bursts.
For example, the peak PRE burst fluxes
from 4U~1728$-$34 
were normally distributed with a
fractional standard deviation of 10\%.
In that case quasi-periodic variations on a timescale of $\approx40$~d
were observed in both the peak PRE burst flux, and the persistent
intensity \cite[measured by the \xte/ASM;][]{gal03b}. 
The residual variation of \fpkre\ for subsets of bursts
observed close together in time (once the $\approx40$~d trend was subtracted)
was consistent with the measurement uncertainties, indicating that the
intrinsic variation of the peak PRE burst luminosity is actually $\la1$\%. 
A correlation between the PRE burst fluence and the peak flux
was attributed to
reprocessing of the burst flux in the
accretion disk. The fraction
of reprocessed flux may vary from burst to burst as a result of varying
projected area of the disk, through precession of the disk possibly
accompanied by radiation-induced warping.
That the persistent flux from
4U~1728$-$34 varies quasi-periodically on a similar time scale to \fpkre\
is qualitatively consistent with such a cause.
It is plausible that comparable variations due to similar mechanisms may
be present in other sources.

Even assuming that the mean peak flux of PRE bursts approaches the
characteristic $F_{\rm Edd}$ value for each source, it
is to be expected that the Eddington luminosities for different
sources are not precisely the same.
Inconsistencies 
are perhaps most likely to arise from
variations in the composition of the photosphere (the hydrogen fraction,
$X$, in equation \ref{ledd}); the neutron star masses, as well as
variations in the typical
maximum radius reached during the PRE episodes (which affects the
gravitational redshift, and hence the observed \ledd) may also contribute.
We can be most confident regarding the photospheric composition in the
ultracompact sources like 3A~1820$-$303 (\S\ref{s1820}), where the lack of
hydrogen in the mass donor rules out any significant abundance in the
photosphere. However, for the majority of bursting sources the uncertainty
in $X$ is the dominant uncertainty in (for example) distance determination
via PRE bursts. One clue as to the composition is provided by the PRE
bursts from 4U~1636$-$536, which reach peak fluxes that are bimodally
distributed \cite[]{gal06a}. The majority of PRE bursts were distributed
normally about the mean with standard deviation of 7.6\%, but two much
fainter bursts reached a peak flux a factor of 1.7 lower than the mean for
the remainder. This distribution suggests that the brighter bursts reach
(approximately) \leddhe, while the two faint bursts reach \leddh.  This
observation implies that, although the source is likely accreting
H-rich material (see \S\ref{diversity}), once a sufficiently strong burst
is triggered, this material is ejected or driven to sufficiently large
radii that it become transparent to X-ray photons \cite[see
also][]{seh84}. The fainter bursts may then arise when the maximum
luminosity exceeds \leddh, but is not sufficient to drive off the outer
layers. If the behaviour of 4U~1636$-$536 is typical, then bursts reaching
\leddh\ are rare, and we can assume with reasonable confidence that PRE
bursts from sources accreting mixed H/He fuel reach the Eddington limit
for essentially pure-He material.

We estimated the distances to the sources with PRE bursts by substituting
the $F_{\rm Edd}$
values into equation
\ref{ledd}. Since the peak flux was typically reached at the end of the
PRE episode, when we presume that the photosphere had touched down on the
NS surface again (see e.g. Fig. \ref{pre_example}), we corrected for
gravitational redshift at the surface of a canonical neutron star with
$R_{\rm NS}=10$~km and $M_{\rm NS}=1.4\ M_\odot$. 
We also estimated the upper limits to the distance for sources without
PRE bursts, from the maximum peak flux of the non-PRE bursts.
We made no correction for the apparently super-Eddington luminosity of
bursts from 4U~2129+12 and 4U~1724$-$307, or the apparently sub-Eddington
luminosity of bursts from 4U~1746$-$37 and GRS~1747$-$312.
We did however calculate the distance to
4U~1636$-$536 in Table \ref{dist} in a different manner to the other
sources with PRE bursts. The listed $\left<F_{\rm pk}\right>$ is for the
high peak flux ($>50\times10^{-9}\ \epcs$) PRE bursts, and we used this
value to calculate the distance assuming that those bursts reach \leddhe\
(fifth column).  For the distance estimates in columns 4 and 6, with
$X=0.7$,
we used only the two PRE bursts with much lower peak fluxes. The
two distances are consistent. 
The distances and limits for all burst sources are both listed in Table \ref{dist}.

\subsection{Burst duration, time scales and fuel composition}
\label{ts}

The characteristic time scale 
$\tau=E_b/F_{\rm pk}$ has long 
served as a measure of the 
lightcurve shape for bursts observed with
low-sensitivity instruments, and the marked differences in the burning
rates for H and He fuel suggests that $\tau$ should also indicate
approximately the fuel composition. 
Here we investigate the variation in $\tau$ as a function of accretion
rate, with a view to identifying systematic trends in the fuel
composition.
Using the high signal-to-noise data
obtained with \xte, we are also able to test this measure against
precisely-determined $\alpha$ values.
For those bursts with sufficient data to resolve the lightcurve, we 
measured additional
decay constant(s) $\tau_1$, $\tau_2$ for the single or broken double
exponential fits to the flux evolution\footnote{ We note that in general $\tau$
and $\tau_1$ were roughly proportional; for
the \burstswithonedecayltfifty\ bursts with both $\tau$ and $\tau_1<50$,
\taurelone\ (rms \taurmsone~s). Similarly for those bursts with two
exponential fits and $\tau<80$~s, we found
\taureltwo\ (rms \taurmstwo~s)}.
A second exponential decay
segment was required for fits
of \burstswithtwodecays\ of the \burstswithonedecay\ lightcurves with
sufficient time-resolved flux measurements to fit at all.

\begin{figure*}
 \plotone{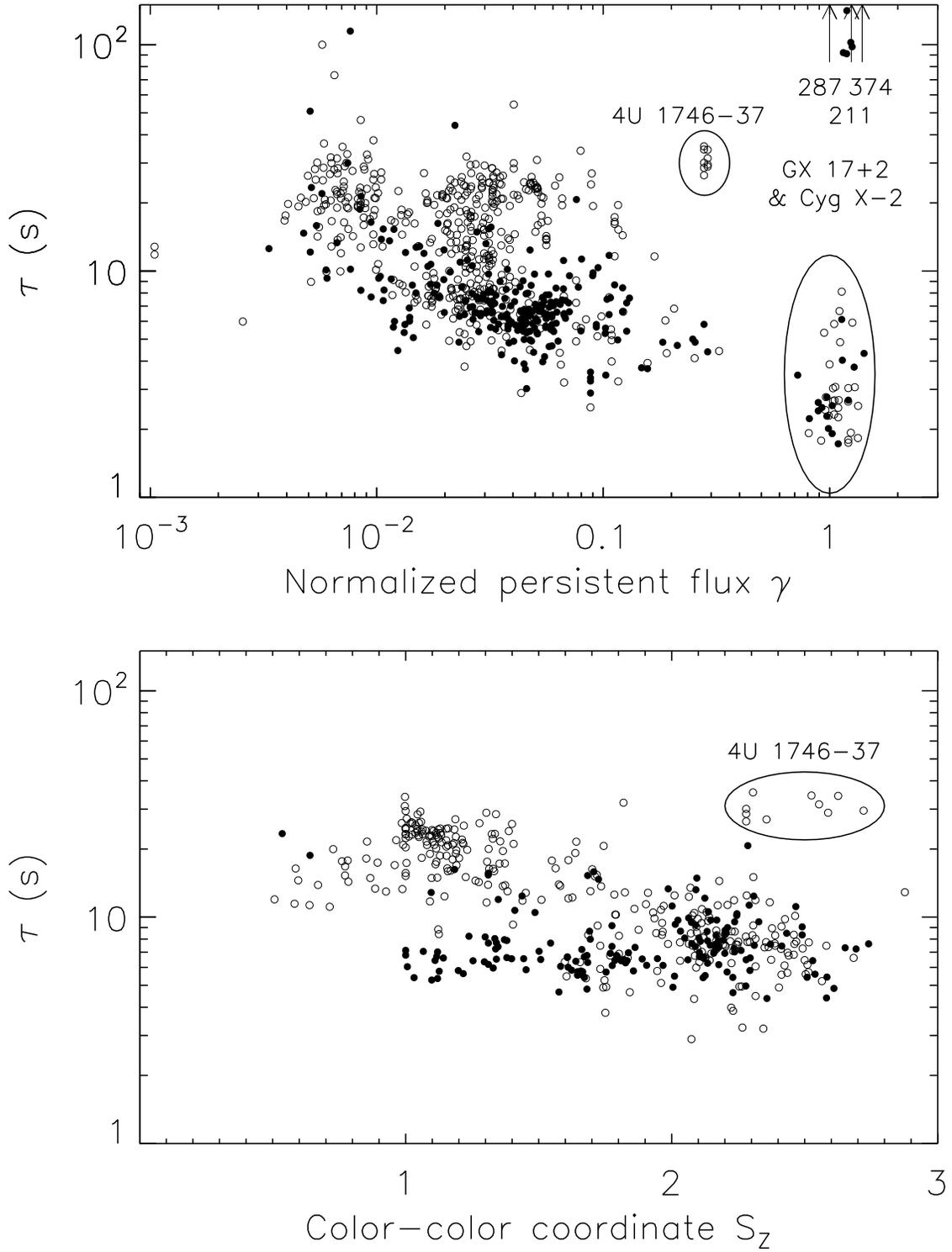}
 \figcaption[tau.eps]{Burst time scale $\tau=E_b/F_{\rm pk}$ as a
function of normalised persistent flux $\gamma$ ({\it top panel}) and
the position $S_Z$ on the color-color diagram ({\it bottom panel}). Open
circles indicate non-radius expansion bursts, while closed circles
indicate PRE bursts. 
Note the markedly different distributions for the PRE and non-PRE bursts,
as well as the presence of groups of outliers in each panel.
In the top panel, the bursts at $\gamma>0.5$ are all from two sources
only, Cyg~X-2 and GX~17+2. Three bursts from GX~17+2 with $\tau>200$
are shown as limits, and labelled by their $\tau$-values. In both the top
and bottom panels, the circled bursts with $\tau\approx30$ at atypically
high $\gamma$/$S_Z$ are almost exclusively from 4U~1746$-$37.
 \label{tauplot} }
\end{figure*}

As has been observed in earlier burst samples \cite[e.g.][]{vppl88}, the
burst timescales vary significantly, and systematically, with accretion
rate, measured by proxy in the first instance as the rescaled persistent
flux $\gamma$ (Fig. \ref{tauplot}, upper panel).
We plot only bursts from sources with at least one radius-expansion
burst (and hence a measurement of $F_{\rm Edd}$) and where
the persistent flux measurements are not confused by other nearby sources;
this subsample is referred to as \sgamma\ (Table \ref{samples}).
The \xte\/ data extend the range in $\gamma$ 
over which timescales can be
measured, 
with extremely short ($\tau\sim2$~s) bursts from Cyg~X-2
and both short and very long bursts from GX~17+2
observed at $\gamma\approx1$.
The long bursts from GX~17+2 are perhaps the most difficult to understand
in the framework of standard burst theory \cite[e.g.][]{kuul02a}, but even
the short bursts from this source and Cyg~X-2 are not predicted by
ignition models at $\dot{M}\approx\dot{M}_{\rm Edd}$.
At this accretion rate both H- and He-burning should be stable, so that
the fuel will burn as it is accreted \cite[e.g.][]{hcw07}.
The long ($\tau\approx30$) bursts from 4U~1746$-$37 at $\gamma\approx0.3$
are the next most signficant outliers from the main sample.
These bursts were
detected during 1998 November when the source was in a
relatively high state; there are other peculiarities regarding the
energetics and recurrence times
(see \S\ref{s1746}).
We omit 75 bursts from sample \sgamma\ in Fig. \ref{tauplot}, top panel,
because the $\tau$ or $\gamma$ values (or the presence or absence of
radius expansion) could not be determined.

As we discuss further in \S\ref{diversity}, the two main groups of
outliers circled in the top panel of Fig. \ref{tauplot} arise from burst
sources (4U~1746$-$37, GX~17+2 and Cyg~X-2) whose behaviour deviates
significantly from the broader population.
Excluding the bursts from these sources reveals trends for the remaining
bursts which differ for the PRE and non-PRE bursts.
For the remaining \preburstswithdecay\ PRE bursts (including those with marginal
evidence for radius expansion; see \S\ref{pre})
the
time scale $\tau$ was 
anticorrelated with $\gamma$
(Spearman's $\rho=\pretaugammacorr$, significant to \pretaugammacorrsig).
For the \nonpreburstswithdecay\ non-PRE bursts from sample \sgamma,
we found a substantially weaker correlation, with
$\rho=\nonpretaugammacorr$, and slightly lower significance.
The distribution of $\tau$-values for the non-PRE bursts in the range
$\gamma=0.03$--0.1 was particularly broad, with short and long bursts
almost equally prevalent.
The
position on the color-color diagram, $S_Z$, for
the subset of sources \ssz\ (Table \ref{samples}) for which this parameter
can be measured (see
\S\ref{pers}), provides an alternative measure of $\dot{M}$ which helps to
understand the variations of $\tau$ with $\gamma$.
Here we find instead
a significant anticorrelation between $\tau$ and $S_Z$ for
the \nonpreburstswithsz\ non-PRE bursts
(\nonpretauszcorr) but no relation for the \preburstswithsz\ PRE bursts
(Fig. \ref{tauplot}, lower panel).
We omit 50 events due to inadequately measured lightcurves and/or $S_Z$
values.
The long 1998 November bursts from 4U~1746$-$37 again fall in a relatively
unpopulated region of the $S_Z$-$\tau$ diagram, this time with
$S_Z$ in the range 2.3--2.8. The other bursts with similar $\tau\approx30$
are all observed at much lower values, $S_Z\approx1$.

With a few exceptions, the \xte\/ data 
confirms the previously-observed decrease (on average) of the burst
timescale $\tau$ with increasing accretion rate (measured by
proxy as either $\gamma$ or $S_Z$). However, there are a number of
aspects which must be considered for a complete understanding of the data.
First, why does the degree of anticorrelation of $\tau$ with accretion
rate (by proxy) differ for the radius-expansion and non-radius expansion
bursts? Selection effects may play a role, and in particular
the lack of correlation between $\tau$ and $S_Z$ for the PRE bursts may be
partly attributed to the fact that almost half (72 of the
\preburstswithsz) PRE bursts
with $S_Z$ values are from
4U~1728$-$34. 
The burst time scales for this source are consistently short
($\tau=6.3$~s on average; see \S\ref{he1728}), possibly because the
accreted fuel is largely helium (see also \S\ref{diversity}).
Furthermore, approximately half of the 20 PRE bursts with large
$\tau$-values (including the three bursts with the largest values) are
from sources that do not have measured $S_Z$ values (and thus are not part
of sample \ssz).

However, there are also systematic effects which bias the
measured $\tau$-values for radius-expansion bursts.
As also noted by \cite{vppl88},
regardless of how large is the fluence for a PRE burst, \fpk\ cannot
significantly exceed the Eddington flux for the source, so that the $\tau$
value will be lower than it would have been were the peak flux not
limited.  
Furthermore, since for the PRE bursts, $F_{\rm pk}\approx F_{\rm Edd}$,
$\tau\approx E_b/F_{\rm Edd}\equiv U_b$.
Thus, the anticorrelation
between $\tau$ and $\gamma$
observed for the PRE bursts
is 
largely a consequence of an anticorrelation
(of almost identical degree)
between the normalised fluence $U_b$ and $\gamma$ ($\rho=\preubgammacorr$,
significant to \preubgammacorrsig; see also \S\ref{secalpha}). For the PRE bursts, the decay
constants $\tau_1$, $\tau_2$ for the exponential fits to the burst
lightcurve following the peak likely serve as a 
less biased measure of the burst time scale, 
and we find flatter correlations between both
$\tau_1$ and $\gamma$
(\preedtgammacorr) and $\tau_2$ and $\gamma$
(\preedttwogammacorr) for the PRE bursts. 

\begin{figure}
 \epsscale{1.2}
 \plotone{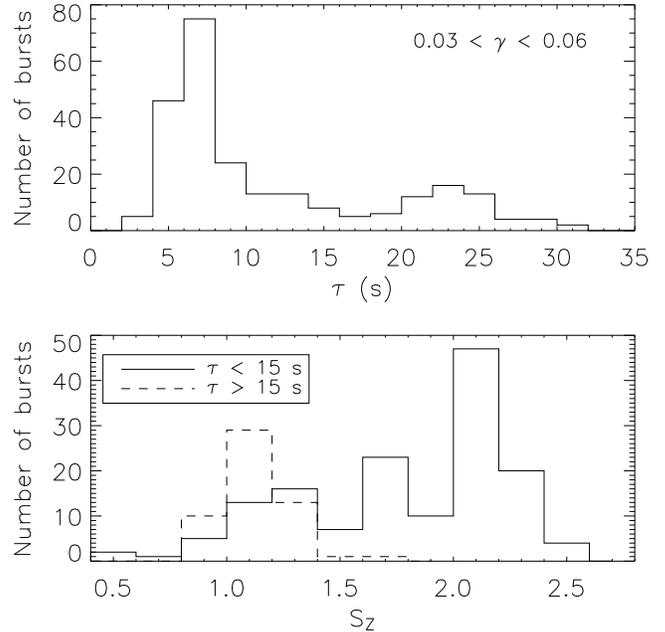}	
 \caption{Relation between $\tau$, $\gamma$ and $S_Z$ for non-radius
expansion bursts.
The $\tau$-distribution ({\it upper panel}) is 
noticeably bimodal at moderate $\gamma=0.03$--0.06.
The distribution of $S_Z$ values for those bursts where it can be
measured, plotted separately for long and short bursts in this
$\gamma$-range ({\it lower
panel}), shows that
the long ($\tau>15$~s) bursts 
occur when the sources are in the ``island'' region of
the color-color diagram
(i.e. $S_Z\approx1$ in Fig. \ref{fig:sz}), while the short bursts 
with $\gamma=0.03$--0.06
largely have $S_Z\ga2$ and occur when the source is in the
``banana'' branch. 
This suggests that the long bursts tend to occur at substantially lower
$\dot{M}$ than the short bursts, despite the comparable
$\gamma$-values. This effect is a consequence of the degeneracy between
$\gamma$ and $S_Z$ (see Fig. \ref{comparison}).
 \label{taugammahist} }
\end{figure}

Second, how do we account for the bimodal $\tau$-distribution for the
non-radius expansion bursts at moderately high $\gamma$ (Fig.
\ref{tauplot}, top panel)? This bimodality is illustrated clearly in Fig.
\ref{taugammahist} (top panel), which also indicates that the 
timescale $\tau$ for bursts observed at $\gamma=0.03$--0.06 is related to
position on the color-color diagram (lower panel).  Bursts with
$\tau>15$~s 
are largely detected in
observations with $S_Z\approx1$, while the short bursts that are also
observed at $\gamma>0.03$ tend to have $S_Z>2$.
Observations with $S_Z\approx1$ correspond to a source in the ``island''
state (e.g. Fig. \ref{fig:sz}), where the accretion rate is thought to be
substantially below that measured when sources are in the ``banana''
state, with $S_Z\ga2$.
Thus,
the bimodal distribution of the $S_Z$ values indicates a mix of high
and low accretion rates, despite the narrow range of $\gamma$.
This ``degeneracy'' between $\gamma$ and $S_Z$
is illustrated clearly in the plot of $S_Z$
as a function of $\gamma$, averaged over each observation (Fig.
\ref{comparison}).
In the range $\gamma=0.01$--0.06, 
$S_Z$ is essentially uncorrelated with $\gamma$, and 
can span almost the full extent of the color-color diagram,
$\approx0.5$--2.5.
Thus, the bimodal distribution of $\tau$-values for bursts at high
$\gamma$ likely does not reflect a bimodal distribution of timescales at
high accretion rates, further supporting the general trend of decreasing
burst timescale as a function of $\dot{M}$.

\begin{figure}
 \epsscale{1.2}
 \plotone{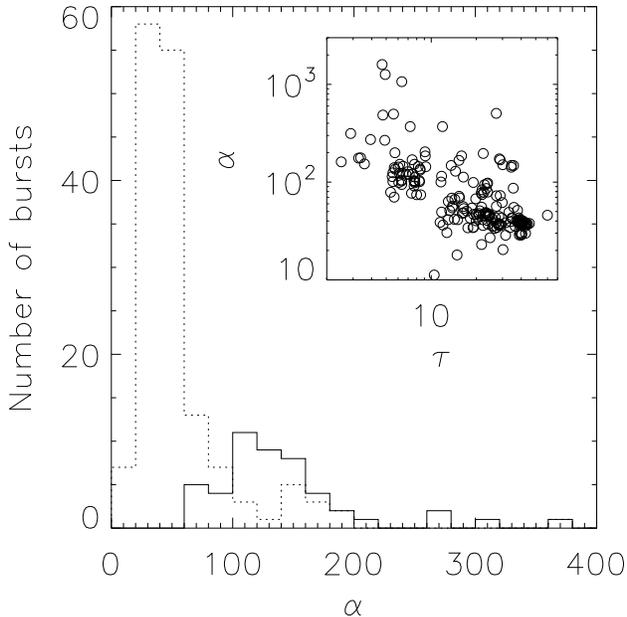}	
 \figcaption[tau-alpha.eps]{Distribution of $\alpha$ for bursts from
sample \sdt\ (Table \ref{samples}) with
$\tau<10$ ({\it solid histogram}) and $\tau>10$ ({\it dotted histogram}). 
Bursts with low $\alpha<60$ are all long with $\tau>10$, while bursts with
$\alpha>60$ are predominantly short with $\tau<10$.
The inset
shows $\alpha$ as a function of $\tau$ for the individual bursts, clearly
showing the correlation between the fuel composition (indicated by
$\alpha$) and the characteristic evolution time-scale for the burst
 \label{taualpha} }
\end{figure}

With the \xte\/ sample we are in a position to directly compare
the measured $\tau$ and $\alpha$ values. From our sample \sdt\ of burst pairs
for which we can precisely measure $\alpha$ (see Table \ref{samples}), we
found that $\tau$ and $\alpha$ were strongly anticorrelated. All the
bursts with $\tau<10$ have $\alpha>70$ (Fig. \ref{taualpha}), which
indicates H-poor bursts in which the H-fraction may have been reduced by
steady burning between bursts. For those
bursts we find a median $\alpha=136$, although we also measured values up
to
1600 (for Ser~X-1). 
For the bursts with $\tau>10$, we instead
found a median $\alpha=43$.
Perhaps most significantly, none of the bursts with low $\alpha$ had
$\tau<10$; 
this highlights the relatively long time required to burn hydrogen via the
rp-process.

When combined with the general trend to shorter timescales with increasing
accretion rate that we deduced from the distributions of $\tau$-values, 
this result argues for a {\it decreasing} H-fraction in bursts as
accretion rate increases, particularly above $\gamma\ga0.07$ and
($S_Z\ga1.8$). As we shall see in \S\ref{global} and \ref{secalpha}, this
trend is difficult to reconcile with theoretical predictions of burst
behaviour as a function of accretion rate.

\subsection{Diversity of burst behaviour}
\label{diversity}

It is well known that the burst behaviour of GX~17+2 and Cyg~X-2 deviate
substantially from that of the majority of burst sources, and this has
also been clear from our analysis so far. Here we address the question of
whether that remaining majority of burst sources also exhibit significant
source-to-source variations in their burst recurrence time and energetics
as a function of accretion rate, or instead follow a consistent global
behavior.
Previous authors analysing burst samples assembled from multiple
sources have generally concluded that ``all sources have the same global burst
behavior as a function of luminosity'' (\citealt{corn03a}; see also
\citealt{vppl88}). However, with the more detailed burst measurements
possible with the \xte\/ sample, it is worthwhile revisiting this issue.

\begin{figure*}
 \plotone{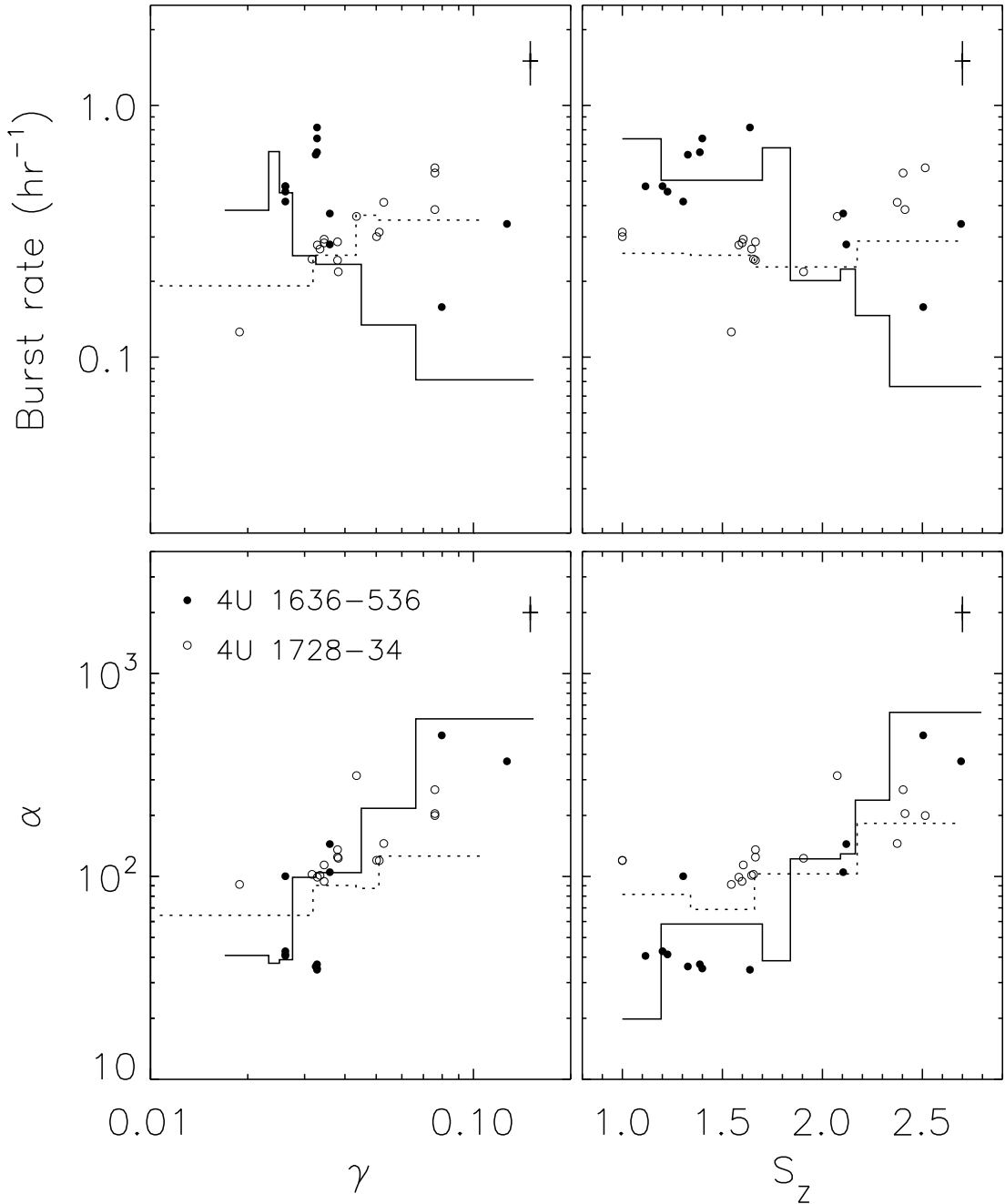}	
 \figcaption[2source_rate.eps]{Burst rates ({\it upper panels})
and $\alpha$-values ({\it lower panels})
for 4U~1636$-$536 
and 4U~1728$-$34 
plotted as a function of $\gamma$ ({\it left-hand panels}) and $S_Z$
({\it right-hand panels}). Bursts are combined in groups of 
$\approx25$
for binning to calculate the mean rates 
(4U~1636$-$536 plotted as a solid histogram, and 4U~1728$-$34 dotted).
The symbol at the top-right of each panel shows the typical $1\sigma$ 
uncertainty on the value in each bin.
Also plotted are the instantaneous burst rates ($1/t_{\rm rec}$) and
$\alpha$-values for selected pairs of bursts from these two sources ({\it
filled/open circles}).
While the $\gamma$/$S_Z$ range spanned by the two sources is almost
identical, the variation of the burst rate and $\alpha$ is distinctly
different.
\label{twosources} }
\end{figure*}

We begin by comparing
the two systems 
with the most bursts in the \xte\/ sample,
4U~1636$-$536 and 4U~1728$-$34.
Bursts were observed from both sources over a similar range of inferred
accretion rate, of about an order of magnitude in $\gamma$ (in the range
0.01--0.1) and over the full extent of the color-color diagrams (see Fig.
\ref{fig:sz}; $S_Z=1.0$--2.5).
The properties of the bursts from 4U~1728$-$34 varied little over these
ranges, with consistently short ($\tau\approx6$~s) bursts that tended to
exhibit radius-expansion. On the other hand, 4U~1636$-$536 showed both
short and long bursts (with $\tau$ up to 30~s), with the latter observed
preferentially at low accretion rates (whether measured by $\gamma$ or
$S_Z$).
For 4U~1728$-$34 
the burst rate 
(calculated as described in \S\ref{energetics})
increased with $\gamma$, as we expect
theoretically, although only at the $1.8\sigma$ level; but no systematic
variation was measured with $S_Z$
(Fig. \ref{twosources}). 
For 4U~1636$-$536, on the other hand, we found a steep {\it decrease}\/ in
burst rate with increasing $\gamma$ and $S_Z$, significant to greater than
$6\sigma$ in either case.
We also found inconsistent behaviour in the measured $\alpha$-values for
these two sources as a function of $\gamma$/$S_Z$ (Fig. \ref{twosources},
lower panels).  For 4U~1728$-$34 
we consistently found 
\malpha\ in the range 60--130
(with measurements from burst pairs
all 
$>90$),
and no significant variation as a function of $\gamma$ or
$S_Z$ ($<2\sigma$ significance in either case).
For 4U~1636$-$536, \malpha\ increased significantly with both $\gamma$ and
$S_Z$, 
increasing from
$\approx40$ to between 600 and 700
(significance of $\ga5\sigma$ in either case).
These two sources thus exhibit very different patterns of variation in
burst timescale, rate and energetics, over the same range of
inferred accretion rate. 

One plausible reason for the difference in burst behaviours between these
two sources is a different accreted composition. The consistently fast
rises and decays, and low $\tau$-values for bursts from 4U~1728$-$34
indicate a consistently low H-fraction at ignition.
In contrast, the at-times long bursts (with $\alpha\approx40$)
observed from 4U~1636$-$536 are characteristic of fuel with a high
proportion of H. While the H-fraction $X$ may be reduced by
steady burning prior to ignition, so that the value inferred from the burst
timescale is a lower limit on the accreted composition $X_0$, the absence of
long bursts characteristic of mixed H/He burning at any $\gamma$ or $S_Z$
value in 4U~1728$-$34\footnote{The ``long'' bursts that have been observed
at times from 4U~1728$-$34 exhibit radius-expansion \cite[e.g][]{bas84},
and are quite distinct from the long bursts which indicate mixed H/He
fuel (exemplified by bursts from GS~1826$-$24) which do not exhibit radius
expansion.} suggests that the H-fraction in the accreted fuel from that
source is significantly lower than in 4U~1636$-$536.
The burst properties of 4U~1728$-$34 in fact closely resemble those of
the ultracompact system 3A~1820$-$303 \cite[]{cumming03}, indicating that
it may also be a He accretor. For 3A~1820$-$303 the measured orbital
period of 11.4~min \cite[]{swp87} requires that the counterpart be evolved
and H-poor; in contrast, the orbital period for 4U~1728$-$34 is unknown.

With orbital periods known for only about 1/3 of bursters in our sample, it
is possible that pure He-accretors may make up a non-neglible fraction of
the remaining sources.
If H-rich and H-poor accretors indeed exhibit systematically different
patterns of burst rate and energetics, as is suggested by the comparison
of 4U~1636$-$536 and 4U~1728$-$34, it is clearly important to
discriminate between these two types of sources when combining bursts.
In the absence of a measured orbital period, the only other guide to
the composition of the accreted material in the remaining sources
is the properties of the bursts.
Consequently, we have reviewed the burst properties of each of the sources 
contributing to our sample, and attempted to classify them
phenomenologically. The sources, ordered by burst behaviour, are listed in
Table \ref{classify},
along with pertinent
burst parameters. Below we briefly describe the salient points of each of
these classifications.

\begin{deluxetable*}{llcccc}
\tabletypesize{\scriptsize}
\tablecaption{Type-I X-ray burst sources arranged by burst behavior \label{classify}
}
\tablewidth{0pt}
\tablehead{
  \colhead{Category}
 & \colhead{Source}
 & \colhead{$\Delta t$ (hr)\tablenotemark{a}}
 & \colhead{$\left<\tau\right> (s)$}
 & \colhead{$\left<U_b\right>$}
 & \colhead{$\left<\gamma\right>$\tablenotemark{b}}
}
\startdata
very low $\dot{M}$ & EXO~0748$-$676 & 2.1 & $24\pm16$ & $4\pm2$ & $0.0075\pm0.0018$
 \\
frequent long bursts & 4U~1323$-$62 & 1.8 & $25\pm6$ & \nodata & ($0.23\pm0.02$)
 \\
 & XTE~J1710$-$281 & 3.3 & $24\pm16$ & $3.3\pm1.6$ & $0.008\pm0.003$
 \\
 & 2E~1742.9$-$2929 & 1.1 & $24\pm8$ & $4.5\pm1.8$ &  \nodata
 \\
 & XTE~J1814$-$338 & 1.7 & $30\pm6$ & \nodata & ($0.44\pm0.08$)
 \\
 \tableline
very low $\dot{M}$ & 4U~0919$-$54 & \nodata & 13, 21 & 1.5, 13 & $0.0033\pm0.0009$
 \\
``giant'' bursts & 1724$-$307 & \nodata & 6.3--44 & 4.8--54 & $0.027\pm0.007$
 \\
 & SLX~1735$-$269 & \nodata & $12.5\pm0.4$ & \nodata & ($0.54\pm0.11$)
 \\
 & GRS~1741.9$-$2853 & \nodata & $16\pm13$ & 4.0--65 &  \nodata
 \\
 & GRS~1747$-$312 & \nodata & 5.4--110 & 1.2--160 & $0.029\pm0.015$
 \\
 & XB~1832$-$330 & \nodata & \multicolumn{2}{c}{$21.4\pm0.8$\tablenotemark{c}} & $0.010\pm0.002$
 \\
 & 4U~2129+12 & \nodata & \multicolumn{2}{c}{$30.0\pm1.2$\tablenotemark{c}} & 0.0074
 \\
 \tableline
low $\dot{M}$ & 4U~0513$-$40 & \nodata & $11\pm5$ & $9\pm6$ & $0.017\pm0.011$
 \\
infrequent short bursts & SAX~J1808.4$-$3658 & 21 & $13\pm2$ & $11\pm5$ & $0.011\pm0.003$
 \\
 & HETE~J1900.1$-$2455 & \nodata & 16, 51 & 15, 55 & $0.007\pm0.003$
 \\
 \tableline
moderate $\dot{M}$ & 1M~0836$-$425 & 2.0 & $22\pm4$ & \nodata & ($1.3\pm0.4$)
 \\
frequent long bursts & KS~1731$-$260\tablenotemark{d} & 2.5 & $23.8\pm0.7$ & $16.1\pm0.5$ & $0.05\pm0.03$
 \\
 & GS~1826$-$24 & 3.2 & $39\pm4$ & \nodata & ($1.6\pm0.2$)
 \\
 \tableline
large $\dot{M}$ range & 4U~1608$-$52 & 3.5 & 5.6--29 & 0.19--24 & 0.0034--0.10
 \\
inhomogeneous bursts & 4U~1636$-$536 & 1.0 & 3.2--32 & 0.13--22 & 0.020--0.15
 \\
 & MXB~1659$-$298 & 1.8 & 3.0--30 & 2.4--9.5 & 0.025--0.055
 \\
 & 4U~1705$-$44 & 0.91 & 4.6--28 & 0.84--11 & 0.017--0.21
 \\
 & KS~1731$-$260\tablenotemark{d} & 6.4 & 4.4--20 & 1.6--15 & 0.0087--0.13
 \\
 & Aql~X-1 & 3.5 & 6.0--34 & 2.1--17 & 0.00011--0.11
 \\
 \tableline
moderate--high $\dot{M}$ & 4U~1702$-$429 & 4.5 & $8.3\pm1.7$ & $6\pm3$ & $0.020\pm0.005$
 \\
consistently short bursts & XTE~J1709$-$267 & \nodata & $5.8\pm0.5$ & \nodata & ($2.8\pm0.3$)
 \\
 & 4U~1728$-$34 & 1.8 & $6.4\pm1.2$ & $5.9\pm1.9$ & $0.041\pm0.013$
 \\
 & 4U~1735$-$44 & 1.1 & $3.7\pm0.8$ & $3.2\pm1.2$ & $0.12\pm0.03$
 \\
 & 3A~1820$-$303 & \nodata & $6.5\pm0.7$ & $6.5\pm0.7$ & $0.061\pm0.009$
 \\
 & Ser~X-1 & 8.0 & $4.8\pm0.6$ & $4.7\pm0.7$ & $0.24\pm0.05$
 \\
 & 4U~1916$-$053 & 6.2 & $7.1\pm1.9$ & $7\pm2$ & $0.014\pm0.005$
 \\
 \tableline
high $\dot{M}$ & GX~17+2 & 5.8 & 5.3, 90--360 & 3.5, 90--370 & $1.21\pm0.13$
 \\
 & Cyg~X-2 & 1.0 & $2.9\pm1.3$ & $2.7\pm1.9$ & $1.05\pm0.19$
 \\
 \tableline
anomalous & EXO~1745$-$248 & 2.9 & $23\pm10$ & $5.1\pm1.9$ & $0.05\pm0.03$
 \\
 & 4U~1746$-$37 & 1.0 & $18\pm11$ & $8\pm3$ & $0.15\pm0.10$
 \\
 \enddata
\tablecomments{Sources not listed have insufficient bursts
to assign them to any of the categories described.}
\tablenotetext{a}{A representative burst recurrence time, which we take
as the shortest burst interval longer than 0.9~hr (thus excluding the
atypical short recurrence time bursts).}
\tablenotetext{b}{Values in parentheses refer to flux (in units of
$10^{-9}\ \epcs$) rather than $\gamma$, for those sources with no
radius-expansion bursts and thus no measured value of $F_{\rm Edd}$.}
\tablenotetext{c}{Where only one burst was observed, and it exhibited
radius expansion, $\tau=U_b$ by definition.}
\tablenotetext{d}{For KS~1731$-$26, we separate out the
regular bursts from the less regular (and generally shorter) bursts
at higher and lower values of $\gamma$}
\end{deluxetable*}

\begin{deluxetable*}{lp{10cm}cc}
\tabletypesize{\scriptsize}
 \tablewidth{0pt}
 \tablecaption{Combined burst samples from the \xte\/ catalog
  \label{samples} }
 \tablehead{
\colhead{Sample} & &
\colhead{No. of} &
\colhead{Total} \\
\colhead{label\tablenotemark{a}} &
\colhead{Description\tablenotemark{b}} &
\colhead{bursts} &
\colhead{duration (Ms)}
}
\startdata
\sgamma  & all sources with at least one radius-expansion burst (see Table
\ref{dist}); 
4U~0513$-$40, EXO~0748$-$676, 4U~0919$-$54, 4U~1608$-$52,
4U~1636$-$536, MXB~1659$-$298, 4U~1702$-$429, 4U~1705$-$44,
XTE~J1710$-$281, 4U~1724$-$307, 4U~1728$-$34, KS~1731$-$260, 4U~1735$-$44,
GX~3+1,
SAX~J1748.9$-$2021, EXO~1745$-$248, 4U~1746$-$37, SAX~J1750.8$-$2900,
GRS~1747$-$312, SAX~J1808.4$-$3658, GX~17+2, 3A~1820$-$303, XB~1832$-$330,
Ser~X-1, HETE~J1900.1$-$2455, Aql~X-1, 4U~1916$-$053, 4U~2129+12, Cyg~X-2
(29) & 834 & 22.9 \\
\ssz & all sources with well-defined color-color diagrams (see
\S\ref{pers}); 4U~1608$-$52, 4U~1636$-$536, 4U~1702$-$429, 4U~1705$-$44,
4U~1728$-$34, KS~1731$-$260, 4U~1746$-$37, Aql~X-1, XTE~J2123$-$058 (9)
& 523 & 10.24 \\
\sdt & sources with burst interval measurements from closely-spaced burst
pairs; EXO~0748$-$676, 1M~0836$-$425, 4U~1254$-$69, 4U~1323$-$62,
4U~1608$-$52, 4U~1636$-$536, 4U~1702$-$429, 4U~1705$-$44, XTE~J1710$-$281,
XTE~J1723$-$376, 4U~1728$-$34, KS~1731$-$260, 4U~1735$-$44,
XTE~J1739$-$285, SAX~J1748.9$-$2021, EXO~1745$-$248, 4U~1746$-$37,
SAX~J1808.4$-$3658, XTE~J1814$-$338, GX~17+2, GS~1826$-$24, 4U~1916$-$053,
XTE~J2123$-$058 (\alphasrc) & \alphadat & \nodata \\
\sgammahe & sources in \sgamma\ with 
consistently fast bursts (see Table \ref{classify});
 4U~1702$-$429, 4U~1728$-$34, 4U~1735$-$44, 3A~1820$-$303, Ser~X-1,
4U~1916$-$053 (6) & 190 & 4.95 \\
\sgammah & sources in \sgamma\ excluding sources in \sgammahe,
  ``anomalous'' bursters EXO~1745$-$248, 4U~1746$-$37, and high-$\dot{M}$
  bursters GX~17+2 and Cyg~X-2
(see Table \ref{classify}; 19) & 525 & 14.5 \\
\sszhe & sources in \ssz\ with consistently fast bursts (see Table
\ref{classify}); 4U~1702$-$429, 4U~1728$-$34 (2) & 153 & 2.64 \\
\sszh & all sources in \ssz\ excluding sources in \sszhe\ and
``anomalous'' burster 4U~1746$-$37 
(6) & 340 & 7.15 \\
\sosc & all sources with detected burst oscillations (see \S\ref{milosc});
  4U~1608$-$52, 4U~1636$-$536, MXB~1659$-$298, 4U~1702$-$429,
  4U~1728$-$34, KS~1731$-$290, GRS~1741.9$-$2853, 1A~1744$-$361,
  SAX~J1750.8$-$2900, SAX~J1808.4$-$3658, XTE~J1814$-$338, 
  4U~1916$-$053 and Aql~X-1 (\nsrcsearched) & 529 & 12.46 \\
\enddata
\tablenotetext{a}{Throughout this paper we refer to these combined
samples as ``${\mathcal S}${\it label}\/'' where {\it label}\/ corresponds
to the selection criteria, as detailed in column 2; see \S\ref{energetics}.}
\tablenotetext{b}{We list the sources and/or the selection criteria for each
sample, followed by the total number of sources in parentheses.}
\end{deluxetable*}

\paragraph{Frequent long bursts at very low $\dot{M}$}

These sources accrete at $\approx1$\%~$\dot{M}_{\rm Edd}$ and below, and
exhibit weak, frequent (1--3~hr), long
($\tau\approx25$~s) bursts with $\alpha\sim40$ characteristic of 
mixed H/He fuel. Orbital periods are 
a few hours, where
known. These sources also characteristically show short ($\Delta
t<30$~min) recurrence time bursts (see \S\ref{dblbursts}).

\paragraph{``Giant'' bursts}

These sources accrete at similar $\dot{M}$ to the above group,
but instead are characterised by extremely energetic bursts, likely arising
from a thick layer of pure He built up over days.  All the bursts exhibit
radius-expansion, and large $\tau$-values (from the large fluence) but
fast rise times. Either H/He accretors in which the accreted H is
exhausted via stable burning before ignition, or systems accreting from
H-poor donors may give rise to such bursts. Orbital periods
are $\ga10$~hr, for the two systems where they are known; in contrast,
4U~0919$-$54 is a candidate ultracompact based on its optical properties.

\paragraph{Infrequent short bursts at low $\dot{M}$}

Also at comparable $\dot{M}$ values as the previous two cases, these
sources typically exhibit fast rise, moderately energetic radius-expansion
bursts with characteristic $\Delta t\ga0.5$~d and $\alpha>100$. Sources
which exhibit
giant bursts commonly also exhibit weaker bursts in this category. For
two cases where the orbital periods are known they are $>80$~min and thus
likely accrete mixed H/He, so that steady H-burning is required to reduce
or exhaust the accreted H prior to burst ignition. This may not be the
case for 4U~0513$-$40, which is a candidate ultracompact.

\paragraph{Frequent long bursts at moderate $\dot{M}$}

Recurrence times are in the range 2--6~hr, varying inversely with
$\dot{M}$, and the long rise times $\approx5$~s, $\tau=20$--40,
$\alpha\approx40$ (as well as detailed model comparisons in the case of
GS~1826$-$24; see \S\ref{s1826}) indicate He-ignition in mixed H/He fuel
where the H-fraction is reduced (but not exhausted) by steady burning
between the bursts.

\paragraph{Inhomogeneous bursts}

These transients routinely span a wide range of $\dot{M}$ during their
outbursts. As a result, the burst properties are highly variable,
at times frequent and long, or infrequent and short, depending upon the
$\dot{M}$. Orbital periods are a few hours or longer, where known; none of
the examples are thought to be ultracompact.

\paragraph{Consistently short bursts}

These sources accrete at $\dot{M}$ levels comparable to the previous
group, but consistently show short ($\tau<10$) bursts with $\alpha>100$
characteristic of pure or almost-pure He fuel. In the case of 3A~1820$-$30
and 4U~1916$-$053, this is consistent with their ultracompact nature,
although this explanation manifestly does not explain 4U~1735$-$44
($P_{\rm orb}=4.65$~hr; see also \S\ref{indbursts}).

\paragraph{High $\dot{M}$}

Both these sources accrete at $\sim1$~$\dot{M}_{\rm Edd}$, and show either
very short bursts (in the case of Cyg~X-2) or a mix of short and extremely
long (GX~17+2) bursts. These bursts represent a considerable challenge for
conventional burst theory (see als \S\ref{hibursts}).

\paragraph{Anomalous bursts}

Two relatively prolific bursters exhibit burst
properties that defy explanation via conventional burst theory. During
it's July 2000 outburst, EXO~1745$-$248
exhibited either frequent, long bursts (accompanied by dramatic dipping
behaviour), or infrequent, short bursts, at roughly comparable $\dot{M}$.
4U~1746$-$37 exhibits faint, frequent, regular long bursts at moderate to
high $\dot{M}$ levels, which are interrupted at times by short bursts with
evidence for radius expansion (see also \S\ref{indbursts}).

It is not unexpected that the first four categories listed in Table
\ref{classify} can each be identified with one of the
theoretically-predicted cases of thermonuclear ignition (see Table
\ref{burstregimes}). The frequent, long
bursts at low-$\dot{M}$ have properties consistent with unstable
H-ignition (case 3); infrequent short and giant bursts with unstable
He-ignition in H-poor fuel (case 2); and the frequent long bursts at
moderate $\dot{M}$ with unstable He-ignition in a mixed H/He environment
(case 1). Additionally, the ``inhomogeneous'' burst samples are
typically accumulated from transients spanning a large range of $\dot{M}$, and thus
likely arise from a mixture of ignition types (although in each case
including at times bursts characteristic of mixed H/He fuel). 
In contrast, the frequent short bursts at moderate $\dot{M}$
($\gamma=0.02$--0.2) {\it cannot}\/ be reconciled with the theoretically
expected ignition conditions if H is present in the accreted fuel at
approximately solar mass fractions $X=0.7$. These bursts all exhibit fast
rise and decay times, despite (at times) insufficiently short recurrence
times to exhaust the accreted H between bursts (unless the accreted
H-fraction is sub-solar, or the CNO abundance above solar). 
Similarly, neither the bursts at high $\dot{M}$ nor the two
anomalous cases in Table \ref{classify} can be easily identified with any
of the theoretically-predicted regimes of bursting. 

In summary, by comparing the two most prolific bursters in our sample, we found
compelling evidence that the burst behaviour as a function of
$\gamma$ or $S_Z$ is not the same for all sources. In a broader sense, the
behaviour of most sources appears consistent with one or more of the
ignition cases expected theoretically for sources accreting mixed H/He.
However, there are several notable exceptions, the most numerous of which
is a significant subgroup of sources which consistently exhibit short
bursts, some of which are confirmed ultracompact systems and thus
He-accretors.

\subsection{Burst frequency as a function of accretion rate}
\label{global}

Assuming ideal conditions (accretion over a constant fraction of the
neutron star surface,
complete consumption of the accreted fuel) and neglecting 
transitions between H- and He-ignition, we expect the burst rate to
increase monotonically with $\dot{M}$.
This increase may be expected to continue until the temperature in the
fuel layer reaches the point where He burning is stabilized (expected
around $\dot{M}_{\rm Edd}$), at which time bursts will essentially cease. 
This pattern is expected both for sources accreting mixed H/He or pure He,
although in the former case steady H-burning between bursts contributes to
higher temperatures in the fuel layer, and thus earlier ignition (and
more frequent bursts).
In order to compare the theoretical predictions with observations, here we
construct burst rate curves as a function of accretion rate (by proxy) for
various subsamples of the bursts detected by \xte.

As a consequence of the diversity in burst behaviour established for
sources contributing to the \xte\/ sample (see \S\ref{diversity}), 
we further restrict the sample of bursters from which we draw our bursts to
construct global curves of burst rate as a function of 
accretion rate.
The presence at times of bursts with long timescales $\tau$ for sources in
the first five categories listed in Table \ref{classify} leads us to
conclude
that these sources accrete mixed H/He, and
thus can be expected to follow broadly consistent patterns of ignition as
a function of accretion rate. We also include in this sample, which we
refer to as \sgammah\ (or \sszh, when binning on $S_Z$) the bursts from
unclassified systems (which likely do not contribute sufficient bursts or
observations 
to significantly affect the combined sample). We {\it
exclude}\/ the sources which exhibit frequent short bursts, bursts at high
$\dot{M}$, and the two anomalous cases in Table \ref{classify}. 
We hypothesize that, like 3A~1820$-$303, the sources which consistently show
short bursts likely accrete H-poor material, and these systems form a
comparison sample which we refer to as \sgammahe\ (or \sszhe; see Table
\ref{samples} for full descriptions of each of these groups).
We further exclude from both samples, bursts with very short recurrence
times $<1$~hr. Such bursts occur episodically (see \S\ref{dblbursts}), and
do not reflect the ``steady'' burst behaviour we are attempting to measure
here.

\begin{figure*}
 \plotone{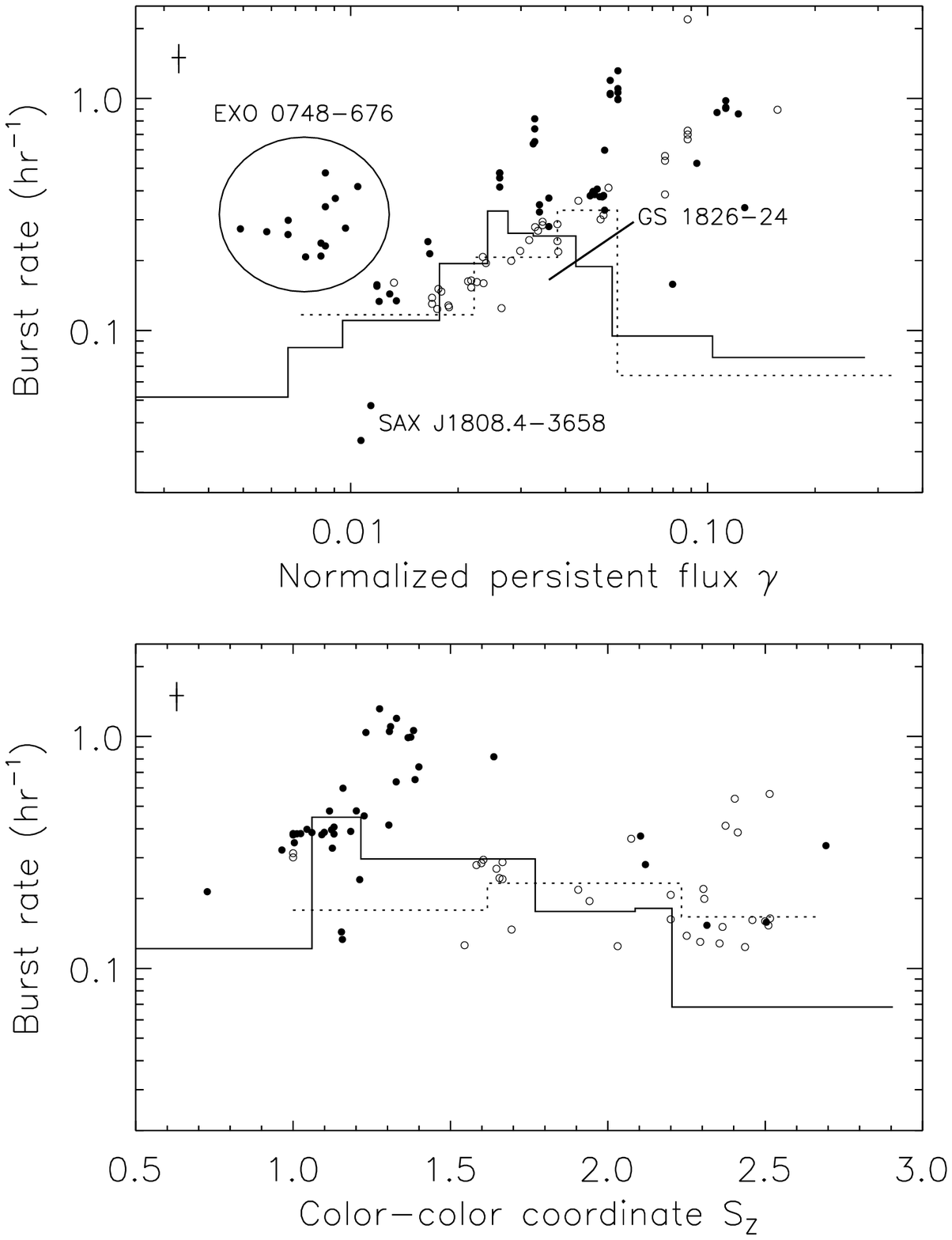}	
 \figcaption[burst_rate.eps]{Burst rates 
measured by \xte\ as a function of normalized persistent flux $\gamma$
({\it upper panel}) and color-color coordinate $S_Z$ ({\it lower panel}). 
We plot separately the mean rates for the sources with evidence of H-rich
fuel (Table \ref{classify}, samples \sgammah/\sszh\ in Table
\ref{samples}; {\it solid histogram}) and the sources with consistently
short bursts (samples \sgammahe/\sszhe, {\it dotted histogram}).
Bursts 
were combined in groups of 
$\approx50$ 
to calculate the ensemble average within each bin;
a representative error bar
indicating the $1\sigma$ uncertainty is shown at the top-left of each
panel.
Burst rates
($1/t_{\rm rec}$) for pairs of bursts 
are also shown ({\it filled/open circles}, for H-rich or H-poor accretors,
respectively).
Individual measurements for notable sources are indicated; we also show 
the approximate $\gamma$-burst rate relationship
derived for GS~1826$-$24 ({\it thick solid line}) by \cite{gal03d}.
For both samples,
the ensemble-averaged burst rate increased with $\gamma$ and reached
a maximum in the range $\gamma\simeq0.02$--0.05 (equivalent to $\approx10^{37}\
\eps$).
The H-accretors appear to reach a peak burst rate at lower $\gamma$ than
the He-accretors.
The behavior as a function of $S_Z$ was less consistent.
  \label{rates} }
\end{figure*}

The mean burst rate for the remaining sample of
\numsrburst\ bursts from
\numsrcrate\ sources 
contributing to \sgammah\
increased with $\gamma$ throughout
the range $\gamma\approx0.001$--0.03 (Fig. \ref{rates}, top panel).
The peak rate of 0.3~hr$^{-1}$ was reached at a $\gamma$ corresponding to
a source luminosity of
0.5--$1\times10^{37}\ \eps$.  
At the lowest accretion rates $\gamma<0.01$ the mean burst rate 
was between 0.04--0.05~hr$^{-1}$.
Above $\gamma=0.03$ the mean
burst rate decreased steadily, reaching a minimum of
0.08~hr$^{-1}$ 
in the range $\gamma=0.1$--0.3. 
No bursts from \sgammah\ were observed at $\gamma>0.3$; the only sources
bursting at such high persistent flux levels are GX~17+2 and Cyg~X-2,
which were excluded from this sample.
The mean burst rate for these two sources
was also in the range 0.07--0.08~hr$^{-1}$.

For comparison, the source which most closely matches predictions of
theoretical burst models for mixed H/He ignition is GS~1826$-$24,
also known as the ``Clocked Burster''
due to its consistently regular bursts (e.g.  \citealt{clock99},
\citealt{corn03a};
see also \S\ref{s1826}).
The burst recurrence time measured by \xte\/ between 1997--2002 decreased
significantly 
in response to a gradually increasing persistent flux \fper\ (i.e.
$\gamma$),
proportional to $F_p^{-1.05\pm0.02}$ \cite[]{gal03d}.
This behaviour is very close to that expected 
assuming $F_p\propto \dot{M}$ and unvarying fuel composition, and a
subsequent comparison of burst lightcurves with time-dependent model
predictions confirm an accreted composition of roughly solar metallicity
and H-fraction $X$ \cite[]{heger07b}.
Without any radius-expansion bursts or a well-defined color-color diagram,
$\gamma$ or $S_Z$ values could not be calculated for 
GS~1826$-$24, and thus the bursts detected from this source by \xte\/ were
not included in 
sample \sgammah\ or \ssz. Instead, we estimated the equivalent
$\dot{M}$ based on a distance of 6~kpc \cite[derived from
comparisons to theoretical ignition models][]{gal03d} at 0.06--$0.10\dot{M}_{\rm
Edd}$, which corresponds to $\gamma=0.04$--0.06 (taking into account the
mean bolometric correction of $c_{\rm bol}=1.678$). The
corresponding variation of burst rate with $\gamma$ observed for
GS~1826$-$24 between 1998--2002 is shown as the solid line in Fig.
\ref{rates} ({\it top panel}). 
Surprisingly, the mean burst rate calculated for sources contributing
to \sgammah\ reaches a maximum below the $\gamma$-range in which
GS~1826$-$24 is active, and in that range the mean rate is in fact
{\it decreasing}\/ rather than increasing.
This discrepancy may be attributed to a systematic error in our
calculation of $\gamma$ for GS~1826$-$24, relative to the sources
comprising \sgammah.
On the other hand, the only other source exhibiting long, regular bursts
similar to those of GS~1826$-$24 in the \xte\/ sample, KS~1731$-$26, also does
so at $\gamma=0.05$.

As we have seen with the variation in burst timescales (\S\ref{ts}), 
the degeneracy between $\gamma$ and $S_Z$ may affect the averaged
burst rates in the 
range $\gamma=0.01$--0.06.
The bursts which fall in this range arise from observations both from the
``island'' state (with $S_Z\la1.5$) and the ``banana'' state ($S_Z\ga2$;
see Fig. \ref{comparison}).
The variation of burst rate as a function of $S_Z$ indicates that these two
groups of bursts have substantially different intrinsic  bursting
frequencies (Fig. \ref{rates}, lower panel),
along with their different timescales (Fig. \ref{taugammahist}).
The long burst timescales suggest that 
GS~1826$-$24 is likely persistently in the ``island'' state,
and the discrepancy between the properties of those
bursts and the broader sample in Fig. \ref{rates} can 
be explained if most of the sources contributing to \sgammah\ in the range
$\gamma=0.04$--0.06 are instead in the ``banana'' state.
This is supported by the distribution of observation-averaged $S_Z$ values
(for the sources contributing to \ssz) with $\gamma$ in this range; 83\% of
the observations (79\% in terms of exposure) have $S_Z\geq2$.
Further highlighting the different intrinsic burst rates is the fact that
the number of bursts arising from ``island'' and ``banana'' state
observations with $\gamma=0.04$--0.06 is approximately equal, despite the
factor of 4 greater exposure in the latter spectral state.
Thus, it is likely that the burst rates in the range
$\gamma=0.01$--0.06  presented here are measured from a sample of bursts
with larger dispersion in accretion rate than their $\gamma$-values would
suggest. For this reason, apparent variations in the burst rates in this
$\gamma$-range must be viewed with some caution. On the other hand, the
burst rates at higher $\gamma\ga0.06$ and lower $\gamma<0.01$ are perhaps
likely to be measured from more uniform samples, since $\gamma$ and $S_Z$
are more closely related in those ranges (see Fig. \ref{comparison}).

The variation of burst rate with $\gamma$ for \numsrbursthe\ bursts
from \numsrcratehe\ sources with
consistently short bursts (sample \sgammahe, excluding the short-recurrence
time bursts; Table \ref{samples}) was 
similar, reaching a comparable maximum rate although at a slightly higher
$\gamma=0.04$--0.05 (Figure \ref{rates}). Above $\gamma=0.05$, however,
the burst rate dropped much more rapidly by a factor of $\approx5$.
Between $\gamma=0.03$ and 0.05 the burst rate for sample \sgammah\ is
decreasing, while for sample \sgammahe\ is increasing. While the variation
is only for a few bins in these data, we note that this result is also
found from the data from 4U~1636$-$536 and 4U~1728$-$34 alone (see Fig.
\ref{twosources}).

The recurrence time for closely-spaced burst pairs from \sdt\ (see Table
\ref{samples}), interpreted as an
instantaneous burst rate (as distinct from a steady recurrence time
measured from a series of regular bursts), are also shown in Fig. \ref{rates}.
Below $\gamma=0.01$, the rates measured from burst pairs were mostly in
the range 0.2--0.6~hr$^{-1}$, well above the mean value; these bursts are
all from EXO~0748$-$676. 
That the mean burst rate 
underestimates the burst pair measurements from EXO~0748$-$676
may be due
to the inclusion in this sample of several
pure He-accretors in the low-$\dot{M}$ ``giant'' burst class (see
\S\ref{diversity}). These sources are likely to exhibit a much lower
burst rate in this range, since H-ignition is not possible.
For most of the burst pair measurements, the recurrence times for the
sources in \sgammah\ are systematically shorter than for the sources in
\sgammahe\ at comparable $\gamma$-values. This is consistent with the
expected effects of steady H-burning to boost the burst rates.
Between $\gamma=0.01$ and 0.03 the rates measured from burst pairs
roughly follow the mean values, although above $\gamma=0.03$ there are
also measurements substantially in excess of the 
mean. 
Sources with frequent ($>0.5$~hr$^{-1}$) bursts in the range $\gamma>0.03$
include 4U~1636$-$536, 4U~1705$-$44, and SAX~J1748.9$-$2021; in addition,
bursts from EXO~1745$-$248
and 4U~1746$-$37 (which are
omitted from Fig. \ref{rates}) also fall in this region of the plot.

We note that there are selection biases
in \sdt\ towards regular bursts with short recurrence
times, and 
in particular it is difficult to unambiguously measure burst intervals
$\ga10$~hr from \xte\/ observations (SAX~J1808.4$-$3658 is a notable
exception; see \S\ref{s1808}).  As a result, the locus of 
measurements from burst pairs cannot be considered
representative.  Nevertheless, 
the extent to which the instantaneous burst rates
can deviate from the ensemble-averaged value clearly indicates that
the statistical errors on the binned measurements significantly
underestimate the true extent of variation.

We also show the burst rate as a function of $S_Z$ for the 
sources
for which it was possible to parametrize the color-color
diagram, and excluding the anomalous sources 
and short-recurrence time bursts, in Fig. \ref{rates} ({\it lower panel}). 
As with sample \sgamma, we divided the bursts from sample \ssz\ based on
evidence for mixed H/He bursts (\numszrate\ sources totalling \numszburst\
bursts, sample \sszh), or lack thereof (\numszratehe\ sources, with
\numszbursthe\ bursts, sample \sszhe; see Table \ref{samples}).
The burst rate for sample \sszh\ reached a maximum of
$\approx0.4$~hr$^{-1}$ around $S_Z=1.2$, and subsequently
decreased steadily as $S_Z$ increased further to around
0.04~hr$^{-1}$.
In contrast, the bursts from sample \sszhe\ exhibited no steady trend over
the range in $S_Z$ in which they were observed. 
As with the mean rates as a function of $\gamma$, the rates measured from
burst pairs as a function of $S_Z$ correspond only loosely to the
mean values. We found variations of up to an order of magnitude compared
to the mean rates over all sources.
The rates largely reflect the behavior of 4U~1636$-$536 and 4U~1728$-$34, which
make up the dominant fraction within each sample (Fig. \ref{twosources}).

It is important to keep in mind that
the bursts contributing to the mean rates in 
the lower panel of Fig \ref{rates} 
are a subset of those contributing to the
upper panel, since the $S_Z$ values could only be determined for a limited
number of sources (see \S\ref{pers}).
There are systematic biases which could be introduced by comparing samples
from different groups of bursts, particularly if one or a few sources
dominate the samples.
However, 
when
binned instead as a function of $\gamma$, samples \sszh\ and \sszhe\
exhibited the same variations as found for samples \sgammah\ and
\sgammahe, used for the top panel of Fig. \ref{rates}.  Thus, the two
samples have comparable variation in burst rate as a function of $\gamma$.

\subsection{Burst energetics and the 
role of steady burning}
\label{secalpha}

The burst rate is perhaps the most straightforward quantity relating to
thermonuclear burning to measure, but the possibility of systematic
variations in the burst properties as a function of $\dot{M}$ means that
the rate alone does not uniquely identify the nature of ignition or the
composition of the burst fuel.
We have previously seen how the rescaled fluence, $U_b$, of PRE bursts is
anticorrelated with $\dot{M}$ (using $\gamma$ as a proxy), 
leading to an anticorrelation between the burst timescale $\tau$ and
$\gamma$ (sec.  \S\ref{ts}). This anticorrelation indicates unambiguously
that, at least
for the PRE bursts, the amount of fuel at ignition becomes systematically
less at higher $\dot{M}$. Since at higher $\dot{M}$ we expect hotter
temperatures in the fuel layer and hence earlier ignition, this effect is
qualitatively consistent with theory. This trend should be independent of
whether or not the burst exhibited radius-expansion; however, we find no
systematic correlation for the non-PRE bursts in \sgamma\ between $U_b$
and $\gamma$. 

This discrepancy between theoretical expectations and our observations
mainly arises from the already-established diversity of our burst sample.
Burst sources accreting mixed H/He may ignite via unstable H- or
He-burning, while He-accretors may only ignite via He burning.
For sample \sgammahe, comprising bursts from sources
with consistently short bursts (which we infer primarily accrete He; Table
\ref{classify}), we find that $U_b$ is significantly anticorrelated with
$\gamma$ both for PRE and non-PRE bursts. However, for sample \sgammah,
which includes bursts from sources which we infer are accreting mixed
H/He, we find a significant anticorrelation of $U_b$ with $\gamma$ (as
before) for the PRE bursts, but a significant {\it correlation}\/ instead
for the non-PRE bursts. That is, for the non-PRE bursts from the sources
which we infer are accreting mixed H/He, the bursts tend to get more
intense as $\dot{M}$ increases.
This correlation arises due to the presence at low $\gamma$ of 
bursts arising from H-ignition. Such events, typified by the bursts from
EXO~0748$-$676 (see \S\ref{s0748}), tend to be much more frequent and less
energetic than bursts at higher $\gamma$ values. As $\dot{M}$ increases
through the range where ignition transitions from H- to He-burning (case
3 to case 2), the burst recurrence time increases dramatically, so that
the amount of fuel accumulated also increases, leading to much more
energetic bursts. This transition, we suggest, is the dominant effect
leading to the correlation between $U_b$ and $\gamma$ uniquely for the
non-PRE bursts from H-rich accretors. It is still possible that within
each of the
two ignition regimes,  the theoretically-expected anticorrelation between
$U_b$ and $\gamma$ may be measured. However, the much smaller ranges of
$\gamma$ spanned by each bursting regime, coupled with the difficulty
discriminating between the two types of bursts (and the
scatter on the measured $U_b$ values at any $\gamma$ value) likely make
such measurements unfeasible.

\begin{figure*}
 \plotone{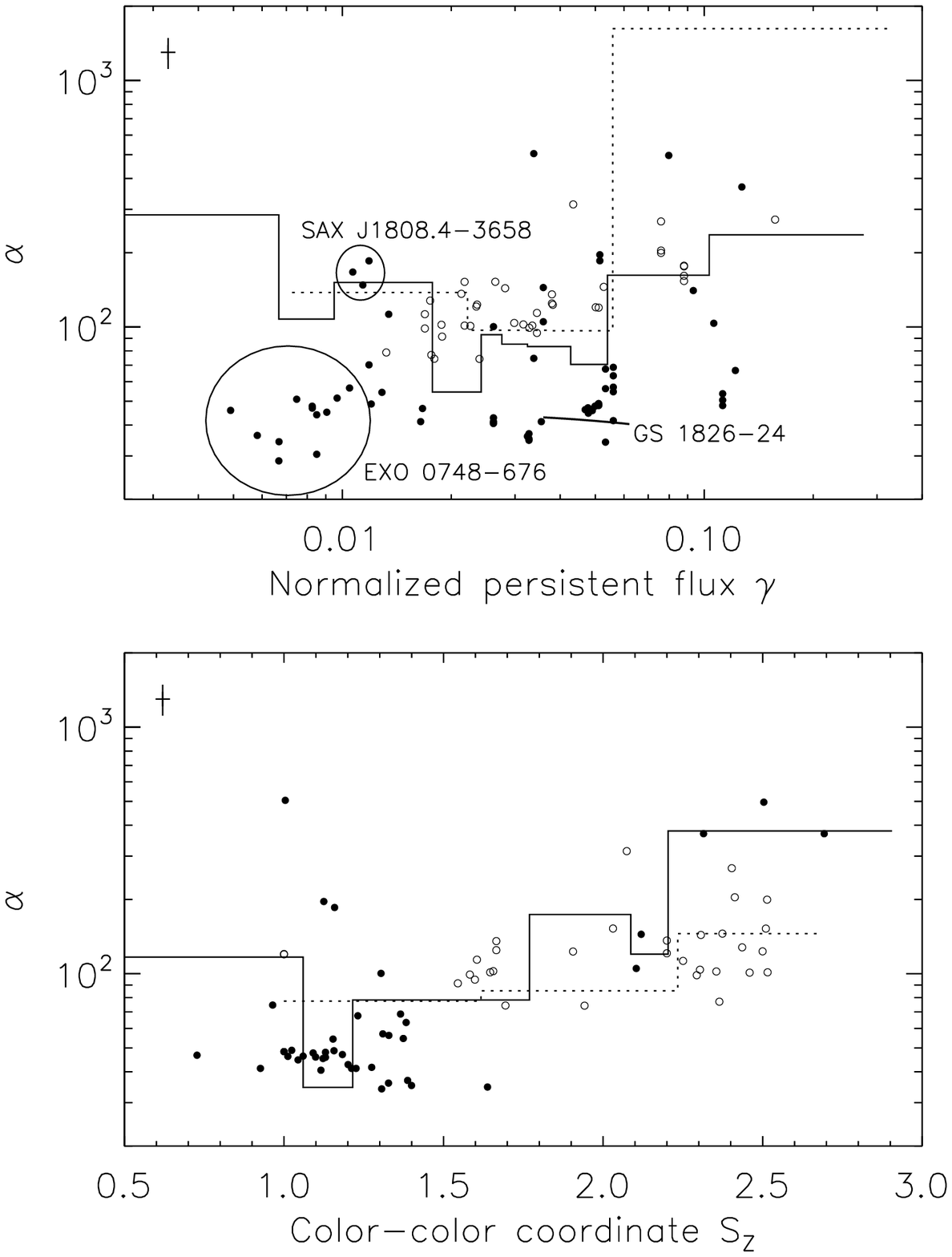}	
 \figcaption[alpha.eps]{Measured $\alpha$-values for bursters observed by
\xte\ as a function of normalized persistent flux $\gamma$ ({\it upper
panel}) and color-color coordinate $S_Z$ ({\it lower panel}), from the
same burst samples as in Fig. \ref{rates}.
The mean \malpha\ for bursts from sample
\sgammah\ and \sgammahe\ are plotted separately as histograms ({\it solid
line, dotted line}, respectively); measurements from burst pairs are
also shown ({\it filled, open circles}).
Bursts 
were combined in groups of 
$\approx50$ 
for binning;
a representative error bar for the \malpha\ values is shown at the
top-left.
Measurements for notable sources are indicated; we also show
the approximate trend of $\alpha$ with $\gamma$
derived for GS~1826$-$24 ({\it thick solid line}) by \cite{gal03d}.
For sample \sgammah, he mean \malpha\ reaches a minium roughly 
where the burst
rate reaches its maximum, around $\gamma\approx0.03$ or $S_Z\approx1$,
decreasing above and below.
For sample \sgammahe, \malpha\ increases by more than an order of
magnitude at $\gamma=0.05$, but exhibits little variation elsewhere, or
with $S_Z$.
  \label{alphagamma} }
\end{figure*}

Neither the burst rates, nor the burst fluences, allow us to unambiguously
determine the energetics of the bursts. For this reason, we here study the
variation in the ratio of burst to persistent fluence, $\alpha$,
calculated for burst pairs and as a mean \malpha\ for the combined
samples (see \S\ref{energetics}).
The binned \malpha\ for \sgammah\ reached a minimum of $\approx50$ 
between $\gamma=0.02$--0.03, 
where the burst rate also reached a maximum
(Fig.  \ref{alphagamma}, upper panel).
Below this $\gamma$ range \malpha\
varied between 80 and 250,
while above $\gamma\approx0.05$ increased steadily up to
$\approx400$.
For the sources with consistently short bursts (sample \sgammahe), \malpha\
exhibited rather different behaviour, $\approx100$ below
$\gamma=0.05$ and
increasing by more than an order of magnitude abruptly above. This step
coincides with an abrupt drop in the burst rate (Fig. \ref{rates}).
As with the burst rates, the $\alpha$
measurements from burst pairs differed from
the binned values by almost an order of magnitude. 
There was no discernable trend in the 
$\alpha$-values as a function of $\gamma$, although the values for the
members of \sgammahe\ were systematically larger in the mean than those of
members of \sgammah.
While \malpha\ agreed well with the $\alpha$-measurements from burst pairs for
SAX~J1808.4$-$3658 at $\gamma=0.01$, \malpha\ was significantly in excess
of the measured values for EXO~0748$-$676 just below.
We also show in the upper panel of Fig. \ref{alphagamma} the measurements
from GS~1826$-$24, which were found to decrease only
slightly as the persistent flux increased \cite[]{gal03d}. This decrease
was attributed to steady H-burning between bursts, which results in a
slightly higher fuel H-fraction (and hence lower $\alpha$) as the burst interval
decreases.
The \malpha-values calculated for sample \sgammah\ (which does not include
GS~1826$-$24) was somewhat in excess of the values for that source.
The variation of $\alpha$ with $S_Z$ for sample \sgammah\ is rather more
consistent between the mean and the measurements from burst pairs (Fig.
\ref{alphagamma}, lower panel),
with $\alpha$ increasing rather steadily from $\sim30$ to 
$\sim400$ as $S_Z$ increases from 1 to 2.5. Below $S_Z=1$,
$\alpha\sim100$. For the bursts from \sgammahe, both the \malpha\ and
burst pair $\alpha$ measurements varied only slightly, between $\approx80$
and 150.
The mean \malpha-values for GX~17+2 and Cyg~X-2 (not shown in Fig.
\ref{alphagamma}) were in the range 1300--4100.

It is worthwhile here to revisit the earlier measurements of $\alpha$ as a
function of accretion rate from a large sample of bursts by \cite{vppl88}.
Those authors found a steep increase in $\alpha$ from $\sim10$ to
$\sim10^3$ as $\gamma$ increased from 0.01 to 0.3.
The earlier sample was assembled from measurements in the literature from
individual burst sources, including representatives from both the
\sgammah\ and \sgammahe\ samples
(Table \ref{samples}). 
As we have seen, these two samples exhibit systematically different burst
behavior as a function of accretion rate; several sources in sample \sgammahe\
are established ultracompact systems, and thus primarily He-accretors,
while those in \sgammah\ exhibit bursts with profiles indicative of mixed
H/He fuel and thus must also accrete hydrogen.
In particular, measurements of $\alpha\sim10$ for bursts from
EXO~0748$-$676 at $\gamma\approx0.01$ solely determined the low-$\gamma$
end of the \cite{vppl88} correlation.  These bursts likely arise from
H-ignition of mixed H/He fuel (i.e. case 1); were 3A~1820$-$303 (another
source contributing to the correlation) accreting at $\gamma=0.01$, it is
not possible that it would exhibit bursts with $\alpha\sim10$, since
3A~1820$-$303 likely accretes pure He from a white dwarf donor (see
\S\ref{s1820}).
Thus, we suggest that the \cite{vppl88} correlation is not indicative of
burst behavior for any one source over the range of $\gamma$ spanned by
the combined sample. In contrast, assuming our subsamples \sgammah\ and
\sgammahe\ contain sources with similar accreted composition (as the
similarity in burst behavior suggests), the
corresponding variation of burst rate and $\alpha$ should go closer to
reflecting realistic behavior for any of the sources in each sample.

The variation of \malpha\ as a function of $\gamma$ differs for the two
samples \sgammah\ and \sgammahe, to a greater extent than the burst rates.
This result lends additional credence to the hypothesis that the sources
contributing to \sgammahe\ (see Table \ref{samples}) exhibit
systematically different burst energetics. That the mean and burst pair
$\alpha$-values for these bursts are consistently $\ga80$ supports the
hypothesis that they are primarily He-accretors, as we inferred from the
consistently short
burst timescales. We note that the range of \malpha\ is significantly different
when binning on $\gamma$ or $S_Z$, which may be attributed to the small
number of sources contributing to the latter sample; only two sources
(4U~1702$-$429 and 4U~1728$-$34) with consistently short bursts have
measured $S_Z$ values in our sample (see Table \ref{samples}; in contrast,
\sgammahe\ includes 190 bursts from 6 sources). 
Perhaps the most remarkable feature
of the \malpha\ variation in this sample is the significant increase
observed at $\gamma=0.05$. Even for pure-He fuel, the maximum $\alpha$
value expected is 120 (equation \ref{alphatheory}), strongly suggesting that
some of the assumptions that enter into the theoretical prediction 
break down. Values of $\alpha\gg120$
indicate
that some process is reducing the energy generated from unstable burning.
One candidate is the onset of steady He-burning, although
$\gamma\approx0.05$ is an accretion rate approximately an
order-of-magnitude lower than where this phenomenon is predicted to
commence theoretically.

\subsection{Boundaries of theoretical ignition regimes}
\label{boundary}

Having measured the variation in the properties of bursts observed by
\xte\/ as a function of accretion rate (by proxy), we here assess how well
these measurements agree with theoretical predictions. 
In particular, while the transition values of $\dot{M}$ between the
different ignition regimes are generally well-reproduced
by different numerical models, there have been few attempts to verify the
values observationally.

We consider the bursts from sources contributing to sample
\sgammah\ (Table \ref{samples}) only, since it is these systems that we
infer accrete the mixed H/He that makes the full range of ignition cases
possible.
Ignition in mixed H/He (cases~1 and 3) and pure He (case~2)
environments are distinguishable by small and large values of
$\alpha$ respectively, since H nuclei contribute much more energy per
nucleon than He (i.e. $Q_{\rm nuc}$ is larger in equation \ref{alphatheory}).
The transition to steady H-burning (between case 3 and 2), expected around
$\dot{M}\sim0.01\dot{M}_{\rm Edd}$, should thus result in an increase in
$\alpha$, as well as a drop in burst rate.
Examination of Figs. \ref{rates} and \ref{alphagamma} indicate that these
expectations are largely unmet. 
The burst rate
for sample \sgammah\ increases steadily from $\gamma\approx0.001$--0.03 (Fig.
\ref{rates}), with no evidence for a decrease. 
We do measure an
increase in the binned $\alpha$ values between the two bins spanning
$\gamma=0.01$, but the increase is only weakly significant. Furthermore, the
lower value is still $\approx100$, which is too large for H-rich fuel.

Given the systematic uncertainties affecting $\gamma$, it is possible that
the transtion may take place somewhat above or below $\gamma=0.01$.
Indeed, the burst rate begins to decrease above $\gamma=0.03$, corresponding
to 0.5--$1\times10^{37}\ \eps$ (based on our uncertainties in the true
value of the Eddington limit; equation \ref{ledd}). This decrease has also
been observed in the behaviour of individual sources observed with {\it
BeppoSAX}/WFC
\cite[]{corn03a},
and was attributed by those authors to 
the onset of stable H burning \cite[i.e. the transition between case 3 to
case
2 of][]{fhm81}.
There are several reasons why 
the \xte\/ data do not support this conclusion.
First, there is no increase in $\alpha$ measured coincidental with
the decrease in burst rate. In fact, between $\gamma\approx0.03$--0.06 the
averaged $\alpha$ values is constant to within the errors, possibly
decreasing slightly (Fig. \ref{alphagamma}). 
Second, at the transition between case~3 and case~2 ignition
we would expect to see only a local decrease in the burst
rate, followed by a subsequent increase at even higher accretion rates,
through the transition from case~2 to case~1. Instead, the burst rate
continues to decrease, up to the limit of $\gamma$ at which we observe
sources contributing to \sgammah. 
Third, detailed analysis of bursts from individual sources confirm the
presence of case~2 or case~1 bursts at comparable or lower accretion
rates.
Analysis of bursts from
SAX~J1808.4$-$3658 at $\gamma\approx0.01$ indicates that the burst fuel is
largely He, so that the accreted hydrogen must have been significantly
reduced by steady burning prior to the bursts \cite[i.e. case~2
ignition;][]{gal06c}.
Additionally,
analysis of the bursts from GS~1826$-$24 which occur in the range
$\gamma=0.04$--0.06 confirm that these arise from case~1 ignition
(\citealt{gal04a}; see also \citealt{zand04b}), suggesting that the
transition 
from case~3 to case~2
must take place at lower $\dot{M}$ ($\gamma$). 

The 
position $S_Z$ on
the color-color diagram (for those sources where it can be measured, i.e.
members of \ssz; Table \ref{samples}) offers an alternative explanation
for the decrease in burst rate above $\gamma=0.03$ which may not be
related to the nuclear physics. The diversity of the burst timescales
and $S_Z$ values for bursts at $\gamma>0.03$ discussed in \S\ref{ts},
coupled with the intrinsic variation in burst rates in the ``island''
($S_Z\approx1$) and ``banana'' ($S_Z>2$) states (Fig. \ref{rates}, lower
panel) indicates that this decrease in burst rate is related to the
transition between these spectral states. That is, above $\gamma=0.03$,
we increasingly (although not exclusively) find sources 
with $S_Z\geq2$, where the bursts are much less frequent.
While there may be a thermonuclear component to this
transition, it is not consistent with the expected behaviour through the
transition between cases~3 and 2 (or cases~2 and 1, for that matter).
As has been suggested earlier, the bursts occuring above $S_Z=2$ may be
the ``delayed mixed bursts'' predicted by \cite{ramesh03} to occur at
higher accretion rates than case~1 bursts (see e.g. Table
\ref{burstregimes}). The tendency for short burst timescales
indicates H-poor fuel, 
which may constrain the extent of steady burning prior to ignition in this
regime.

It seems more probable that the transition in burst behaviour around $S_Z=2$
is related to the onset (or increase in the rate) of stable He-burning,
even though the inferred accretion rate is much lower than predicted from
models \cite[e.g][]{vppl88}.  Millihertz oscillations observed around the
transition to the ``banana'' state \cite[]{rev01} may arise from
marginally stable burning, as suggested by \cite{hcw07}. If fuel is
accreted onto some fraction of the neutron star at a high enough (local)
rate for He-burning to stabilize, the $\dot{m}$ onto the remainder may be
small enough to still permit infrequent bursts, where the accreted
H-fraction is reduced by stable H-burning. 
The variation in effective gravity between the equator and higher
latitudes, which depends upon the spin rate, may also contribute to such
effects \cite[]{cn07b}.
A detailed
study linking the predictions of theoretical models for accretion
\cite[e.g.][]{is99} with time-dependent ignition models
\cite[e.g.][]{woos03,ramesh03} may help to establish the validity of this
hypothesis.

Another contributing factor to the lack of evidence for a transition
between case~3 and 2 is the apparent scarcity of bursts
from H-ignition at low $\gamma$. At the lowest accretion rates a
significant fraction of sources exhibit infrequent, energetic bursts
consistent with largely-He fuel (Table \ref{classify}). It is difficult to
determine whether these bursts arise from case~2 ignition following
exhaustion of the accreted hydrogen by steady burning, as in the case of
SAX~J1808.4$-$3658 \cite[][see also \S\ref{s1808}]{gal06c}, or from
accretion and ignition of intrinsically H-poor material, as might be
expected from an ultracompact system with an evolved donor. Indeed,
several of the systems with the lowest accretion rates are candidate
ultracompacts, based on their X-ray to optical luminosity, or other
indirect evidence.

The effect of including these systems in \sgammah\ 
(which we have done in the absence of evidence precluding them accreting
hydrogen) will be to reduce the mean burst rate at low $\gamma$, and
correspondingly increase the mean $\alpha$-values.
The best candidates for H-ignition bursts in this
$\gamma$-range are the frequent, weak,
long-timescale bursts from EXO~0748$-$676 observed
in the range $\gamma=0.005$--0.01.
Many of these bursts are separated by no
more than 5 hours, and both the long timescales and the typical
$\alpha\approx40$ indicate 
a large fraction of H in the burst fuel. While these properties are also
shared by case~1 ignition bursts, exemplified by bursts from GS~1826$-$24,
the bursts from EXO~0748$-$676 are around a factor of 4 less intense (on
average).
As suggested by
\cite{boirin07a}, and earlier by \cite{gottwald86}, these properties
indicate that the long bursts are ignited by unstable hydrogen
ignition\footnote{Bursts arising from H-ignition and occurring as often as
every few hours are predicted by the two-zone model of \cite{cn07a},
although in this accretion rate range the He is expected to accumulate and
ignite in an energetic pure-He burst every few days. Such intense
He-bursts interrupting trains of much weaker and more frequent H-bursts
have not been observed to date.}
(i.e. case~3), rather than helium ignition \cite[which likely
triggers the PRE bursts from this source at higher $\dot{M}$; e.g. ][]{wolff05}.
As we have seen, the individual rates for burst pairs from EXO~0748$-$676
are in excess of the mean rate (Fig. \ref{rates}), and the measured
$\alpha$-values lower than the mean (Fig. \ref{alphagamma}).
We note that the upper $\gamma$-limit for the long-timescale, frequent
bursts from EXO~0748$-$676 is $\gamma\approx0.01$, which coincides with
the $\gamma$-value at which the short, infrequent case~2 bursts from
SAX~J1808.4$-$3658 are observed (e.g. Fig. \ref{rates}). If the case~3
ignition is indeed giving rise to the bursts from EXO~0748$-$676, and
these sources can be taken as representative of the larger group which
accrete mixed H/He fuel, then this appears to confirm the prediction that
the case~3 to 2 transition takes place close to $\dot{M}/\dot{M}_{\rm
Edd}=0.01$.

While the case~3 to 2 transition may be expected to result in a fairly
sharp change in burst properties with $\gamma$, the case~2 to 1 likely is more
subtle.  As the
accretion rate increases, the burst recurrence time is expected to
steadily decrease so that eventually steady H-burning will not be
complete,
leading to a corresponding steady decrease in $\alpha$.
This is roughly as observed. The burst rate increases steadily up to
$\gamma=0.03$, while the mean $\alpha$-value decreases (on average) from
$\approx150$ at $\gamma=0.01$ down to 70 at $\gamma=0.05$.
The bursts from
GS~1826$-$24 represent the best-studied example of case~1 ignition, and
exhibit a steep \cite[almost 1:1;][]{gal03d} increase in burst rate with
persistent flux (i.e.  $\gamma$) although with only a slight decrease
in $\alpha$. 
The mean \malpha-values for sample \sgammah\ show little
variation in the inferred $\gamma$-range in which the bursts from
GS~1826$-$24 are observed, but are systematically higher by a factor of
$\approx1.5$ (Fig.
\ref{alphagamma}). However, as already noted, the mean burst rate above
$\gamma=0.03$ is already decreasing, likely because the observations in
this range are a mix of ``island'' ($S_Z\sim1$) and ``banana'' ($S_Z>2$)
states, with their corresponding distinctive bursting behaviours.
We conclude that the transition from case~2 to case~1 burning
occurs between $\gamma=0.01$ and 0.03, 
but note that the weakness of this transition likely makes any further
improvement on observational constraints unlikely.

The variations of burst rate and $\alpha$ with $S_Z$ are even harder to match
with theoretical expectations. With fewer sources contributing to sample
\sszh\ (Table \ref{samples}),
the measured variations in $S_Z$ are on a much
coarser grid of binned values.
Similar to the burst rate dependence on $\gamma$, we seen a maximum rate
at $S_Z=1.1$, and lower burst rates above and below. However, the
data permits only one bin at lower values of $S_Z$, so that we do not
resolve the increase to reach the maximum burst rate, and
likely masking the transition between
case~3 and case~2.  Also at this accretion rate, \malpha\ reaches a
minimum of $\approx30$, increasing above and below. 
It seems likely that the transition to case~1
ignition must also take place near $S_Z=1$, since the 
subsequent decrease in burst rate and increase in \malpha\ as $S_Z$
increases through $S_Z=2$ is the same effect largely
contributing to the drop in burst rates above $\gamma=0.03$. As we have
discussed above, these variations in burst rate likely
do not correspond to one of the transitions in ignition conditions that
are predicted by models.

\subsection{Millisecond oscillations}
\label{milosc}

Millisecond oscillations during thermonuclear bursts have been detected to
date in \noscsrc\ sources,
including two that also show
persistent pulsations (SAX~J1808.4$-$3658 and XTE~J1814$-$338).
The properties of burst oscillations detected in \xte\/ data have
previously been described in detail in a number of papers \citep[see][for a
review]{sb03}.  Several key questions remain.
One of the most puzzling aspects is that the burst oscillations 
are frequently detectable far into the decays of the bursts
\citep{stroh97, smb97}.  During the burst rise, it is expected that the
burning will spread to cover the entire surface of the neutron star, so
that subsequent anisotropy in the emission will be small. Nevertheless,
oscillations are frequently detectable as much as 10~s after the burst
peak.
Thus, for the combined sample of bursts observed with \xte, we have
examined the detectability of oscillations, where in the bursts they
occur, and the properties of the bursts that produce them. 

\begin{figure*}
 \plotone{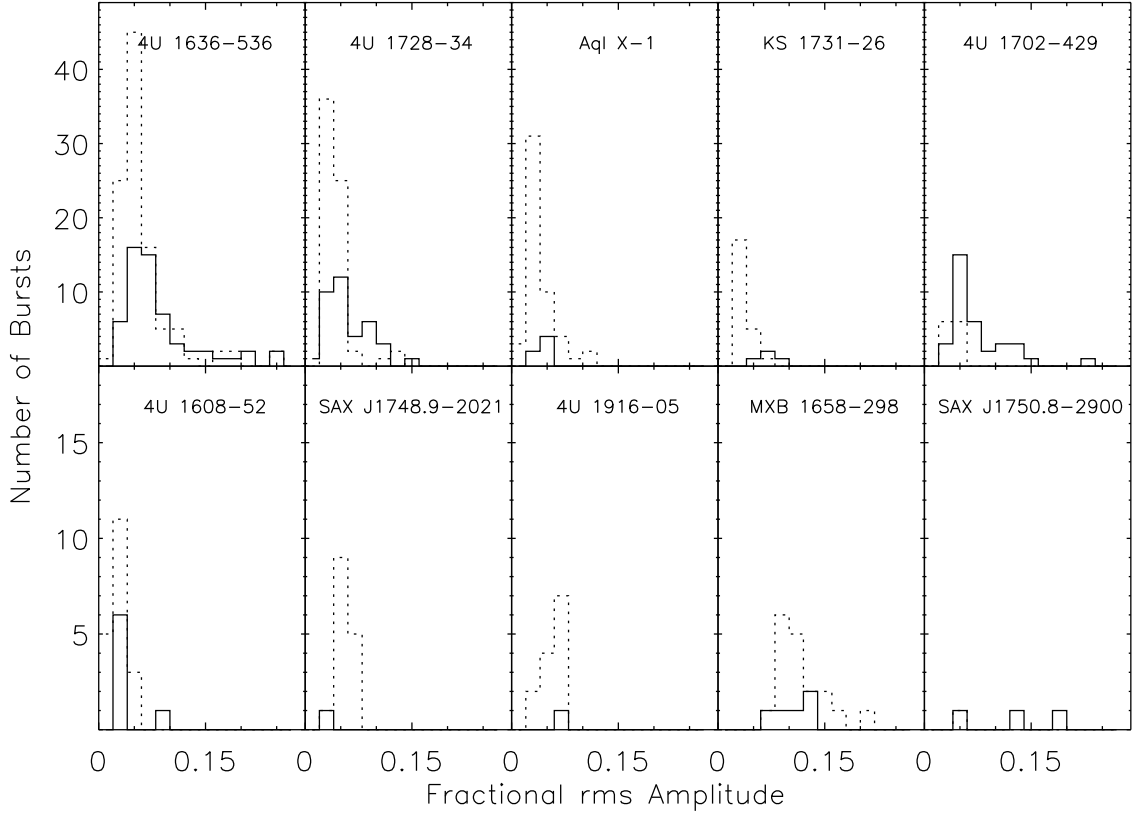}	
\caption{Histograms of the largest fractional rms amplitudes of detected
oscillations ({\it solid lines}), and of the upper limits on the rms 
amplitude
in the burst decay when oscillations are not detected ({\it dotted 
line}).
\label{fig:rmshist}}
\end{figure*}

We list the numbers of bursts with oscillations by source in
Table~\ref{tab:osc}.  We omitted from our search the bursts from
EXO~0748$-$676, in which the 45~Hz oscillations cannot be detected in
single bursts \cite[]{villarreal04}, and the burst from IGR~J17191$-$281
\cite[]{mark07a},
for which the \xte\/ data 
was not available at the time of writing.
We searched for, but did not detect, oscillations near the 377.3~Hz spin
frequency in the two bursts observed by \xte\/ from the millisecond
pulsar, HETE~J1900.1$-$2455.  The persistent pulsations in this system
were present only in the first two months of the outburst, which is
unusual \cite[]{gal07a}.
No bursts
have been detected by \xte\/ from 4U~0614+09, preventing a search for the
oscillations detected in that source by {\it Swift}/BAT \cite[]{stroh08a}.

%
\begin{deluxetable*}{lccccccccccc}
\tablecolumns{11}
\tabletypesize{\scriptsize}
\tablewidth{0pc}
\tablecaption{Summary of Burst Oscillations\label{tab:osc}}
\tablehead{
\colhead{} & \colhead{$\nu_{\rm spin}$} & \colhead{Number} &
\multicolumn{9}{c}{Number of Oscillations} \\
\colhead{Source} & \colhead{(Hz)} & \colhead{of Bursts\tablenotemark{a}} & \colhead{Total} &
\colhead{4~s} & \colhead{Rise} & \colhead{R$+$P} & 
\colhead{Peak} &
\colhead{P$+$D} & \colhead{Decay} & \colhead{R$+$D} & 
\colhead{Cont.}
}
\startdata
4U~1916$-$053 & 270 & 14(14) &   1 & \nodata & \nodata & \nodata & \nodata & \nodata & \nodata & \nodata &   1 \\
\hspace{0.5cm}with PRE & & 12(12) &   0 & \nodata & \nodata & \nodata & \nodata & \nodata & \nodata & \nodata & \nodata \\
XTE~J1814$-$338\tablenotemark{b} & 314 & 28(28) &  28 &   3 & \nodata & \nodata & \nodata &   2 & \nodata &   3 &  20 \\
4U~1702$-$429 & 329 & 47(47) &  35 &   1 &   2 &   5 &   3 &   9 &   8 &   2 &   5 \\
\hspace{0.5cm}with PRE & & 5(5) &   0 & \nodata & \nodata & \nodata & \nodata & \nodata & \nodata & \nodata & \nodata \\
4U~1728$-$34 & 363 & 106(104) &  38 &   2 &   5 &   1 &   3 &  11 &   9 &   1 &   6 \\
\hspace{0.5cm}with PRE & & 69(69) &  18 &   2 &   1 & \nodata & \nodata &   6 &   8 & \nodata &   1 \\
SAX~J1808.4$-$3658\tablenotemark{b} & 401 & 6(4) &   4 & \nodata &   1 & \nodata & \nodata & \nodata &   1 &   2 & \nodata \\
KS~1731$-$260 & 524 & 27(27) &   4 & \nodata &   1 & \nodata & \nodata &   1 &   1 &   1 & \nodata \\
\hspace{0.5cm}with PRE & & 4(4) &   3 & \nodata & \nodata & \nodata & \nodata &   1 &   1 &   1 & \nodata \\
1A~1744$-$361 & 530 & 1(1) &   1 & \nodata &   1 & \nodata & \nodata & \nodata & \nodata & \nodata & \nodata \\
Aql~X-1 & 549 & 57(55) &   6 & \nodata & \nodata &   1 &   2 &   3 & \nodata & \nodata & \nodata \\
\hspace{0.5cm}with PRE & & 9(9) &   5 & \nodata & \nodata & \nodata &   2 &   3 & \nodata & \nodata & \nodata \\
MXB~1659$-$298 & 567 & 26(25) &   6 &   1 &   2 & \nodata &   1 & \nodata &   1 &   1 & \nodata \\
\hspace{0.5cm}with PRE & & 12(12) &   5 &   1 &   2 & \nodata & \nodata & \nodata &   1 &   1 & \nodata \\
4U~1636$-$536 & 581 & 172(169) &  59 &   6 &   5 &   5 &   1 &   4 &  17 &  12 &   9 \\
\hspace{0.5cm}with PRE & & 46(45) &  38 &   4 &   1 & \nodata & \nodata &   4 &  14 &  10 &   5 \\
GRS~1741.9$-$2853\tablenotemark{c} & 589 & 8(8) &   2 &   2 & \nodata & \nodata & \nodata & \nodata & \nodata & \nodata & \nodata \\
\hspace{0.5cm}with PRE & & 6(6) &   2 &   2 & \nodata & \nodata & \nodata & \nodata & \nodata & \nodata & \nodata \\
SAX~J1750.8$-$2900 & 601 & 4(4) &   3 & \nodata &   2 & \nodata & \nodata & \nodata & \nodata &   1 & \nodata \\
\hspace{0.5cm}with PRE & & 2(2) &   2 & \nodata &   1 & \nodata & \nodata & \nodata & \nodata &   1 & \nodata \\
4U~1608$-$52 & 620 & 31(29) &   7 & \nodata &   2 & \nodata &   1 &   1 &   1 & \nodata &   2 \\
\hspace{0.5cm}with PRE & & 12(12) &   7 & \nodata &   2 & \nodata &   1 &   1 &   1 & \nodata &   2 \\
\tableline
Total & & 527(515) & 194 & 15 & 21 & 12 & 11 & 31 & 38 & 23 & 43 \\
PRE & & 182(180) &  84 &  9 &  8 &  0 &  3 & 15 & 26 & 15 &  8 \\
\enddata
\tablenotetext{a}{The first number is the number of bursts observed from
this source, while the number in parentheses is the number of bursts
which were searched for oscillations.}
\tablenotetext{b}{These sources also exhibit persistent pulsations
at the listed frequency. Note that 
for XTE~J1814$-$338, only the last burst observed exhibited marginal
evidence for PRE.}
\tablenotetext{c}{We attributed bursts with oscillations from the
Galactic center region to this source; see \S\ref{gcbo}. }
\end{deluxetable*}


Excluding the millisecond pulsars (and sources with
less than five bursts total), between
7 and 75\% of bursts exhibited
oscillations. Two of the millisecond pulsars with bursts (SAX~J1808.4$-$3658
and XTE~J1814$-$314) exhibited oscillations in every burst detected;
oscillations were detected in the only burst from 1A~1744$-$361, and three
of the four bursts observed from SAX~J1750.8$-$2900.
Excluding these four, the next most frequent burst oscillation source was
4U~1702$-$429, with 75\% of bursts exhibiting oscillations.

Also in Table~\ref{tab:osc} we
summarize the number of bursts with oscillations detected in each part of the
lightcurve: rise (R), peak (P) and decay (D; see \S\ref{ibo}). For many non-PRE bursts,
oscillations were detected in all three phases, i.e. continuously (last
column).
In the 4th column of Table \ref{tab:osc} we list the number of bursts for
which we detected no oscillations in the 1-s FFTs, but did detect
oscillations in the 4-s intervals.
We found that
55\% of the oscillations were detected in the rises of bursts,
54\% in the peaks, 
75\% in the tails, 
and 8\% only in the 4-s FFTs (for which the location
could not be determined). 
Even if all the oscillations which were only detected in the 4-s FFTs
were actually present only during the burst rise or peak, the most
frequent part of the burst in which oscillations were detected remains the
burst tail.
We confirm the tendency for oscillations to be interrupted during the
burst peak for radius-expansion bursts; of the bursts with oscillations
detected at the peak, 69\% were from non-radius expansion bursts.
Oscillations in the burst rise were also preferentially (65\%) found in
non-radius expansion bursts; only oscillations in the burst tails showed
no preference for the presence or absence of radius expansion, being
equally prevalent in each type of burst.

In Table \ref{osctbl} we list the properties of the oscillations for
individual bursts from each source: where the oscillations occurred, the
maximum (Leahy-normalized) power, and the mean \% rms.
The bursts in which oscillations were detected were 
unremarkable, compared to the entire sample.
We found oscillations in long bursts (with $\tau$ up to 39.4~s), as well as
in short; although the proportion of bursts with $\tau>10$ exhibiting
oscillations was, at 25\%, rather lower than the proportion of all bursts
(from the burst oscillation sources) with $\tau>10$ (43\%). Previously we
saw that bursts with $\tau>10$ are associated with small $\alpha$-values,
indicating mixed H/He fuel (see \S\ref{ts}).
The distribution of burst separations were similar; the shortest
wait time to a burst with oscillations was 13.6~min; the shortest overall
for any of the bursts from sources with oscillations was 
between $4.3<\Delta t<6.4$~min (see \S\ref{dblbursts}). The distribution
of (normalized) fluences $U_b$ was also similar.
Histograms of the
amplitudes of the detected oscillations are displayed by source in
Figure~\ref{fig:rmshist} ({\it solid lines}).  The rms amplitudes 
are typically between
2\% and 20\%, with a median amplitude of about 5\%.
When oscillations
were not detected, we report the upper limit on the rms amplitude from the
first 5~s of the decay in Figure~\ref{fig:rmshist} ({\it dotted lines}).
The median values of the upper limits are typically lower than the
detected oscillations. This indicates that the failure to detect oscillations
is generally a consequence of lower amplitudes, and not a result of a lack 
of sensitivity in the relevant bursts.

It has previously been noted that the properties of the 
bursts in which strong oscillations are observed are correlated 
with the spin frequency of the neutron star 
\citep{muno00, franco01, vs01, muno01}. Specifically, \citet{muno01}
claimed, based on the sample of bursts taken through 2001 March, that 
oscillations appeared preferentially in bursts without radius expansion
when the neutron star was spinning slowly ($<$400 Hz), and in bursts with 
radius expansion when the neutron star was spinning rapidly ($>$400 Hz).
A re-examination of the trend using data taken through 2003 August 
revealed that the division between slow and rapid rotators was not 
absolute, in that oscillations were often observed in bursts both 
with and without radius expansion in both classes of source \citep{muno04a}. 
This is 
also evident in our full sample from Table~\ref{tab:osc}, in which only 
18\%
of oscillations appear in bursts with radius expansion in slow rotators, 
whereas 
73\%
of oscillations in bursts with radius expansion in 
rapid rotators. Therefore, a trend is present, although 
the properties of the bursts are not the only determinant
of whether or not oscillations occur. 

\begin{figure}
 \epsscale{1.2}
 \plotone{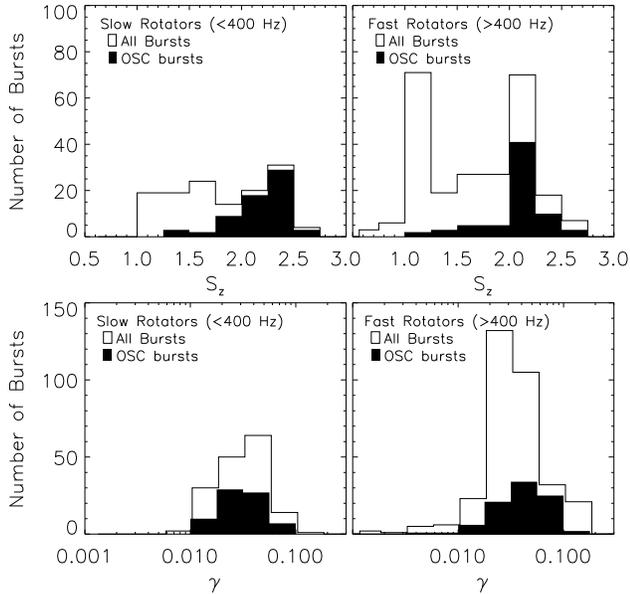}	
\caption{Illustration of how the detection of oscillations depends 
upon accretion rate, for the sources in Figure~4. In each panel, the
open histogram displays the number of bursts as a function of $S_Z$
({\it top panels}) or $\gamma$ ({\it bottom panels}), and the solid
histogram indicates the number of those bursts in which oscillations
were detected. The {\it left panels} are for slow rotators
($\nu$$<$400 Hz), and the right panel for fast rotators ($\nu$$>$400
Hz). For both slow and fast rotators, oscillations are more likely to
be detected in bursts at large values of $S_Z$. No trend is seen as a
function of $\gamma$.
\label{fig:whereosc}}
\end{figure}

We point out that the sample of burst sources with slow ($<400$~Hz)
oscillations, i.e.  4U~1916$-$053, XTE~J1814$-$338, 4U~1702$-$429
and 4U~1728$-$34 
is dominated by sources for
which we consistently observe short bursts (see Table \ref{classify}).
Conversely, none of the burst sources with consistently short bursts are
included in the sample of systems with fast ($>400$~Hz) oscillations.
Thus, the previously identified differences between these systems
may instead arise from the mechanisms that give rise to consistently short
bursts (or not). 
Indeed,
\cite{muno04a} suggested that the apparent trend was actually a
consequence of two separate tendencies as 
sources move through the color-color diagram ($S_Z$): 
first, that
oscillations tend to be observed only at high $S_Z$, and
second, that 
as $S_Z$ increased, bursts from slow rotators (i.e. predominantly sources
with consistently short bursts) became
less likely to exhibit radius expansion, whereas bursts from
rapid rotators (sources with evidence for mixed H/He accretion) became
more likely to exhibit radius expansion.  For the six sources with
well-defined $S_Z$ values (Fig.~\ref{fig:sz}), we display when
oscillations are detected in Figure~\ref{fig:whereosc}.  
Of the bursts with $S_Z$$>$1.75, 
57\%
exhibited oscillations. However, only 
9\% of bursts with 
$S_Z$$\le$1.75 exhibited oscillations. In fact, 
87\%
of the bursts 
with oscillations had $S_Z$$>$1.75. Similar trends are seen for both 
slow and fast rotators. 
We found no comparable preference for oscillations to be found in bursts
at high $\gamma$, likely a consequence of the degeneracy between this
parameter and $S_Z$, as explored earlier (see \S\ref{global};
\S\ref{pers}).

\begin{figure}
 \epsscale{1.2}
 \plotone{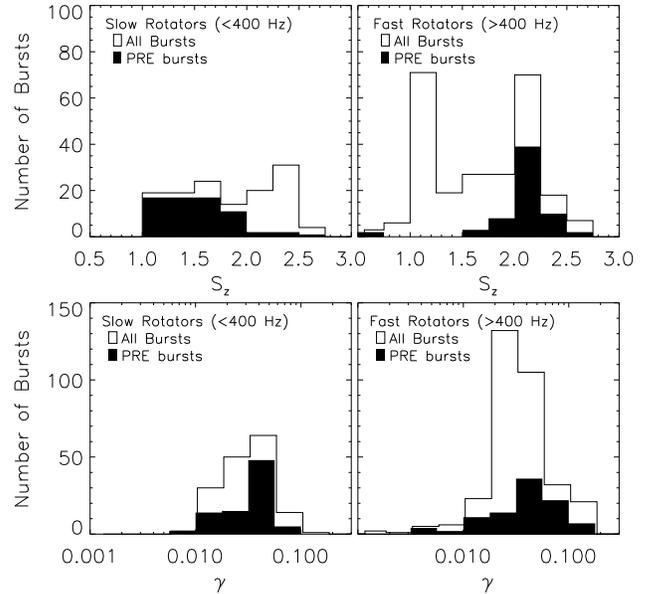}	
\caption{
Illustration of how the occurrence of radius expansion depends 
upon accretion rate for the sources in Figure~4. 
In each panel, the open histogram displays 
the number of bursts as a function of $S_Z$ ({\it top panels}) 
or $\gamma$ ({\it bottom panels}), and the solid histogram indicates
the number of those bursts in which radius expansion was observed. The 
{\it left panels} are for slow rotators ($\nu$$<$400 Hz), and the right
panel for fast rotators ($\nu$$>$400 Hz). For slow rotators, radius 
expansion tends to occur in bursts at low $S_Z$. For fast rotators, 
radius expansion occurs at high $S_Z$. No trends are observed as a
function of $\gamma$.
\label{fig:wherepre}}
\end{figure}

We examine when radius expansion occurs in Figure~\ref{fig:wherepre}. 
If we consider all 
six neutron stars with $S_Z$ values (regardless of their rotational
period), radius expansion is 
found in 
41\% of bursts with $S_Z$$>$1.75 and 
35\% of bursts with 
$S_Z$$\le$1.75. However, if we separate the sources into slow and fast
rotators, we find different behaviors. 
For slow rotators (i.e. 4U~1702$-$429 and 4U~1728$-$34), radius expansion
is found in only 
24\% of bursts with $S_Z$$>$1.75, but in 
86\% of bursts
with $S_Z$$\le$1.75. 
For fast rotators, the opposite trend is evident: 
radius expansion is found in 
49\% of bursts with
$S_Z$$>$1.75, but in only 
6\% of bursts with $S_Z$$\le$1.75. 
The vast majority 
(90\%) of radius-expansion bursts occurred in the fast rotators at
$S_Z$$>$1.75, while for the slow rotators, only 23\% did.

The predominance of systems with consistently short bursts in the
group of slow ($<400$~Hz) oscillators suggests a link between the
physics which determines the burst properties and which sets the
neutron-star rotation speed.
We consider two possible explanations.
As suggested previously, it is plausible that the slow rotators are
largely accreting from degenerate companions, so that the accreted
material is H-deficient.  XTE~J1814$-$338, at 314~Hz a slow oscillator, is
the exception to this rule, since it shows consistently long, weak bursts
possibly arising from H-ignition of mixed H/He fuel (see \S\ref{s1814}).
If this explanation is valid, it would suggest that a difference in the
evolutionary history of systems with and without degenerate companions leads
to the difference in spin periods.

The second possibility is that the rotation rate of the neutron stars
determines how much mixing occurs in the layer of accreted fuel before a
burst, and hence the burst properties. \cite{pb07} suggested that the 
larger shear between the disk and the neutron star for the slow rotators
may cause the CNO-rich ashes to be mixed into freshly accreted fuel,
leading to rapid exhaustion of the accreted H via steady hot-CNO burning
and hence helium-rich (i.e.  consistently fast) bursts. Since the group of
fast oscillators contains at least one ultracompact source
(4U~1916$-$053), this cannot be the only mechanism. 
Additional numerical calculations are required to trace the evolution in
composition of the fuel layer for mixed H/He accretion, in order to
establish whether this effect should occur, and whether it is strong
enough to contribute significantly the trends above.
Observationally, determining the orbital periods for the remaining systems
with slow spins (4U~1702$-$429 and 4U~1728$-$34) can also help to
constrain the accreted composition and hence the role of shear-mediated
mixing.

\subsection{Theoretical challenges}
\label{challenges}

Having attempted a genuinely global study of burst behaviour covering all
the sources observed by \xte, it is appropriate to summarise here the
remaining phenomenological aspects revealed by our analysis which cannot
be understood by current burst theories.

\subsubsection{Bursts at high accretion rate}
\label{hibursts}

There are two aspects to the thermonuclear burst behaviour observed at
high accretion rates (above $\sim10^{37}\ \eps$, or $\gamma\ga0.03$) that
are contrary to the predictions of burst theory. First, that the burst
rate decreases for most sources with increasing $\dot{M}$, despite the
more rapid accumulation of fuel. Previous authors have noted that the
bursts tend to become shorter as well as less frequent as the apparent
$\dot{M}$ increases \cite[e.g.][]{bil00}. The picture from the
\xte\/ sample is somewhat more complex. Within a range of accretion rates
($\gamma=0.03$--0.06) there is a mix of frequent (long) and infrequent
(short) bursts. These bursts largely appear to be distinguished by the
source's position on the color-color diagram at the time they occur; the
long bursts occur when the sources are still in the ``island'' state
($S_Z<2$, where it can be measured) while the short bursts occur only once
the sources transition to the ``banana'' state ($S_Z>2$). It is appealing
to try to resolve the degeneracy between the accretion rate
parametrisation based on $\gamma$ or $S_Z$, e.g. by referring to our set
of derived bolometric correction values ($c_{\rm bol}$; see \S\ref{pers}).
However, we find that $c_{\rm bol}$ is inversely correlated with $S_Z$
(for the observations where we measure both parameters), so that
attempting to correct the $\gamma$ (which is based on the 2.5--25~keV
flux) using this parameter will tend to increase the degree of overlap
between the two samples, rather than separate them. 

Second, although bursting behaviour ceases for most sources at
$\approx10^{38}\ \eps$ ($\gamma\approx0.3$), there are two well-known
outliers exhibiting at times frequent burst behaviour at much higher
$\dot{M}$, GX~17+2 and Cyg~X-2. The range of $\gamma$ spanned by both the
bursts and observations provides some additional details here. No bursts
are observed at all in the range $\gamma=0.3$--0.7; GX~17+2 and Cyg~X-2
are the only sources with bursts at $\gamma>0.7$. The gap in observations
is smaller, although still present; no sources are observed at
$\gamma=0.3$--0.5. We note that there is no shortage of sources accreting
at these levels \cite[e.g.][]{grimm02}, although it is difficult to
completely rule out selection effects since we only analyse sources that
are already known as burst sources.

The bursts that are observed at $\dot{M}/\dot{M}_{\rm
edd}\approx\gamma\approx1$ are intriguing in themselves. The \xte\/ sample
provides examples of both the very short (down to $\tau=1$~s) bursts from
both GX~17+2 and Cyg~X-2, and the intermediate-duration bursts ($\tau=100$--400~s)
from GX~17+2 \cite[cf. with][]{twl96}. It is puzzling why the bursts are so diverse at these
accretion rates, let alone why they occur at all. Furthermore, GX~17+2 also
exhibits superbursts \cite[]{zand04a}, while as yet no such long or
intermediate duration bursts have been detected from Cyg~X-2.

\subsubsection{``Double'' bursts}
\label{dblbursts}

Thermonuclear bursts with extremely short recurrence times (``double'' or
''prompt'' bursts) have long presented a challenge to our understanding of
burst physics.  Their recurrence times of $\ga5$~min are too short for
sufficient fuel to accumulate to allow ignition by unstable thermonuclear
burning \cite[see e.g.][]{lew93}.  
The ``classical'' double burst consists of an initial bright burst
followed by a much fainter secondary, although bursts which are more similar
in fluence are also observed, as well as groups of three closely-spaced
bursts \cite[e.g.][]{boirin07a}.  
We note also that there are cases of weak burst-like events {\it preceding}
otherwise normal bursts by only a few seconds, in 4U~1636$-$536 (see
\S\ref{s1636}), 4U~1709$-$267 \cite[see \S\ref{s1709} and][]{jonk04} and
SAX~J1808.4$-$3658 \cite[]{bhatt06d}. It is presently not clear whether
these events can be attributed to the same processes that give rise to
burst pairs with separations $\ga5$~min, or if they are instead more
closely related to double-peaked bursts \cite[see e.g.][]{bhatt06c}.

An important question is whether the fuel for the second (and possible
subsequent) bursts
is residual material left unburnt by the first,
or if it is newly accreted.
Studies of bursts from EXO~0748$-$676 indicate
that some small fraction
($\approx10$\%) of the fuel left over from the previous burst contributes
to the fluence of subsequent bursts \cite[]{gottwald87}, 
perhaps reaching ignition by
mixing deeper into the NS atmosphere \cite[]{ww84,fuji87}.

\begin{figure}
 \epsscale{1.2}
 \plotone{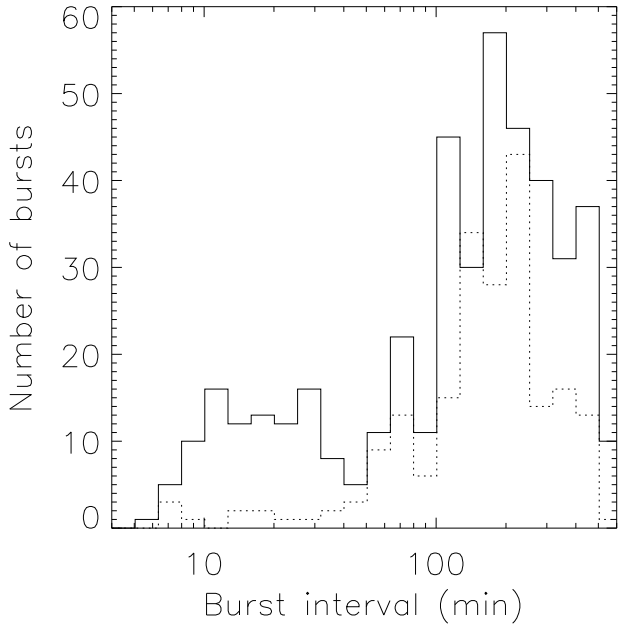}	
 \caption{Distribution of observed burst recurrence times $\Delta t$ below
10 hours. The solid histogram shows the distribution for all the bursts
observed by \xte; the dotted histogram shows the distribution for those
pairs of bursts with measured $\alpha$-values (i.e. in sample \sdt; see
Table \ref{samples} and \S\ref{energetics}).
The majority of the latter subset of bursts are regular, and so represent
conventional burst behaviour. Regular bursting with recurrence times
$\la30$~min is not observed; pairs of bursts this close in time are only
observed episodically.
 \label{tdeldist} }
\end{figure}

Good instrumental sensitivity is required to detect the typically much
weaker secondary (and sometimes, tertiary) events, and the \xte\/ sample
likely contains a greater number of multiple events than other large burst
samples (for example, accumulated by the {\it BeppoSAX}/WFC).
We plot the small-$\Delta t$ end of the burst recurrence time distribution
in Fig. \ref{tdeldist}. The distribution exhibits a deficit of bursts with
$\Delta t=40$--100~min; the 90~min satellite orbit falls within this
range, so that this deficit is likely a consequence of the regular data
gaps (the typical duty cycle is $\approx0.65$) due to Earth occultations of
bursters. For $\Delta t>100$~min it is possible 
that intermediate bursts have been missed in data gaps, so that the actual
recurrence time is $1/2$ or $1/3$ the measured value.
The burst pairs with $\Delta t\la30$~min occur within uninterrupted
stretches of data, and so
we extract this subsample
for our analysis. 
These burst pairs occurred primarily while the normalised persistent flux
of the bursting sources were between $\gamma=0.02$ and 0.04, or (where it
could be measured; see \S\ref{pers}) $S_Z=0.8$ and 1.8.

Of the \npair\ pairs of bursts 
with $\Delta t\leq30$~min, we found \ntriple\ cases
where short-recurrence time pairs immediately followed each other, i.e. a
series of three bursts with recurrence times $<30$~min. Sources
contributing to this subset were 4U~1705$-$44, Rapid Burster (4),
4U~1636$-$536, 2E~1742.9$-$2929 (3), EXO~1745$-$248 and 4U~1608$-$52. 
Previously, the only known source to exhibit triple bursts was
EXO~0748$-$676 \cite[]{boirin07a}. We
also found two instances of four bursts following each other, all with
recurrence times $<30$~min, from the Rapid Burster and 4U~1636$-$536.

\begin{figure}
 \epsscale{1.2}
 \plotone{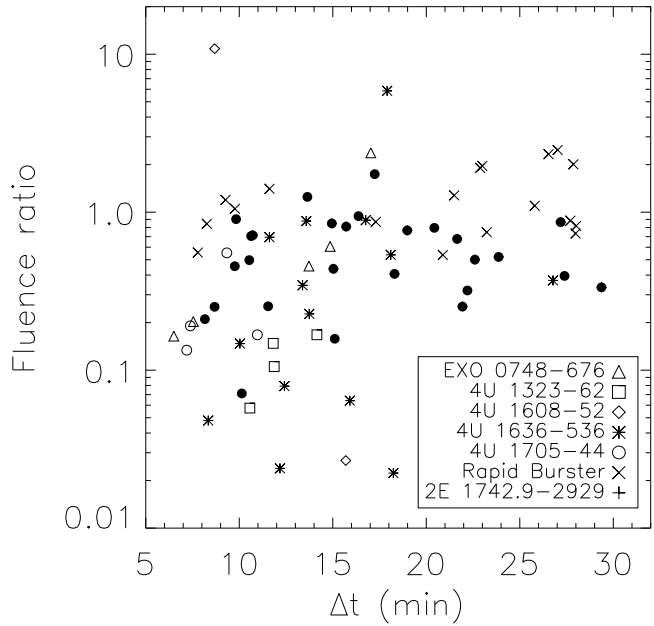}	
 \caption{Ratio of fluences $E_{b,1}/E_{b,0}$ of pairs of bursts plotted
as a function of their separation $\Delta t$.  We plot the values for
various sources of interest contributing to the sample with different
symbols; the remaining measurements come from XTE~J1710$-$281 (1),
4U~1735$-$44 (1), EXO~1745$-$248 (2), 4U~1746$-$37 (5), Aql~X-1 (6) and
Cyg~X-2 (3) ({\it filled circles}).
The shortest measured ratio plotted here was for a pair of bursts from
EXO~0748$-$676 separated by just 6.5~min, in which
the first was more than 5 times more energetic than the one which
followed (i.e. fluence ratio of 0.164). These ``classical'' double bursts
(i.e. with $E_{b,1}\la0.2E_{b,0}$) are not found with recurrence times
longer than $\approx20$~min.
 \label{tdelratio} }
\end{figure}

We show the ratio of fluences for pairs of bursts with $\Delta t<30$~min
in Fig. \ref{tdelratio}. 
For \npairbad\ pairs we could not measure the fluence for one of the
pairs, because either the high time-resolution data did not cover the
burst, or because the burst was too faint to reliably measure the fluence.
The pair of bursts with possibly the shortest measured recurrence time, from
4U~1608$-$52 on 2001 November 21 15:29, fell into the former category; we
can only limit the separation to $4.3<\Delta t<6.4$~min, since we did not
observe the start of the second burst due to an unexplained data gap. The
next shortest recurrence time was from a pair of bursts from
EXO~0748$-$676 on 2003 July 1, at 6.5~min.
With examples of three and even four bursts following each other
in rapid succession, we hypothesize that many of the short-recurrence time
bursts from the Rapid Burster may be type-II rather than
type-I bursts.  
In few of the short-recurrence time bursts from the Rapid
Burster are temperature variations present at a significance of more than
$3\sigma$, so that we cannot confirm these bursts as arising from
thermonuclear ignition.
For the remaining sources, less than half of the bursts have highly
significant ($>5\sigma$) temperature variations. However, the distribution
of burst intervals and fluence ratios is not substantially different if
only these bursts are considered.
A significant fraction of the bursts pairs had
comparable fluences, i.e. $E_{b,1}/E_{b,0}\approx1$. 
The ``classical'' double bursts were observed with fluence ratios of
$\la0.2$, and recurrence times of 
$\Delta t= 6.5$--18~min (Fig. \ref{tdelratio}). 

Double bursts appear to occur primarily (but not exclusively) from sources
in the H-ignition (i.e. case 1) regime \cite[the best example is
EXO~0748$-$676; see][]{boirin07a}. Furthermore, we note only one case
of a burst with $\Delta t<30$~min from a source with consistently short
bursts (see Table \ref{classify}); this burst is from 4U~1735$-$44, with
$\Delta t=27.4$~min. This system incidentally has a measured orbital
period of 4.65~hr (see \S\ref{s1735}), and an optical counterpart
remarkably similar to that of 4U~1636$-$536 \cite[e.g.][]{august98}
extending to the presence of emission lines from hydrogen. Thus, it seems
that short ($\la30$~min) recurrence time bursts are strongly associated
with sources accreting hydrogen. While the effects of sedimentation at low
accretion rates \cite[]{pbt07} may play a role in igniting these bursts,
it remains to be seen if a more general mechanism can explain the
properties of the entire sample.

\subsubsection{Individual sources: 4U~1746$-$37 
  and 4U~1735$-$44}
\label{indbursts}

There are a number of sources whose behaviour revealed by the \xte\/
observations is particularly difficult to reconcile with current burst
theory. 
4U~1746$-$37, the bright bursting source in NGC~6441 (see
\S\ref{s1746}), exhibits long, regular bursts characteristic of mixed H/He
burning at apparent accretion rates (parametrised by either $\gamma$ or
$S_Z$; see \S\ref{pers}) much higher than other sources (Fig.
\ref{tauplot}). Sequences of regular bursts are occasionally interrupted by
short, radius-expansion bursts which are ``out of phase'', but which
do not otherwise interrupt the regular bursting \cite[]{gal04a}. The
unusually high persistent flux and X-ray colors, as well as the burst
behaviour, suggest that at times two sources may be bursting independently
in the cluster; however, the lack of spatial information with \xte\/ makes
it impossible to localise the two types of bursts. Additionally, the peak
flux of the (apparent) radius-expansion bursts are significantly below the
expected Eddington flux for a source at the cluster distance (see
\S\ref{pflux}). The spectral variation of these bursts were substantially
different to the other radius-expansion bursts in the \xte\/ sample, and
may require an alternative explanation. On the other hand, the atypical
spectral evolution could be due to geometrical effects common to sources
with high system inclination, as inferred from the presence of regular
X-ray dips attributed to partial eclipses of the neutron star by material
in the outer disk \cite[e.g.][]{gal08a}.

4U~1735$-$44
displayed episodic bursting behaviour in the \xte\/ observations 
with consistently short bursts (rise times
$\lesssim 2$~s, and mean timescale of $\tau=3.7\pm0.8$~s)
characteristic of very H-poor fuel (see Table \ref{classify};
\S\ref{s1735}). A significant fraction (73\%) of these bursts exhibited
radius-expansion.
The other sources identified as consistently exhibiting short bursts are
either known ultracompacts (3A~1820$-$30, 4U~1916$-$053) or do not have a
measured orbital period (4U~1728$-$34, 4U~1702$-$429; Table \ref{classify}). 
4U~1735$-$44 is thus the only system in the group with a measured orbital
period longer than 80~min; at 4.65~hr, it is the only system for
which the companion, V926~Sco, may be a main-sequence star and thus likely
accretes mixed H/He.
While the spectral properties of the optical
counterpart to 4U1735$-$44\ bear a remarkable similarity to 4U~1636$-$536
\cite[e.g.][]{august98}, the burst behaviour is markedly
different; 4U~1636$-$536\ shows at times long bursts, characteristic of H-rich
fuel.
The inferred accretion rate of 4U~1735$-$44 at 0.16--$0.5\dot{M}_{\rm
Edd}$ was typically higher than that
of 4U~1636$-$536, so that the short bursts may arise via the transition
to short bursts noted by \cite{bil00} to take place at $\approx10^{37}\
\eps$. However, the bursts were instead frequent, with typical recurrence
times of $\approx1.5$~hr, rather than the longer recurrence times usually
associated with the burst behaviour transition at high flux.
The difference may instead be related to some physical property
of the sources, perhaps the spin rate. \cite{pb07} studied the effect of mixing between the fuel
and ash layers in He-accretors, and found that these
layers will become mixed preferentially at high accretion rates and low spin frequencies.
Low neutron-star spin frequency results in a larger shear between the
innermost Keplerian-orbiting material in the disk and the neutron-star
surface, enhancing mixing. The ashes are rich in CNO nuclei, which if
mixed into the fuel layer may allow more rapid exhaustion of the accreted
hydrogen through steady burning, perhaps explaining the short time-scales of
bursts from 4U1735$-$44.
While the spin rate in 4U~1636$-$536 has been measured at 581~Hz via burst
oscillations \cite[\S\ref{milosc}; see also][]{stroh02b}, no such
oscillations or pulsations have been detected in 4U1735$-$44. The latter
source may be a key system for future studies of rotation-mediated mixing
and its effect on thermonuclear bursts.

\section{Summary}
\label{summary}

We have analysed a catalog of \burstnum\ thermonuclear (type-I) bursts
observed by \xte\/ from \numbursters\ LMXBs. Although this is not the
largest sample of bursts accumulated by recent satellite missions, the
unparalleled PCA sensitivity offers the best signal-to-noise for precise
measurements of burst flux, fluence, time-scales, and energetics.
Furthermore, the high timing capability has allowed a detailed comparison
of the properties of the bursts and the burst oscillations, where
detected. Below we summarize the principal results obtained through our
study of this sample.

\begin{itemize}
\item We found significant variations in the peak flux of photospheric
radius-expansion bursts from most sources with more than one such burst.
Most of the variation can be attributed to three of the four distinct
classes of PRE bursts; faint symmetric bursts, reaching significantly
sub-Eddington luminosities (in 4U~1746$-$37 and GRS~1747$-$312);
rare, hydrogen-limited bursts (much weaker than normal PRE bursts, which
likely reach the He limit) in 4U~1636$-$536; 
and ``giant'' bursts reaching
fluxes in excess of the Eddington limit, typically the most energetic
bursts from each source.
\item 
We confirmed the previously-observed tendency for the burst duration $\tau$
(the ratio of the burst fluence to peak flux) to
decrease as accretion rate increases \cite[e.g.][]{vppl88}. However,
short-duration bursts at high persistent flux levels were largely
associated with sources in the ``banana'' spectral state, while
long-duration bursts were still observed at similar persistent flux levels
so long as sources remained in the ``island'' state.
As expected from time-dependent ignition models, the timescale $\tau$ was
strongly anticorrelated with the ratio of integrated persistent flux to
burst fluence $\alpha$, and in particular all bursts
with $\tau<10$ had $\alpha>70$ (indicating He-rich fuel), while the
majority of bursts with $\tau>10$ had $\alpha\approx40$ (H-rich).
\item 
We found 
evidence for distinctly different burst behavior as a function of
accretion rate for the two sources with the largest number of
bursts in the sample, 4U~1636$-$536 and 4U~1728$-$34.
The absence of long-duration bursts indicative of H-rich fuel for
4U~1728$-$34 (for which the orbital period is
presently unknown) suggests that the neutron star accretes primarily He
from an evolved donor. 4U~1636$-$536 on the other hand likely accretes
mixed H/He at roughly solar fraction, like the majority of LMXBs.
We identified a number of systems with consistently fast bursts, similar
to
4U~1728$-$34, including two known ultracompact systems: 3A~1820$-$303 and
4U~1916$-$053, and several candidates.
\item We estimated the mean burst rates as a function of accretion rate,
using our qualitative assessment of the burst properties as a guide for
combining samples of bursts from different sources.
We measured a peak burst rate of around 0.3~hr$^{-1}$ at
$\dot{M}\approx0.03\dot{M}_{\rm Edd}$ ($\approx10^{37}\
\eps$) for the systems with evidence for mixed H/He accretion, and
approximately the same maximum rate but at a factor of two larger $\dot{M}$
for the sources with consistently fast bursts.
The burst rates for both samples decreased significantly above this level,
more steeply for the latter group.
The derease in burst rate appears largely due to an increasing fraction of
observations with X-ray colors consistent with the ``banana'' spectral
state; this transition does not appear to be related to any of the
theoretically predicted ignition regimes.
Above $\approx0.3\dot{M}_{\rm Edd}$,
only bursts from GX~17+2 and Cyg~X-2 were detected.
\item We also calculated the mean $\alpha$-values for the two samples of
bursts as a function of accretion rate.
We found 
significantly different behaviour for 
sources which exhibit at times long bursts and sources with
consistently fast bursts,
further evidence for the different compositions of the accreted material
in those two classes.
Perhaps most remarkably, for the systems with consistently fast bursts, we
found an abrupt increase in the mean $\alpha$ of an order of magnitude
(from $\approx100$ to $>1000$) at persistent flux levels corresponding to
an accretion rate of $5\%\dot{M}_{\rm Edd}$.
\item We qualitatively compared the observed burst behaviour with that
predicted by numerical models, and found that the mean behaviour largely
did not match theoretical expectations. Even so, we identified bursts in
the sample corresponding to all three cases of ignition presently
understood
theoretically, as described by
\cite{fhm81}. Case~3
bursts, likely arising from H-igition of mixed H/He fuel were observed from
EXO~0748$-$676 in a range of
accretion rate $\approx0.5$--1\%~$\dot{M}_{\rm Edd}$. At slightly higher
$\dot{M}$ ($\approx2$--5\%~$\dot{M}_{\rm Edd}$) we found infrequent,
short bursts from SAX~J1808.4$-$3658 which have been shown to arise from
He-ignition in an almost pure He enviroment (case 2). 
These observations support the
theoretical prediction that the boundary between case 3 and 2 occurs
around 1\%~$\dot{M}_{\rm Edd}$.
Earlier analyses of the long bursts from
GS~1826$-$24 confirm that these arise from He-ignition in a H/He
environment, where the burst interval is insufficient to completely
exhaust the accreted H (case~1).
\item We analysed all the public bursts from the sources with burst
oscillations and characterised them based on the peak rms and where in the
burst the oscillations are present. 
The previously noted differences in the type of bursts which exhibit
oscillations between the ``fast'' ($>400$~Hz) and ``slow'' ($<400$~Hz) 
oscillators may be related to the prevalence of systems with consistently
fast bursts (for which we infer H-poor accretion) in the latter group. The
data suggests a correspondence between the neutron star spin and the
composition of the accreted fuel (and hence the evolutionary history).
\item 
Finally, we identified four key theoretical challenges highlighted by the
results of our analyses. First, the 
decrease in burst rate above $0.1\dot{M}_{\rm Edd}$, as well as the
occurrence of bursts at $\ga\dot{M}_{\rm Edd}$, remain significant puzzles.
Second, we examined the properties of bursts with very short ($\la30$~min)
recurrence times, identifying burst triplets (three closely-spaced bursts)
from five systems, and instances of burst quadruplets from two
systems. Such bursts appear to occur only in those systems with evidence
for mixed H/He fuel, particularly at very low accretion rates.
Third, the ``out-of-phase'' bursts in 4U~1746$-$37, which
also exhibit radius-expansion at sub-Eddington luminosities. Fourth, 
the case of 4U~1735$-$44, a system with a 4.65~hr
orbital period (and hence not ultracompact) as well as an optical spectrum
similar to that of 4U~1636$-$536, and yet consistently short bursts,
similar to known and suspected ultracompacts including 3A~1820$-$303.
\end{itemize}

\acknowledgments

We are grateful to Pavlin Savov, who created the first version of the
catalogue on which this paper is based for an undergraduate project.
%
We are also indebted to the referee team, whose extensive and detailed
comments helped to significantly improve the presentation and content of
this paper.
This research has made use of data obtained through the High Energy
Astrophysics Science Archive Research Center Online Service, provided by
the NASA/Goddard Space Flight Center.  This work was supported in part by
the NASA Long Term Space Astrophysics program under grant NAG 5-9184 (PI:
Chakrabarty).

\appendix

\section{Thermonuclear (type-I) burst sources}
\label{sources}

In the following sections we discuss the properties of bursts observed from
individual sources by \xte.
For each source, we list the Galactic coordinates, the first detection of
thermonuclear bursts, the system properties (where known, including the
orbital period and the method of measurement and the presence of burst
oscillations),
and the properties of bursts from previous observations. 
We describe the type of source (persistent/transient) and
the persistent flux history throughout \xte's lifetime.
The burst properties include the presence of PRE and/or double peaked
bursts, the corresponding distance, recurrence times and $\alpha$-values,
whether short recurrence time bursts have been found, and the patterns of
variation in burst properties with
source state ($F_p$, position on the color-color diagram etc.).

\subsection{4U~0513$-$40 in NGC 1851}

Thermonuclear bursts from the persistent source
\cite[]{clark75} at $l=244\fdg51$, $b=-35\fdg04$ were first detected by \sas\/
\cite[]{clark77} and also possibly \uhu\/ \cite[]{fj76}.  \chandra\/
observations of the host cluster NGC~1851 allowed an optical
identification \cite[]{homer01}; while the orbital period has not been
measured, the large $L_{\rm X}/L_{\rm opt}$ ratio indicates an
ultracompact (i.e. $P_{\rm orb}<80$~min) system \cite[e.g.][]{vpmcc94}.
Additionally, the lack of UV and X-ray modulation at time scales within
the expected orbital period
range ($\la1$~hr) suggests low inclination. Only a handful of bursts has
ever been observed (see \citealt{lew93} and \citealt{kuul03a} for
summaries).

Throughout the \xte\/ mission 4U~0513$-$40 has been active at a
flux of approximately $1.4\times10^{-10}\ \epcs$ (2.5--25~keV), reaching
as high as $8.5\times10^{-10}\ \epcs$ during a $\approx50$~d transient
outburst in 2002. For $d=12.1$~kpc \cite[the distance to the host cluster,
NGC~1851;][]{kuul03a}, this corresponds to an isotropic accretion rate of
2--12\%~$\dot{M}_{\rm Edd}$ (adopting a bolometric correction of $c_{\rm
bol}=1.34$).
We found 
seven widely-separated bursts in
the public \xte\/ data, the first six with rather consistent properties:
peak fluxes of
$(14\pm2)\times10^{-9}\ \epcs$, 
$\tau=10\pm2$~s in the mean and no evidence of PRE.
The most recent burst, on 2006 November 4, occurred at $\approx60$\% lower
persistent flux level than the earlier bursts, and reached a peak flux of
$(19.8\pm0.5)\times10^{-9}\ \epcs$, exhibiting moderate radius-expansion.
This burst was more intense than the earlier non-PRE bursts,
with a 8-s interval during which the flux was close to maximum,
contributing to a significantly larger $\tau=22.0\pm0.6$~s.
The peak flux implies a source distance of 8.2~(11)~kpc, depending upon
whether the burst reached the Eddington limit for H-rich ($X=0.7$ in
equation \ref{ledd}) or pure He ($X=0$) material.
An earlier 
PRE burst observed by \sax\/ reached an essentially identical peak flux
\cite{kuul03a}.
That the distance calculated from the PRE bursts is closest to the 
cluster distance of $12.1\pm0.3$~kpc for $X=0$ suggests that these bursts
reach the Eddington limit in a pure-He atmosphere, \leddhe, rather than in
mixed H/He.
The short $\tau$-values for the non-PRE bursts, as well as the short rise
times ($1.5\pm0.6$~s in the mean) are all consistent with ignition of
primarily He fuel, as might be expected from a neutron star accreting from
an evolved (and hence H-poor) donor.

\subsection{EXO~0748$-$676} 
\label{s0748}

This transient at $l=279\fdg98$, $b=-19\fdg81$ was discovered during
\exo\/ observations in 1995 \cite[]{parmar86}, which also revealed
the first thermonuclear bursts from the source. EXO~0748$-$676 exhibits
synchronous X-ray and optical eclipses \cite[]{crampton86} once every
3.82~hr orbit, and also shows X-ray dipping activity. The bursting
behaviour was studied in detail with \exo, which revealed that the burst
rate was inversely correlated with persistent flux (\citealt{gottwald86};
see also \citealt{lew93}). In addition, the burst properties varied
significantly with \fper; both \fpk\ and \fluen\ increased, and $\tau$
decreased when \fper\ exceeded $7.5\times10^{-10}\ \epcs$ (0.1--20~keV).
That the $\alpha$-values increased from $\sim10$ to $\sim100$ confirms
that the changes in burst properties resulted from a transition from H/He
to He-dominated burst fuel as the persistent flux (accretion rate)
increases. Bursts with the highest peak fluxes (3--$4\times10^{-8}\
\epcs$, 0.1--20~keV) exhibited PRE, leading to an estimated distance of
$8.3/(1+X)$~kpc.

The combined spectra of bursts observed by {\it XMM-Newton}\/ exhibited
discrete spectral features \cite[]{cott01,bb01}, some of which appeared to
be redshifted features from near the neutron star surface
\cite[$z=0.35$;][]{cott02}. Subsequent followup studies have failed to
confirm this result \cite[e.g.][]{cott07}. \cite{boirin07a} studied the
energetic properties of these same bursts, identifying for the first time
several examples of burst triplets separated by typically 12~min, in
addition to closely spaced burst pairs \cite[which have been observed
previously; e.g.][]{gottwald86}. These
bursts occurred at an accretion rate of about 1\%~$\dot{M}_{\rm Edd}$, and
likely arose from H-ignition in a mixed H/He environment \cite[i.e. case 3
of][]{fhm81}. It seems probable that the presence of closely-spaced burst
pairs and triplets is linked to hydrogen (case 3) ignition; the
characteristic wait time between the bursts may be associated with a
nuclear $\beta$-decay timescale.

The source was also in a low flux state throughout most of the \xte\/
observations, at a mean level of $2.7\times10^{-10}\ \epcs$ (2.5--25~keV).
In 2004 May ongoing monitoring of the source with \xte\/ revealed a
radius-expansion burst, 
which reached a peak flux of $5.2\times10^{-8}\ \epcs$ \cite[]{wolff05}.
Two additional PRE bursts were detected in subsequent observations in 2005
June and August, and the overall range of peak fluxes 
(lower due to the reduced instrumental effective areas used by {\sc
lheasoft} version \lheasoftver) was (3.8--$4.7)\times10^{-8}\ \epcs$. The
short rise times (2~s) and low $\tau=9.2\pm0.9$~s suggest ignition in a
He-rich environment, so that the distance inferred from the peak burst
fluxes was $7.4\pm0.9$~kpc. At this distance the accretion rate over the
course of the \xte\/ observations was
$\approx2.2$\%~$\dot{M}_{\rm Edd}$ (adopting a bolometric correction of
$1.93\pm0.02$).
We observed 
94 
bursts in total, most of which did not exhibit PRE and had properties
typical for the low-state bursts observed previously: separations
consistent with a recurrence time of
2--5~hr, \fpk
$=(5$--15$)\times10^{-9}\ \epcs$, varying fluences typically
$\sim0.2\times10^{-6}\ \epc$, and long durations ($\tau\sim20$~s in the
mean).  We also found five burst pairs with much shorter recurrence
times in the range 6.5--17~min (see \S\ref{dblbursts}).
Several bursts occurred partially or wholly within dips.
Three bursts (on 1999 Oct 17, 15:11:15 04:10:57
2000 Dec 17 23:53:34, and 2000 Dec 18) occurred close to, or coincident
with, a dip egress, so that the measured peak flux was achieved while the
source was still obscured.
For these bursts
the peak flux and fluence measurements (and derived parameters including
$\tau$) are to be treated with caution.

\subsection{1M~0836$-$425}
\label{s0836}

A Galactic-plane ($l=261\fdg95$, $b=-1\fdg11$) transient discovered during
\osos\/ observations between 1971--74 \cite[]{markert77,com78},
1M~0836$-$425 was first observed to exhibit thermonuclear bursts by \gin\/
during an outburst between 1990 November and 1991 February
\cite[]{aoki92}. 
The typical recurrence time for the 28 bursts detected was $\approx
2$~hr, but on one occasion was 8~min. No bursts exhibited radius
expansion, leading to an upper limit on the distance of 10--20~kpc.
The optical counterpart has not been identified, since several stars down
to a limiting $R$-band magnitude of $\sim23.5$ are present in the $9''$ \ros\/
error circle \cite[]{belloni93}.

A new outburst was detected by the \xte/ASM on 2003 January, 
lasting until May.
A second interval of activity
commenced around 2003 September, and continued throughout the remainder of
2003--04.  Seventeen bursts were observed in total, with rather homogeneous
properties; peak fluxes of
$(13\pm3)\times 10^{-9}\ \epcs$, fluence
$(0.27\pm0.07)\times10^{-6}\ \epc$ and time-scale $\tau=22\pm4$.
\cite{chelov05} analysed 15 of these bursts, in addition to 24
bursts observed by \igr; their maximum peak flux of $1.5\times10^{-8}\
\epcs$ (3--20~keV) led to a distance upper limit of 8~kpc.
The maximum estimated bolometric flux from our analysis was
$2\times10^{-8}\ \epcs$, which leads to a more conservative upper limit
(for $X=0$) of $d<11$~kpc.
No kHz oscillations were detected in any of the bursts.
The peak persistent flux was $1.96\times10^{-9}\ \epcs$ (2.5--25~keV),
corresponding to an accretion rate of $\la33$\% ~$\dot{M}_{\rm Edd}$
(adopting a bolometric correction
of $1.82\pm0.02$).

Near the end of 2003 January two pairs of bursts were observed, each
separated by approximately 2~hr. The other bursts in the early part of the
outburst occurred at much longer intervals, 
although consistent with a steady recurrence time of $\sim2$~hr
(but where the intervening bursts were missed due to data gaps).
Assuming the bursts occurred quasi-periodically, the inferred recurrence time
was $2.20\pm0.18$~hr on average, similar to that in
previous observations \cite[]{aoki92}.  The measured $\alpha$-values
were betwen 40 and 100, although the persistent flux measured from the
field may be contaminated by emission from the nearby
($\Delta\theta=0\fdg4$) 12.3~s X-ray pulsar GS~0834$-$340.
Thus, the derived broad-band flux (and hence $\alpha$) may be
overestimated. Even so, 
$\alpha\la40$, the long duration of the bursts, and the $\sim2$~hr
recurrence times \cite[too short to exhaust the accreted H by steady
burning between the bursts, except for an extremely low accreted H fraction;
e.g.][]{fhm81} all indicate ignition of mixed H/He fuel.

\subsection{4U~0919$-$54}
\label{s0919}

This weak, persistent source at $l=275\fdg85$, $b=-3\fdg84$ has been
observed by all the major early satellites (\uhu,{\it OSO-7, Ariel-5,
SAS-3, HEAO-1, Einstein}). A $V=21$ star has been identified as the
optical counterpart \cite[]{ci87}; although the orbital period is unknown,
the system is a candidate ultracompact based on the optical properties,
and X-ray spectroscopic measurements indicate enhanced abundances
(\citealt{ucb01,ucb02}; see also \citealt{zand05a}). Observations by
\xte\/ led to the first detection of a thermonuclear burst from the
source, as well as the discovery of a 1160~Hz QPO \cite[]{jon01}.

Our analysis of the lone PRE burst, detected by \xte\/ on 2000 May 12
19:50:17 UT 
(Fig. \ref{profiles})
leads to a distance estimate of 4.0 (5.3)~kpc assuming that the burst
reaches \leddh\ (\leddhe).
We found 
three other burst candidates, the brightest of which
(on 2004 June 18 23:38:17 UT) 
peaked at $(5.7\pm0.2)\times10^{-9}\ \epcs$ and exhibited no evidence for
PRE. The other two were sufficiently faint that our spectral analysis
results did not allow a test for cooling in the burst tail, and thus these
three events must remain as merely burst candidates.
The 2.5--25~keV PCA flux was between 0.7--$4.5\times10^{-10}\ \epcs$,
which for $d=5.3$~kpc corresponds to an accretion rate of
$\la1.2$\%~$\dot{M}_{\rm Edd}$.	

\subsection{4U~1254$-$69} 

Thermonuclear X-ray bursts and periodic dips were discovered in this
3.9~hr binary ($l=303\fdg48$, $b=-6\fdg42$) during \exo\/ observations in
1984 \cite[]{c3p86,motch87}. Just two thermonuclear bursts were detected,
with rise times of $\sim1$~s; durations of $\sim20$~s; peak fluxes of
$\sim1.1\times10^{-8}\ \epcs$; and no evidence of PRE \cite[see also][]{lew93}.
\cite{zand03b} reported a superburst from this source, as detected by
\sax\/ on 1999 January 9. 
Possible PRE during the type-I burst precursor to the superburst
leads to a distance estimate of $13\pm3$~kpc.
Analysis of \xte\/ observations confirm that the system is an atoll
source, and provided weak evidence for 95~Hz oscillations in one burst
\cite[]{bhatt07a}.

\xte\/ observations in 1996--7 and 2001 detected the source at a
persistent level of $8\times10^{-10}\ \epcs$ (2.5--25~keV), which for
$d=13$~kpc corresponds to an accretion rate of 12\%~$\dot{M}_{\rm Edd}$
(for a bolometric correction of $1.13\pm0.03$).
Four of the five thermonuclear bursts occurred between 2001 December 6--7
and exhibited rather unusual evolution of the blackbody radius.
The radius was elevated during the rise and peak of the burst, and was
accompanied by a decrease in the blackbody temperature, but not to a
sufficient degree to be confirmed as PRE bursts (see Fig. \ref{profiles}).
Typically for these bursts the peak flux ($5.6\pm0.7\times10^{-9}\ \epcs$
in the mean) was reached at or before the peak radius.
The bursts were short (mean $\tau=5.5\pm0.7$~s) and weak, with mean
$E_b=(0.028\pm0.003)\times10^{-6}\ \epc$. The recurrence times were 10.7,
4.04 and 9.16~hr; the observations had sufficient coverage over the final
interval to exclude an intermediate burst, although a dip (which
could potentially have prevented detection of a burst) was observed at
approximately the halfway mark. The measured $\alpha$-values were
$1260\pm60$, $490\pm20$ and $1070\pm60$.

\subsection{4U~1323$-$62}
\label{s1323}

Bursts with a steady recurrence time of 5.3~hr were discovered
from 4U~1323$-$62 ($l=307\fdg03$, $b=+0\fdg46$) in 1984 during \exo\/
observations \cite[]{vdk84}.
The bursts were homogeneous, with
rise times of $4.0\pm0.6$~s, \fpk\ of $(5.2\pm0.9)\times10^{-9}\ \epcs$,
$\tau=14\pm2$~s, and no evidence of PRE \cite[]{lew93}.
``Double'' bursts, with extremely short ($\la10$~min) recurrence
times, are commonly observed from this source \cite[see e.g.][]{balucinska99}.
4U~1323$-$62 also exhibits 2.93~hr periodic
intensity dips \cite[]{parmar89}, and is associated with a faint
($K'=17.05\pm0.20$, $J\sim20$) IR counterpart \cite[]{smale95}.

\xte\/ observations found 4U~1323$-$62 at a persistent, steady flux level of
(2--$2.5)\times10^{-10}\ \epcs$ (2.5--25~keV), somewhat higher than the
$\sim1.7\times10^{-10}\ \epcs$ of the \exo\/ observations.
A total of 
30 
X-ray bursts were found, in observations taken early in 1997,
1999, late 2003 and late 2004.  Seven bursts were observed between 1997
April 25--28,
5 with a regular recurrence time of $2.7\pm0.3$~hr.  Two of these regular
bursts were closely ($\approx10$~min) followed by extremely faint
secondary bursts
\cite[see also][]{barnard01}.
Two other examples of double bursts were detected by \xte\/ in 1999
January and 2004 December.
Neglecting the bursts with short recurrence times, the burst properties were 
roughly consistent with previous observations:
\fpk~$=(3.8\pm1.5)\times10^{-9}\ \epcs$, $\tau=27\pm4$ and
$E_b=(0.10\pm0.02)\times10^{-6}\ \epc$.
The $\alpha$-values were $38\pm4$ in the mean, which, combined with the
long burst duration, strongly suggests mixed H/He fuel in the bursts.
Without any radius-expansion bursts, we have only upper limits on the
distance, of 11~(15)~kpc assuming the bursts do not exceed \leddh\
(\leddhe). For a distance of 11~kpc, the persistent flux indicates an
accretion rate of a few percent of $\dot{M}_{\rm Edd}$. At this level, we
expect either He-rich bursts with long recurrence times resulting from
He-ignition (e.g. SAX~J1808.4$-$3658, \S\ref{s1808}), or mixed H/He bursts
with short recurrence times resulting from H ignition (e.g.
EXO~0748$-$676, \S\ref{s0748}). The burst properties of 4U~1323$-$62 (as
well as the presence of faint secondary bursts), strongly
suggests the latter regime.
We note that the regular occurence of double bursts
provides an opportunity to
constrain the fuel source for the secondary bursts
(see \S\ref{dblbursts}).

\subsection{4U~1608$-$52} 
\label{s1608}

Bursts from this Galactic plane ($l=330\fdg93$, $b=-0\fdg85$) transient were
first detected by the two {\it Vela-5}\/ satellites \cite[]{bce76}.
\uhu\/ observations confirmed the link between the bursting and persistent
source \cite[]{tananbaum76}, and a more precise X-ray position obtained with
HEAO-1 \cite[]{fabbiano78} permitted identification of the $I=18.2$ optical
counterpart QX~Nor \cite[]{gl78}. Variations in the persistent X-ray flux
as measured by {\it Vela-5B}\/ have suggested an orbital period of either
4.10 or 5.19~d \cite[]{lr94}.
Along with 4U~1728$-$34, 4U~1608$-$52 was one of the first sources for
which the intensity dips at the burst peak were identified as PRE episodes
\cite[]{fg89}.  \hak\/ observations revealed recurrence times as short as
10~min, as well as two separate bursting modes (in terms of recurrence time
and profile shape) depending upon the persistent flux level
(\citealt{murakami80a,murakami80b}; see also \citealt{lew93}). At high
\fper, the bursts are bright and fast, with rise times of $\sim2$~s and
$\tau\approx8$~s; at low \fper, the rise times exceeded 2~s and the bursts
were much longer with $\tau=10$--30~s.
A superburst has been identified in \xte/ASM data; this source is the
first frequent transient in which such an event has occurred
\cite[]{keek07}.

Two large and several smaller outbursts have occurred during \xte's
lifetime; the most recent observed by \xte\/ in 2005 March. The large
outbursts both peaked at $\sim2.7\times10^{-8}\ \epcs$ (2.5--25~keV),
corresponding to an accretion rate of 62\%~$\dot{M}_{\rm Edd}$ (for a
distance of 4.1~kpc, see Table \ref{dist}; and a bolometric correction of
$1.77\pm0.04$). The lowest measured fluxes of $\approx2.2\times10^{-11}\
\epcs$ provide an upper limit for the quiescent accretion rate of
0.5\%~$\dot{M}_{\rm Edd}$.
We found 
31 
bursts in public \xte\/ data, of which 
12 exhibited
PRE.  At the lowest \fper\ values, we found five PRE bursts
with long time scales ($\tau=16\pm4$ in the mean).  In five of the other 6 PRE
bursts we detected burst oscillations at 619~Hz, making this source the
most rapidly spinning neutron-star LMXB known to date (Hartman et al.
2008, in preparation).
Three bursts exhibited oscillations both before and after the PRE episode,
and showed increases in frequency with time of up to 6.6~Hz (1.1\%) over
8~s.  The oscillations during the burst rise had the largest amplitudes,
up to 13\% (rms), while detections during the tails were up to 10\% but
more typically 3--5\%.

The bursts from 4U~1608$-$52 were some of the brightest seen in the entire
\xte\/ sample, and peaked between
1.2--$1.5\times10^{-7}\ \epcs$.  As with previous observations, $\tau$
appeared to decrease significantly with persistent flux, although there
was substantial scatter. For four pairs of closely-spaced non-PRE bursts
with $\tau=16$--22~s we estimated the recurrence time at between
4.14--7.5~hr, so that $\alpha=41$--54.
We observed two instances of extremely short recurrence times for bursts
from 4U~1608$-$52. On 1996 March 22 we observed three bursts in quick
succession; two brighter bursts separated by 24~min, with an extremely
faint burst (\fpk$=(1.0\pm0.2)\times10^{-9}\ \epcs$) inbetween, just
16~min after the first burst. Such burst triplets have only been observed
from a few other sources, including EXO~0748$-$676 \cite[]{boirin07a} and
4U~1705$-$44 (see \S\ref{s1705}). None of the
three bursts exhibited PRE. On 2001 November 21 we observed two bursts
separated by at most 6~min (the start of the second burst fell within a
data gap). The persistent flux level between these two observations varied
by a factor of 2, and suggest an accretion rate of 3--5\%~$\dot{M}_{\rm
Edd}$, in the range where we expect bursts to ignite via H-ignition.  The
bursts
comprising the burst triplet and pair all had long $\tau=19\pm7$ in the
mean, and the second and third burst of the triplet had atypically low
fluences of 0.025 and $0.27\times10^{-6}\ \epc$.
These properties are similar to other short-recurrence time
bursts such as those observed from EXO~0748$-$676 (see \S\ref{s0748}).

\subsection{4U~1636$-$536} 
\label{s1636}

4U~1636$-$536 ($l=332\fdg9$, $b=-4\fdg8$) is a well-studied LMXB in a
3.8~h orbit with an 18th magnitude blue star, V801~Ara \cite[]{1636orb}.
X-ray bursts were first detected from a region containing the
previously-known persistent source by \osoe\ \cite[]{swank76a}, and were
subsequently studied in great detail with observations by {\it EXOSAT}\/
\cite[see][for a review]{lew93}.
\xte\/ observations revealed burst oscillations at 579.3~Hz
\cite[]{stroh98b,stroh98}, as well as a possible first detection of a
harmonic \cite[which has not been confirmed in other bursts from the
source;][]{mill99}.  The properties of the oscillations have been analysed
by \cite{muno01} and \cite{gil02}.  Three ``superbursts'' have also been
detected, separated by 2.9 and 1.75~yr
\cite[]{wij01b,kuul04}.
In one of the superbursts,
which was observed with the PCA, \cite{stroh02b} found an $\approx800$~s
interval during which oscillations were consistently detected.

The persistent flux of the source varied between 4--$6\times10^{-9}\
\epcs$ (2.5--25~keV) between 1996--2000, but since then it 
declined steadily, reaching $1.25\times10^{-9}\ \epcs$
during 2004 January. For a distance of 6~kpc (see Table \ref{dist}), this
corresponds to an accretion rate between 3--16\%~$\dot{M}_{\rm Edd}$.
We detected 
123 bursts from public \xte\ observations, of which \preburst\ exhibited
PRE. The peak fluxes were bimodally distributed, as in previous
observations \cite[e.g.][]{seh84}, although we found both PRE and non-PRE
bursts with \fpk\ falling within the gap noted by those authors. 
While the majority of the \preburst\ PRE bursts had peak fluxes which
varied significantly about a mean of \highmean, with a standard deviation
of \highsig\ \cite[in contrast to earler studies which measured consistent
PRE burst peak fluxes;][]{eb87} we also found two PRE bursts with much
smaller peak fluxes of $38\times10^{-9}\ \epcs$.  The ratio of the mean
peak flux of the bright to faint PRE bursts is 1.7, suggesting that the
bright bursts reach the Eddington limit for pure He, while the faint
bursts reach the limit for mixed H/He at approximately solar composition
\cite[][see also \S\ref{pflux}]{gal06a}.
At the highest \fper\ range the
bursts were fast, with $\tau=6.4\pm1.1$~s in the mean; at lower \fper\
the $\tau$ values became both larger and more variable. The burst rate
decreased significantly as \fper\ increased (see Fig.
\ref{twosources} and \S\ref{global}).
Typical $\alpha$-values
for bursts with $\Delta t= 1.2$--2.2~hr at low \fper\ were
$\sim40$, although on occasion were as high as 100 (for somewhat
longer $\Delta t=2.4$~hr); however, at the
highest flux range we measured $\alpha=500$ for a pair of bursts with
$\Delta t=6.3$~hr.

We found 4 bursts with distinct double peaks in the bolometric flux,
separated by 4--5~s. For three of these bursts (on 2001 September  5
08:15:04, 2001 October  3 00:22:18, and 2002 February 28 23:42:53 UT) the
second peak was larger than the first, by 70, 40, and 230\% respectively
\cite[the lattermost case bore a striking resemblence to a burst from
4U~1709$-$267; see \S\ref{s1709} and][]{jonk04}. For the other burst, on
2002 January  8 12:22:44 UT, the first peak was very slightly greater than
the second, and between the two peaks the flux reached a minimum of around
45\% \fpk. This burst was also analysed by \cite{bhatt06a}, who
interpreted the variation of the blackbody radius as arising from
two-phase spreading of the nuclear burning following ignition near the
pole.

\subsection{MXB~1659$-$298} 

Regular ($\Delta t=2.1$--2.6~hr) X-ray bursts were first detected 
from MXB~1659$-$298 ($l=353\fdg83$, $b=7\fdg27$) by \sas\ in 1976
\cite[]{lhd76}. An upper limit on the persistent flux at this time led to
a constraint on $\alpha<25$. Persistent emission at $\sim5\times10^{-10}\
\epcs$ attributed to the bursting source was detected by \sas\ two years
later \cite[]{lewin78}, although no bursts were detected at that time
\cite[see also][]{lew93}. An improved X-ray position from observations by
{\it HEAO-1}\/ led to the identification of the optical counterpart,
V2134~Oph \cite[]{doxsey79}.  Irregular X-ray dips as well as 15~min
eclipses at the 7.1~hr orbital period were discovered in {\it HEAO}\/ A-1
scanning observations \cite[]{cw84}.

A new active period which lasted around 2.5~yr
began in 1999 April, during which the source was observed extensively by
\xte\ and \sax/WFC. The PCA flux peaked at $10^{-9}\ \epcs$ (2.5--25~keV)
in 1999 April, but was 4--$6\times10^{-10}\ \epcs$ throughout the
remainder of the outburst. For $d=12$~kpc (see Table \ref{dist}) this
corresponds to a range of accretion rates of 
6--15\%~$\dot{M}_{\rm Edd}$.	
Burst oscillations at 567~Hz were detected in most of the PRE bursts
observed by \xte\/ \cite[]{wij01}; a detailed study of 14 of the 26 bursts
observed in total was presented in \cite{wij01c}.

Those authors found no clear correlations between their properties and the
accretion rate, although only a limited range of accretion rates were
sampled. In the full set of bursts observed by \xte\/ there is a marked
division between the PRE bursts, which had $\tau=4.6\pm1.0$~s in the mean
and were generally observed at higher \fper, and the non-PRE bursts, for
which $\tau=19\pm5$~s (with one exception) and which were observed at
somewhat lower \fper. We also found unusually large variations in the peak
flux of the PRE bursts (see \S\ref{pflux}); this may be a consequence of
the high system inclination, as evidenced from the presence of eclipses.
The burst intervals measured by \xte\/ were all $>14$~hr, with just two
exceptions; for one of those intervals ($\Delta t=1.82$~hr) we could not
exclude intermediate bursts due to a data gap, while for the other
(0.53~hr) the high-resolution datamodes did not cover the second burst.
Thus, we could not reliably estimate $\alpha$ for any of the burst pairs.

MXB~1659$-$298 became undetectable by the PCA on 2001 September 7, after which
the cooling of the NS was monitored by \chandra\ observations
\cite[]{wij04b}.

\subsection{4U~1702$-$429} 
\label{s1702}

Bursts from this Galactic plane ($l=343\fdg89$, $b=-1\fdg32$) source were
first detected by \osoe\/ \cite[]{swank76}. \cite{mak82} detected 14
bursts with characteristic recurrence times of 9--12~hr from the region in
\hak\/ observations in 1979, and attributed them to a persistent \uhu\/
source. The peak fluxes were
(18--$30)\times10^{-9}\ \epcs$ (3--10~keV); rise times a few seconds or
less; and $\tau=10$--15 \cite[see also][]{lew93}.  A distance of
$\sim10$~kpc was inferred from the similarity of the peak fluxes with
those of other Galactic centre sources.
A $K=16.5$ star at the edge of the \chandra\/ error circle has been
suggested as the counterpart \cite[]{wachter04}.
\xte\/ observations in 1997 July 29--30 revealed kHz QPOs as well as 6
type-I bursts with coherent oscillations near 330~Hz \cite[]{mss99}.

4U~1702$-$429 was persistently active at low levels
(0.7--$2.3\times10^{-9}\ \epcs$, 2.5--25~keV) throughout the 
\xte\/ mission to date. For a distance of 5.5~kpc (derived from 
the mean peak flux of 
five 
PRE bursts; Table \ref{dist}) the accretion
rate was 2--6\%~$\dot{M}_{\rm Edd}$ (where we adopt a bolometric
correction for the 2.5--25~keV flux which averages $1.12 \pm 0.04$ over 6
selected observations).
We found 
47 
bursts in total,
with peak fluxes 
which were strongly correlated with the burst fluence.
The correlation was most striking for the bursts with fluences
$<0.6\times10^{-6}\ \epc$ (primarily non-PRE bursts), and 
the scatter about a linear fit was only about 10\%.
The relation saturated at about $75\times10^{-9}\ \epcs$, and for bursts
with fluences larger than
$0.6\times10^{-6}\ \epc$ (all PRE bursts) the peak fluxes were all quite
similar.  We note that this behavior is distinct from that of
4U~1728$-$34, which shows a correlation between peak flux and fluence most
strongly for PRE bursts \cite[]{gal03b}.
The non-PRE burst profiles from 4U~1702$-$429 were quite homogeneous, with 
$\tau=7.8\pm0.7$~s in the mean,
and rise times
consistently $<2$~s (see Fig. \ref{profiles}).
The PRE bursts had substantially larger $\tau$-values 
($\tau>10$) than the non-PRE bursts.

The shortest burst interval was 
4.5~hr;
on
16 separate occasions 
between 1997 July and 2004 April
we detected a pair of bursts separated by no more than
8.1~hr; in the mean, the burst separations were $6.4\pm1.1$~hr.
We tentatively identify this value as the characteristic burst recurrence
time for the source during the \xte\/ observations. While it is possible that 
intervening bursts were missed in data gaps (so that the
actual recurrence times were one-half or less of the measured intervals),
we note
that these intervals are already significantly lower than the 9--12~hr
measured in previous observations \cite[]{mak82}, suggesting that an even
shorter recurrence time is less likely.
The source was quite variable on timescales
of a few days, so that the persistent 2.5--25~keV flux during these
observations varied between (1.3--$2.3)\times10^{-9}\ \epcs$.
We estimate
$\alpha$-values in the range 74--153, with the
wide range arising mainly from the 
large variations in burst fluence $E_b$.
This inconsistency between the burst fluences for comparable recurrence
times and $\dot{M}$ perhaps indicates incomplete burning of the accreted
fuel in the low-fluence bursts.
We found the three smallest values of
$\alpha$ clustered tightly about a mean of
 $75.3\pm1.5$,
suggesting a H-fraction at ignition of less than half the solar value.
A low H-fraction is also consistent with the short rise times
and low $\tau$-values for the bursts.

\subsection{4U~1705$-$44}
\label{s1705}

This persistently bright, variable 
source 
($l=343\fdg33$, $b=-2\fdg33$)
was detected initially in
\uhu\/ observations \cite[]{4ucat}.  Bursts were first discovered by \exo\
in 1985, and the bursting behaviour was subsequently studied in detail by
that satellite \cite[]{langmeier87,gottwald89}.  The burst profile was
observed to change from slow (with decay $\tau\sim100$~s) to 
fast ($\tau\sim25$~s) as the intensity increased, in a similar fashion to
EXO~0748$-$676 \cite[see also \S\ref{s0748},][]{lew93}. Typical recurrence
times were
1.9--2.5~hr, except for occasional faint bursts following brighter events
by just
500--1000~s.  Model atmosphere fits to spectra of non-PRE bursts
observed by \exo\/ 
suggest a distance to the source of $7.5^{+0.8}_{-1.1}$~kpc
\cite[]{hab95}, and also indicate a H-rich atmosphere; on the other hand,
\cite{cs97} derive a distance from the peak flux of PRE bursts of 11~kpc.
One (or possibly two) high-frequency QPOs were discovered through \xte\/
observations \cite[]{ford98}.
The optical counterpart is unknown.

The persistent flux from the source varies substantially in a
quasi-periodic manner 
on a time-scale of several hundred days
\cite[e.g.][ see also fig. \ref{asmlc}]{pried86}. The 2.5--25~keV flux measured by
\xte/PCA varied between 0.18--$10.7\times10^{-9}\ \epcs$.
We found a total of 
47 
type-I bursts in the public PCA data, including
three exhibiting PRE.
The distance derived from the PRE bursts is
consistent with the distance estimate of \cite{hab95} at between
5.8--7.6~kpc, depending upon the composition (Table \ref{dist}).
The persistent flux range is thus equivalent to an accretion rate of
1.1--70\%~$\dot{M}_{\rm Edd}$ (for a bolometric correction to the
2.5--25~keV flux which ranged between 1.29--1.75 for 3 selected
observations, and was 
1.48 in the mean).
All but one of the bursts was observed when
$F_p\la2.5\times10^{-9}\ \epcs$ (i.e. $\dot{M}\la16$\%~$\dot{M}_{\rm
Edd}$).
The non-PRE bursts had $\tau=19\pm5$~s in the mean, while the PRE bursts,
which were all observed at higher persistent flux levels
($F_p=2.25$--2.5$\times10^{-9}\ \epcs$),
were much faster with $\tau=5.0\pm0.4$~s. The single burst observed at
high $F_p=8.1\times10^{-9}\ \epcs$ on 2002 June 29 19:54:43 UT also had a
low $\tau=6.8\pm0.4$~s but did not exhibit PRE.

We found four bursts with short recurrence times of between 7--11~min, two
of which were themselves part of a closely-spaced burst triplet, on 2000
February 7. Such triplets have only been observed in a handful of sources,
including EXO~0748$-$676 \cite[]{boirin07a} and 4U~1608$-$52 (see
\S\ref{s1608}). Each of the short recurrence time bursts
had unusually small fluences, and occurred when the persistent flux
indicated an accretion rate of 5--10\%~$\dot{M}_{\rm Edd}$.
For the remaining bursts, taking into
account the possibility of missed bursts, the recurrence times were
0.76--3.1~hr.  Between 1997 May 16--19 ten bursts were detected off-axis
during observations of PSR~B1706$-$44, just $0\fdg41$ away.
Assuming that the X-ray flux from the pulsar is negligible, the majority
of the bursts for which the recurrence time could be confidently measured
had $\Delta t\approx1$~hr and $\alpha=34$--75. For three other
pairs of bursts with somewhat longer $\Delta t=1.6$--3~hr, the
$\alpha$-values ere
much higher, at $196\pm16$, $186\pm14$, and, for an unusually weak burst,
$500\pm50$. 

\subsection{XTE~J1709$-$267}
\label{s1709}

XTE~J1709$-$267 ($l=357\fdg47$, $b=+7\fdg91$) went into a large outburst in
1997 and was discovered by the \xte/PCA during a satellite maneuver
\cite[]{marshall97}. 
An improved position was obtained from \sax\/
observations \cite[]{cocchi98}, also during which bursting behaviour was
first observed.  Three bursts were observed in total by \sax, with
absorption-corrected bolometric peak fluxes of $16\times10^{-9}\ \epcs$,
leading to a distance upper limit of $10\pm1$~kpc.
A second outburst, very similar to the first, was observed with \xte\/
beginning 2001 December
\cite[]{jonk04}. The maximum PCA flux was $3.2\times10^{-9}\ \epcs$
(2.5--25~keV), slightly smaller than for the previous outburst
(which peaked at $4.2\times10^{-9}\ \epcs$). For $d=10$~kpc this
corresponds to a peak accretion rate of 
33--43\%~$\dot{M}_{\rm Edd}$.	
PCA observations detected 3 bursts during the 2001--02 outburst with peak
fluxes of 
$(11.6\pm0.6)\times10^{-9}\ \epcs$ and $\tau=5.8\pm0.5$~s in the mean. All
three bursts exhibited low-level flux for 2--3~s prior to the burst, and
one (on 2002 January 30 04:16:02 UT) exhibited a precursor event peaking
at $\approx20$\% of the subsequent maximum, remarkably similar to a burst
observed from 4U~1636$-$536 on 2002 February 28 23:42:53 UT (see Fig.
\ref{profiles}).
While none of the bursts exhibited evidence for significant radius
expansion, the third burst, on 2002 Feb  7 01:12:03 UT, exhibited marginal
evidence (although it did not reach a peak flux level significantly
higher than the other two). The distance limits derived from these bursts
is consistent with that of the \sax\/ observations, at 11 (14)~kpc for
$X=0.7$ (0.0).

\subsection{XTE~J1710$-$281}

This nearby source ($l=356\fdg36$, $b=+6\fdg92$) was also detected in
outburst by the PCA in 1998, and was identified with the \ros\/ All-Sky
Survey Bright Source Catalog source 1RXS~J171012.3$-$280754
\cite[]{mmst98}. Low-level activity continued throughout 1999, and
recurring $<800$~s eclipses with a periodicity of 3.27~h (assumed to
be the orbital period) were detected in
subsequent observations
\cite[]{msm98,ms02c}.
Bursting activity was also discovered during these \xte\/ observations,
with a total of 
19 
bursts detected since, including one PRE burst and one
pair of bursts separated by 10~min.  The distance range implied by the
peak flux of 
$(9.2\pm0.2)\times10^{-9}\ \epcs$ for the single PRE burst is 12--16~kpc
\cite[Table \ref{dist}; see also][]{ms02c}.  The source was also active
during 2001--02, at an overall flux range of 0.4--$1.4\times10^{-10}\
\epcs$ (2.5--25~keV). For $d=16$~kpc this corresponds to
1--4\%~$\dot{M}_{\rm Edd}$ (assuming a bolometric correction of
$1.421\pm0.126$).
The non-PRE bursts were exceedingly faint, with peak fluxes of
0.22--$3.6\times10^{-9}\ \epcs$.
The bursts observed when $F_p>1\times10^{-10}\ \epcs$ had a range of
$\tau=5$--17~s, with the PRE burst having the smallest value of
$\tau=5.7$~s. At lower persistent fluxes the bursts were longer, with
(typically) $\tau\approx30$.
The pair of closely spaced bursts occurred when the persistent flux
indicated an accretion rate of 1.6\%~$\dot{M}_{\rm Edd}$.
On three occasions we measured burst intervals of $\sim7$~hr, 
while in 2005 November we found two bursts separated by 3.3~hr. Thus, we
tentatively identify the typical recurrence time as
$\approx3.5$~hr.
The corresponding $\alpha$-values varied widely between 22--190.

\subsection{XTE~J1723$-$376}

This transient ($l=350\fdg18$, $b=-0\fdg87$) underwent a moderate outburst
on 1999 January and was discovered during an \xte/PCA scan of the region
\cite[]{marshall99a}.  An \asca\ observation on 1999 March 4 led to an
improved position \cite[]{marshall99b}, and the
source activity continued throughout March \cite[]{msm98}.  Thermonuclear
bursts were first observed during the \xte\/ observations, and a total of
three bursts were observed. 
The brightest burst reached $(14.1\pm0.1)\times10^{-9}\ \epcs$,
implying a source distance of at most 13~kpc.
The persistent flux reached $1.5\times10^{-9}\ \epcs$ (2.5--25~keV) in
outburst, which for $d\la13$~kpc corresponds to $\la20$\%~$\dot{M}_{\rm Edd}$
(for a bolometric correction to the 2.5--25~keV flux of $1.05\pm0.02$).
The first of the pair of bursts on 1999 February 3--4 ($\Delta t=2.7$~hr)
exhibited evidence for a double peak in the bolometric flux, and had
$\tau=12$~s. The second burst was faster and somewhat flat-topped, with
$\tau=7.5$~s and $\alpha=170$.

\subsection{4U~1724$-$307 in Terzan 2}
\label{s1724}

A long thermonuclear burst was observed from 4U~1724$-$307 ($l=356\fdg32$,
$b=+2\fdg30$) by \osoe\ \cite[]{swank77}, reaching a peak flux of
$6.2\times10^{-8}\ \epcs$. The burst source was identified with the
globular cluster Terzan~2 \cite[]{grindlay78b,grindlay80}, 7.5--12~kpc
away (\citealt{kuul03a}; although \citealt{obb97} estimated a range of
5--8~kpc). No optical counterpart is known.
Bursts are observed from the source relatively infrequently
\cite[e.g.][]{lew93}, and the most detailed studies to date have been with
\sax/WFC monitoring observations of the Galactic center region which found
24 bursts with inferred PRE reaching 5.4--$8.4\times10^{-8}\ \epcs$
\cite[]{kuul03a}.

4U~1724$-$307 was persistently bright although declining during \xte\/
observations; in 1996--98 the persistent flux was $F_p=1.2\times10^{-9}\
\epcs$ (2.5--25~keV, see also \citealt{olive98}), while in 2001--02 it
had declined to $7\times10^{-10}\ \epcs$. For a distance of 9.5~kpc this
corresponds to an accretion rate of 
6--11\%~$\dot{M}_{\rm Edd}$.	
A 
burst observed by \xte\/ on 1996 November 8 07:00:31 UT
exhibited sufficiently intense PRE that the color temperature fell below
$\sim0.5$~keV 
shortly after the start, so that the burst flux
effectively dropped out of the PCA band \cite[see Fig. \ref{b1724} and
][]{mgl00,kuul03a}. Apart from this episode the burst flux was approximately
constant at $6\times10^{-8}\ \epcs$ for the first 25~s; this
is substantially above the expected Eddington flux for a source at
9.5~kpc.
A second PRE burst observed on 2004 February 23 exhibited much less
extreme radius expansion, and reached a peak flux around 30\% lower at
$4\times10^{-8}\ \epcs$. The implied distance from this burst is 7.4~kpc,
consistent with the lower limit of the estimated distance range for
Terzan~2.  The weaker PRE burst lasted only $\approx20$~s, and the fluence
was less than a tenth of that of the brighter. The only other burst was
similar in profile to the 2004 February burst, reaching a peak of
$5\times10^{-8}\ \epcs$, but exhibiting only marginal radius expansion.
Intense, long duration PRE bursts, as also observed from 4U~2129+12 (see
\S\ref{s2129}) and GRS~1747$-$312 (\S\ref{sc1747}), appear to consistently
exceed the expected Eddington limit (see \S\ref{pflux}).

\subsection{4U~1728$-$34 ($=$ GX 354+0)}
\label{he1728}

4U~1728$-$34 (GX~354+0; $l=354\fdg3$, $b=-0\fdg15$) was
first resolved by {\it Uhuru}\/ scans of the galactic center region
\cite[]{ftj76}. 
The position of a possible radio counterpart
suggested identification with a $K=15$ infrared counterpart
\cite[]{marti98}.
Thermonuclear X-ray bursts were discovered during {\em SAS-3}\/
observations of the Galactic center region \cite[]{lcd76,hoff76}.  The
bursting behaviour was subsequently studied in detail using extensive {\em
SAS-3\/} observations, which included 96 bursts in total. The burst
intervals were moderately regular, varying by a factor of $\sim2$, and ---
along with the burst properties --- were apparently not correlated with
$F_p$ \cite[]{lew93}. 
The average $\tau=7.8\pm2.4$~s, while $\alpha=110$
with only $\sim15$\% variation between observations.  \cite{bas84} found
evidence for a narrow distribution of peak burst fluxes, as well as a
correlation between peak flux and the burst fluence.  Assuming that the
maximum burst flux is the Eddington limit, the distance to the source is
4.2--6.4~kpc \cite[see also][]{vp78};
other measurements are all around these values \cite[e.g.
6~kpc;][]{kam89}.
Early \xte\/ observations of the source led to the
discovery of nearly coherent 363~Hz oscillations during the bursts
\cite[]{stroh96} that were subsequently observed in 12 other sources
(see \S\ref{bosc}). Subsets of the bursts observed during the PCA observations
have been studied by \cite{vs01}, \cite{franco01}, and
\cite{muno01,muno04a}, with particular attention to the relationship
between the appearance of burst oscillations and the mass accretion rate.

4U~1728$-$34 was persistently bright during \xte\/ observations at
$F_p=1$--$7\times10^{-9}\ \epcs$ (2.5--25~keV). We note that these
persistent flux levels may be affected by the presence of the nearby
transient 4U~1730$-$335 (the Rapid Burster; see \S\ref{srb} and
\S\ref{sc1728rb}).  For a distance of 5.2~kpc \cite[see also][]{gal03b}
this is equivalent to an accretion rate of 3--18\%~$\dot{M}_{\rm Edd}$
(for a bolometric correction to the 2.5--25~keV flux between 1.05--1.55,
or $1.24\pm0.20$ in the mean).
We found 
106 
bursts in total attributable to 4U~1728$-$34 in public \xte\/
observations. The bursts were all rather homogeneous, with short rise times
($\approx1$~s) and time scales ($\tau=6.3\pm1.3$). The shortest measured burst
interval was 1.77~hr; \cite{corn03a} found evidence for clustering of
recurrence times in bursts observed by \sax\/ between 2.5--5~hr, which is
consistent with the \xte\/ measurements. The $\alpha$-values for the
bursts varied between 91--310, or $150\pm70$ in the mean. The burst
properties bear a remarkable resemblence to those of 3A~1820$-$30 \cite[see
\S\ref{s1820};][]{cumming03}, which is an ultracompact binary with an
evolved, H-poor mass donor. 
Model fits to the \xte\/ spectra during PRE also suggest an atmosphere
dominated by helium \cite[]{sth03}.
It seems likely that the mass donor
in 4U~1738$-$34 is also H-poor.

A significant fraction ($\approx2/3$) of the bursts observed by \xte\
showed evidence for PRE episodes.
The peak flux \fpkre\ of these bursts 
varied with a standard deviation of 9\%, and was correlated with the
persistent emission, both varying quasi-periodically with a time scale of
$\approx40$~d \cite[]{gal03b}.  The peak PRE burst flux and fluence \fluen\
were also strongly
correlated, suggesting
reprocessing of the burst flux, perhaps by
a precessing, warped accretion disk 
to give the $\approx40$~d time scale.
\cite{sth03} suggests instead that the variations in \fpkre\ arise from
increased visibility of the neutron star following the atmospheric
contraction,
and estimate a system inclination of $\sim50\arcdeg$.

\subsection{Rapid Burster ($=$ MXB 1730$-$335) in Liller 1}
\label{srb}

This remarkable Galactic bulge ($l=354\fdg84$, $b=-0\fdg16$) transient was
discovered during \sas\ observations \cite[]{lewin76b} to exhibit
unusually frequent X-ray bursts, with intervals of 6~s to 5~min.  An
apparent second class of ``anomalous'' bursts, with 3~s rise times and
lower peak intensities \cite[]{ulmer77}, were later identified as
thermonuclear (type-I) bursts;
the brighter, more frequent (type II) bursts were attributed instead to
episodic accretion \cite[][]{hoff78}.
Thermonuclear bursts occur in the Rapid Burster at intervals of $\sim1.5$--4~hr
\cite[]{lew93}, in the same range as other bursters, and are almost
always observed alongside type-II bursts.  No PRE bursts have been
reported; the peak flux of the brightest thermonuclear bursts is
$1.7\times10^{-8}\ \epcs$ (\citealt{kuul03a}; see also
\citealt{lew93}). The type-I bursts have long durations, suggesting that
substantial amounts of H is present at ignition. Historically, the source
has exhibited approximately periodic outbursts every $\sim200$~d
\cite[]{guerr99}; thermonuclear bursts are seen preferentially in the
first 15--20~d of the outburst.
Located in the globular cluster Liller 1 \cite[]{liller77}, no conclusive
optical counterpart has been found despite \chandra\/ and {\it HST}\/
observations of the field \cite[]{homer01c}, and a confirmed radio
counterpart \cite[]{moore00}.

\xte\/ observations have detected $\approx25$ outbursts (as of 2006
March), and indicate that the outburst recurrence time has decreased to
$\sim100$~d (\citealt{masetti02}; see also Fig. \ref{asmlc}). The peak
2.5--25~keV flux typical for outbursts prior to 2000 was
$1.2\times10^{-8}\ \epcs$, which for a distance to the host cluster of
8.8~kpc \cite[]{kuul03a} corresponds to an accretion rate of
95\%~$\dot{M}_{\rm Edd}$.	
Outbursts occuring in 2000 and later appeared to
peak at a significantly smaller flux of $\approx5\times10^{-9}\ \epcs$, or
$\approx40$\%~$\dot{M}_{\rm Edd}$.	
The minimum detectable \fper\ was
$1.6\times10^{-10}\ \epcs$, or 
1.2\%~$\dot{M}_{\rm Edd}$.	
We note that these \fper\ measurements may include a contribution from the
nearby source 4U~1728$-$34 (see \S\ref{he1728}, \S\ref{sc1728rb}).
In the public \xte\/ observations of the Rapid Burster we found 66 type-I
bursts, none of which exhibited PRE.
Analysis of a subset of the thermonuclear bursts
by \cite{fox01} revealed a possible burst oscillation at 306.5~Hz. The
signal was detected by combining the power density spectra of 31
individual bursts, and is not detected in any single burst.
As also noted by \cite{fox01}, the bursts exhibit a range of profiles,
with $\tau=5$--40; some bursts appear to last for more than 100~s.
A few bursts appear to
be followed by an increase in the persistent flux level, which makes the
fluence difficult to constrain.
The peak flux was $\sim10^{-8}\
\epcs$, similar to previous observations.
The burst rate was $0.43\pm0.06$~hr$^{-1}$ on average, and increased
significantly with persistent flux (although it is possible that some
faint thermonuclear bursts are in fact mis-identified type-II bursts).
Interestingly, the Rapid Burster appears to be ``rapid'' both in the sense
of type-I and type-II bursts.

\subsection{KS~1731$-$260}
\label{s1731}

A transient located near the Galactic center ($l=1\fdg07$, $b=+3\fdg66$),
KS~1731$-$26 was discovered in August~1989 using the imaging spectrometer
aboard the {\it Mir-Kvant}\/ observatory \cite[]{sun90}.  Type~I X-ray
bursts were also first seen during these observations, lasting 10--20~s
and reaching $\sim0.6$~Crab ($3\times10^{-8}\ \epcs$).
An improved X-ray position from \chandra\/ observations \cite[]{wij01e} led to
the identification of the $J=17.32\pm0.2$, $K' = 16.36\pm0.18$ counterpart
\cite[]{rs02,mign02}.
\xte\/ observations revealed 524~Hz burst oscillations which occurred
preferentially in PRE bursts (\citealt{smb97,muno00})

The source was active at 1--$6\times10^{-9}\ \epcs$ (2.5--25~keV) although
declining steadily in intensity between 1996--2000, before transitioning
to quiescence early in 2001 \cite[]{wij01d}. After this time the source
became undetectable by \xte, and no more bursts were detected 
We found 27 bursts from KS~1731$-$26, with 4 PRE bursts observed at
relatively high $F_p\ga3.9\times10^{-9}\ \epcs$ \cite[see
also][]{muno04a}; oscillations were found preferentially in the PRE bursts
(\citealt{muno01}; see also \S\ref{milosc}).
The PRE bursts reached peak fluxes indicating a source distance of
$7.2\pm1.0$~kpc (assuming the bursts reach \leddhe; Table \ref{dist}).
Thus, the accretion rate while the source was active was
6--38\%~$\dot{M}_{\rm Edd}$ (for a bolometric correction averaging 1.62
over two observations). The PRE bursts were of short duration, with
$\tau=8.7\pm1.4$~s in the mean. In observations between 2000
August--September, during which $F_p= (2.12\pm0.07)\times10^{-9}\ \epcs$
(i.e. 14\%~$\dot{M}_{\rm Edd}$), the bursts were much longer duration
($\tau=23.8\pm0.7$~s, rise time 
$4.8\pm0.6$~s) 
and occurred regularly at
$\Delta t=2.59\pm 0.06$~hr.  The $\alpha$-values were $46.9\pm1.4$ in the
mean; in many respects these bursts were remarkably similar to those
observed from GS~1826$-$24 (\citealt{gal03d}; see also \S\ref{s1826},
\S\ref{global}). At both higher and lower \fper\ this regular bursting
ceased \cite[also noted by][]{corn03a}.

\subsection{SLX~1735$-$269}

This persistent Galactic-center ($l=0\fdg79$, $b=+2\fdg40$) source was
discovered in observations with a coded-mask X-ray telescope aboard {\it
Spacelab 2}\/ \cite[]{skin87}, and a single X-ray burst detected 
later by \sax/WFC was attributed to the source \cite[]{bazz97}. The burst
lasted 30~s, and had a peak flux of $1.8\times10^{-8}\ \epcs$.
\cite{molkov04} detected six bursts in {\it INTEGRAL}\/
observations between 2003 April and September, one with a brief precursor,
an unusually long duration of $\sim600$~s, and indications of PRE. The peak
flux of $6\times10^{-8}\ \epcs$ for the long burst suggests a distance of
5--6~kpc, depending upon the composition. The five bursts in 2003 September
were consistent with a steady recurrence time of $12.3$~hr; the estimated
$\alpha$-values were 100--200, and the inferred accretion rate was
$\approx1.7$\%~$\dot{M}_{\rm Edd}$ (for $d=8.5$~kpc).
An improved X-ray position has been determined from \chandra\/
observations by \cite{wilson03}, although this position did not match that
of any IR counterpart (to an upper limit of $J>19.4$).

The persistent flux measured by the \xte/PCA during 1997 and 2001--02 was
2.5--$6.5\times10^{-10}\ \epcs$ (2.5--25~keV); the timing behaviour during
these observations was studied by \cite{wij99b}. We found just one
burst, on 2002 January 23, with a peak flux of $(43.0\pm1.3)\times10^{-9}\
\epcs$ and $\tau=12.5$, and with no evidence for PRE.
This implies an upper limit to the distance of 7.3~kpc (Table \ref{dist}).
The inferred accretion rate during the \xte\/ observations was then
1--4\%~$\dot{M}_{\rm Edd}$.	

\subsection{4U~1735$-$44} 
\label{s1735}

This bright,
persistent atoll source at $l=17\fdg7$, $b=17\fdg5$ was first detected by
\uhu, while thermonuclear bursts were discovered during \sas\/
observations \cite[]{lewin77}. The 20--$30\arcsec$ position from \sas\/
\cite[]{jern77} led to the identification of the $V\sim17.5$ optical
counterpart, V926~Sco \cite[]{mcc77}, with optical and UV properties very
similar to those of the counterpart of Sco~X-1 \cite[e.g][]{mcb78}.
Delayed optical bursts have been observed from the counterpart
\cite[]{grindlay78a}, and periodic photometric variations indicate an
orbital period of 4.65~hr (\citealt{mp81,corbet86b}; note that
\citealt{lew93} erroneously lists the $P_{\rm orb}$ as 3.65~hr). The
bursts are notable for their rapid time scales ($\tau=4.4$~s), irregular
recurrence times and consequently extremely variable $\alpha$-values (from
250 to $\sim8000$; \citealt{vp88,lew93}).
The source is one of a growing number which exhibit extremely long
``superbursts'', detected in \sax\ observations \cite[]{corn00}.

The persistent flux measured by \xte\/ varied between 3--$8\times10^{-9}\
\epcs$ (2.5--25~keV), with evidence for a long ($\sim1000$~d) time scale
(see Fig. \ref{asmlc}). We detected 
11	
bursts in total from the PCA
observations, 6 of which exhibited evidence for PRE with a mean peak flux
of $(31\pm5)\times10^{-9}\ \epcs$. Assuming these bursts reach \leddhe,
the distance to the source is 8.5~kpc, implying a range of accretion rates
of 19--50\%~$\dot{M}_{\rm Edd}$ (adopting a bolometric correction of
1.137, the mean of values from two observations).
All the bursts were observed when $F_p\la5.4\times10^{-9}\
\epcs$, i.e. $\dot{M}\la34$\%~$\dot{M}_{\rm Edd}$. The majority 
were fast, with $\tau=3.4\pm0.4$~s.
We found four pairs of bursts with recurrence times of 1.1--1.5~hr, and
one pair with $\Delta t=0.46$~hr; the alpha values varied from 150--270.
These properties are all consistent with pure He fuel.

\subsection{XTE~J1739$-$285}

This Galactic-center region ($l=359\fdg71$, $b=+1.30$) transient was first
detected in monitoring \xte\/ observations in 1999 October
\cite[]{mmsw99}. At that time, no bursts or high-frequency variability
were detected. Little more was learned about XTE~J1739$-$285 until it
became active once again in 2005 August \cite[]{bodaghee05}, when it was
detected by {\it INTEGRAL}\/ as well as several other X-ray instruments.
Despite a {\it Chandra}\/ position with estimated uncertainty of
$0\farcs6$ (90\%), the optical counterpart has not been identified
\cite[]{torres06}.
Thermonuclear bursts were first detected also in {\it INTEGRAL}\/
observations \cite[]{brandt05}, and subsequent \xte\/ observations led to
detection of six additional bursts. Evidence for oscillations at 1122~Hz
during one of the bursts \cite[]{kaaret07a} suggests this system may
harbour the fastest-spinning neutron star yet known.

The persistent flux was greatest during the observations at the time of
discovery in 1999, at between 3 and $5\times10^{-9}\ \epcs$ (2.5--25~keV).
For the observations in 2005--6 (and during an earlier active phase in
2001)
the flux was typically in the range 0.3--$1.5\times10^{-9}\ \epcs$. At
the maximum possible distance of 10~kpc \cite[based on the peak flux of the
brightest burst, since none of the 6 bursts observed with \xte\/ exhibited
radius expansion; see also][]{kaaret07a} this corresponds to a range of
accretion rates of 3--50\%~$\dot{M}_{\rm Edd}$ (for a bolometric
correction of $1.30\pm0.06$). We note that bursts were
only observed at peak fluxes between 0.9 and $1.3\times10^{-9}\ \epcs$,
i.e. an estimated range of accretion rates of 9--13\%~$\dot{M}_{\rm
Edd}$.

The burst peak fluxes varied significantly, reaching between 9 and
$25\times10^{-9}\ \epcs$. Four of the six bursts had comparable fluences,
while the other two were much fainter, about 1/3 the mean of the brighter
bursts. The peak burst flux was significantly correlated with the burst
fluence. The burst rise times and timescales also varied significantly, in the
range 1--4.5~s and 6--12~s respectively.
One of the two faint bursts (with rise time 2~s and $\tau=6$~s) was the
only one that occurred within 24~hr of the previous burst; assuming no
bursts were missed in the (single) data gap inbetween, the recurrence time
was 1.95~hr. The corresponding $\alpha$-value was $143\pm8$, indicating
H-poor fuel, which is also consistent with the fast rise time and low
$\tau$. The longer rise times and higher $\tau$ for the other, brighter
bursts suggest a larger contribution of H in the burst fuel.

\subsection{KS~1741$-$293 ($=$ AX~J1744.8-2921)}

This Galactic center ($l=359\fdg55$, $b=-0\fdg07$) source was discovered
in 1989 August with TTM/Kvant aboard {\it Mir}\/
\cite[]{zand91}; it is within the
error boxes of both MXB~1742$-$29 \& MXB~1743$-$29 \cite[]{lhd76b}. Two
single-peaked X-ray bursts were observed, with estimated peak fluxes of
12 and $16\times10^{-9}\ \epcs$.
KS~1741$-$293 was visible to \xte\/ in all the Galactic center fields in
which bursts were observed (see \S\ref{gcf}), as well as the field centered
on GRO~J1744$-$28 (\S\ref{gcbo}), although none of those
bursts were conclusively attributable to this source. Because the source
density is so high in this region, it was not possible to independently
measure the source flux with the PCA observations. Furthermore, because
the source position is only known to $\approx1'$, KS~1741$-$293 is not
included in the list of sources for which the ASM routinely provides
intensity measurements.

An extremely
faint burst was observed on 1998 Sep 24 00:22:24 UT 
during pointings towards 1E~1740.7$-$2942, with an intrinsic peak flux of
$(1.6\pm0.3)\times10^{-9}\ \epcs$.  KS~1741$-$293, 2E~1742.9$-$2929 and
SLX~1744$-$300 are all $\approx1\arcdeg$ from the center of this field,
with KS~1741$-$293 the closest by a small margin.  Since the scaling
factor due to the collimator response is extremely sensitive to offset
angles around this value, we attribute this burst to the closest source,
KS~1741$-$293.  The resulting scaled peak flux was $(41\pm9)\times10^{-9}\
\epcs$, which is a factor of $\approx2$ larger than previously observed
bursts from the source.  While 2E~1742.9$-$2929 exhibited other bursts
soon after, on 1998 September 30, the rescaled flux assuming the burst
originated from that source instead would be almost a factor of two
higher, which would be in excess of the Eddington limit for a Galactic
center source.

\subsection{GRS~1741.9$-$2853 ($=$ AX~J1745.0$-$2855)}
\label{s17419}

This Galactic center ($l=359\fdg96$, $b=+0\fdg13$)
transient was discovered during observations with 
the ART-P coded-mask X-ray telescope aboard the
{\it GRANAT}\/ observatory \cite[]{pgs94}.
Bursts (with indications of PRE) were first detected from this source by
\sax\/ in August and September 1996 \cite[]{cocchi99}; the peak flux
indicates a distance of $\sim8$~kpc, consistent with the distance to the
Galactic center.
The source has been detected in outburst several times by {\it ASCA}\/ and
{\it Chandra}\/ \cite[]{muno03b}; the {\it Chandra}\/ observations also
revealed an extremely weak burst (peak flux $6\times10^{-10}\ \epcs$, 2--8~keV).

We attributed 8 bursts in observations covering GRS~1741.9$-$2853 to this
source; we note that \cite{stroh97} reported millisecond oscillations in 3
of the 8, originally attributed to MXB~1743$-$29 (see also \S\ref{gcbo}).
The two shortest burst intervals were 35.6~hr and 39.4~hr, similar to the
recurrence time of 1.46~d observed for MXB~1743$-$29 
\cite[]{lew93}, although it is possible that intermediate bursts
were missed during the \xte\/ observations.
Six of the 8 bursts exhibited PRE, with peak fluxes varying significantly
between 22--$52\times10^{-9}\ \epcs$. This range of peak PRE burst fluxes
implies a range of distances of 5--10~kpc, consistent with the distance to
the Galactic center (see also Table \ref{dist}). The two brightest bursts,
on 1996 July 8 01:57:47 UT and July 23 04:13:56 UT, had unusual profiles
with broad maxima, long durations ($\tau=20.8$ and 46.4~s, respectively;
see Fig. \ref{profiles}) and PRE to large radii, similar as has been
observed recently for GRS~1747$-$312 \cite[]{zand03a} as well as a few
other sources. The remaining bursts had much shorter durations,
$\tau=11\pm2$~s in the mean.

\subsection{2E~1742.9$-$2929 ($=$ GC X-1/1A 1742$-$294)}

Bursts from this Galactic center ($l=359\fdg56$, $b=-0\fdg39$) source were
probably first observed with \sas, and attributed to a source designated
MXB~1742$-$29 \cite[]{lhd76b}. 2E~1742.9$-$2929 was subsequently observed with
\arv\/ and \ein, among others. The most detailed study of the bursts 
were of 26 observed with ART-P/{\it Granat}\/ \cite{lgps01}. Both ``weak''
and ``strong'' bursts were observed; the brightest of the latter class
reached a maximum of
3.5--$4\times10^{-8}\ \epcs$ (3--20~keV).

2E~1742.9$-$2929 was the most active Galactic center burster during the period
covered by the \xte\/ observations. In the ASM the source was persistently
bright at $\approx2\ {\rm counts\,s^{-1}}$ (25~mCrab, or $6\times10^{-10}\
\epcs$ in 2--10~keV; see Fig. \ref{asmlc}).
We attributed more than 80 bursts to this source in total, the majority
from 2001 September 26 to October 8 (see \S\ref{gcften}).  All but two of
the bursts were faint, with inferred peak fluxes of $(7\pm3)\times10^{-9}\
\epcs$ in the mean, $\tau=10$--40~s, and no evidence for PRE.  The other
two bursts, observed in 
the field centered on $\alpha=17\fh44\fm02.6$,
$\delta=-29\arcdeg43\arcmin26\arcsec$ (J2000.0; see \S\ref{gcfone}) on
1997 March 20 and 1998 November 12, were of shorter duration with
$\tau\approx8$~s and exhibited strong PRE with rescaled peak fluxes of
$4\times10^{-8}\ \epcs$, consistent with the brightest bursts previously
observed from this source by {\it Granat}. About 20\% of the bursts had
recurrence times $\la0.5$~hr ($0.31\pm0.09$~hr in the mean), while 35\%
had $\Delta t=1.5$--3~hr.

\subsection{SAX~J1747.0$-$2853}
\label{sax1747}

This transient was first detected
in 1998 \cite[]{zand98b} at a position ($l=0\fdg21$, $b=-0\fdg24$)
consistent with a source detected in the 1970s by rocket-borne
coded-mask/\arv\/ observations, GX~.2$-$.2 \cite[also known as
1A~1743-288;][]{psw78}.
Bursting behaviour was first observed by \sax\ \cite[]{sidoli98}.
Followup observations revealed at least one radius-expansion burst,
indicating a source distance of $\sim9$~kpc \cite[]{nat00}.

The source appeared in outburst on several subsequent occasions, including
2000 February-June and 2001 September \cite[]{nat04}. During \xte\/
observations in 2001 September--October, 15 bursts were detected which
we attributed to this source (see \S\ref{gcften}). The bursts were
relatively long, with $\tau=11\pm2$~s in the mean; 10 of the 15
exhibited PRE, and 7 bursts also exhibited a distinct double-peaked
morphology. If the PRE bursts reached \leddhe, the estimated mean
peak flux implies a distance of 6.7~kpc (see Table \ref{dist}). Assuming
that SAX~J1747.0$-$2853 was the only active source in the field during the
2001 September observations, the persistent flux was
1.5--$1.7\times10^{-9}\ \epcs$, equivalent to an accretion rate of
7--8\%~$\dot{M}_{\rm Edd}$.	
The shortest recurrence times measured for
the bursts was 3--4.2~hr.

\subsection{IGR~J17473$-$2721 ($=$ XTE~J1747$-$274)}
\label{xmmu1747}
\label{igrj17473}

The transient IGR~J17473$-$2721 
($l=1\fdg410$, $b=0\fdg425$)
was first detected with {\it INTEGRAL}\/ in 2005 March
and April \cite[]{greb05b}.
One month later, a previously unknown source designated
XTE~J1747$-$274 was reported in \xte\/ observations of the Galactic bulge
\xte\/ observations \cite[]{mark05d}. Subsequent {\it Swift}\/ and {\it
Chandra}\/ observations of the field detected a single active source,
indicating that IGR~J17473$-$2721 and XTE~J1747$-$274 were the same
source \cite[]{kennea05a,juett05}.

Two X-ray bursts were detected in pointed \xte\/ observations, on 2005 May
24 and 31, which we attributed to this source (see \S\ref{igrsrc}).
The corresponding peak fluxes, taking into account the offset between the
pointing direction and the source position, were 5.5 and $4.5\times10^{-8}\
\epcs$. Neither burst exhibited oscillations or indications of
radius-expansion. The correspoinding
upper limit on the distance (from the brightest burst, and adopting the
Eddington limit for pure He material) is 6.4~kpc. The rise times were
5 and 7~s, which in addition to the relatively long $\tau=20.7$ and 17.4~s
indicate H-rich fuel.

\subsection{SLX~1744$-$299/300}
\label{slx1744}

This close ($2\farcm8$ separation) pair of Galactic center ($l=359\fdg26$,
$b=-0\fdg91$) sources was discovered during mapping observations with the
SL2-XRT instrument aboard {\it
Spacelab-2}\/ \cite[]{skin87}.
Their dual nature was revealed
when a burst was observed from the southern source, with a peak flux of
$1.4\times10^{-8}\ \epcs$ \cite[including a 20\% correction for
absorption;][]{skin90}. Bursts from this region were also observed by
\exo, TTM/{\it Kvant} and \sas\/ \cite[]{lew93}.

The source was persistently active in the ASM at $\approx2\ {\rm
counts\,s^{-1}}$ (equivalent to 25~mCrab, or $6\times10^{-10}\ \epcs$ in
2--10~keV; see Fig. \ref{asmlc}). The source was only observed with the
PCA well off-axis in a field containing a number of other sources (see
\S\ref{gcfone}), so it was not possible to measure the flux more
precisely.  We found 3 bursts in public observations attributable to this
pair of sources.  The bursts were short, with $\tau=6$--7.6~s, and with
peak fluxes between 1.4--$1.9\times10^{-8}\ \epcs$, roughly consistent
with earlier observations.

\subsection{GX~3+1}

This persistent Galactic center 
source ($l=2\fdg29$, $b=+0\fdg79$)
was discovered during a rocket flight in 1964 \cite[]{bowyer65}, and
subsequently proved to be one of the brightest persistent sources in the
Galaxy.
A detailed study of low-frequency QPOs in this and other sources led to
the introduction of the atoll/Z-source classification \cite[]{hvdk89}.
No optical counterpart is known \cite[e.g.][]{ncl91}.
Thermonuclear X-ray bursts were discovered during \hak\/ observations
\cite[]{mak83}. The bursts were observed at a particularly low \fper\
level, and reached peak fluxes of 4--$8\times10^{-8}\ \epcs$ \cite[see
also][]{lew93}.
\cite{kuul02b} also found evidence for 
a possible superburst from \xte/ASM monitoring; an
intermediate-duration event, lasting $\approx30$~min, was detected by {\it
INTEGRAL}\/ \cite[]{chenevez06}.
The most detailed study of the thermonuclear bursts to date was by
\cite{hartog03}, who detected 61 bursts with \sax/WFC and found them
remarkably homogeneous with a weighted mean $\tau=3.63\pm0.10$~s. The
burst rate dropped by a factor of $\approx6$ as the persistent flux
increased through a relatively small range.

Long-term flux measurements suggest that \fper\ varies on a time scale of a
few years (e.g. Fig. \ref{asmlc}). At the peak in 2001--03 the PCA flux
was 8--$12\times10^{-9}\ \epcs$, while during the minimum was
4--$7\times10^{-9}\ \epcs$.
\xte\/ observations revealed the first radius expansion burst from the
source on 1999 August 10 18:35:54 UT \cite[]{kuul00}, leading to a
distance estimate of $\sim4.5$~kpc with estimated uncertainty of up to
30\%. 
The rapid rise (0.75~s) and short duration ($\tau=4.96\pm0.13$~s) of the
burst suggests He-rich fuel, and assuming the flux reaches \leddhe\ a
somewhat larger value of 6.5~kpc is indicated (see Table \ref{dist} and
\citealt{hartog03}). In that case, the persistent flux range corresponds
to accretion rates of 
17--50\%~$\dot{M}_{\rm Edd}$.	
Just one other burst was observed, on
2001 Aug  7 16:38:46 UT, with $\tau=11.6\pm1.0$~s and no evidence for PRE.
While the \fpk\ was significantly lower than for the PRE burst, the
fluence was significantly larger.

\subsection{1A~1744$-$361}	

This variable source ($l=354\fdg14$, $b=-04\fdg20$) was discovered in {\it
Ariel V}\/ observations in 1976 \cite[]{carp77}. 
A single X-ray burst was observed from the source in 1989 August by the
TTM/COMIS instrument onboard {\it Mir}\/ \cite[]{emelyanov01}.
A new period of
activity beginning 2003 November was detected initially by the \xte/ASM
\cite[]{remillard03a}. \xte\/ and {\it INTEGRAL}\/ detected the source
(briefly designated XTE~J1748$-$361) again in 2004 April, at which time
the optical counterpart was also identified \cite[]{steeghs04}.
More extensive \xte\/ observations followed a subsequent outburst in 2005
July, revealing energy-dependent dips separated by 97~min, and a
thermonuclear burst with 530~Hz burst oscillations \cite[]{bhatt06b}. The
correlated spectral and timing variations were similar to those of atoll
sources \cite[]{bhatt06e}.

The X-ray flux from the \xte\/ observations was in the range
0.5--$2\times10^{-9}\ \epcs$ (2.5--25~keV).  The single X-ray burst was
observed when the persistent flux was $1.1\times10^{-9}\ \epcs$, and
reached a peak flux of $(19.0\pm0.6)\times10^{-9}\ \epcs$. With no
evidence for radius expansion, the peak burst flux is a lower limit to the
Eddington flux, implying an upper limit on the distance of 11~kpc. At this
distance, the X-ray flux range implies an accretion rate between 5 and
18\%~$\dot{M}_{\rm Edd}$. The burst was fairly fast, with a 1~s rise time
and $\tau=5.8\pm0.2$~s. These properties suggest H-poor fuel.

\subsection{SAX J1748.9$-$2021 in NGC 6440}

SAX J1748.9$-$2021 is one of an estimated 4--5 LMXBs \cite[]{pooley02} in the
globular cluster NGC~6440 \cite[$l=7\fdg73$, $b=+3\fdg80$,
$d=8.4^{+1.5}_{-1.3}$~kpc;][]{kuul03a}.  Transient X-ray emission was
detected from the cluster in 1971 \cite[]{markert75,fjt76}, 1998
\cite[]{zand99b}, and 2001 \cite[]{zand01b}, although it was not certain
initially that the outbursts were all from the same source.  
An observation with {\it Chandra}\/ in 2001 also allowed the optical
counterpart to be securely identified \cite[]{zand01b}.
Thermonuclear bursts were first
detected during the 1998 outburst \cite[]{zand99b}. Three bursts were detected
in the WFC, with recurrence times of $\approx2.8$~hr. A similar burst
observed with the narrow-field instruments reached a peak flux of
$1.7\times10^{-8}\ \epcs$.
No bursts were detected by \xte\/ during the 1998 outburst, but
observations during the 2001 outburst revealed 16 bursts, one of which
exhibited weak evidence for burst oscillations at
$409.7$~Hz \cite[]{kaaret03}.
A subsequent outburst was detected in PCA scans of the Galactic bulge region on
2005 May 12--16 \cite[]{mark05a}.
Intermittent persistent pulsations at 442~Hz were detected
during the 2001 and 2005 outbursts \cite[]{gavriil07,altamirano07}. The
pulsations exhibited Doppler shifts from an 8.7~hr orbit.

Six of the bursts observed by \xte\/ exhibited PRE, most with a pronounced
double-peaked maximum (see Fig. \ref{profiles}) which varied in flux
between
28--$40\times10^{-9}\ \epcs$.  The inferred distance (assuming, based on
the short PRE burst duration of $\tau=6.9\pm1.3$~s, that the bursts
reached \leddhe) is 8.1~kpc (see Table \ref{dist}). The range of \fper\/
measured by \xte, from $6\times10^{-10}\ \epcs$ prior to the 2001 outburst
to $4.4\times10^{-9}\ \epcs$ (2.5--25~keV) at the peak, thus translates to
an inferred accretion rate range of 3--25\%~$\dot{M}_{\rm Edd}$ (with a
bolometric correction averaging $1.157\pm0.015$ over four observations).

The non-PRE bursts bursts had longer durations, $\tau=15\pm4$ on average,
and peaked in the range 
1.9--$2.2\times10^{-8}\ \epcs$
\cite[roughly consistent with the burst observed by the
\sax/NFI;][]{zand99b}. The inferred burst recurrence times varied between
1.02--1.90~hr, rather faster than in previous observations
\cite[]{zand99b}. While the measured $\alpha$ was not correlated with
\fper\ (which only varied by about 9\% rms over the observations with
bursts), it was strongly anticorrelated with $\tau$, with $\alpha=100$--150
for the fast bursts and $\alpha=50$--65 for the slow bursts.
This is consistent with the fast bursts arising primarily from He burning,
while in the slow bursts the fuel is a mixture of H/He (see also Fig.
\ref{taualpha}).  Perhaps most interestingly, the long and short bursts
appeared to alternate 
independently of the 
persistent flux.

\subsection{EXO~1745$-$248 in Terzan 5}
\label{s1745}

EXO~1745$-$248, in the Galactic bulge ($l=3\fdg84$, $b=+1\fdg46$) globular
cluster Terzan~5,
was discovered
during \hak\/ observations of the region \cite[]{mak81}. The source has
been noted for episodic burst behaviour, as well as burst intervals as
short as 8~min (\citealt{inoue84}; see also \citealt{lew93}).  PCA/\xte\/
scans of the bulge detected a new transient outburst in 2000 July
\cite[]{markwardt00}. Followup pointed observations initially revealed 15 X-ray
bursts
with an average separation of 25~min, as well as dipping activity and QPOs
around 65 and 134~mHz
\cite[]{markwardt00b}.

Prior to the outburst peak, the source was active at a flux level of
1--$5\times10^{-9}\ \epcs$ (2.5--25~keV).
During this time we detected 21 type-I bursts with peak fluxes of
3--$19\times10^{-9}\ \epcs$, and no evidence for PRE.  For those
observations where we saw more than one burst, the recurrence
times were between 17 and 49~min.  
The estimated $\alpha$-values were in the range
20--46
which (along with the long burst durations $\tau\approx25$~s) 
indicates H-rich fuel. 
Following the outburst peak (between August
15--18) the frequent bursting ceased, and just two more bursts were
observed, on September 24 and October 2.
These two bursts were of markedly different character to
those prior to outburst maximum, with peak fluxes of
$\approx6\times10^{-8}\ \epcs$,
shorter durations of $\tau=6.6$ and 7.3~s,
and both exhibiting strong PRE \cite[see Fig. \ref{profiles};][]{kuul03a}. For
the distance of 8.7~kpc for Terzan~5 derived by \cite{kuul03a}, 
we expect an Eddington flux
for cosmic abundances of $1.7\times10^{-8}\ \epcs$, or $3\times10^{-8}\
\epcs$ for pure He; thus, the two PRE bursts appear to be
super-Eddington by a factor of at least 2.
However, the burst peak fluxes are consistent with the 
more recent distance of $5.5\pm0.9$~kpc
\cite[][cf. with Table \ref{dist}]{ortolani07}.
\cite{kuul03a}
also noted that the peak fluxes for PRE bursts from this source measured
by different instruments exhibited a large (factor of $\sim3$) variation.

\subsection{4U~1746$-$37 in NGC 6441}
\label{s1746}

Persistent emission from 4U~1746$-$37 ($l=353\fdg53$, $b=-5\fdg01$) was
first recorded in the 3rd \uhu\/ catalog \cite[]{3ucat};  thermonuclear
X-ray bursts were probably first observed during \sas\/ observations
\cite[]{lc77}.  PRE bursts have previously been observed by \exo\/
\cite[]{szt87} with peak fluxes of $(1\pm0.1)\times10^{-8}\ \epcs$
\cite[see also][]{lew93,kuul03a}.
Periodic intensity dips every 5.7~hr were reported from {\it Ginga}\/
observations \cite[]{sansom93}; more recent analysis of \xte\/ data
indicate a somewhat shorter dip period of $5.16\pm0.01$~hr
\cite[]{balucinska03}.
The optical counterpart, identified from an \hst\/ image following
\chandra\/ observations, also shows variations which are consistent with a
period of $\approx5$~hr \cite[]{homer02}.

4U~1746$-$37 was active but variable at between 0.16--$1.6\times10^{-9}\
\epcs$ (2.5--25~keV) throughout the \xte\/ observations. For the distance
to the cluster of $11.0_{-0.8}^{+0.9}$~kpc \cite[]{kuul03a} this
corresponds to a range of accretion rates of 2--16\%~$\dot{M}_{\rm Edd}$
(for a bolometric correction of between 
1.09--1.45,
depending upon the
epoch).
The catalog contains a total of 
30 bursts from 4U~1746$-$37. The burst
properties were clustered into three groups, depending upon the persistent
flux level; at $F_p\approx1.6\times10^{-10}\ \epcs$ (2.5--25~keV), we
detected long-duration bursts with $\tau=13\pm2$~s in the mean, while at
higher \fper\ the bursts fell into two groups, one even longer duration
with $\tau=31\pm3$~s and the other very short with $\tau=4.5\pm0.9$~s.
This latter group included three PRE bursts, which reached peak fluxes a
factor of two lower than previous PRE bursts from the source, at
$(5.3\pm0.9)\times10^{-9}\ \epcs$ (see also \S\ref{pflux} and Table
\ref{dist}). We note that \cite{kuul03a} did not categorize these three bursts
as PRE.
Overall the peak fluxes were approximately bimodally distributed, with 15
bursts reaching fluxes between (0.4--$2.8)\times10^{-9}\ \epcs$, and the
rest peaking at between (3.8--$6.3)\times10^{-9}\ \epcs$. The
characteristic $\alpha$-values for the bright bursts ($\tau\approx12$~s)
was 35--50, while for the faint bursts ($\tau\approx32$~s) was 140--180
(we note that 4U~1705$-$44 was the only other source with bursts with
$\tau>20$~s and $\alpha>100$).

On two separate occasions (1996 October
25--27 and 1998 November 7th), a train of regular bright (faint) bursts
was interrupted by an out-of-phase faint (bright) burst. As discussed by
\cite{gal04a}, such interrupted regular bursting has not been previously
observed, and is difficult to understand in the context of standard burst
models. An alternative possibility is
the presence of {\it two}\/ sources in the cluster, bursting
independently. From the stellar encounter rate for the host cluster
NGC~6441 (which is the second-highest of all Galactic globular clusters),
we expect around 6 LMXBs in the cluster \cite[]{pooley03,heinke03b}.
High spatial resolution observations
during intervals of burst activity are required in order to independently
localize individual bursts and confirm this hypothesis.

\subsection{SAX~J1750.8$-$2900}
\label{sax1750}

This Galactic center ($l=0\fdg45$, $b=-0\fdg95$) source 
was first detected by \sax\/ in 1997 as a weak, bursting transient
\cite[]{nat99}.
The source was detected in outburst once more in 2001 March, exhibiting an
initial rise and steep fall early in March, followed by a second peak
around April 21.

We found four bursts between 2001 April 6--15 attributable to the source,
three of which were bright with $F_{\rm peak}\approx5\times10^{-8}\ \epcs$
and two of those with evidence for PRE. The corresponding distance
estimate (assuming, given the short durations $\tau=5$--7.3~s of the
bursts, that they reach \leddhe) is $6.79\pm0.14$~kpc (see Table
\ref{dist}). 
Burst oscillations at 600.75~Hz were detected in the second burst on 2001
April 12, which also exhibited PRE implying a
distance $6.3\pm0.7$~kpc \cite[]{kaaret02}.
The peak flux measured by the PCA during the 2001 outburst
was $2.7\times10^{-9}\ \epcs$ (2.5--25~keV), which corresponds to
13\%~$\dot{M}_{\rm Edd}$;	
during the secondary peak the source reached
$2.4\times10^{-9}\ \epcs$, while the persistent level was
$\approx3\times10^{-10}\ \epcs$ 
($1.4$\%~$\dot{M}_{\rm Edd}$,	
although the flux may contain contribution from other souces in the field;
see \S\ref{sax12}). The shortest recurrence time measured, between one of
the PRE bursts and the final, faint burst ($F_{\rm peak}=7.2\times10^{-9}\
\epcs$, 2.5--25~keV) was 1.58~hr.

\subsection{GRS~1747$-$312 in Terzan 6}
\label{sc1747}

GRS~1747$-$312 ($l=358\fdg56$, $b=-2\fdg17$), in the globular cluster
Terzan 6, was discovered by 
ART-P/{\it GRANAT}\/ in 1990--1992 \cite[]{pgs94}.  \xte\
observations revealed quasi-periodic outbursts every $\approx4.5$~months,
as well as thermonuclear bursts, eclipses and dips \cite[]{zand03}. The
orbital period is 12.36~hr.

\xte/PCA observations of two successive outbursts in 2001 May-June and
October found a maximum persistent flux of 8--$9\times10^{-10}\ \epcs$
(2.5--25~keV); the minimum flux measured following the first outburst was
$0.6\times10^{-10}\ \epcs$. At the distance to the host cluster of
$9.5_{-2.5}^{+3.3}$~kpc \cite[]{kuul03a}, this corresponds to a range
of accretion rates of 
0.6--8\%~$\dot{M}_{\rm Edd}$.	
Of the 
seven 
bursts from public \xte\/ observations towards GRS~1747$-$312,
four have been previously discussed by \cite{zand03}. The bursts had short
durations, of $5.5\pm1.2$~s on average. Two of the
bursts
exhibited PRE, and despite their similar profiles reached distinctly
different peak fluxes of
$1.0$ and $1.7\times10^{-8}\ \epcs$ respectively \cite[see also
\S\ref{pflux} and][]{kuul03a}.
The first PRE burst may have been fainter because it actually originated from
the nearby ($\Delta\theta=0\fdg485$) source SAX~J1752.3$-$3138
\cite[]{cocchi01b} instead.  If that was the case, the corrected peak flux
would be consistent with the earlier PRE burst observed from that source
by \sax\/. With only 4 PCUs on during that observation, and rather low
count rate at the peak of the burst, it was not possible to rule out
either source as the origin. Furthermore, for none of the other 3 bursts
can we rule out an origin at GRS~1747$-$312, so that we can attribute none
of the bursts concusively to SAX~J1752.3$-$3138.  Thus, we attribute all
the bursts to GRS~1747$-$312.

One additional PRE burst was observed in the field of the millisecond X-ray
pulsar XTE~J1751$-$305, but was subsequently attributed to GRS~1747$-$312
\cite[]{zand03a}.  This burst exhibited approximately constant flux
for $\approx50$~s, interrupted between 10--30~s by an excursion up to a
maximum flux of almost
a factor of two higher (see Fig. \ref{profiles}). During this excursion,
however, the blackbody radius reached a maximum, and the color temperature
reached a minimum of $\approx0.6$~keV, at which level extrapolating the
blackbody spectra outside the PCA bandpass becomes particularly
error-prone. Thus, for this burst we exclude the data during the radius
maximum for the purposes of calculating the peak flux, and instead adopt
the mean peak flux between 5--10 and 30--50~s as the peak, i.e.
$(22.4\pm0.7)\times10^{-9}\ \epcs$. Even with this correction, the peak
flux significantly exceeds than that of the other two PRE bursts, leading
to a fractional standard deviation of peak PRE burst flux of \fstdevmax,
the largest of any of the sources with PRE bursts (see \S\ref{pflux}).
The estimated fluence for the brightest PRE burst was almost thirty times
larger than the next most energetic PRE burst; the profile was similar to
other extreme bursts from 4U~1724$-$307 (see \S\ref{s1724}) and 4U~2129+12
(\S\ref{s2129}).

\subsection{XTE J1759$-$220}
\label{s1759}

This quasi-persistent source towards the Galactic bulge ($l=7\fdg58$,
$b=0\fdg78$) was detected by \igr\/ between 2003 March--April
\cite[]{lutovinov03}, and was subsequently identified with a new source
detected by \xte\/ since 2001 February \cite[]{markwardt03c}.
Significant spectral variability was measured during the \igr\/
observations between 2003 and 2004, suggestive of transitions between
low/hard and soft/high states \cite[]{lutovinov05}. In addition, there was
evidence of dipping behaviour in the \xte\/ observations, suggesting high
inclination.
A single X-ray burst was detected in an \xte\/ observation on 2004
September 13.  The burst was faint, reaching a peak flux of just
$(5.07\pm0.16)\times10^{-9}\ \epcs$. Both the slow (4~s) rise and long
$\tau=24.8$~s indicate a H-rich burst; the upper limit on
the distance (assuming $X=0.7$) is 16~kpc. Assuming that the source is
equidistant with the Galactic bulge, the distance is $\sim8.5$~kpc.
The X-ray flux measured by \xte\/ during 2004 Mar--September was between
2--$4\times10^{-10}\ \epcs$ (2.5--25~keV). For a distance of 8.5~kpc, this
corresponds to an accretion rate of a few percent $\dot{M}_{\rm Edd}$. 

\subsection{SAX J1808.4$-$3658} 
\label{s1808}

The first accreting millisecond X-ray pulsar, SAX J1808.4$-$3658
($l=355\fdg38$, $b=-8\fdg15$) was discovered during \sax/WFC observations
\cite[]{zand98c}. Two bright thermonuclear bursts were also observed from
the source, separated by 14~hr. \xte\/ observations during a subsequent
outburst in 1998 revealed persistent millisecond pulsations at 401~Hz
\cite[]{wij98b}, modulated by Doppler shifts arising from a 2.1~hr binary
orbit \cite[]{chak98d}. \cite{gil99} made observations of the $V\sim20$
(in quiescence) optical counterpart as it faded following the outburst
peak.  Reanalysis of the 1996 \sax\/ discovery observations revealed a
third, previously undetected, brighter burst, leading to a revised
distance estimate of 2.5~kpc \cite[]{zand01}. The source has continued to
exhibit outbursts every $\sim2$~yr; the latest was in 2005 June
(\citealt{mark05b}; see also \citealt{wij04a}). At the peak of
the October 2002 outburst four bursts were observed by \xte/PCA, each with
burst oscillations also at 401~Hz, confirming the link with the NS spin
\cite[]{chak03a}.
One of these bursts exhibited a faint precursor event, 1~s
the burst, also exhibiting oscillations \cite[]{bhatt06d}; variations in the 
observed oscillation frequency have been interpreted as arising from
spreading of the burning front following ignition at mid-latitudes of the
neutron star \cite[]{bhatt06f}.

The 2002 outburst was the best sampled so far by \xte, and reached a
maximum persistent
flux of $2.6\times10^{-9}\ \epcs$ (2.5--25~keV). The four bursts were
all observed within a 100~hr interval just after the outburst peak, and
the last three were separated by 21.1 and 29.8~hr, respectively. The
bursts were quite homogeneous, all exhibiting strong PRE; the fluence
increased steadily by 30\%, and $\tau$ by 20\% (total) as \fper\
decreased. The peak fluxes exhibited little variation and indicate a
distance of 2.77 (3.61) kpc assuming the bursts reach \leddh\ (\leddhe;
see also Table \ref{dist}). For $d=3.61$~kpc the peak persistent flux
corresponds to a maximum accretion rate of just 5.5\%~$\dot{M}_{\rm Edd}$
(adopting a bolometric correction averaging $2.12\pm 0.04$ over four
observations near the peak).
The estimated $\alpha$-values
for the last two burst intervals were
$\alpha=148$ and 167, respectively.

The $\alpha$-values, as well as the fast rise
times ($\approx0.5$~s) suggest almost pure He fuel. A comparison of the
burst properties with an igition model \cite[]{cb00}
indicates that the mean H-fraction at ignition is $\approx0.1$ \cite[]{gal06c}.
These are the first He-rich bursts which have been securely
observationally identified. The ignition model comparison allowed an
estimate of the distance, which was consistent with the estimate
derived by equating the long-term time-averaged X-ray flux with the expected
mass transfer rate due to gravitational radiation, as well as the peak
flux of the bursts. The derived distance range for the source is
3.4--3.6~kpc.

\subsection{XTE~J1814$-$338}
\label{s1814}

XTE~J1814$-$338 ($l=358\fdg75$, $b=-7\fdg59$) was discovered in outburst
during \xte/PCA scans of the Galactic center region \cite[]{markwardt03a}.
Subsequent PCA observations revealed persistent pulsations at 314.4~Hz,
making this source the fifth known accretion-powered millisecond pulsar
\cite[]{stroh03a}.  Doppler variations in the persistent pulsation
frequency indicate an
orbital period of 4.28~hr.
A total of 
28 bursts were observed throughout the outburst, all with burst
oscillations at the pulsar frequency \cite[see e.g.][]{watts05}, and all
without conclusive evidence of PRE.  From the maximum peak flux of the
bursts, an upper limit to the
distance of $\approx 8$~kpc is derived.

The persistent flux level while the source was bursting was
0.4--$0.5\times10^{-9}\ \epcs$,
equivalent to 3.6--4.5\%~$\dot{M}_{\rm Edd}$ averaged over the NS surface
(for $d=8$~kpc and a bolometric correction of $1.86\pm0.3$).
We found five bursts separated by $<10$~hr. The burst times were not
consistent with a constant $\Delta t$, and instead suggest irregular
reccurence times of 4--6~hr (with longer intervals resulting from missed
bursts in data gaps).  The two bursts with shorter recurrence times (1.7
and 2.3~hr) both had fluences around $1\times10^{-7}\ \epc$ (as did
three others), while the remainder had fluences of
$(2.6\pm0.3)\times10^{-7}\ \epc$ in the mean. Thus, the burst behaviour
appears to consist of irregular bursts with recurrence times of 4--6~hr
and roughly constant fluence interrupted occasionally by bursts with
approximately half the fluence, occurring after approximately half the
usual interval.
The measured alpha values from the bursts with recurrence times $\la10$~hr
ranged between 55--100.
The low $\alpha$ values, coupled with the relatively long
burst time scales of $\tau=30\pm6$ in the mean indicate that mixed H/He
makes up the burst fuel.

The burst behaviour of XTE~J1814$-$338 is in marked contrast to the
infrequent, He-rich bursts observed at similar accretion rates from the
other accretion-powered millisecond X-ray pulsars, SAX~J1808.4$-$3658 (see
\S\ref{s1808}) and HETE~J1900.1$-$2455 (\S\ref{h1900}). 
The long bursts in XTE~J1814$-$338 may arise instead from H-ignition, as
is seen in EXO~0748$-$676 (\S\ref{s0748}).

\subsection{GX~17+2}
\label{sgx17p2}

One of the first cosmic X-ray sources ever detected \cite[e.g.][]{bradt68},
GX~17+2 ($l=16\fdg43$, $b=+1\fdg28$) is one of the few Z-sources which
exhibits thermonuclear bursts \cite[]{hvdk89}.
Despite a precise position from radio detection
\cite[]{hjellming78} the optical counterpart long eluded observers; 
HST observations finally led to identification of a variable IR
counterpart \cite[]{dma99,callanan02} with $\sim4$~mag modulation on
a time scale of days to weeks \cite[]{bandy02}.
Bursting behaviour was discovered with \hak\/ \cite[]{oda81}; the unique
features of the characteristic long bursts unique to this source (rise
time 1.5~s, duration 3--15~min) were subsequently discussed by
\cite{tawara84}. Difficulties for the ``standard'' burst analysis
presented by these long (as well as short $\sim10$~s) bursts were
explored by \cite{szt86}, who concluded they were indeed type-I
(thermonuclear) bursts.
A search for superbursts in \sax/WFC data was presented by \cite{zand04a}.

GX~17+2 was persistently bright during the \xte\/ observations at
15--$33\times10^{-9}\ \epcs$.
\xte\/ observed 
12	
thermonuclear bursts from the source, 10 of which were
studied in detail previously by \cite{kuul02a}. One
additional event, on 1998 Nov 19 03:39:10 UT, was not discussed by those
authors;
however, the lack of spectral softening during this event appears to rule
out a thermonuclear burst.  Instead, this event, along with the four other
``flares'' noted by \cite{kuul02a} may be type-II bursts (i.e. accretion
instability events), analogous to those observed in the Rapid Burster and
previously observed from GX~17+2 during \ein\/ observations \cite[]{kg84}.
Two of the short ($\tau\la10$~s) and six of the long ($\tau\sim100$--300~s)
bursts exhibited indications of PRE, with peak fluxes of
$14.8\times10^{-9}\ \epcs$ in the mean. Neglecting the persistent
emission, this suggests a distance of 9.8 (12.8)~kpc, assuming the bursts
reach \leddh\ (\leddhe). 
In GX~17+2, as distinct from almost all the other bursters, the
persistent flux is comparable to the peak burst flux; even for a distance
of 10~kpc, the persistent flux level suggests accretion rates consistently
$\ga \dot{M}_{\rm Edd}$ (for a bolometric correction of $1.083\pm0.017$).
Thus, the estimated distance will be significantly closer if we sum the
two contributions for our estimate of the Eddington flux.
However, we note that detailed
spectral studies seem to indicate that the two are truly independent, so
that combining them may not be correct \cite[]{kuul02a}.
We found three pairs of bursts with relatively short intervals of 5.77,
13.0 and 11.5~hr, and assuming that this represents
the recurrence time, we derive $\alpha=7200\pm600$, $6000\pm1000$ and
$580\pm60$, respectively. The first two bursts were of short duration,
while the third was long with $\tau=92.1$~s, and followed another long
($\tau=113$~s) burst.

\subsection{3A~1820$-$303 ($=$ Sgr X-4) in NGC 6624}
\label{s1820}

Thermonuclear bursts from this globular cluster source at $l=19\fdg06$,
$b=18\fdg81$ were first discovered by \ans\/ \cite[]{grindlay76}, although
some bursts were observed earlier but not initially detected by \sas\/ 
\cite[]{clark76}.  The $B=18.7$ UV/optical counterpart detected
by \cite{king93} was later confirmed by the detection of periodic
variations at $P_{\rm orb}=685$~s (\citealt{kw86,swp87}; see also
\citealt{anderson97}).  The NS is thus in an ultracompact binary with an
evolved, H-poor companion, and one
of the shortest orbital periods known.
The source is also notable for a steady long-term (176~d) periodicity in the
persistent X-ray intensity, detected initially with \vfb\/ observations
\cite[][see also Fig.  \ref{asmlc}]{pt84}.  Long-term observations
indicate that regular motions throughout the color-color diagram of this
atoll source also reflect the 176~d period \cite[]{bloser00b}.
Some authors have suggested that this periodicity indicates that the
source is in fact a heirarchical triple \cite[e.g.][]{chou01}.

X-ray burst activity appears to be confined to within $\pm23$~d of the
minima in the long-term periodicity. A 20~hr \exo\/ observation found 7
extremely regular ($\Delta t=3.21\pm0.04$~hr) PRE bursts
\cite[]{haberl87}; PRE bursts were also detected by \sas\/
\cite[]{vlvp86}. The $\Delta t$ was found to decrease with increasing
$F_p$ \cite[]{clark77b}, up to a critical level of around $2\times10^{-9}\
\epcs$ (i.e. $\sim9$\%~$\dot{M}_{\rm Edd}$ for $d=7.6$~kpc;
\citealt{kuul03a}) at which the bursts stopped completely.  A comparison
of the burst properties with theoretical ignition models indicates pure He
fuel, which is consistent with the expected H-poor nature of the donor
\cite[]{cumming03}.
\xte\/ observations also revealed 
a ``superburst'' with 3~hr duration, following (by $<20$~s) a normal (type I)
thermonuclear burst \cite[]{stroh02}.

The \xte\/ observations of 3A~1820$-$303 were almost always made when the
source was above the critical threshold for burst activity; the persistent
flux level was 3--$16\times10^{-9}\ \epcs$ (2.5--25~keV). For $d=7.6$~kpc
\cite[the host cluster distance;][]{kuul03a}.
this is equivalent to
18--95\%~$\dot{M}_{\rm Edd}$.	
As a result, only 
five 
thermonuclear bursts were detected, when the source was between
$F_p=2.7$--$3.7\times10^{-9}\ \epcs$.
All five bursts exhibited extreme PRE and reached fluxes of around
$54\times10^{-9}\ \epcs$,
leading to a distance estimate of 6.4~kpc (for $X=0$, based on the
H-poor nature of the mass donor). This value is somewhat lower than the
distance to NGC~6624, indicating that the bursts are slightly
under-luminous.

\subsection{GS~1826$-$238} 
\label{s1826}

This quasi-persistent source ($l=9\fdg27$, $b=-6\fdg09$) was discovered 
during \gin\/ observations \cite[]{tanaka89}.  Thermonuclear bursts were
first conclusively detected by \sax\/ \cite[]{ubert97}, although this
source may also have been the origin of X-ray bursts observed much earlier
by \osoe\/ \cite[]{becker76b}.  
Optical photometry of the $V\approx19$ counterpart
\cite[]{motch94,barret95} revealed a 2.1~hr modulation, as well as optical
bursts \cite[]{homer98}. The delay time measured between the X-ray and
optical bursts is consistent with the binary separation for a 2.1~hr orbit
\cite[see also][]{kong00}.  Based on optical measurements, the distance to
the source is at least 4~kpc \cite[]{barret95}; since no PRE bursts have
been observed, an upper limit of 8~kpc has been derived from the peak
fluxes of bursts measured by \sax, \asca\/ and \xte\/
\cite[]{zand99,kong00}, placing the source just outside the Galactic
bulge.
Analysis of the $\approx260$ bursts observed by the \sax/WFC revealed that
the source consistently exhibits approximately periodic bursts, with a
recurrence time which decreases significantly as the persistent flux
increases \cite[]{clock99,corn03a}.

\xte\/ observations revealed that the source intensity steadily increased
between 1997--2003, from 1.1--$1.9\times10^{-9}\ \epcs$ (2.5--25~keV). For
$d=6$~kpc, this corresponds to a range of accretion rates of
5--9\%~$\dot{M}_{\rm Edd}$ (for a bolometric correction of
$1.653\pm0.009$). We detected a total of 
54 
remarkably homogeneous, long
($\tau=39\pm3$~s, rise time $6.0\pm0.8$~s) bursts in the \xte\/
observations, with regular recurrence times that decreased proportionately
with the increase in \fper\ (\citealt{gal03d}; see also \S\ref{global}).
The mean $\alpha$-value was $37.5\pm 1.2$\footnote{Note that this is
slightly smaller than the value quoted by \cite{gal03d} of $41.7\pm1.6$,
due to an improved estimate of the burst fluence. The fractional
variation in $\alpha$ with \fper\ is unchanged.},
indicating a high proportion of H in the burst fuel.
None of the bursts exhibited any evidence for PRE.

\subsection{XB~1832$-$330 in NGC 6652}

This globular-cluster source at $l=1\fdg53$, $b=-11\fdg37$ was discovered
by \ros\/ 
during a probable transient outburst \cite[]{phv91}; the first
thermonuclear bursts were observed by \sax\ \cite[]{zand98}. The bursts
were long, with exponential decay times of 16 and 27~s, and peak fluxes of
$\sim8\times10^{-9}\ \epcs$ (bolometric). A third burst was detected by
\asca, reaching a peak flux of $2\times10^{-9}\ \epcs$ \cite[]{ms00}.  A
\chandra\ observation reavealed 3 new sources in the cluster, and the
improved position for XB~1832$-$330 allowed identification of the blue
variable $M_V=3.7$ optical counterpart from archival {\it HST}\/
observations \cite[]{heg01}.  Sparse optical data suggest a 43.6~min
periodic intensity modulation with semiamplitude 30\% \cite[]{dma00},
although no periodic modulation of the X-rays was seen in 2001 by \sax\
\cite[]{parmar01}. A 43.6~min period would indicate an ultracompact
binary with an evolved, likely H-poor companion similar to 3A~1820$-$30
(see \S\ref{s1820}).

\xte/PCA measurements in 1998 and 2001--2 indicate a flux of
2--$3.5\times10^{-10}\ \epcs$ (2.5--25~keV), although this may include
contributions from the other (typically quiescent) LMXBs in the cluster.
For
$d=9.6\pm0.4$~kpc \cite[]{kuul03a}, this gives an upper limit to the
accretion rate in XB~1832$-$330 of 
2--3\%~$\dot{M}_{\rm Edd}$.	
We found just one burst in public \xte\/ observations, on 1998
November 27 05:45:15 UT. 
The blackbody radius reached local maxima during the rise and near the
flux maximum, and the simultaneous inflection of the blackbody temperature
suggests that this burst may have experienced modest PRE \cite[although
see][]{kuul03a}.
The peak flux suggests a distance
consistent with that of the host cluster, assuming the burst reached
\leddhe\ (see Table \ref{dist}).
The burst exhibited a steep initial decay, but then a long $\approx100$~s
tail (Fig. \ref{profiles}), so that the overall $\tau$ was long at 21.4~s.

\subsection{3A~1837+049 ($=$ Ser~X-1)} 

This persistent source at  $l=36\fdg12$, $b=+4\fdg84$ was first detected
in early rocket flights \cite[]{bowyer65}.  A more precise position from
\sas\/ observations \cite[]{doxsey75} led to a suggested optical
counterpart \cite[]{davidsen75}; later observations revealed that this
candidate was actually two stars, one of which (with He{\sc ii} 4686~\AA\
emission) was the counterpart
\cite[]{tbc80}. The inferred $L_X/L_O$ ratio is $>100$.
X-ray bursts were discovered more or less simultaneously by \osoe\/
\cite[]{swank76b} and \sas\/ \cite[]{li77}. The bursts exhibited irregular
recurrence times of 1--38~hr, average $\tau=6.8\pm2.1$~s, and showed no
indications of PRE (\citealt{szt83}; see also \citealt{lew93}). The
variations in burst interval were apparently independent of \fper,
although \fpk\ increased with \fper.
A ``superburst'' was detected by \sax\/ \cite[]{corn02}, after which 
regular thermonuclear bursts were not detected for 34~d.

The source was persistently bright at 4--$6\times10^{-9}\ \epcs$
(2.5--25~keV) in \xte/PCA observations. We found 7 bursts in public data,
two of which exhibited weak PRE indicating a distance of 7.7 (10)~kpc
assuming the bursts reached \leddh\ (\leddhe; see Table \ref{dist}). We
note that one other burst exceeded the peak flux of the two PRE bursts by
$\approx30$\% but did not itself exhibit PRE. The
corresponding accretion rate range is 38--56\%~$\dot{M}_{\rm Edd}$ (for a
bolometric correction of $1.24\pm0.08$).
The bursts were of short duration, with mean $\tau=4.8\pm0.6$~s. We found
one pair of bursts separated by 7.99~hr, from which we derived
$\alpha=1590\pm150$ (although there may have been intermediate bursts
which were missed during Earth occultations).

\subsection{HETE J1900.1$-$2455}
\label{h1900}

This 
source ($l=0\fdg00$, $b=-12\fdg87$) was
discovered on 2005 June 14 when a strong thermonuclear (type-I) burst was
detected 
by {\it HETE-II}
\cite[]{vand05a}. A subsequent PCA observation of the field on 2005 June
16 revealed 2\% rms pulsations at 377.3~Hz, confirming the bursting source
as the seventh accretion-powered millisecond pulsar \cite[]{morgan05}. A
series of followup PCA observations allowed measurements of Doppler shifts
of the apparent pulsar frequency on the is 83.25~min orbital period
\cite[]{kaaret05a}.
The optical counterpart was identified by its brightening to $R\sim18.4$
during outburst \cite[]{fox05}.
Based on the peak flux of the burst observed by {\it HETE-II}\/, the
distance was estimated at 5~kpc \cite[]{kawai05}.

Two bursts were detected during followup PCA observations, the first on
2005 July 21.  The burst profile was complex, with a
precursor lasting 2~s, followed by a slower rise to a maximum of
$(110.7\pm1.5)\times10^{-9}\ \epcs$ lasting approximately 20~s.  While the
burst flux was close to maximum, the blackbody radius reached two
successive local maxima, each accompanied by local minima in the blackbody
temperature.
The second burst was much less energetic, reaching a peak flux $\approx20$\%
lower, and with a total fluence only a quarter of the first burst.
Assuming both bursts reached the Eddington limit for pure He material, the
distance to the source is $4.7\pm0.6$~kpc, consistent with the earlier
estimate from the first burst observed by {\it HETE-II}\/ by \cite{kawai05}.

The source activity continued for
more than 
1~yr after the outburst began \cite[e.g.][]{gal05d}.
This is
much longer than the typical outburst duration for the other accretion-powered
millisecond pulsars ($\approx2$~weeks), and a factor of three longer than the
previous record-holder, XTE~J1814$-$338, at 50~d (see \S\ref{s1814}). The
inferred accretion rate (for a bolometric correction of $1.96\pm0.02$) was
2--3\%~$\dot{M}_{\rm Edd}$. Should activity persist at this level
indefinitely, HETE~J1900.1$-$2455 will have the highest time-averaged
accretion rate of all the millisecond pulsars.

\subsection{Aql~X-1} 

One of the earliest cosmic X-ray sources detected \cite[e.g.][]{fbc67},
Aql~X-1 ($l=35\fdg72$, $b=-4\fdg14$) is a recurrent transient with a
quasi-regular
outburst interval variously reported as $\approx230$~d
\cite[e.g.][]{khbs77}, 122--125~d \cite[]{pt84b}, or 309~d \cite[]{kit93}.
Optical photometry of the highly variable ($B=20$--17) K0 counterpart
\cite[]{tcb78} throughout an outburst revealed a 19~hr period, assumed
initially to be the binary period \cite[]{ci91}. The $I$-band periodicity
is twice this value
\cite[]{shah98a}.
Thermonuclear bursts were probably first detected by \sas\
\cite[]{lhd76c}, but were confirmed during \hak\ observations in the
declining phase of an outburst \cite[]{koy81}. The bursts reached peak
fluxes between 7--$11\times10^{-8}\ \epcs$, with time scales
$\tau\sim10$--18~s \cite[e.g.][]{lew93}.
\xte\/
observations during an outburst in 1997 February--March revealed a QPO in
the frequency range 740--830~Hz as well as burst oscillations around
549~Hz \cite[]{zhang98}.
More recently, \cite{casella07} reported detection of persistent
pulsations at a frequency just above the burst oscillation frequency, in
an otherwise unremarkable 150~s stretch of data.
Note the nearby source 1A~1905+00 ($\Delta\theta=0\fdg82$), which has also
exhibited bursts \cite[][ see also Table \ref{nobursts}]{lhd76c}. While
bursts were detected during \xte\/ observations centered on this source,
we attributed them all to Aql~X-1 instead (see
\S\ref{aqlsc}).

\xte\/ has observed around 8 outbursts 
since 1996
(Fig. \ref{asmlc}). The 2.5--25~keV flux reached 2--$18\times10^{-9}\
\epcs$ at the peak of these outbursts.
The 
57 
bursts detected by \xte\/ occurred at \fper\ levels which span more
than an order of magnitude.
The burst properties were correspondingly diverse, and indicate three
approximately distinct groups: one with short time scales ($\tau=5$--10)
and rather low fluences, another of non-radius expansion bursts with
$\tau\approx15$--30
and a third group which have
$\tau\approx8$--15. It is this third group in which all the bursts which
exhibit PRE and oscillations occur. 
The peak flux of the PRE bursts indicates a distance of 
3.5 (4.5)~kpc, assuming the bursts reach \leddh\ (\leddhe).  The peak
accretion rate reached during the outbursts is thus 6--56\%~$\dot{M}_{\rm
Edd}$ (for $d=5$~kpc and a bolometric correction of $1.65\pm0.05$).
We also found 
six 
instances of short recurrence times, between 8--22~min, including a burst
triplet on 2005 April 16--17. Triplets of closely-spaced bursts have been
observed only from a handful of sources, including EXO~0748$-$676
\cite[]{boirin07a}, 4U~1705$-$44 (see \S\ref{s1705}), and 4U~1608$-$52
(\S\ref{s1608}). As
is typical for short-$\Delta t$ bursts, the fluence of these bursts was
significantly smaller than the mean value, and the bursts occurred at
low persistent flux levels, in the range 0.2--8\%~$\dot{M}_{\rm Edd}$
(see also \S\ref{dblbursts}).

\subsection{4U~1916$-$053} 

This source at $l=31\fdg36$, $b=-8\fdg46$
was discovered by the {\it Uhuru}\/ satellite \cite[]{1ucat}.
{\it EXOSAT} observations revealed irregular X-ray dipping behaviour with
a period of $\approx50$~min \cite[]{walter82,white82}, which optical
observations of the $V=21$ companion confirmed was approximately the
orbital period \cite[]{grin88}. The source is thus an ``ultracompact''
system, which cannot accommodate a H-rich mass donor.
Bursts were first observed from the source by \osoe\/ \cite[]{becker77},
and were subsequently detected by \sas, {\it HEAO-1}\/ and \exo\/
\cite[see][]{lew93}. The typical burst interval is 4--6~hr, although
bursts may sometimes occur at longer intervals or not at all.
Bursts exhibiting PRE suggest a source distance of 8.4--10.8~kpc
\cite[]{smale88}. Measured $\alpha$-values vary between 120--170; burst
durations are typically $\tau\sim5$~s, but may be up to a factor of two
longer.
A burst oscillation at 270~Hz was discovered in a single burst observed by
\xte, on 1998 August 1 \cite[]{1916burst}. This source is the only one in
which the burst oscillation frequency is significantly below the kHz QPO
peak separation.

The persistent flux of the source was between 0.2-$1\times10^{-9}\ \epcs$
(2.5--25~keV) throughout the \xte\/ observations, although we note that
these values are not corrected for the presence of dips.
We found a total of 14 bursts from the source, with 12 exhibiting 
PRE. The inferred distance is 7--9~kpc (see Table \ref{dist}), giving
a range of accretion rates of 1.5--8\%~$\dot{M}_{\rm Edd}$ (for $d=9$~kpc
and a bolometric correction of $1.37\pm0.09$).
The bursts were short, with $\tau=6.5\pm1.3$~s, although one burst
(which also had the largest fluence) exhibited a much broader peak,
resulting in $\tau=10.2$~s.
Just one pair of bursts were separated by $<10$~hr, on 1998 July
23, with $\Delta t=6.33$~hr; from these two bursts
we calculate $\alpha=78.8\pm0.3$, equivalent to a mean H-fraction at
ignition of $X\approx0.2$ (equation \ref{alphatheory}). 

\subsection{XTE~J2123$-$058} 

This source ($l=46\fdg48$, $b=-36\fdg20$) was discovered as an X-ray
transient by \xte\/ in late June 1998 \cite[]{lss98}. The optical
counterpart was identified and found to have a 
5.96~hr periodic optical modulation, identical to the
spectroscopic period \cite[]{tomsick99}.  The counterpart was monitored
extensively throughout the outburst \cite[e.g.][]{swg99}, and into
quiescence.  Keck measurements resulted in a narrowing of the distance
range to $8.5\pm2.5$~kpc (\citealt{tomsick01}; see also
\citealt{tomsick02}).
\xte\/ observations revealed thermonuclear X-ray bursts and high-frequency QPOs
\cite[]{homan99b}; optical bursts have also been detected.

The peak PCA flux during the 1998 outburst was $1.74\times10^{-9}\ \epcs$
(2.5--25~keV). At $d=8.5$~kpc, this corresponds to 11\%~$\dot{M}_{\rm
Edd}$ (for a bolometric correction of $1.19\pm0.06$).
We found a total of 6 weak bursts from the source, the two brightest (on
1998 July 22) reaching a peak of just $\approx6\times10^{-9}\ \epcs$.
The remaining four bursts all reached peak fluxes below $3\times10^{-9}\
\epcs$.  None of the bursts exhibited PRE; the peak fluxes were all well
below the expected value for bursts reaching \leddh\ at 8.5~kpc.
The two brightest bursts were separated by just
6.5~hr, which for the persistent flux level measured during the
observation leads to an $\alpha=370\pm40$.
Such a large value suggests that intervening bursts may have been missed
in data gaps, or that only a fraction of the accreted material
was burned.

\subsection{4U~2129+12 ($=$ AC 211) in M15}
\label{s2129}

This source ($l=65\fdg01$, $b=-27\fdg31$) is one of two bright LMXBs in
the globular cluster M15 \cite[$d=10.3\pm0.4$~kpc;][]{kuul03a}, separated
by just $2\farcs7$ \cite[]{wa01}. Originally the x-ray source was
identified with the 17.1~hr binary AC~211
\cite[]{ilovaisky93,alft84,cjn86}; the other source, M15~X-2, is the
suggested origin of the strong PRE bursts observed by \gin\/
\cite[]{dotani90,vpd90}, \xte/PCA \cite[]{smale01}, \sax/WFC
\cite[]{kuul03a} and \xte/ASM \cite[]{ccvz02}.  The latter work also
identified 15
burst candidates in the ASM data, leading to a lower limit on the burst
recurrence time of 1.9~d.

The PCA flux of the source in observations in 1997 and 2000 was
2--$4\times10^{-10}\ \epcs$ (2.5--25~keV). Although this flux contains
contributions from both LMXBs, \cite{wa01} found M15~X-2 (the suggested
origin of the bursts) to be 2.5 times brighter than M15~X-1, so that the
inferred range of accretion rate of 
2--4\%~$\dot{M}_{\rm Edd}$ 	
should be approximately correct. The inferred $\dot{M}$ is also consistent
with the long burst recurrence times of $\ga1.9$~d. The single burst
observed by \xte/PCA, on 2000 September 22, peaked at $4\times10^{-8}\
\epcs$
(note that the higher value of $5\times10^{-8}\
\epcs$ quoted by \citealt{smale01} was derived using the older response
matrices), which agrees well with the peak flux of the burst observed by
\gin\/ of $4.2\times10^{-8}\ \epcs$. Although the burst duration was long,
with $\tau=30$~s (see Fig. \ref{profiles}), the \gin\/ burst was even
longer. 
The burst exhibited very strong radius expansion, similar to that seen in
the bursts from 4U~1724$-$307 (\S\ref{s1724}, Fig. \ref{b1724}) and
GRS~1747$-$312 (\S\ref{sc1747}) although insufficient to
drive the emission at the radius peak completely out of the PCA band (the
minimum blackbody temperature reached was 0.8~keV).
As noted by \cite{kuul03a}, the peak fluxes of these bursts are
substantially in excess of the expected range of \leddhe\ for
$d=10.3$~kpc.

\subsection{Cyg~X-2} 
\label{scygx2}

This source ($l=87\fdg33$, $b=-11\fdg32$) was detected in the very first
observations which indicated the existence of cosmic X-ray sources
\cite[]{giacc62}.  A $V=14.7$ optical counterpart was identified shortly
afterwards \cite[]{giacc67}; the binary orbit is very wide, with $P_{\rm
orb}=236.2$~hr \cite[]{cow79}.
An event resembling a thermonuclear burst was first observed during \ein\
observations (\citealt{kg84}; see also \citealt{lew93}).  In their
analysis of an event observed by \xte, \cite{smale98b} detected a decrease
in color temperature late in the burst, following an apparent PRE episode.
This appeared to confirm the thermonuclear nature of these events, as well
as allowing a distance estimate of $11.6\pm0.3$~kpc to be made.
At accretion rates comparable to the Eddington limit, which is typical for
this Z source, it is expected that bursts should be extremely infrequent
or absent altogether since the temperature in the accreted layer may be
sufficient for the accreted fuel to burn stably, instead. 
That the bursts have such short
time-scales presents an additional puzzle, since in other
sources such bursts are identified with pure He fuel, whereas at the high
$\dot{M}$ typical for Cyg~X-2 it is expected that a substantial H-fraction
remains at ignition \cite[see also][]{kuul02a}.

Cyg~X-2 was persistently bright in \xte/PCA observations at
6--$21\times10^{-9}\ \epcs$, with a mean level of $11\times10^{-9}\
\epcs$. For $d=11.6$~kpc, this corresponds to accretion rates
of 
$>0.8\ \dot{M}_{\rm Edd}$, and for much of the time well in excess of
$\dot{M}_{\rm Edd}$.
We found 
55 
burst-like events from Cyg~X-2 in public data from \xte,
including 
8 
apparently exhibiting PRE \cite[similar to the burst on
1996 Mar 27 14:29:07 UT analysed by][]{smale98b}. The mean peak flux of
these bursts suggests a distance of 11 (14)~kpc, assuming the bursts
reach \leddh\ (\leddhe; see Table \ref{dist}).
However, some of the PRE bursts did not exhibit any decrease in $T_{\rm
bb}$ following the peak. Furthermore, only a handful of the other events
showed a decrease in $T_{\rm bb}$ following the maximum flux. In many
bursts, $T_{\rm bb}$ was constant or even {\it increased} with time
throughout the burst.
Thus, we consider there to be some doubt yet as to the thermonuclear
explanation for these bursts.

\section{Determining the origin of bursts}
\label{localize}

The mechanical collimators on each of the PCUs aboard \xte\/ admit photons
over a relatively large field of view ($\approx1\arcdeg$ radius). The
collimator response decreases approximately $\propto1/\Delta\theta$, where
$\Delta\theta$ is the angle between the source position and the nominal
pointing direction, in degrees.  The wide field of view means that
correctly attributing bursts to sources in crowded fields (particularly
the Galactic center) is problematic. 

Where bursts were observed in fields containing multiple sources, we
attempted to match the bursts with the known characteristics of individual
sources (see Table \ref{gcbursters}). 
We also exploited the fact that the 5 PCUs are not
perfectly aligned. As a result, the ratio of observed count rates in each
PCU depends upon the position of the source within the field of view.
From the modeled collimator responses for each PCU we have deduced the most
probable origin for each burst. We first determined an interval covering the
burst over which the count rate was greater than $\approx10$\% of the
maximum (neglecting the pre-burst persistent emission), and accumulated all
the counts observed in each PCU over this interval. We then stepped over a
grid of positions covering the field of view and performed a linear fit to
test the hypothesis that the variations in the PCU-to-PCU total count rates
arose solely from differences in the collimator responses at each
position. Although we used the same set of collimator responses over all gain
epochs, we renormalised the responses based on observed count rates for
Crab observations close in time to each burst. We then identified the source
at which position we found the minimum goodness of fit statistic ($\chi^2$)
as the most probable origin of the burst.

Naturally, this calculation can most easily distinguish between sources
which are widely separated in the field of view. For more crowded fields,
we may only be able to narrow down the possible origin as one of a few
nearby sources.
While this method works best for very bright bursts, where we observe only
faint bursts (whether intrinsically faint or originating from sources
that are far off-axis), we can combine counts from multiple bursts, so long
as the pointing and spacecraft orientation is consistent, to improve the
localization.

Once the burst origin was identified with confidence, for bursts observed
$\ga0\fdg1$ off-axis we scaled the measured burst flux and fluence by the
ratio of the collimator response (averaged over those PCUs that were
operating) at the aimpoint, to the response at the source location.

Below we describe close pairs of bursting sources, and the attribution of
bursts in each case. Where a burst from an observation centered on one
source is later attributed to another, we flag that burst as uncertain in
origin in Table \refbursts.

\subsection{4U~1728$-$34 and the Rapid burster ($\Delta\theta=0\fdg56$)}
\label{sc1728rb}

4U~1728$-$34 is one of the most prolific bursters (see
\S\ref{pflux}), and also produces a large proportion of PRE bursts
\cite[]{gal03b}. The Rapid Burster, on the other hand, tends to produce
preferentially non-PRE thermonuclear bursts \cite[]{fox01}, in addition to
the much more frequent type-II bursts \cite[e.g.][]{lew93}. The peak
fluxes of the majority
of bursts observed in the 4U~1728$-$34 field were bimodally distributed,
with non-PRE bursts peaking at around $4\times10^{-8}\ \epcs$ on average,
and PRE bursts peaking at $9\times10^{-8}\ \epcs$. We also observed six
bursts with peak fluxes around $5\times10^{-9}\ \epcs$ and no evidence of
PRE, each close to the time of one of the semi-regular transient Rapid
Burster outbursts.
Four of these bursts (on 1996 May 3 13:56:30, 13:57:49, 13:59:15 and
14:00:16 UT; obsid \#10410-01-01-00) had no evidence of decreasing
blackbody temperature with time, and also exhibited recurrence times much
shorter than expected for thermonuclear bursts ($\sim100$~s).  Thus, we
identified these as type-II bursts from the Rapid Burster.
The other two bursts (on 2001 May 27 09:15:59 and May 29 09:05:57 UT) did,
however, show weak evidence of cooling. The first of these bursts was
consistent with an origin at either 4U~1728$-$34 or the Rapid Burster, but
the second was consistent with an origin only at the latter source
(4U~1728$-$34 was excluded at the $>5\sigma$ level). Based on this
evidence, and the similar long $\tau\approx11$~s for these bursts, we
identified them as type-I bursts from the Rapid Burster.

The majority of the bursts observed in the field centered on the Rapid
Burster, on the other hand, peaked at $\approx1\times10^{-8}\ \epcs$ with
relatively long rise times ($\ga2$~s) and time scales ($\ga10$).
At least 13 bursts were notable exceptions, with 
profiles much more similar to bursts from 4U~1728$-$34. The Rapid Burster
was excluded as an origin for these bursts at (typically) the 3--$5\sigma$
level. Thus, we attribute these bursts to 4U~1728$-$34, instead.  We note
that the
corrected peak fluxes were
similar to the other 
bursts observed from 4U~1728$-$34.

\subsection{SAX~J1750.8$-$2900 and SAX~J1747.0$-$2853
($\Delta\theta=0\fdg750$)}
\label{sax12}

Four bursts were observed in the field of SAX~J1750.8$-$2900 (see
\S\ref{sax1750}) and SAX~J1747.0$-$2853 (see \S\ref{sax1747}) by \xte, all
during the rise to the second peak of the SAX~J1750.8$-$2900 outburst.
Three of the bursts reached
apparent peak fluxes of $\approx5.5\times10^{-8}\ \epcs$, and two of those
exhibited PRE, while the fourth reached just $9\times10^{-9}\ \epcs$.
The three bright bursts exhibited count rate variations between the PCUs
inconsistent to a high level of confidence with an origin at 
SAX~J1747.0$-$2853; furthermore, their corrected peak fluxes (assuming
they arose instead from that source) would also be inconsistent with the
distance to the Galactic center.
While the origin of the faint burst cannot be constrained within the
$1\arcdeg$ field of view, if it originated from SAX~J1747.0$-$2853 the
corrected peak flux would be a factor of two larger than that of bursts
observed previously from the source (Table \ref{gcbursters}). Thus, we
attribute all these bursts to SAX~J1750.8$-$2900 \cite[see
also][]{kaaret02}.

\subsection{Aql~X-1 and 1A~1905+00 ($\Delta\theta=0\fdg82$)}
\label{aqlsc}

Bursts from 1A~1905+00 were discovered by \sas\ \cite[]{lhd76c},
and were attributed to a previously known persistent source
\cite[]{seward76}.  The apparent burst recurrence time was 8.9~hr; one
long radius-expansion burst was observed in a 17~hr observation by \exo\/
\cite[]{ci90}, with a peak flux of $(2.4\pm0.2)\times10^{-8}\ \epcs$.
We found three bursts from observations towards 1A~1905+00 (on 1996 July 24
08:56:40, July 24 09:07:18 and 2002 February 15 22:56:28 UT), with peak
fluxes 0.8--$1.0\times10^{-8}\ \epcs$. All three bursts were observed
during periods of transient activity by Aql~X-1.
The corrected peak fluxes for these three bursts (assuming they originated
from Aql~X-1) were consistent with those of other bursts from that source.
Furthermore, the most probable origin for two of the three bursts (given
the variations in count rate between the PCUs) were within $<0\fdg1$
of Aql~X-1, although we can formally exclude an origin at 1A~1905+00 at
only the 2.5--3.1$\sigma$ confidence level. For the third burst, the most
probable origin is within $0\fdg25$ of Aql~X-1, and we can exclude
1A~1905+00 at only the $1.8\sigma$ level. Given the lack of evidence of
bursting behaviour from the latter source during the \xte\/ observations,
we attribute all three to Aql~X-1.

\subsection{IGR~J17473$-$2721 
and IGR~J17464$-$2811
  ($\Delta\theta=0\fdg84$)}
\label{igrsrc}

Following the discovery by {\it INTEGRAL}\/ of the transient
IGR~J17473$-$2721 \cite[also known as
XTE~J1747$-$274;][]{greb05b,mark05d}, pointed \xte\/ observations were
made throughout 2005 May--June. The
\xte\/ field of view also covers XMMU~J174716.1$-$281048
\cite[also known as IGR~J17464$-$2811;][]{sidoli04}, which
was also active during 2005 and exhibited a burst on May 22 with an
estimated peak flux of $2.6\times10^{-7}\ \epcs$ \cite[]{delsanto07}.
Two bursts were detected shortly after in \xte\/ observations, on 2005 May
24 and 31.
The burst on May 24 was detected in an observation offset by just
$0\fdg085$ from IGR~J17473$-$2721; the rescaled peak flux assuming instead
an origin at IGR~J17464$-$2811 would be around $2.5\times10^{-7}\ \epcs$,
similar to that measured by {\it INTEGRAL}. However, the burst on May 31
was detected in an observation pointed inbetween the two sources,
so that the rescaled peak flux for either origin was only around
$4\times10^{-8}\ \epcs$.
Furthermore, the variation in observed count rates between
different PCUs 
during both bursts indicates a more likely origin with IGR~J17473$-$2721.
Thus, we attributed the two bursts to that source. 

\subsection{Galactic center fields}
\label{gcf}

Observations towards the Galactic center were categorized based on the
pointing direction into 10 (generally overlapping) fields, with pointing
directions separated by $>0\fdg1$. Below we describe the fields for which
bursts were observed, and which sources we attribute them to.
 
\subsubsection{Galactic center field 1}
\label{gcfone}

This field, centered on $\alpha=17\fh44\fm02.6$,
$\delta=-29\arcdeg43\arcmin26\arcsec$ (J2000.0), includes the known burst
sources KS~1741$-$293 ($\Delta\theta=0\fdg41$ from the center of the field),
	2E~1742.9$-$2929 ($\Delta\theta=0\fdg49$),
	1A~1742$-$289 ($\Delta\theta=0\fdg78$),
	SLX~1744$-$299/300 ($\Delta\theta=0\fdg80$),
and GRS~1741.9$-$2853 ($\Delta\theta=0\fdg85$), 
as well as the Bursting
Pulsar GRO~J1744$-$28 at the edge of the field ($\Delta\theta=0\fdg99$).
We observed 13 bursts with most of the peak fluxes below $1\times10^{-8}\
\epcs$.
Six of the faint bursts observed in 1997, on Feb 26 01:12:24, Feb 27
01:21:04 and 21:25:59, Mar 21 20:13:01, Mar 23 12:15:07 and Mar 25
17:32:13, had roughly symmetric profiles and no evidence of a temperature
decrease in the burst tail (although the first three were affected by data
gaps following the peak). The ratio of the integrated count rate in
different PCUs indicates that the best candidate for the burst origin was
GRO~J1744$-$28; we note also that the bursts were coincident with the
second outburst observed by \xte\/ from this source, beginning around
December 1996.
The burst profiles were similar to other type-II bursts
observed previously from that source \cite[]{giles96}. Thus, we attributed
all six to GRO~J1744$-$28.
Of the two brighter bursts, one (on 1997 March 21 20:07:28 UT) 
exhibited two distinct peaks at $1\times10^{-8}\ \epcs$, separated by
$\approx3$~s, while the other (on 1997 Mar 20 17:07:16 UT) featured strong
PRE and peaked at $2.3\times10^{-8}\ \epcs$. Both of these bursts were
consistent with an origin at 2E~1742.9$-$2929 ($\approx0.6\sigma$), with
the next best candidates inconsistent at the 2--$3\sigma$ level.
Two more bursts (on 1997 Apr 19 23:52:43 and 1997 Apr 21 00:13:27 UT) were
also consistent with an origin at this source, although we cannot rule out
other sources (KS~1741$-$293, 1A~1742$-$289) at quite as high confidence
due to the lower flux of these bursts.

The remaining three bursts were all consistent with an origin at
SLX~1744$-$299/300 (see \S\ref{slx1744}), which was close to the edge of
the field, and we can rule out the alternatives at between
2.5--$5.4\sigma$
confidence.
SLX~1744$-$299 has previously exhibited at least one very bright, long
burst, with exponential decay time 43.3~s 
(Table \ref{gcbursters}), while the three
bursts detected by \xte\/ were all much shorter, with decay times of
3--5~s.
Thus, we attribute the bursts to SLX~1744$-$300, although it is possible
they actually originated from SLX~1744$-$299.

\subsubsection{Galactic center field 3}

Centered approximately on the position of SAX~J1747.0$-$2853
($\Delta\theta=0\fdg02$),
GC field 3 also includes the bursters 
	1A~1742$-$289 ($\Delta\theta=0\fdg36$),
	GRS~1741.9$-$2853 ($\Delta\theta=0\fdg47$),
	2E~1742.9$-$2929 ($\Delta\theta=0\fdg68$),
	KS~1741$-$293 ($\Delta\theta=0\fdg69$), and 
	SAX~J1750.8$-$2900 ($\Delta\theta=0\fdg73$),
the recently-discovered bursting transient
	XMMU~J174716.1$-$281048 ($\Delta\theta=0\fdg70$),
as well as the Bursting
Pulsar GRO~J1744$-$28 ($\Delta\theta=0\fdg59$).
We found four faint (\fpk$<5\times10^{-9}\ \epcs$) bursts in observations
of this field, on 2000 Mar 12 06:22:25 and 2001 Oct  8 13:06:03, 17:32:56
and 17:51:14 UT.
While these bursts exhibited variations between the count rates for each
PCU consistent with an origin in a number of sources, their low peak
fluxes and long time scales suggest the most likely origin was
2E~1742.9$-$2929. Thus, we attribute them to that source.

\subsubsection{Galactic center field 10}
\label{gcften}

This field, centered on $\alpha=17\fh45\fm12.0$,
$\delta=-28\arcdeg48\arcmin18\arcsec$ (J2000.0), includes the burst
sources
	GRS~1741.9$-$2853 ($\Delta\theta=0\fdg11$),
	1A~1742$-$289 ($\Delta\theta=0\fdg23$),
	SAX~J1747.0$-$2853 ($\Delta\theta=0\fdg41$),
	KS~1741$-$293 ($\Delta\theta=0\fdg55$), and
	2E~1742.9$-$2929 ($\Delta\theta=0\fdg74$),
as well as the Bursting
Pulsar GRO~J1744$-$28 ($\Delta\theta=0\fdg16$). This field was observed
intensely for 355~ks between 2001 September 26 and 2001 October 6.

We found 80 bursts from these observations, with three-quarters of
the bursts reaching apparent peak fluxes $<10^{-8}\ \epcs$. For these
faint bursts the mean $\tau=25\pm11$~s, and the median delay time was
2.6~hr. These properties, once the off-axis angle is taken into account,
suggest that the bursts arose from 2E~1742.9$-$294 \cite[aka
1A~1742$-$294;][]{lgps01}. While the ratio of count rates from different
PCUs were not particularly constraining in determining the bursts location,
due to their faintness, only 6 bursts were inconsistent with an origin in
2E~1742.9$-$294, and then only at the 3--$5\sigma$ level.

Several of the brighter bursts exhibited PRE, often with a pronounced
double-peaked structure in the bolometric flux. Variations in the count
rates in different PCUs could not distinguish between a number of
closely-spaced sources as origins for these bursts. However, all the
bursts were observed over a short time interval, between 2001 September
26--29 and October 3--6. The only source in the field of view which was
active around this time was SAX~J1747.0$-$2853 \cite[]{wij02a}, from which
bursts were also observed during 2001 September by \sax\/ \cite[]{werner04}.
Thus, we attribute the bright bursts from this field to that source.

\subsection{GRO J1744$-$28}
\label{gcbo}

We found 19 type-I (thermonuclear) bursts in observations of the field of
the ``Bursting Pulsar'', GRO~J1744$-$28, which also includes the sources
  KS~1741$-$293 ($\Delta\theta=0\fdg62$),
  1A~1742$-$289 ($\Delta\theta=0\fdg36$),
  2E~1742.9$-$2929 ($\Delta\theta=0\fdg85$),
  SAX~J1747.0$-$2853 ($\Delta\theta=0\fdg56$),
  XTE~J1739$-$285 ($\Delta\theta=1\fdg05$) and
  GRS~1741.9$-$2853 ($\Delta\theta=0\fdg19$).
These bursts were comparatively easily distinguished from the much more
frequent type-II bursts from GRO~J1744$-$28 by the fast rise and exponential
decay profile, as well as the detection of falling blackbody temperature
in the burst tail. GRO~J1744$-$28 is not known to exhibit
thermonuclear bursts, and all the type-I bursts we observe from the field
we attribute to nearby sources.

Eleven of the bursts had low measured peak fluxes of $\la5\times10^{-9}\
\epcs$, and long time scales. Each of these bursts were consistent with an
origin at  2E~1742.9$-$2929, although only for the brighter bursts could
we exclude other sources in the field. We attributed all the faint bursts to
that source.

Millisecond oscillations at 589~Hz were previously detected in three of
the eight brighter bursts from the field, on 1996 May 15 19:32:23, Jun 4
14:41:12 and Jun 19 09:55:44 UT \cite[]{stroh97}. These bursts were
attributed by the latter authors to MXB~1743$-$29, which is in turn
thought to be identified with either KS~1741$-$293 or 1A~1742$-$289.
Another candidate source which was not considered at the time is
GRS~1741.9$-$2853 (see \S\ref{s17419}).
We could only conclusively exclude
KS~1741$-$293 or 1A~1742$-$289
as the origin for one of the 8 bursts, on 1996 Jul  8 01:57:47 UT.
Since the \sax\/ observations indicate bursting
activity shortly after the bursts observed by \xte, and the scaled peak
fluxes for the bursts assuming an origin in GRS~1741.9$-$2853 are
consistent with those observed by \sax\/ \cite[]{cocchi99}, we assume that
source as the origin.


\clearpage

\LongTables
\begin{deluxetable*}{lcccccc}
\tabletypesize{\scriptsize}
\tablecaption{Bursts with oscillations detected by \xte
  \label{osctbl}
}
\tablewidth{0pt}
\tablehead{
  \colhead{}
 & \colhead{Burst}
 & 
 & 
 & 
 & \colhead{Maximum}
 & \colhead{Mean}
\\
  \colhead{Source and frequency}
 & \colhead{ID}
 & \colhead{Start time}
 & \colhead{PRE?}
 & \colhead{Location}
 & \colhead{power}
 & \colhead{\% RMS}
}
\startdata
4U~1916$-$053 (270 Hz) &   9 & 1998 Aug  1 18:23:49 & N &  R P D & 28.3 & $7.5\pm1.4$ \\
\colrule
XTE~J1814$-$338 (314 Hz) &   1 & 2003 Jun  6 00:58:30 & N &  R -- D & 21.2 & $15\pm3$ \\
                       &   2 & 2003 Jun  7 06:26:52 & N &  -- P D & 40.3 & $13\pm2$ \\
                       &   3 & 2003 Jun  7 21:12:21 & N &  R P D & 37.1 & $11.0\pm1.8$ \\
                       &   4 & 2003 Jun  9 04:42:35 & N &  R P D & 41.0 & $10.0\pm1.6$ \\
                       &   5 & 2003 Jun 10 02:23:00 & N &  R P D & 37.8 & $11.7\pm1.9$ \\
                       &   6 & 2003 Jun 11 00:42:02 & N & \nodata & 36.0 & $9.5\pm1.6$ \\
                       &   7 & 2003 Jun 12 11:11:37 & N &  -- P D & 44.5 & $13\pm2$ \\
                       &   8 & 2003 Jun 12 13:31:06 & N &  R -- D & 30.2 & $17\pm3$ \\
                       &   9 & 2003 Jun 13 01:25:31 & N & \nodata & 49.0 & $8.9\pm1.3$ \\
                       &  10 & 2003 Jun 13 17:37:13 & N &  R P D & 26.4 & $10.3\pm2.0$ \\
                       &  11 & 2003 Jun 14 00:20:54 & N &  R P D & 38.9 & $11.8\pm1.9$ \\
                       &  12 & 2003 Jun 15 18:56:50 & N &  R P D & 34.0 & $12\pm2$ \\
                       &  13 & 2003 Jun 16 17:56:22 & N &  R P D & 49.0 & $19\pm3$ \\
                       &  14 & 2003 Jun 16 19:37:54 & N & \nodata & 40.0 & $10.2\pm1.6$ \\
                       &  15 & 2003 Jun 17 16:00:21 & N &  R P D & 36.0 & $13\pm2$ \\
                       &  16 & 2003 Jun 18 19:05:18 & N &  R P D & 35.5 & $15\pm2$ \\
                       &  17 & 2003 Jun 19 18:45:29 & N &  R P D & 37.3 & $14\pm2$ \\
                       &  18 & 2003 Jun 20 01:38:10 & N &  R P D & 37.4 & $10.3\pm1.7$ \\
                       &  19 & 2003 Jun 20 21:38:02 & N &  R -- D & 20.9 & $17\pm4$ \\
                       &  20 & 2003 Jun 21 15:20:59 & N &  R P D & 27.0 & $12\pm2$ \\
                       &  21 & 2003 Jun 22 21:19:41 & N &  R P D & 44.2 & $14\pm2$ \\
                       &  22 & 2003 Jun 23 11:15:00 & N &  R P D & 42.0 & $14\pm2$ \\
                       &  23 & 2003 Jun 27 16:38:50 & N &  R P D & 42.3 & $14\pm2$ \\
                       &  24 & 2003 Jun 27 21:12:01 & N &  R P D & 38.5 & $15\pm2$ \\
                       &  25 & 2003 Jun 28 20:29:15 & N &  R P D & 42.1 & $17\pm3$ \\
                       &  26 & 2003 Jul  7 06:05:03 & N &  R P D & 41.5 & $13\pm2$ \\
                       &  27 & 2003 Jul  8 19:02:33 & N &  R P D & 35.7 & $16\pm3$ \\
                       &  28 & 2003 Jul 17 17:42:46 & ? &  R P D & 25.1 & $8.0\pm1.6$ \\
\colrule
4U~1702$-$429 (329 Hz) &   2 & 1997 Jul 19 18:40:58 & N &  -- P D & 28.5 & $5.6\pm1.0$ \\
                       &   3 & 1997 Jul 26 09:03:23 & N &  -- P D & 120 & $11.7\pm1.1$ \\
                       &   4 & 1997 Jul 26 14:04:18 & N &  R P D & 127 & $10.7\pm1.0$ \\
                       &   5 & 1997 Jul 30 07:22:37 & N &  -- P D & 49.9 & $7.4\pm1.0$ \\
                       &   6 & 1997 Jul 30 12:11:56 & N &  R -- D & 87.3 & $12.1\pm1.3$ \\
                       &   7 & 1999 Feb 21 23:48:32 & N &  -- P D & 28.9 & $5.1\pm0.9$ \\
                       &   8 & 1999 Feb 22 04:56:05 & N &  R P D & 22.6 & $22\pm5$ \\
                       &   9 & 2000 Jun 22 11:57:46 & N &  R -- -- & 25.2 & $4.1\pm0.8$ \\
                       &  10 & 2000 Jul 23 07:09:42 & N &  -- -- D & 71.3 & $10.6\pm1.3$ \\
                       &  14 & 2001 Feb  4 03:50:29 & N &  -- P D & 27.2 & $4.3\pm0.8$ \\
                       &  15 & 2001 Apr  1 15:47:17 & N &  R P -- & 32.0 & $4.8\pm0.8$ \\
                       &  16 & 2001 Apr  1 21:55:53 & N &  -- P D & 115 & $12.3\pm1.2$ \\
                       &  18 & 2001 Nov 16 17:02:10 & N &  -- -- D & 21.7 & $4.7\pm1.0$ \\
                       &  20 & 2004 Jan 18 01:17:56 & N &  -- P -- & 15.2 & $3.1\pm0.8$ \\
                       &  21 & 2004 Jan 18 21:12:37 & N &  -- -- D & 18.5 & $4.9\pm1.1$ \\
                       &  22 & 2004 Jan 20 00:30:01 & N &  -- P -- & 29.7 & $5.2\pm0.9$ \\
                       &  23 & 2004 Feb 29 01:59:38 & N &  -- -- D & 34.1 & $13\pm2$ \\
                       &  24 & 2004 Feb 29 06:32:16 & N &  R P D & 19.4 & $5.5\pm1.2$ \\
                       &  26 & 2004 Mar  1 23:26:40 & N &  R -- -- & 31.7 & $6.2\pm1.1$ \\
                       &  27 & 2004 Mar  2 07:27:52 & N &  R P -- & 19.2 & $4.2\pm0.9$ \\
                       &  28 & 2004 Apr  8 22:12:46 & N &  -- P D & 39.4 & $6.4\pm1.0$ \\
                       &  29 & 2004 Apr  9 06:18:07 & N &  -- -- D & 68.0 & $14.4\pm1.7$ \\
                       &  30 & 2004 Apr  9 21:32:20 & N &  -- -- D & 27.0 & $5.5\pm1.1$ \\
                       &  31 & 2004 Apr 11 08:44:45 & N &  R -- D & 22.0 & $5.7\pm1.2$ \\
                       &  32 & 2004 Apr 12 06:56:34 & N &  R P -- & 18.5 & $3.9\pm0.9$ \\
                       &  35 & 2004 Apr 13 05:40:02 & N &  R P D & 25.8 & $ 9.7\pm1.9$ \\
                       &  36 & 2004 Apr 13 12:11:53 & N &  -- -- D & 27.5 & $9.1\pm1.7$ \\
                       &  37 & 2004 Apr 13 18:17:23 & N &  R P -- & 36.3 & $5.6\pm0.9$ \\
                       &  38 & 2004 Apr 14 18:32:01 & N &  -- -- D & 31.9 & $7.1\pm1.2$ \\
                       &  39 & 2004 Apr 15 00:43:14 & N &  R P D & 28.1 & $5.5\pm1.0$ \\
                       &  40 & 2004 Apr 16 02:08:57 & N & \nodata & 30.0 & $3.8\pm0.7$ \\
                       &  41 & 2004 Apr 16 20:29:36 & N &  -- P D & 56.5 & $8.0\pm1.1$ \\
                       &  42 & 2004 Apr 17 02:45:18 & N &  -- P D & 67.8 & $7.3\pm0.9$ \\
                       &  46 & 2004 Nov  3 03:28:32 & ? &  R P -- & 14.9 & $32\pm8$ \\
                       &  47 & 2005 Sep 17 10:26:25 & N &  -- P -- & 22.9 & $4.6\pm0.9$ \\
\colrule
4U~1728$-$34 (363 Hz) &   2 & 1996 Feb 15 21:10:21 & Y &  -- -- D & 22.8 & $3.1\pm0.6$ \\
                       &   3 & 1996 Feb 16 03:57:11 & N &  R -- -- & 14.2 & $3.4\pm0.9$ \\
                       &   4 & 1996 Feb 16 06:51:10 & Y &  -- -- D & 37.3 & $4.9\pm0.8$ \\
                       &   5 & 1996 Feb 16 10:00:47 & Y &  R P D & 55.0 & $9.1\pm1.2$ \\
                       &   6 & 1996 Feb 16 19:27:13 & Y &  -- -- D & 21.9 & $4.4\pm0.9$ \\
                       &   7 & 1996 Feb 18 17:31:52 & Y &  -- -- D & 30.4 & $4.2\pm0.8$ \\
                       &  13 & 1997 Jul 17 05:14:34 & Y &  -- P D & 21.0 & $5.2\pm1.1$ \\
                       &  14 & 1997 Sep 19 12:32:58 & N &  R -- -- & 25.1 & $4.5\pm0.9$ \\
                       &  15 & 1997 Sep 20 10:08:52 & N &  R P D & 73.1 & $8.7\pm1.0$ \\
                       &  16 & 1997 Sep 21 15:45:31 & N &  R P -- & 19.3 & $4.0\pm0.9$ \\
                       &  17 & 1997 Sep 21 18:11:07 & N &  R P D & 30.0 & $5.1\pm0.9$ \\
                       &  18 & 1997 Sep 22 06:42:53 & N &  -- P D & 94.2 & $ 9.9\pm1.0$ \\
                       &  19 & 1997 Sep 26 14:44:12 & N &  R -- -- & 18.2 & $15\pm3$ \\
                       &  20 & 1997 Sep 26 17:29:50 & N &  -- P D & 25.1 & $5.9\pm1.2$ \\
                       &  22 & 1997 Sep 27 15:54:06 & Y &  -- P D & 32.3 & $3.7\pm0.6$ \\
                       &  39 & 1998 Nov 16 16:09:06 & Y &  -- -- D & 30.2 & $3.8\pm0.7$ \\
                       &  41 & 1998 Nov 17 13:44:09 & Y &  -- -- D & 25.8 & $3.4\pm0.7$ \\
                       &  46 & 1999 Jan 23 23:49:02 & Y &  -- -- D & 19.9 & $6.4\pm1.4$ \\
                       &  47 & 1999 Jan 24 08:35:55 & ? &  -- -- D & 22.5 & $9.0\pm1.9$ \\
                       &  51 & 1999 Jan 28 03:22:34 & Y & \nodata & 20.0 & $1.6\pm0.4$ \\
                       &  53 & 1999 Jan 31 22:02:00 & Y &  -- P D & 80.6 & $6.1\pm0.7$ \\
                       &  54 & 1999 Feb  1 01:58:43 & Y &  -- P D & 15.2 & $2.9\pm0.7$ \\
                       &  55 & 1999 Feb  4 22:31:25 & Y &  -- P D & 46.3 & $4.3\pm0.6$ \\
                       &  59 & 1999 Mar  3 01:07:06 & Y &  R -- -- & 19.4 & $42\pm9$ \\
                       &  64 & 1999 Aug 19 09:33:48 & N &  R -- D & 23.8 & $5.8\pm1.2$ \\
                       &  65 & 1999 Aug 19 12:09:22 & N &  R P D & 86.2 & $ 9.8\pm1.1$ \\
                       &  66 & 1999 Aug 19 14:00:44 & N &  -- P -- & 21.7 & $6.0\pm1.3$ \\
                       &  67 & 1999 Aug 19 15:46:58 & N &  R P D & 40.9 & $11.4\pm1.8$ \\
                       &  68 & 1999 Aug 20 05:54:45 & N &  R P D & 43.9 & $8.5\pm1.3$ \\
                       &  70 & 1999 Sep 22 05:14:11 & Y &  -- -- D & 20.5 & $12\pm3$ \\
                       &  85 & 2001 Apr  8 14:42:54 & Y & \nodata & 21.0 & $2.0\pm0.4$ \\
                       &  86 & 2001 Apr  9 02:05:24 & Y &  -- P D & 21.0 & $3.5\pm0.8$ \\
                       &  93 & 2001 Sep 17 22:15:24 & N &  -- P D & 25.8 & $5.4\pm1.1$ \\
                       &  94 & 2001 Oct 18 03:33:29 & ? &  -- P D & 35.0 & $5.8\pm1.0$ \\
                       &  95 & 2001 Oct 18 09:26:09 & N &  -- P -- & 19.9 & $3.7\pm0.8$ \\
                       &  96 & 2001 Oct 27 23:53:44 & N &  R -- -- & 31.0 & $6.5\pm1.1$ \\
                       & 105 & 2004 Mar 12 01:41:13 & N &  -- P D & 40.7 & $10.5\pm1.6$ \\
                       & 106 & 2005 Apr  5 06:19:54 & ? &  -- P -- & 21.4 & $4.0\pm0.9$ \\
\colrule
SAX~J1808.4$-$3658 (401 Hz) &   1 & 2002 Oct 15 09:55:37 & Y &  R -- -- & 143 & $7.4\pm0.6$ \\
                       &   2 & 2002 Oct 17 07:19:24 & Y &  -- -- D & 63.1 & $4.3\pm0.5$ \\
                       &   3 & 2002 Oct 18 04:25:20 & Y &  R -- D & 40.9 & $3.3\pm0.5$ \\
                       &   4 & 2002 Oct 19 10:14:33 & Y &  R -- D & 123 & $18.8\pm1.7$ \\
\colrule
KS~1731$-$260 (524 Hz) &   1 & 1996 Jul 14 04:16:13 & Y &  -- P D & 41.3 & $6.8\pm1.1$ \\
                       &   7 & 1999 Feb 23 03:09:01 & N &  R -- -- & 18.9 & $6.7\pm1.5$ \\
                       &   8 & 1999 Feb 26 17:13:09 & Y &  R -- D & 27.9 & $4.3\pm0.8$ \\
                       &   9 & 1999 Feb 27 17:25:08 & Y &  -- -- D & 93.2 & $8.9\pm0.9$ \\
\colrule
1A~1744$-$361 (530 Hz) &   1 & 2005 Jul 16 22:39:56 & N &  R -- -- & 38.3 & $11.3\pm1.8$ \\
\colrule
Aql~X-1 (549 Hz) &   4 & 1997 Mar  1 23:26:36 & Y &  -- P D & 71.5 & $4.7\pm0.6$ \\
                       &   5 & 1997 Sep  5 12:33:57 & Y &  -- P -- & 21.1 & $2.3\pm0.5$ \\
                       &  10 & 1999 Jun  3 18:43:03 & Y &  -- P -- & 53.3 & $4.1\pm0.6$ \\
                       &  19 & 2000 Nov  8 03:45:56 & Y &  -- P D & 33.6 & $4.4\pm0.8$ \\
                       &  24 & 2001 Jul  1 14:18:37 & N &  R P -- & 16.7 & $5.4\pm1.3$ \\
                       &  28 & 2002 Feb 19 23:46:23 & Y &  -- P D & 30.6 & $3.1\pm0.6$ \\
\colrule
MXB~1659$-$298 (567 Hz) &   2 & 1999 Apr  9 14:47:34 & Y &  R -- -- & 19.0 & $6.1\pm1.4$ \\
                       &   3 & 1999 Apr 10 09:49:36 & Y &  R -- D & 28.9 & $13\pm2$ \\
                       &   4 & 1999 Apr 14 11:48:56 & Y &  -- -- D & 20.8 & $28\pm6$ \\
                       &   9 & 1999 Apr 21 11:45:57 & Y & \nodata & 45.0 & $8.3\pm1.2$ \\
                       &  12 & 1999 May  5 11:09:51 & Y &  R -- -- & 22.3 & $12\pm3$ \\
                       &  21 & 2001 Apr 30 17:41:33 & N &  -- P -- & 19.7 & $13\pm3$ \\
\colrule
4U~1636$-$536 (581 Hz) &   1 & 1996 Dec 28 22:39:24 & Y &  R -- D & 37.4 & $21\pm3$ \\
                       &   2 & 1996 Dec 28 23:54:04 & N &  R P D & 40.8 & $8.5\pm1.3$ \\
                       &   3 & 1996 Dec 29 23:26:47 & Y &  R -- D & 58.1 & $5.9\pm0.8$ \\
                       &   4 & 1996 Dec 31 17:36:52 & Y &  -- -- D & 23.5 & $4.8\pm1.0$ \\
                       &   6 & 1998 Aug 19 11:44:39 & Y &  R -- D & 69.7 & $7.4\pm0.9$ \\
                       &   7 & 1998 Aug 20 03:40:09 & Y &  -- -- D & 42.2 & $6.1\pm0.9$ \\
                       &   8 & 1998 Aug 20 05:14:12 & N &  R P D & 92.8 & $10.5\pm1.1$ \\
                       &   9 & 1999 Feb 27 08:47:29 & Y &  R P D & 34.9 & $5.1\pm0.9$ \\
                       &  10 & 1999 Apr 29 01:43:39 & Y & \nodata & 21.0 & $2.3\pm0.5$ \\
                       &  12 & 1999 Jun 10 05:55:30 & Y &  -- -- D &  98 & $9.3\pm0.9$ \\
                       &  13 & 1999 Jun 18 23:43:04 & Y &  R -- D & 28.1 & $4.8\pm0.9$ \\
                       &  14 & 1999 Jun 19 17:30:58 & Y &  -- -- D & 76.8 & $8.0\pm0.9$ \\
                       &  15 & 1999 Jun 21 19:05:53 & Y &  R -- D & 23.3 & $4.4\pm0.9$ \\
                       &  16 & 1999 Sep 25 20:40:49 & Y &  R -- -- & 22.2 & $32\pm7$ \\
                       &  20 & 2000 Jun 15 05:05:44 & Y &  -- P D & 137 & $8.7\pm0.7$ \\
                       &  21 & 2000 Aug  9 01:18:40 & Y &  -- -- D & 62.8 & $5.6\pm0.7$ \\
                       &  22 & 2000 Aug  9 08:56:53 & Y &  -- -- D & 38.4 & $5.4\pm0.9$ \\
                       &  23 & 2000 Aug 12 23:32:21 & Y &  -- -- D & 68.8 & $6.3\pm0.8$ \\
                       &  24 & 2000 Oct  3 23:32:48 & Y &  R -- D & 20.8 & $18\pm4$ \\
                       &  25 & 2000 Nov  5 04:21:59 & Y &  R -- D & 60.2 & $6.0\pm0.8$ \\
                       &  26 & 2000 Nov 12 18:02:28 & Y &  R P D & 40.1 & $6.1\pm1.0$ \\
                       &  27 & 2001 Jan 28 02:47:13 & Y &  -- -- D & 47.5 & $6.1\pm0.9$ \\
                       &  28 & 2001 Feb  1 21:00:50 & Y &  -- -- D & 37.8 & $4.6\pm0.7$ \\
                       &  29 & 2001 Feb  2 02:24:20 & Y &  -- -- D & 21.9 & $3.7\pm0.8$ \\
                       &  30 & 2001 Apr  5 17:07:05 & Y & \nodata & 23.0 & $3.8\pm0.8$ \\
                       &  31 & 2001 Apr 30 05:28:34 & Y &  R P D & 30.5 & $4.1\pm0.7$ \\
                       &  34 & 2001 Jun 15 03:14:04 & Y &  -- -- D & 65.0 & $11.9\pm1.5$ \\
                       &  37 & 2001 Aug 23 00:50:33 & N &  R P -- & 19.3 & $6.5\pm1.5$ \\
                       &  38 & 2001 Aug 28 06:41:20 & Y &  -- P D & 46.9 & $5.4\pm0.8$ \\
                       &  39 & 2001 Sep  5 05:42:07 & N &  -- -- D & 21.7 & $7.1\pm1.5$ \\
                       &  45 & 2001 Sep 30 14:47:17 & Y &  R P D & 43.7 & $24\pm4$ \\
                       &  49 & 2001 Nov  1 07:38:18 & Y &  -- P D & 40.2 & $5.8\pm0.9$ \\
                       &  57 & 2001 Dec 31 04:05:41 & N &  -- -- D & 20.3 & $7.8\pm1.7$ \\
                       &  61 & 2002 Jan  9 00:26:38 & Y &  R -- D & 32.1 & $21\pm4$ \\
                       &  62 & 2002 Jan  9 12:48:24 & Y &  -- P D & 56.0 & $7.3\pm1.0$ \\
                       &  66 & 2002 Jan 11 16:52:27 & N &  -- -- D & 21.2 & $6.1\pm1.3$ \\
                       &  70 & 2002 Jan 12 01:53:57 & N &  -- P -- & 39.6 & $14\pm2$ \\
                       &  71 & 2002 Jan 12 02:10:43 & N & \nodata & 21.0 & $4.9\pm1.1$ \\
                       &  72 & 2002 Jan 12 13:18:42 & Y & \nodata & 29.0 & $2.6\pm0.5$ \\
                       &  75 & 2002 Jan 12 21:35:34 & N &  R P -- & 53.9 & $9.4\pm1.3$ \\
                       &  77 & 2002 Jan 13 01:29:03 & N &  R P -- & 110 & $15.0\pm1.4$ \\
                       &  80 & 2002 Jan 13 12:47:26 & N &  R -- -- & 26.9 & $24\pm5$ \\
                       &  84 & 2002 Jan 14 01:22:36 & N &  R P -- & 62.3 & $13.1\pm1.7$ \\
                       &  86 & 2002 Jan 14 12:20:36 & Y &  -- -- D & 15.9 & $5.5\pm1.3$ \\
                       & 102 & 2002 Jan 22 07:07:20 & N &  R P D & 58.8 & $12.0\pm1.6$ \\
                       & 105 & 2002 Jan 25 03:58:06 & N &  R -- -- & 15.1 & $8\pm2$ \\
                       & 109 & 2002 Jan 30 23:06:55 & ? &  R -- D & 23.4 & $6.2\pm1.3$ \\
                       & 110 & 2002 Feb  5 22:21:51 & Y &  R -- D & 43.3 & $28\pm4$ \\
                       & 111 & 2002 Feb 11 17:35:07 & Y &  R P D & 39.1 & $4.7\pm0.7$ \\
                       & 113 & 2002 Feb 28 23:42:53 & N &  R -- D & 21.1 & $8.2\pm1.8$ \\
                       & 115 & 2002 Apr 26 05:07:18 & ? &  R P D & 49.9 & $6.7\pm0.9$ \\
                       & 127 & 2005 Mar 23 05:27:58 & N &  R -- -- & 18.8 & $17\pm4$ \\
                       & 136 & 2005 May 26 07:30:53 & Y &  -- -- D & 20.5 & $7.6\pm1.7$ \\
                       & 137 & 2005 Jun  3 09:19:54 & Y & \nodata & 27.0 & $2.5\pm0.5$ \\
                       & 138 & 2005 Jun 11 02:42:04 & N &  R -- -- & 15.3 & $11\pm3$ \\
                       & 146 & 2005 Aug  4 04:34:09 & N & \nodata & 22.0 & $7.0\pm1.5$ \\
                       & 148 & 2005 Aug 10 05:36:36 & N &  R P -- & 31.9 & $3.9\pm0.7$ \\
                       & 150 & 2005 Aug 16 01:45:36 & Y &  R -- D & 33.0 & $5.9\pm1.0$ \\
                       & 168 & 2005 Nov 14 22:50:45 & Y &  -- -- D & 19.9 & $4.0\pm0.9$ \\
\colrule
GRS~1741.9$-$2853 (589 Hz) &   4 & 1996 May 15 19:32:24 & Y & \nodata & 20.0 & $3.4\pm0.8$ \\
                       &   6 & 1996 Jun 19 09:55:40 & Y & \nodata & 21.0 & $3.6\pm0.8$ \\
\colrule
SAX~J1750.8$-$2900 (601 Hz) &   2 & 2001 Apr 12 14:20:31 & Y &  R -- D & 20.3 & $5.1\pm1.1$ \\
                       &   3 & 2001 Apr 15 17:02:25 & Y &  R -- -- & 24.1 & $20\pm4$ \\
                       &   4 & 2001 Apr 15 18:37:07 & N &  R -- -- & 20.8 & $12\pm3$ \\
\colrule
4U~1608$-$52 (620 Hz) &   5 & 1998 Mar 27 14:05:19 & Y &  -- -- D & 19.4 & $3.9\pm0.9$ \\
                       &   8 & 1998 Apr 11 06:35:31 & Y &  R -- -- & 18.0 & $2.6\pm0.6$ \\
                       &  10 & 2000 Mar 11 01:42:36 & Y &  R -- -- & 15.9 & $2.8\pm0.7$ \\
                       &  21 & 2002 Sep  7 02:26:15 & Y &  R P D & 24.2 & $2.6\pm0.5$ \\
                       &  22 & 2002 Sep  9 03:50:29 & Y &  R P D & 49.5 & $3.3\pm0.5$ \\
                       &  23 & 2002 Sep 12 04:18:15 & Y &  -- P D & 126 & $8.8\pm0.8$ \\
                       &  25 & 2002 Sep 19 07:38:20 & Y &  -- P -- & 49.1 & $3.7\pm0.5$ \\
 \enddata
\end{deluxetable*}

\clearpage

\begin{deluxetable}{lcccccc}
\tabletypesize{\scriptsize}
\tablecaption{ Mean peak fluxes and estimated distances 
  from PRE bursts observed by \xte
  \label{dist} }
\tablewidth{0pt}
\tablehead{
  & \colhead{Number}
  & \colhead{$\left<F_{\rm peak}\right>$}
  & \multicolumn{2}{c}{Distance (kpc)\tablenotemark{a}}
  & \multicolumn{2}{c}{Max. distance (kpc)\tablenotemark{b}} \\
  \colhead{Source}
  & \colhead{of bursts}
  & \colhead{$(10^{-9}\ \epcs)$}
  & \colhead{$X=0.7$}
  & \colhead{$X=0$}
  & \colhead{$X=0.7$}
  & \colhead{$X=0$}
}
\startdata
4U~0513$-$40 &   1 &  19.8  & 8.2 & 11 &  \nodata & \nodata \\
EXO~0748$-$676 &   3 &  $41\pm5$  & $5.7\pm0.7$ & $7.4\pm0.9$ & $<5.9$ & $<7.8$ \\
1M~0836$-$425 &  \nodata & \nodata  & $<8.2$ & $<11$ & \nodata & \nodata \\
4U~0919$-$54 &   1 &  81.9  & 4.0 & 5.3 &  \nodata & \nodata \\
4U~1254$-$69\tablenotemark{c} &   4 & $5.6\pm0.7$ & $15.5\pm1.9$ & $20\pm2$ & $<17$ & $<22$ \\
4U~1323$-$62 &  \nodata & \nodata  & $<11$ & $<15$ & \nodata & \nodata \\
4U~1608$-$52 &  12 &  $132\pm14$  & $3.2\pm0.3$ & $4.1\pm0.4$ & $<3.4$ & $<4.4$ \\
4U~1636$-$536\tablenotemark{d} &  46 &  $64\pm5$  & $5.95\pm0.12$ & $6.0\pm0.5$ & $<6.0$ & $<6.6$ \\
MXB~1659$-$298 &  12 &  $17\pm4$  & $9\pm2$ & $12\pm3$ & $<11$ & $<14$ \\
4U~1702$-$429 &   5 &  $76\pm3$  & $4.19\pm0.15$ & $5.46\pm0.19$ & $<4.3$ & $<5.5$ \\
4U~1705$-$44 &   3 &  $39.3\pm1.7$  & $5.8\pm0.2$ & $7.6\pm0.3$ & $<5.9$ & $<7.8$ \\
XTE~J1709$-$267\tablenotemark{c} &   1 & 11.0 & 11 & 14 & \nodata & \nodata \\
XTE~J1710$-$281 &   1 &  9.25  & 12 & 16 &  \nodata & \nodata \\
XTE~J1723$-$376 &  \nodata & \nodata  & $<10$ & $<13$ & \nodata & \nodata \\
4U~1724$-$307 &   2 &  $53\pm17$  & $5.0\pm1.6$ & $7\pm2$ & $<5.7$ & $<7.4$ \\
4U~1728$-$34 &  69 &  $84\pm9$  & $4.0\pm0.4$ & $5.2\pm0.5$ & $<4.6$ & $<5.9$ \\
Rapid~Burster &  \nodata & \nodata  & $<8.9$ & $<12$ & \nodata & \nodata \\
KS~1731$-$260 &   3 &  $43\pm6$  & $5.6\pm0.7$ & $7.2\pm1.0$ & $<6.0$ & $<7.8$ \\
SLX~1735$-$269 &  \nodata & \nodata  & $<5.6$ & $<7.3$ & \nodata & \nodata \\
4U~1735$-$44 &   6 &  $31\pm5$  & $6.5\pm1.0$ & $8.5\pm1.3$ & $<7.6$ & $<10$ \\
XTE~J1739$-$285 &  \nodata & \nodata  & $<7.3$ & $<10$ & \nodata & \nodata \\
KS~1741$-$293 &  \nodata & \nodata  & $<5.7$ & $<7.5$ & \nodata & \nodata \\
GRS~1741.9$-$2853 &   6 &  $38\pm10$  & $6.0\pm1.6$ & $8\pm2$ & $<7.8$ & $<10$ \\
2E~1742.9$-$2929 &   2 &  $40.07\pm0.08$  & $5.770\pm0.011$ & $7.523\pm0.015$ & $<5.8$ & $<7.5$ \\
SAX~J1747.0$-$2853 &  10 &  $50\pm8$  & $5.2\pm0.8$ & $6.7\pm1.1$ & $<5.7$ & $<7.5$ \\
IGR~17473$-$2721 &  \nodata & \nodata  & $<4.9$ & $<6.4$ & \nodata & \nodata \\
SLX~1744$-$300 &  \nodata & \nodata  & $<8.4$ & $<11$ & \nodata & \nodata \\
GX~3+1 &   1 &  53.0  & 5.0 & 6.5 &  \nodata & \nodata \\
1A~1744$-$361 &  \nodata & \nodata  & $<8.4$ & $<11$ & \nodata & \nodata \\
SAX~J1748.9$-$2021 &   6 &  $34\pm5$  & $6.2\pm1.0$ & $8.1\pm1.3$ & $<6.9$ & $<9.0$ \\
EXO~1745$-$248 &   2 &  $59.6\pm1.1$  & $4.73\pm0.09$ & $6.17\pm0.11$ & $<4.8$ & $<6.2$ \\
4U~1746$-$37 &   3 &  $5.3\pm0.9$  & $16\pm3$ & $21\pm4$ & $<17$ & $<22$ \\
SAX~J1750.8$-$2900 &   2 &  $49.2\pm1.0$  & $5.21\pm0.11$ & $6.79\pm0.14$ & $<5.2$ & $<6.8$ \\
GRS~1747$-$312 &   3 &  $16\pm6$  & $9\pm3$ & $12\pm4$ & $<12$ & $<15$ \\
XTE~J1759$-$220 &  \nodata & \nodata  & $<16$ & $<21$ & \nodata & \nodata \\
SAX~J1808.4$-$3658 &   5 &  $174\pm8$  & $2.77\pm0.11$ & $3.61\pm0.14$ & $<2.9$ & $<3.7$ \\
XTE~J1814$-$338\tablenotemark{c} &   1 & 21.3 & 7.9 & 10 & \nodata & \nodata \\
GX~17+2 &   2 &  $14.8\pm1.7$  & $ 9.8\pm0.4$ & $12.8\pm0.6$ & $<10$ & $<13$ \\
3A~1820$-$303 &   5 &  $54.8\pm1.9$  & $4.94\pm0.17$ & $6.4\pm0.2$ & $<5.0$ & $<6.6$ \\
GS~1826$-$24 &  \nodata & \nodata  & $<6.7$ & $<8.8$ & \nodata & \nodata \\
XB~1832$-$330 &   1 &  29.7  & 6.7 & 8.7 &  \nodata & \nodata \\
Ser~X-1 &   2 &  $23\pm3$  & $7.7\pm0.9$ & $10.0\pm1.1$ & $<8.0$ & $<10$ \\
HETE~J1900.1$-$2455 &   3 &  $ 99\pm10$  & $3.6\pm0.5$ & $4.7\pm0.6$ & $<3.8$ & $<4.9$ \\
Aql~X-1 &   9 &  $89\pm15$  & $3.9\pm0.7$ & $5.0\pm0.9$ & $<4.6$ & $<6.0$ \\
4U~1916$-$053 &  12 &  $29\pm4$  & $6.8\pm1.0$ & $8.9\pm1.3$ & $<7.5$ & $<10$ \\
XTE~J2123$-$058 &  \nodata & \nodata  & $<14$ & $<19$ & \nodata & \nodata \\
4U~2129+12 &   1 &  39.6  & 5.8 & 7.6 &  \nodata & \nodata \\
Cyg~X-2 &   8 &  $12\pm2$  & $11\pm2$ & $14\pm3$ & $<13$ & $<17$ \\
 \enddata
\tablenotetext{a}{Where no radius expansion bursts have been observed,
the upper limit on the distance is calculated from the peak flux of the
brightest burst observed by \xte.
}
\tablenotetext{b}{Upper limits on the distance calculated from the peak
flux of the faintest burst exhibiting radius expansion}
\tablenotetext{c}{Only marginal cases of radius-expansion were
available for this source}
\tablenotetext{d}{For 4U~1636$-$536, the peak flux distribution is
bimodal, with a separation factor of $\approx1.7$. The lower flux
radius-expansion bursts are thus identified with the Eddington
limit for material with cosmic abundances ($X=0.7$), while the
brigher bursts are assumed to reach the Eddington limit for
He-only. The two distances are thus calculated using the
appropriate group for each of the two abundance options}
\tablecomments{The distances here are derived assuming a canonical
neutron star with $M=1.4M_\sun$ and $R=10$~km. The corresponding
distances for a $2M_\sun$ neutron star will be a factor of 9.3\%
larger.
}
\end{deluxetable}

\clearpage

\begin{deluxetable}{llcccccl}
\tablecaption{Type-I X-ray bursters within $1\fdg5$ of the Galactic center
  \label{gcbursters}
}
\tablewidth{0pt}
\tablehead{
  & & \multicolumn{2}{c}{Position}
  & \colhead{Peak burst flux} & \colhead{Energy} & & \\
 \colhead{Source}
 & \colhead{Alt. name}
 & \colhead{$l$} & \colhead{$b$}
 & \colhead{($10^{-9}\ \epcs$)} 
 & \colhead{range}
 & \colhead{PRE}
 & \colhead{Ref.}
}
\startdata
 XTE~J1739$-$285 & \nodata & 359.725 & 1.30 & 
10--28 & bolometric & N & [1] \\
 SLX 1737$-$282 & 2E~1737.5$-$2817 & 359.973 & 1.25 & 
$60\pm5$ & bolometric & Y & [2] \\
 KS~1741$-$293 & AX~J1744.8-2921 & 359.554 & -0.0677 & 
    $14\pm3$ & 2--30~keV & single peak & [3] \\
 & & & & $17\pm2$ &  & N &  \\
 \raisebox{6pt}[0pt]{GRS~1741.9$-$2853} & 
 \raisebox{6pt}[0pt]{AX~J1745.0$-$2855} & 
\raisebox{6pt}[0pt]{359.960} & \raisebox{6pt}[0pt]{0.132} &
	
	 $31\pm2$ & \raisebox{6pt}[0pt]{3--28~keV} & Y & 
 \raisebox{6pt}[0pt]{[4]} \\
 1A~1742$-$289 & AX~J1745.6$-$2901   & 359.929 & -0.0421 & 
   $9.2\pm0.8$ & bolometric & N & [5] \\
 & & & & $13\pm5$\tablenotemark{a} & & N? & \\
 \raisebox{6pt}[0pt]{2E~1742.9$-$2929} & 
   \raisebox{6pt}[0pt]{GC X-1/1A 1742$-$293/4} & 
   \raisebox{6pt}[0pt]{359.558} & \raisebox{6pt}[0pt]{-0.393} &
    $38\pm3$ & \raisebox{6pt}[0pt]{3--20~keV} & Y? & \raisebox{6pt}[0pt]{[6]} \\
& & & & $22_{-8}^{+6}$ & bolometric & N & [7] \\
 \raisebox{6pt}[0pt]{SAX~J1747.0$-$2853} & 
 \raisebox{6pt}[0pt]{GX +0.2,$-$0.2} &
 \raisebox{6pt}[0pt]{0.207} & \raisebox{6pt}[0pt]{-0.238} &  
$31.8\pm2.7$ & bolometric & Y & [8] \\
 XMMU~J174716.1$-$281048 & IGR~J17464$-$2811 & 0.834 & 0.0837 & 
$260^{+170}_{-100}$ & 1--30~keV & double-peaked & [9] \\
 SLX~1744$-$299 & AX J1747.4$-$3000  & 359.296\tablenotemark{b} & -0.889\tablenotemark{b} & 
   25 & 3--20~keV & ? & [10] \\
 SLX~1744$-$300 & AX J1747.4$-$3003  & 359.260 & -0.911 & 
   $35\pm5$ & 2--30~keV & ? & [11] \\
 SAX~J1750.8$-$2900 & AX~J1750.5$-$2900 & 0.452651 & -0.948 & 
   $39\pm11$\tablenotemark{c} & 2--26~keV & single peaks & [12] \\
 \enddata
\tablenotetext{a}{Weighted mean and standard deviation of peak flux from 19
fainter bursts (mean 2--60~keV flux $\la5.8\times10^{-9}\ \epcs$)}
\tablenotetext{b}{R.A. = $17^h 47^m 25.9^s$, dec. = $-29^{\circ}59'58''$}
\tablenotetext{c}{Mean and standard deviation of peak flux from 7 bursts
which were not affected by atmospheric attenuation}
\tablerefs{
1. \cite{kaaret07a}, see also \cite{brandt05};
2. \cite{zand02};
3. \cite{zand91}; 4. \cite{cocchi99}; 5. \cite{maeda96}; 6. \cite{lgps01};
7. \cite{sidoli98}; 8. \cite{nat00}; 
9. \cite{delsanto07};
10. \cite{pgs94}; 11. \cite{patt89}; 12. \cite{nat99} }
\end{deluxetable}


\end{document}